\documentclass[times,12pt,thesis]{uom_thesis_deniaud} 

\usepackage{graphicx}              
  \graphicspath{ {./images/} }
\usepackage{amsmath}               
  \allowdisplaybreaks[1]           
\usepackage{amssymb}               
\usepackage{url}                   
\usepackage{rotating}              
\usepackage{multirow}              
\usepackage{lscape}                
\usepackage{tabularx}              
\usepackage{verbatim}              
\usepackage{footnote}              
\usepackage{float}                 
\usepackage{booktabs}              
\usepackage{lipsum}                
\usepackage[base]{babel}           
\usepackage{subcaption}            
\usepackage{siunitx}               

\newcommand{\degree}{\ensuremath{^\circ}}

\newcolumntype{Z}{>{\centering\arraybackslash}X}  

\usepackage[style=ieee,backend=biber,backref=true,hyperref=auto,maxbibnames=3,minbibnames=1]{biblatex}
\renewcommand*{\multicitedelim}{\addcomma}
\DeclareCiteCommand{\cites}
  {\printtext{[}\usebibmacro{prenote}}
  {\usebibmacro{citeindex}%
   \printtext[bibhyperref]{\printfield{labelnumber}}}
  {\multicitedelim}
  {\usebibmacro{postnote}\printtext{]}}

\DefineBibliographyStrings{english}{backrefpage = {cited on p\adddot},  backrefpages = {cited on pp\adddot}}

\begin{document}


  


\makeatletter
\title{Design and study of the electron beam-line at the European XFEL (ELBEX) for strong field quantum electro-dynamics}
\author{Marin Deniaud}
\makeatother

\faculty{Science and Engineering}                  
\school{School of Natural Sciences} 
\departmentordivision{Department of Physics and Astronomy} 
\submitdate{2026}                                  
\maketitle


\uomtoc 
\uomlof 
\uomlot 

\begin{uomlop} 
  \begin{refsection}[publications.bib]
    \nocite{*}
    \printbibliography[heading=none]
  \end{refsection}
\end{uomlop}

\begin{uomterms}
  \begin{table}[!ht]
    \centering
    \small
    \begin{tabular}{| c | c |}
       \hline
       Abbreviation & Explanation \\
       \hline
       \hline
       LUXE & Laser und XFEL experiment \\
       QED & Quantum electrodynamics \\
       SFQED & Strong-field quantum electrodynamics \\
       EM & Electromagnetic \\
       FEL & Free electron laser \\
       XFEL & X-ray free electron laser \\
       EuXFEL & European X-ray free electron laser \\
       DESY & Deutsches Elektronen-Synchrotron \\
       CDR & Conceptual design report \\
       TDR & Technical design report \\
       ICS & Inverse Compton scattering \\
       IP & Interaction point \\
       SASE & Self-amplified spontaneous emission \\
       ELBEX & Electron beamline at the European XFEL \\
       MAD & Methodical Accelerator Design \\
       RF & Radio-frequency \\
       PWFA & Plasma wakefield acceleration \\
       
       CW & Continuous wave \\
       
       BDSIM & Beam delivery simulation \\
       LMA & Locally monochromatic approximation \\
       LCFA & Locally constant field approximation \\
       TM & Transverse magnetic \\
       BPM & Beam position monitors \\
       BLM & Beam loss monitors \\
       BAM & Bunch arrival monitor \\
       MLO & Master laser oscillator \\
       EOM & Electro-optic modulator \\
       CRISP & Coherent radiation intensity spectrometer \\
       CoDR & Coherent diffraction radiation \\
       EBPM & Energy beam position monitor \\
       VM & Virtual measurement \\
       SVD & Singular value decomposition \\
       SR & Synchrotron radiation \\
       CSR & Coherent synchrotron radiation \\
       DBA & Double bend achromat \\
       MBA & Multi bend achromat \\
       SC & Superconducting \\
       
       DAC & Digital-to-analogue converter \\
       MDCS & Mott differential cross section \\
       
       PFH & Primary first hit \\
       CDF & Cumulative distribution function \\
       CLARA & Compact linear accelerator for research and applications \\
       QRR & Quantum radiation reaction \\
       VELA & Versatile electron linear accelerator \\
       FEBE & Full energy beam exploitation \\
       FEC & FEBE experimental chamber \\
       \hline
    \end{tabular}
\end{table}
\end{uomterms}

\begin{abstract} 
    Quantum electro-dynamic (QED) processes have been thoroughly studied in the perturbative regime, well below the electric field limit of $\mathcal{E}_{\rm cr} = 1.32 \times 10^{18}\ {\rm V/m}$. Exploring QED in the non-perturbative regime above this limit would require electric fields not obtainable in the lab frame. However, an electromagnetic field encountered by a relativistic particle would be increased by the Lorentz factor. 
Laser Und XFEL Experiment (LUXE) is an international project that aims to collide the \SI{17.5}{\giga\electronvolt} electrons of the European XFEL with a high power laser (\SI{40}{\tera\watt} in phase 0 and \SI{350}{\tera\watt} in phase 1). The electromagnetic field perceived by the electrons will exceed the Schwinger limit, allowing the study of non-linear Compton scattering, non-linear Breit-Wheeler pair creation and non-linear Trident process. 
The electron beam line at the European XFEL (ELBEX) project aims to provide the electron beam for LUXE, as well as a plasma acceleration experiment and test beam applications. This project will be located in the currently unused XTD8 tunnel of the EuXFEL. To deliver electron bunches at ELBEX, a new transport beam line must be designed. From the existing beam path, it will extract specific bunches in the bunch train and focus the transverse beam sizes at the needed locations for the experiments. Using the python based accelerator tracking code Xsuite, an achromatic and isochronous beam line was designed.
There are two main phenomena that can impact the measured signal at LUXE. The first one is the background particles reaching the detectors, coming from the interactions between the electron beam and both the residual gas and the beam pipe. Background simulations realized with the Geant4 based software BDSIM showed that there was not significant impact on the LUXE IP background rates.
The second phenomena is the variations in electron bunches properties, i.e. beam jitter. Measurements at the EuXFEL are performed to evaluate the jitter for various beam parameters. Using the transfer matrices of the new lattice, these variations can be propagated to ELBEX. Electron-laser collision simulations performed with Ptarmigan showed that beam size jitter is the only parameter with major impact on the signal at LUXE.
The compact linear accelerator for research and applications (CLARA) facility at Daresbury is also looking at the possibilities of QED interactions by colliding a \SI{250}{\mega\electronvolt} electron beam with \SI{100}{\tera\watt} laser pulses. 
The CLARA beam energy is not sufficient to probe non-linear QED effects, but the beam energy spread is sufficiently low to measure radiation reaction of the electron beam.
\end{abstract}%
\clearpage

\begin{uomlay} 
    In the $\rm 19^{th}$ century, Maxwell's equations unified the concepts of electricity and magnetism. They became a thoroughly tested and broadly used tool to understand the behaviour of electromagnetic fields and charged particles. Later, the early $\rm 20^{th}$ century saw the arrival of the quantum theory which stipulates that the exchanges in energy only happens by a discrete amount, a quanta. This theory redefined the fundamental understanding of multiple physics domains, including electrodynamics. After the second world war, the theory of quantum electrodynamics (QED) was progressively laid down by Hans Bethe, Richard Feynman, Julian Schwinger and others. 
Among the predictions of QED was the interaction between particles when they collide. Some observations were already made on the linear regime of QED, where a single collision makes a fixed number of product particles. This regime occurs at a relatively low energy. However, the behaviour of those collision at higher energies, what is referred as strong-field QED, should result in a non-linear regime, 
where the number of produced particles is not linearly proportional to the number of interactions.
This non-linear regime have not been yet measured in an experiment and that is the goal of Laser und XFEL experiment (LUXE). 
This experiment will be installed at the European XFEL (EuXFEL), an linear accelerator at the DESY facility in Hamburg, Germany. 
The high energy electron beam produced by the EuXFEL, colliding with a high power laser, should produce QED interactions in the non-linear regime. 
The first aim of this study is to design a new section of the accelerator. A particle accelerator uses magnets to transport charged particles in pipes with a high vacuum. The sequence of those magnets, called a lattice, must be carefully designed to deliver an electron beam with the right properties at LUXE. This is done with the help of different accelerator simulation software like MAD-8 or Xsuite.
The next goal is to insure the electron beam arrives at LUXE with the required beam properties. Residual gas in the beam pipe and collisions between electrons an said pipe can degrade the quality of the beam. The study also include an exhaustive analysis on the stability of the EuXFEL beam and how the variation of specific parameters affect the experiment. From all the simulations ran, the conclusion is that all the considered have a minimal impact on the experiment.
The compact linear accelerator for research and applications (CLARA) is another accelerator located at the Daresbury laboratory, in the United Kingdom. This accelerator can also provide opportunities for electron-laser collision and the study of strong-field QED. Complementing the study on LUXE, this thesis shows that only one of the QED process could be observed at CLARA in its current state.
\end{uomlay}

\begin{uomoriginality} 
  \uomoriginalitydeclaration 
\end{uomoriginality}

\uomcopyrightstatement 

\begin{uomacknowledgements} 
This PhD thesis started in March 2022, almost four and a half years ago. At the end of this journey, I realise how much I have learned as a physicist and also as a person. This accomplishment would not have been possible without the help and support of many people. Through these few words, I wish to express how thankful I am to those who contributed to my work and believed in my capabilities, sometimes more than I did. 

To my supervisor, Stewart T. Boogert. I struggle to find the right words to fully express my gratitude. Beneath all my struggles and doubts, you saw my strengths and you knew that I belonged in this thesis. Now I realise how much I have grown thanks to you. I can say without hesitation that you are a brilliant physicist and an incredible teacher.

To Laurence Nevay, William Shields, Siobhan Alden and all the other members of the BDSIM group. Our weekly meeting was always a valuable moment to discuss my difficulties, receive precious advice and improve the way I present my work. As we continue working together on BDSIM, I hope to be as helpful as you have been.

To Noam Tal Hod, Stuart Walker, Louis Helary, Ruth Jacobs and all the other members of the LUXE collaboration. Despite my anxious nature, I have always felt welcome, listened to and helped, especially during my trips at the European XFEL. I am honoured to contribute to a project like LUXE, in an international group with such nice people.

To Angus, William and all the other colleagues at the Cockcroft Institute. Thank you for the spontaneous help and kindness, the silly lunchtime discussions and the overall positive energy in our office. Working in such an environment was a real comfort to me.

To all my friends and family, who have always been a source of cheerful support. Even if this accelerator physics thesis sounds a bit like nonsensical gibberish, you have always been there whenever I needed to talk about it or about something else. Maybe one day I will be able to explain what all of this is about with the simplicity and confidence I wish I had.

To Sloann who, from the other side of the Channel, believed in me like no one else. You have supported my journey long before the beginning of this thesis, regardless of the distance between us. In the hardest periods and the moments of greatest doubt, you were always present. I am certain I could not have succeeded without you and your love.
\end{uomacknowledgements}


\uomstartmainbody 

\chapter{Introduction} \label{chap:intro}

\section{Structure of the thesis}

This thesis is divided in three main parts. The first two chapters, Chapter~\ref{chap:intro} and Chapter~\ref{chap:theory} lay down the context of the experiment and the physics theory. Both are derived from previous work on which I have based my studies. From Chapter~\ref{chap:measurments} to Chapter~\ref{chap:clara} is the actual core of my work, it contains measurement on an currently operating machine, the design of a new beam line and some simulations on said beam line. The two last chapters, Chapter~\ref{chap:next} and Chapter~\ref{chap:conclu}, discuss the future steps of those studies and summarise the results in the conclusion. In this section, a complete description for each chapter can be found. 

Chapter~\ref{chap:intro} contains a description of the LUXE experiment and a brief explanation on the quantum electrodynamics processes that this experiment aims to observe. This chapter introduces the layout of the European XFEL accelerator and the properties of the electron beam it produces. The ELBEX project is also presented in this chapter as LUXE will be one of its parts.

Chapter~\ref{chap:theory} contains the accelerator theory needed as a basis for the thesis. There is both a section on transverse beam dynamics and longitudinal beam dynamics. This chapter also presents the different simulation software that are used in different parts of the study. Some of them are used for accelerator design and others for collisions and beam interactions.

Chapter~\ref{chap:measurments} contains measurements of the electron beam parameters at the European XFEL. In the control room of the accelerator, at DESY, I was able to retrieve measurements on various parameters of the electron beam. The chapter features a description of each diagnostic device that was used. The rest of the chapter has analyses of the data and extracts, for each parameter, the correlated variations along multiple bunches (the jitter) as well as the individual or collective resolution for each devices (the noise). 

Chapter~\ref{chap:design} contains a walk-through of the design of the new electron beam line needed for ELBEX. It details the requirements for this new line, the different sections, and how Xsuite is used to match and optimize each section. The chapter starts from an initial design and progressively refines it to meet the requirements and add important features such as the beam diagnostics, and the trajectory correctors. The effect of coherent synchrotron radiation is then tested on the lattice to evaluate the impact on the twiss functions. Some modifications of the base design are considered, either for the radioprotection of a downstream experiment or the possibility to include permanent magnets. 

Chapter~\ref{chap:impact_luxe} contains extensive simulations on the different effects that could impact the signal measured at LUXE. It is divided into two main sections, one on the background particles and the other on the beam jitter impact. The first section uses the BDSIM software the study the background particles from residual gas in the beam pipe and the background particles from interactions with the beam pipe itself, i.e. the beam halo. The second section uses the jitter and noise measured at an upstream location (described in Chapter~\ref{chap:measurments}) and propagates them to LUXE's location. The Ptarmigan software is then used to simulate the collision between electron bunches and laser pulses. With multiple Ptarmigan simulations, the impact of every measured parameter on LUXE's signal is evaluated.

Chapter~\ref{chap:clara} contains a study of the quantum electrodynamics experiment possibilities at CLARA, an accelerator at the Daresbury laboratory. As there were some uncertainties in the development of LUXE, QED at CLARA appeared as an interesting study with a closely related subject. This chapter presents the CLARA facility and its experimental beam line FEBE which features a \SI{100}{\tera\watt} laser that can collide with a \SI{250}{\mega\electronvolt} electron beam. This study details the clear possibility of a non-linear Compton experiment as well as the impossibility of an electron-positron pair production experiment without a significant electron beam energy upgrade.

Chapter~\ref{chap:next} contains a discussion on the future steps that could be investigated in the continuity of this thesis. Some of them are related to additional features in the lattice design and how to integrate it to a more realistic CAD design. There is a section on the missing background from the electron dump. As I made an attempt at this study, it might require a new final state splitting tool in BDSIM. Then, the question of energy upgrade at CLARA is raised. The current setup should allow for plasma wakefield acceleration in one of the FEBE chambers. A new simulation study is needed to evaluate what energies can be reached and if it is sufficient for non linear pair-creation.

Chapter~\ref{chap:conclu} contains a conclusion on the thesis, it summarises the results and the major takeaways from each previous chapter.

\section{LUXE experiment}

\subsection{Scientific objectives}

Laser Und XFEL Experiment~\cites{luxe_cdr,luxe_tdr} (LUXE) is an international project that aims to study quantum electrodynamics~\cite{qed} (QED) processes in the strong field regime. In a weak electromagnetic (EM) field, QED has already been tested with a high level of precision~\cites{qed_classical_meas_1,qed_classical_meas_2,qed_classical_meas_3}. In this regime, the coupling between particles is expressed with a Taylor expansion of the form $\Pi = a\alpha + b\alpha^2 + ...$ where the coupling factor $\alpha$ is the fine structure constant, expressed as

\begin{equation}
    \alpha = \frac{e^2}{4\pi \epsilon_0\ \hbar c} \approx \frac{1}{137},
\end{equation}

where $e$ is the electron charge, $\epsilon_0$ is the permittivity of vacuum, $\hbar$ is the reduced Plank constant and $c$ is the speed of light in vacuum. Because $\alpha \ll 1$, the Taylor expansion converges quickly, QED is then a weakly coupled model and it can be described as using perturbative model~\cite{coupling}.
But at high energy (order of the ``Landau pole"~\cite{landau} at \SI{e286}{eV}), it becomes a non-perturbative model. In this highly coupled regime, the coupling factor $\alpha$ is not expressed as the fine structure constant and tends to infinity.
The fields at which non-linear effects occurs can be estimated by finding field strengths where particles can be created. In an electric field $\mathcal{E}$, the typical work $W$ done by the electric force $F$ over a distance $d$ is

\begin{equation}
    W=Fd=e\mathcal{E}\lambda_c,
\end{equation}

where $\lambda_c$ is the Compton wavelength, the typical distance for QED interactions. It is expressed as

\begin{equation}
    \lambda_c=\frac{\hbar}{m_ec}\approx400\ \rm fm,
\end{equation}

where $m_e$ is the electron mass. Reaching the pair production threshold of $m_ec^2$~\cite{pair_threshold} implies that

\begin{equation}
    m_ec^2 = \frac{e\mathcal{E}\ \hbar}{m_ec} \Rightarrow \mathcal{E}_{\rm cr} = \frac{m_e^2 c^3}{e\ \hbar} \approx 1.32 \times 10^{18}\ {\rm V/m}.
\end{equation}

This $\mathcal{E}_{\rm cr}$ electric field is defined as the Schwinger limit~\cite{schwinger} and is the threshold to perform the predicted strong field QED experiments. Generating a field strength of that magnitude in the lab frame is not possible with our current technology. However an electric field $\mathcal{E}$, perceived from a relativistic particle reference frame, will experience a relativistic boost~\cite{lorentz_boost} and become

\begin{equation}
    \mathcal{E}^*=\gamma\mathcal{E}(1+\cos\theta),
\end{equation}

where $\mathcal{E}^*$ is the boosted field, $\gamma$ is the Lorentz factor and $\theta$ the collision angle between the particle axis and the electric field propagation direction. The Lorentz boosted electric field can reach values beyond the Schwinger limit and allow for the study of strong field QED processes. But even with this Lorentz boosting, a high initial electric field is needed to reach the Schwinger limit. This can be done using a high power laser whose peak electric field $\mathcal{E}_L$ is expressed as

\begin{equation}
    \mathcal{E}_L=\sqrt{\frac{2I_0}{c\epsilon_0}},
\end{equation}

where $I_0$ is the peak laser intensity and $\epsilon_0$ is the dielectric permittivity of the vacuum. Rather than $\mathcal{E}_L$, lasers are often characterised by their power $P_L$ expressed as

\begin{equation}
    P_L=\frac{\pi I_0 w_0^2}{2},
\end{equation}

where $I_0$ is the same peak laser intensity and $w_0$ is the laser waist. The electric field can then be expressed as

\begin{equation}
    \mathcal{E}_L=\sqrt{\frac{4P_L}{c\epsilon_0\pi w_0^2}}.
\end{equation}

Increasing the electric field produced by the laser can be done by increasing the laser power or by reducing the waist. 
At a typical laser power of \SI{100}{\tera\watt}, focused at a \SI{5}{\micro\meter} waist, the peak electric field is around \SI{4.38e13}{\volt/\meter}. Assuming this laser is used in a head-on collision ($\theta=0$) with a \SI{10}{\giga\electronvolt} electron beam, the Lorentz factor is around $\gamma=2\times10^4$ and the boosted electric field is around \SI{1.71e18}{\volt/\meter} which is above the Schwinger limit.
In LUXE's design, the goal is to collide \SI{16.5}{\giga\electronvolt} electrons with high intensity laser pulses at a collision angle $\theta = 17.2^{\circ}$. The expected electic field is then around \SI{2.77e18}{\volt/\meter}. The electrons are provided by the European XFEL (EuXFEL), a linear accelerator in the DESY laboratory (Germany). In phase 0 of LUXE, the \SI{40}{TW} JETI (JEna TItanium:sapphire) laser~\cite{laser} will be employed, and in phase 1, it will be replaced by a \SI{350}{TW} laser. The strength of this laser can be expressed in term of the classical non-linearity parameter, 

\begin{equation}
    \xi = \frac{|e| \mathcal{E}_L}{m_e c \omega_L}
    = \frac{e \mathcal{E}_L \lambda_c}{\hbar \omega_L}
    = \frac{m_e c^2 \mathcal{E}_L }{\hbar \omega_L \mathcal{E}_{\rm cr}},
    \label{eq:xi}
\end{equation}

where $\omega_L$ is the laser angular frequency and $\mathcal{E}_L$ is the instantaneous laser field strength. A more powerful laser will increase this $\xi$ factor and result in increased rate and intensity for the processes studied. For a near head-on collision ($\theta = 0$), a relativistic factor of $\gamma \sim 10^4$ (\SI{10}{\giga\electronvolt} electrons) would be needed to get $\xi \sim 1$. With the real angle $\theta$ planed for LUXE, the required electrons energy is approximately \SI{16.5}{GeV} to achieve a unitary $\xi$. The three main aims of LUXE are

\begin{itemize}
    \item measure the interactions of real photons with electrons and photons at field strengths where the coupling to charges becomes non-perturbative;
    \item make precision measurements of electron-photon and photon-photon interactions in a transition from the perturbative to the non-perturbative regime of quantum electrodynamics (QED);
    \item use strong-field QED (SFQED) processes to design a sensitive search for new particles beyond the Standard Model that couple to photons.
\end{itemize}

The conceptual design report~\cite{luxe_cdr} (CDR) for LUXE was completed in 2021 and the technical design report~\cite{luxe_tdr} (TDR) was completed in 2023.

\subsection{Strong-field QED processes}

In the strong-field regime provided at LUXE, there are three main QED processes expected, non-linear inverse Compton~\cites{compton,compton_pair}, non-linear Breit-Wheeler~\cite{compton_pair} and non-linear trident~\cite{trident}.

\subsubsection{Non-linear inverse Compton}

\begin{figure}[!ht]
    \centering
    \includegraphics[width=4cm]{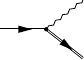}
    \caption{Feynman diagram of the non-linear inverse Compton scattering. Double lines indicate fermions in a laser background.}
    \label{fig:non_lin_compton}
\end{figure}

In a linear Compton interaction, i.e. $\xi \ll 1$, an electron absorbs a number $n$ of photons and emits the same number of photons. From $\xi \sim 1$ the interaction becomes non-linear. For $n$ photons absorbed, a single high energy photon is produced as shown in Figure~\ref{fig:non_lin_compton}.

\begin{figure}[!ht]
    \centering
    \includegraphics[width=\textwidth]{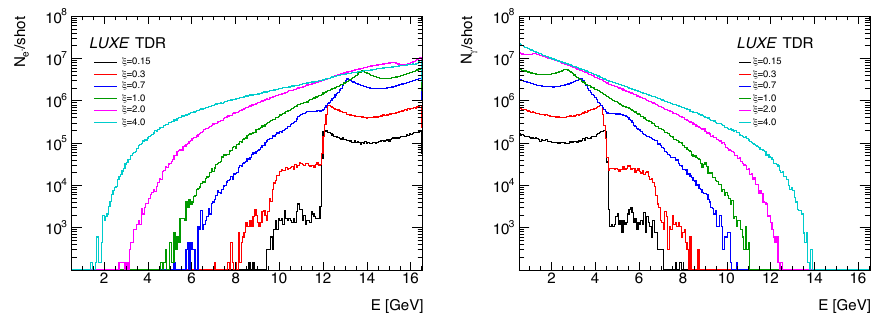}
    \caption{Plots of the effect of the non-linearity parameter $\xi$ on non-linear Compton scattering. Plots are reproduced from LUXE TDR~\cite{luxe_tdr}. a) On the left is the electron spectrum for one bunch crossing. b) On the right is the photon spectrum for one bunch crossing.}
    \label{fig:compton_spectums}
\end{figure}

The formula for non-linear Compton is then

\begin{equation}
    e^{\pm} + n\gamma_L \rightarrow e^{\pm} + \gamma
\end{equation}

where $\gamma_L$ are the laser photons. The Compton scattering interaction of an electron or positron depends on the quantum non-linearity parameter $\chi_e$ which is expressed as
 
\begin{equation}
    \chi_e = \frac{\mathcal{E}^*}{\mathcal{E}_{\rm cr}},
\end{equation}

where $\mathcal{E}^*$ is the Lorentz boosted electric field. At $\chi_e \ll 1$ the Coulomb scattering follows a classical non-linear model. With higher values of $\xi$, the highest energy a photon can carry after a Compton scatter. Preliminary simulations of the non-linear Compton interaction were realised in LUXE's TDR using Ptarmigan, a Monte-Carlo simulation software described in Section~\ref{sec:software} and used in Chapter~\ref{chap:impact_luxe}. Expected spectrums for electrons and photons are shown in Figure~\ref{fig:compton_spectums}.

\begin{figure}[!ht]
    \centering
    \includegraphics[width=12cm]{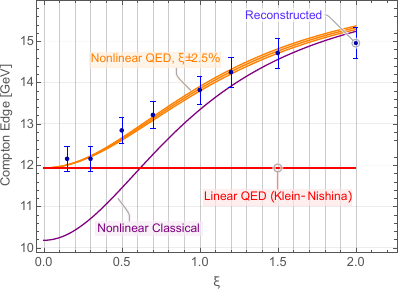}
    \caption{Plot of Compton edge energy as a function of the non-linearity parameter $\xi$. In orange is the non-linear QED predicted band with a \SI{2.5}{\%} error on $\xi$. The blue points with error bars are the reconstructed results from Ptarmigan. The red line is the linear QED theory. The purple line is the non-linear classical theory.}
    \label{fig:compton_edge}
\end{figure}

In the electron spectrum (Figure~\ref{fig:compton_spectums} a), for each histogram, the maximum energy an electron can carry after a Compton scattering interaction is called the ``Compton edge". This is clearly visible for $\xi=0.15$, with a Compton edge at around \SI{12}{\giga\electronvolt}. As $\xi$ increases, so does the Compton edge energy. The relation between $\xi$ and the Compton edge in the electron spectrum is predicted by strong-field QED and is an expected result at LUXE. In Figure~\ref{fig:compton_edge}, another Ptarmigan simulation from LUXE's TDR shows the Compton edge values for different values of $\xi$.

At low values of $\xi$ LUXE should observe higher Compton edge energies than the non-linear classical prediction. At high values of $\xi$, a difference should be observe between non-linear QED and the linear QED prediction of the Klein-Nishina formula~\cite{klein_nishina}.

\subsubsection{Non-linear Breit-Wheeler}

\begin{figure}[!ht]
    \centering
    \includegraphics[width=4cm]{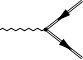}
    \caption{Feynman diagram of the non-linear Breit-Wheeler pair creation. Double lines indicate fermions in a laser background.}
    \label{fig:non_lin_bw}
\end{figure}

The Breit-Wheeler interaction creates an electron-positron pair from a incoming photon as shown in Figure~\ref{fig:non_lin_bw}

With $n$ laser photons $\gamma_L$ the formula is then

\begin{equation}
    \gamma + n\gamma_L \rightarrow e^{+} + e^{-}
\end{equation}

Similar to $\chi_e$ the quantum nonlinearity parameter for the photon $\chi_\gamma$ is expressed as 

\begin{equation}
    \chi_\gamma = \frac{\mathcal{E}_L}{\mathcal{E}_{\rm cr}}\frac{\hbar\omega_\gamma}{\gamma m_ec^2}(1+\cos\theta).
\end{equation}

\begin{figure}[!ht]
    \centering
    \includegraphics[width=\textwidth]{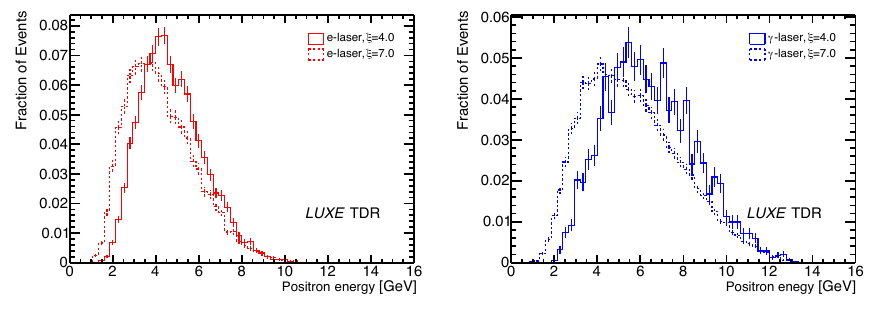}
    \caption{Histograms of the positron energy spectrum for $\xi=4$ and $\xi=7$. On the left plot are the histograms in electron-laser configuration. On the right are the histograms in gamma-laser configuration.}
    \label{fig:pair_spectum}
\end{figure}

With $\chi_\gamma \ll 1$, the Breit-Wheeler pair-creation process follows a linear regime. The number of interactions is linearly proportional to $\xi$. The non-linear behaviour of Breit-Wheeler is expected at $\chi_\gamma \sim 1$. Contrary to the Compton scattering, there is no classical analogue for this process. Preliminary simulation were realised with Ptarmigan and the resulting spectrums are shown in Figure~\ref{fig:pair_spectum}.

\begin{figure}[!ht]
    \centering
    \includegraphics[width=12cm]{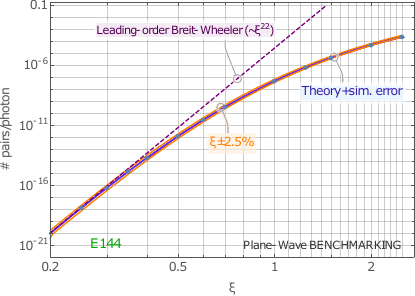}
    \caption{Plot of the number of electron-positrons pairs per photon as a function of the non-linearity parameter $\xi$. In orange is the non-linear QED predicted band with a \SI{2.5}{\%} error on $\xi$. The blue points are the results simulated with Ptarmigan. The purple line is the linear QED theory. The green band represent what could be measured at the E144 experiment~\cite{E144}.}
    \label{fig:pair_curves}
\end{figure}

At $\xi=7$, there is a higher number of positron produced and with higher energy than at $\xi=4$. In the gamma-laser configuration, i.e. where non-linear Breit-Wheeler is the only QED process, there are less positrons overall. However, they reach energies around \SI{13}{\giga\electronvolt} whereas the maximum energy in the electron-laser configuration is below \SI{11}{\giga\electronvolt}. In Figure~\ref{fig:pair_curves}, another Ptarmigan simulation from LUXE’s TDR shows the number of electron-positron pairs for different values of $\xi$.
At low values of $\xi$, the number of pairs follows a linear regime. This range of $\xi$, up to \SI{0.35}{}, was accessible at the E144 experiment~\cite{E144} at SLAC~\cite{slac}. At higher $\xi$, this number of pairs diverges form the linear regime. It is expected to observe less pair creation than what is predicted by the linear theory.

\subsubsection{Non-linear trident}

\begin{figure}[!ht]
    \centering
    \includegraphics[width=4cm]{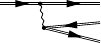}
    \caption{Feynman diagram of the non-linear inverse trident interaction. Double lines indicate fermions in a laser background.}
    \label{fig:non_lin_trident}
\end{figure}

The non-linear trident can be described as a combination of both a non-linear Compton and a non-linear Breit-Wheeler. As shown in Figure~\ref{fig:non_lin_trident} an incoming electron interacts first through a non-linear Compton scattering. The produced high energy photon then creates an electron-positron pair through non-linear Breit-Wheeler.

\begin{figure}[!ht]
    \centering
    \includegraphics[width=\textwidth]{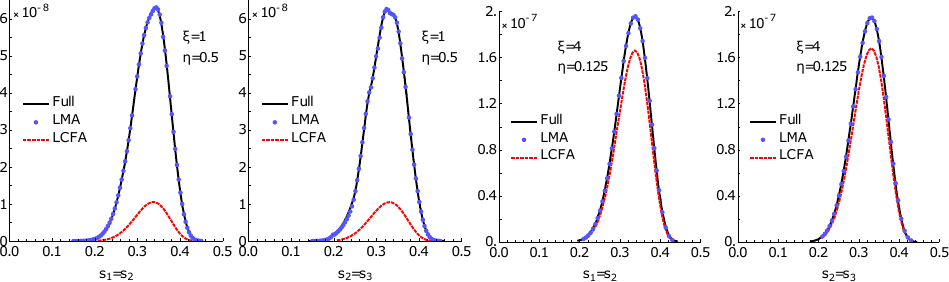}
    \caption{Plots of the momentum spectrum for the non-linear trident products. The black lines are derived from the full QED plane-wave calculations. The blue dots are the results from Ptarmigan using LMA. The red dashed line are the results form Ptarmigan using LCFA.}
    \label{fig:trident}
\end{figure}

The combined formula for non-linear trident is then

\begin{equation}
    e^{\pm} + n\gamma_L + n'\gamma_L \rightarrow e^{\pm} + e^{+} + e^{-}
\end{equation}

where $n$ is the number of laser photons involved in the non-linear Compton and $n'$ the number of laser photons involved in the non-linear Breit-Wheeler. This process have been measured for the first time in the non-perturbative regime at NA63~\cite{na63}. Simulations of this process in Ptarmigan requires the introduction of the locally monochromatic approximation~\cite{ptarmigan} (LMA) instead of the locally constant crossed field approximation~\cite{ptarmigan} (LCFA). A comparison of those two approximation for non-linear trident was done in LUXE's CDR, as shown in Figure~\ref{fig:trident}.

At $\xi=1$, which is the expected value for LUXE, the simulations with LMA follows the full plane-wave model with high precision, the LCFA simulation produces values around six times below the full simulation. Even at $\xi=4$, the LCFA model is better but not as good as LMA. Those two approximations are detailed in Section~\ref{sec:software}.

\subsection{Design of the experiment}

\begin{figure}[!ht]
    \centering
    \includegraphics[width=12cm]{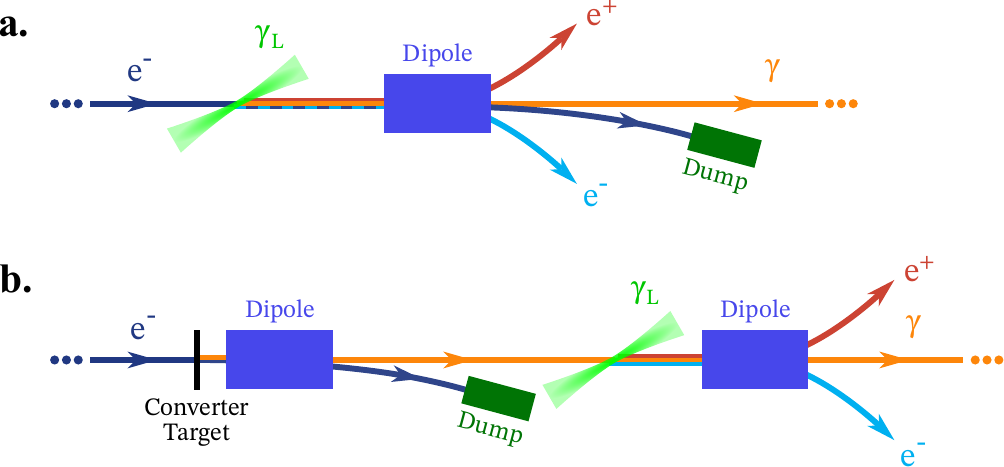}
    \caption{Schematic of the two setups for LUXE. On the top (a) the electron-laser setup and on the bottom (b) the gamma-laser setup.}
    \label{fig:LUXE_setups}
\end{figure}

LUXE is designed to measure the QED processes described in the previous section and is setup to function in two configurations. One where the EuXFEL electron beam collides directly with the laser and one where the electron are converted to photons which then interacts with the laser. Both configurations are shown in Figure~\ref{fig:LUXE_setups}. 

In the electron-laser configuration (shown in Figure~\ref{fig:LUXE_setups} a), the electrons interact directly with the laser pulse and produce electrons, positrons and photons through the three processes described above. The collision location is called the interaction point (IP). A bending magnet is then used to separate the three types of produced particles. Through the same magnet, the EuXFEL electron bunch is sent to a dump. 

In the gamma-laser configuration (shown in Figure~\ref{fig:LUXE_setups} b), the electrons are used to create a gamma photon beam which will interact with the laser. Upstream of the IP, a tungsten converter target can be inserted in the beam path. Electrons colliding with the target will generate photons through bremsstrahlung~\cite{brem_target}. The incoming electrons are immediately sent to a dump after the target by another bending magnet. The interactions at the IP between the photon beam and the laser are then exclusively Breit-Wheeler pair creations. 
Another possible method to create high energy gammas is an inverse Compton scattering (ICS) chamber~\cite{ics}. Here the electrons interact with a laser to create photons instead of a solid target. For LUXE, both methods are considered for the gamma-laser configuration, so the upstream chamber will be able to host either the tungsten target or the laser focus. In the following studies this location will mostly be referred as the ICS chamber (ICSC).

\begin{figure}[!ht]
    \centering
    \includegraphics[width=\textwidth]{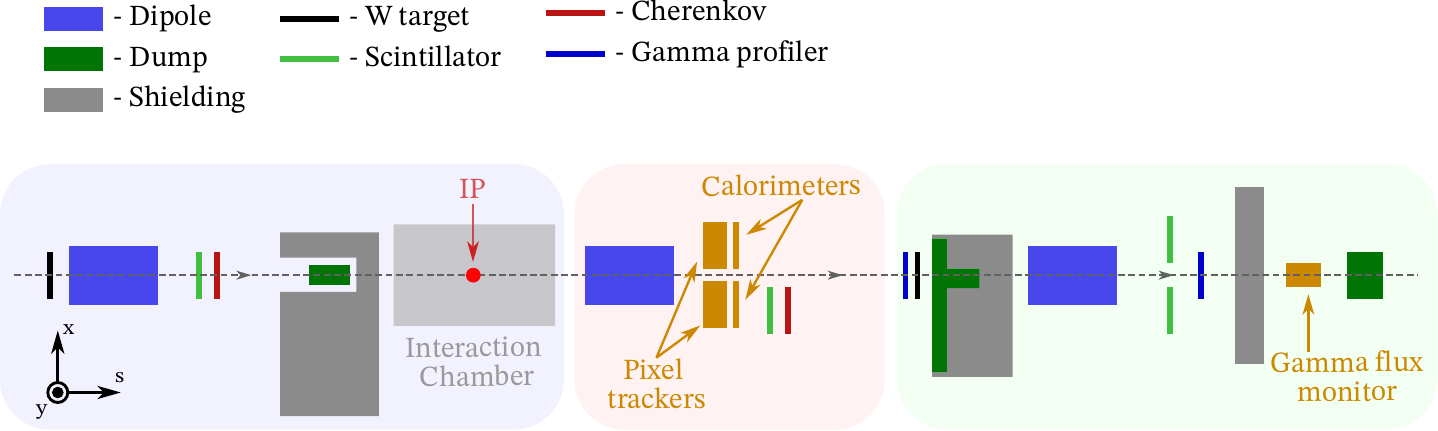}
    \caption{Schematic of the elements and detectors of LUXE viewed from above. The electron beam passes from left to right. The blue shaded section is the interaction section. The red shaded section is the diagnostics section for electrons and positrons. The green shaded section is the diagnostics sections for gamma photons.}
    \label{fig:LUXE_top}
\end{figure}

After the IP chamber, a horizontal dipole separates the QED products (electrons, positrons and photons) before the detector section. Electrons and positrons are bent on each sides of the beam path and the photons are sent further downstream. A complete layout of LUXE's elements and detectors can be found in Figure~\ref{fig:LUXE_top}.

The complete layout of LUXE is composed of three section. The first one is the interaction section whose two configurations are described in Figure~\ref{fig:LUXE_setups}. 
After the interaction chamber, there is the diagnostics section for electrons and positrons. In Figure~\ref{fig:LUXE_top} the positron diagnostics are shown above beam axis and the electron diagnostics are below the beam axis. For the electron path, the detectors used depends on the configuration of LUXE. In the electron-laser configuration only the scintillator~\cite{scintillator} and the Cherenkov detector~\cite{cherenkov} are used. The scintillators in LUXE are made of Gadolinium Oxysulfide (GadOx) doped with Terbium, $\rm Gd_2O_2S$:Tb, and can respond to electrons of only \SI{10}{\kilo\electronvolt}. The Cherenkov detectors are based on a design proposed at the international linear collider~\cite{ilc} (ILC). In the gamma-laser configuration, a pixel tracker~\cite{pixel} and a calorimeter~\cite{calorimeter} are inserted before the scintillator. The pixel trackers are made of silicon and are directly derived from the ``ALPIDE" pixel sensors~\cite{alpide} used at the ALICE experiment~\cite{alice}. The calorimeter is based on a design developed by the FCAL collaboration~\cite{fcal}. 
FCAL is a worldwide collaboration that aims to develop compact calorimeters with fast reading and in the very forward region of a future electron-positron collider.
On the positron path, only the pixel tracker and the calorimeter are used. The combination of those detectors is used to reconstruct the electrons and positrons path and energy. 

The last section is dedicated to the diagnostics for the gamma photons. A first set of gamma profilers~\cite{profiler} is placed after the electrons and positrons detectors. Those profilers are made of sapphire strips and allow a measurement of the gamma photons beam size with a \SI{5}{\micro\meter} resolution~\cite{luxe_tdr}. Just after this profiler, a gamma spectrometer section is used to analyse the energy of the gammas. In this spectrometer, the gamma photon beam first interacts with a tungsten target and produce electron-positron pairs. The pairs passes through a third dipole where the electrons and positrons are sent to scintillators and the residual gamma beam is sent to another gamma profiler. After the spectrometer, behind a third shielding, a gamma flux monitor~\cite{luxe_tdr} measures a signal proportional to the energy of particles back-scattered from the photon beam dump. This monitor is composed of eight blocks of lead-glass arranged cylindrically and symmetrically around the beam axis.

\subsection{Previous location in XTD20}

\begin{figure}[!ht]
    \centering
    \includegraphics[width=\textwidth]{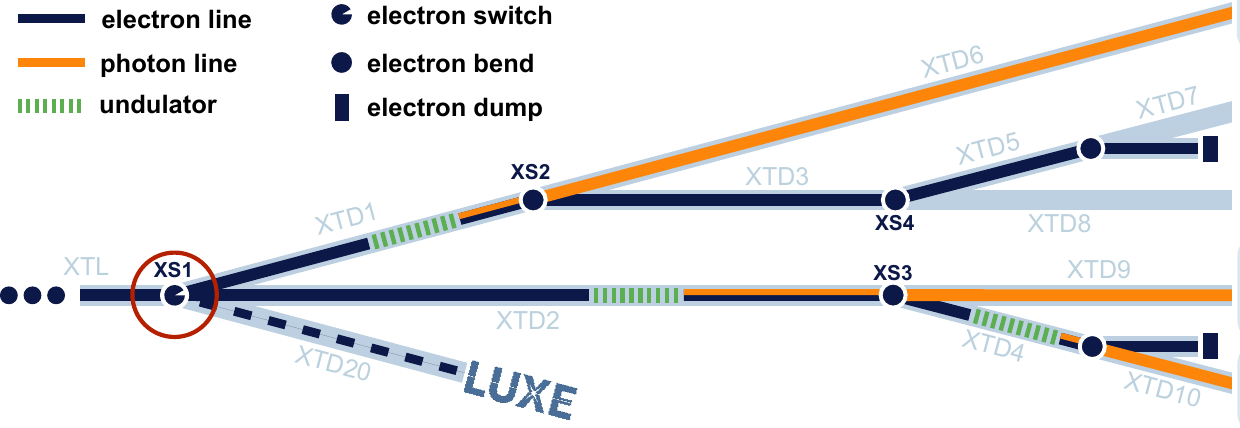}
    \caption{Schematic of the EuXFEL lines after the main linac with LUXE in the previously planned location. The red circle indicates the location of the main switch yard where the new lattice begins.}
    \label{fig:euxfel_lines_luxe_t20}
\end{figure}

\begin{figure}[!ht]
    \centering
    \includegraphics[width=12cm]{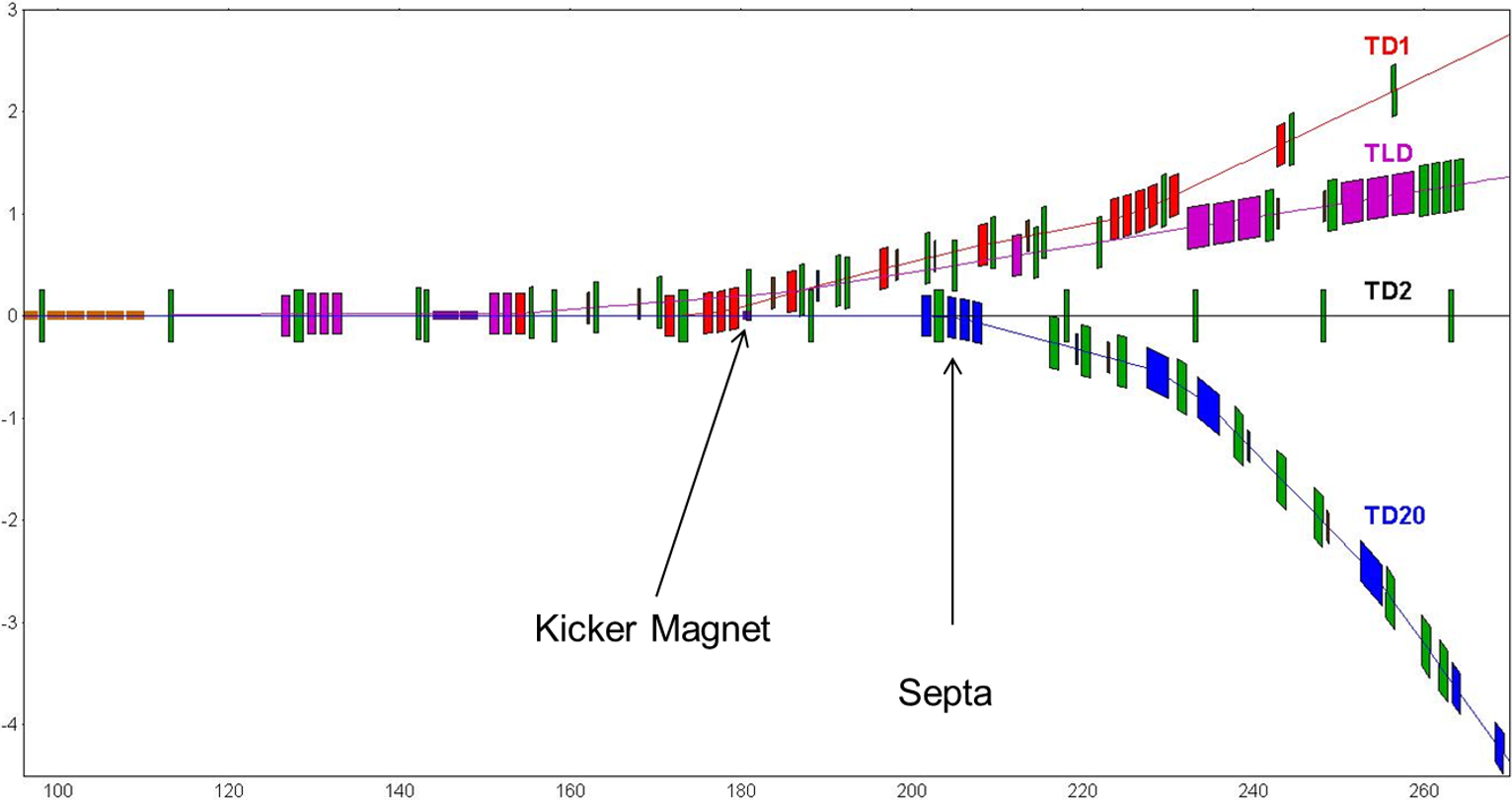}
    \caption{Schematic of the switch yard magnets with the previously planned kicker and septa dipoles for XTD20. This plot is reproduced from LUXE CDR. The dipoles are coloured differently for each beam line. Purple dipoles correspond to the extraction into the dump line XTLD. Red dipoles correspond to the extraction the main line XTD1. Blue dipoles correspond to the extraction into the XTD20 line. Green elements are the quadrupoles and magnets of higher order.}
    \label{fig:switchyard_luxe_t20}
\end{figure}

The location for LUXE at the EuXFEL was recently changed with the opportunity of the ELBEX project that will be presented in Section~\ref{sec:elbex}. However, from the early conception of LUXE to its TDR, the installation of LUXE was proposed in the currently unused XTD20 tunnel, as shown in Figure~\ref{fig:euxfel_lines_luxe_t20}.
The XTD20 tunnel starts at the XS1 building where the main switch yard distributes electron bunches between the two main undulator tunnels, XTD1 and XTD2. This building is upstream of the the FEL lines. As an addition to the current switch yard, a new set of kicker and septa dipoles was designed to send electrons to the XTD20 tunnel as shown in Figure~\ref{fig:switchyard_luxe_t20}

\begin{figure}[!ht]
    \centering
    \includegraphics[width=\textwidth]{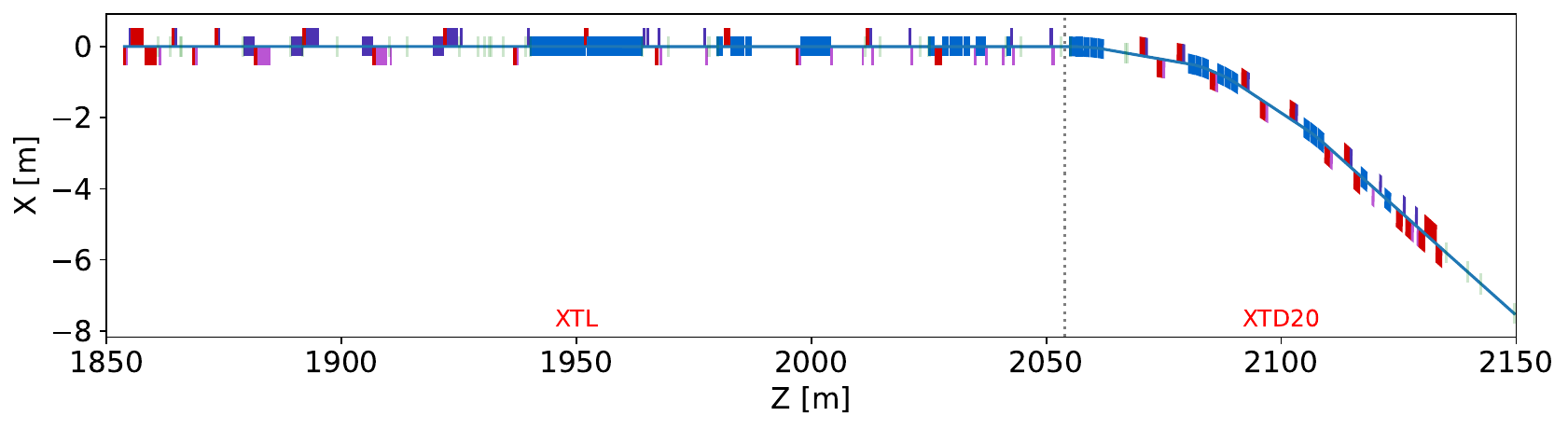}
    \caption{Magnet survey in $X$ and $Z$ of the lattice section used in the simulations for the previous location. It contains both the existing XTL transfer line and the proposed XTD20. In the BDSIM conventions, blue elements are the dipoles, red are quadrupoles, yellow are sextupoles, dark green are octupoles, purple and magenta are horizontal and vertical correctors, light green lines are markers.}
    \label{fig:XTL_XTD20}
\end{figure}

The extraction and the transport line are designed to be commissioned with LUXE. Since the experiment is not planned to be permanent, the designed transport line will be the beginning of the future fan of FEL lines. After the switch yard, a \SI{70}{\meter} beam line transports the electrons to LUXE. The layout of this lattice can be found in Figure~\ref{fig:XTL_XTD20}

This lattice was extracted from MAD-8 files publicly available on the DESY website~\cite{desy_mad8} and contains both the XTD20 line and the XTL line. Some of the studies presented later were originally performed with this lattice. In 2025 LUXE was associated to the ELBEX project which takes place further downstream in the XTD8 tunnel of the EuXFEL.

\section{The European XFEL}

\begin{figure}[!ht]
    \centering
    \includegraphics[width=\textwidth]{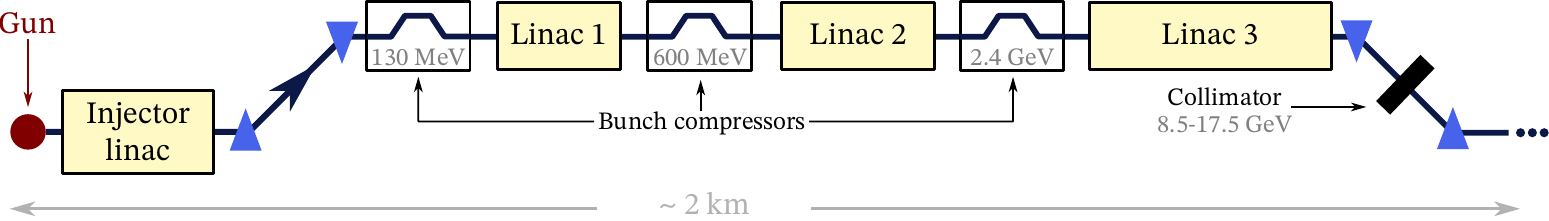}
    \caption{Schematic of the EuXFEL linear accelerator. The blue triangles represent dipole magnets.}
    \label{fig:euxfel_accel}
\end{figure}

The EuXFEL~\cite{euxfel_tdr} is a linear accelerator located at the Deutsches Elektronen-Synchrotron (DESY) facility in Hamburg. The goal of this accelerator is to generate high energy X-rays in short pulses of around \SI{40}{\femto\second}. Those pulses are coherent in wavelengths down to \SI{0.1}{\nano\meter}. The facility is \SI{3.4}{\km} long and is composed of two main sections. The first one is a \SI{2}{\km} accelerator section, referred as the linac~\cite{euxfel_linac}, whose layout is shown in Figure~\ref{fig:euxfel_accel}.

At the beginning of the accelerator is a laser-driven photo-cathode gun~\cite{euxfel_gun}. As a laser pulse hits the solid cathode, an electron bunch is created by the photo-electric effect. The gun generates electron bunch trains at a rate of \SI{10}{\hertz}. Each train contains up to \SI{2700}{} bunches each separated by around \SI{220}{\nano\second}. Each bunch position in the train is referred as a bunch ID. Bunches of the train are accelerated to \SI{120}{\mega\electronvolt} by the RF cavity of the injector linac. Along the remaining \SI{1.6}{\kilo\meter} of linac, the electron bunch is compressed three times in bunch compressors~\cite{euxfel_bc}. Each compressor is designed as a four bend chicane that applies a rotation in the time-energy phase space. This notion of bunch compression is developed in the design of the new beam line for LUXE in Chapter~\ref{chap:design}. The main linear accelerator (from linac 1 to linac 3) is composed of 96 superconducting (SC) accelerator modules. Each module contains eight \SI{1.3}{\giga\hertz} TESLA-type cavities~\cite{tesla}, for an average design gradient of \SI{23.6}{\mega\volt/\meter}.

\begin{table}[!ht]
\centering
\begin{tabular}{| c | c | c |}
\hline
Parameter & Unit & Value \\
\hline
\hline
Energy & GeV & 8.5-17.5 \\
\hline
Bunch charge & nC & 0.1-1 \\
\hline
Normalized RMS emittance & mm.mrad & 0.42-1.05 \\
\hline
RMS energy spread & MeV & 2-3 \\
\hline
RMS pulse duration & fs & 12-60 \\
\hline
\end{tabular}
\caption{Properties of the EuXFEL electron bunches. RMS stands for root mean square.}
\label{tab:euxfel_elec_param} 
\end{table}

\begin{figure}[!ht]
    \centering
    \includegraphics[width=\textwidth]{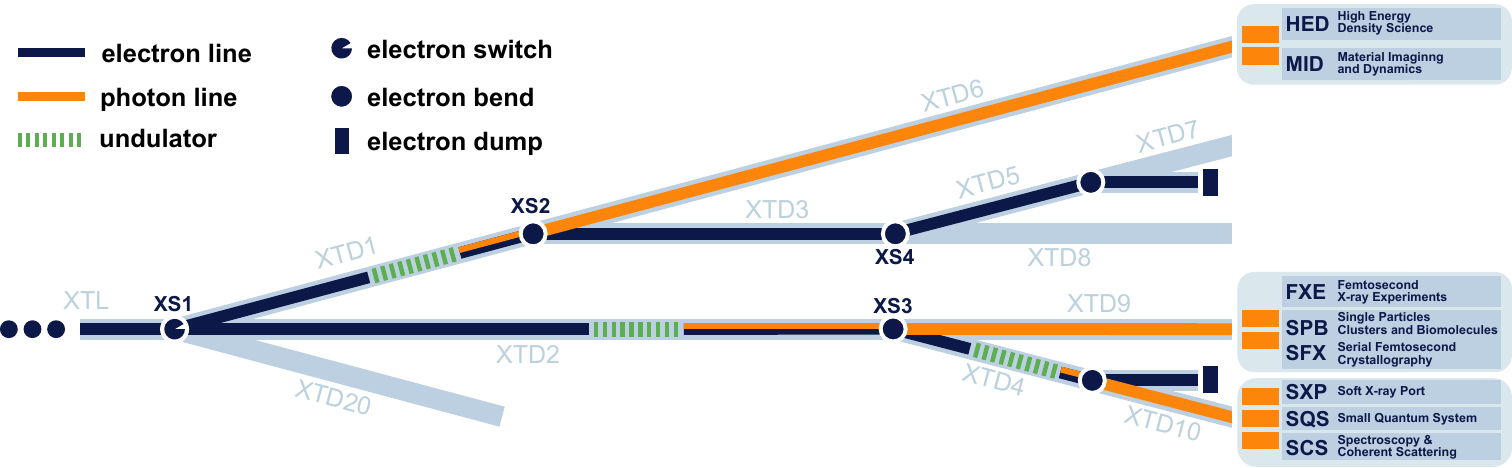}
    \caption{Schematic of the EuXFEL lines. On the right are the explanations of the different experiments using the high energy photons. The linear accelerator described previously is further to the left.}
    \label{fig:euxfel_lines}
\end{figure}

A summary of the electron bunch properties is in Table~\ref{tab:euxfel_elec_param}.
In the second section of the EuXFEL, the high energy electron bunches are distributed by the main switch yard and transported to undulators. Those undulators generates the X-ray photons beams used in various experiments. This section is \SI{1.4}{\kilo\meter} long and the different lines and experiments are shown in Figure~\ref{fig:euxfel_lines}.

\begin{figure}[!ht]
    \centering
    \includegraphics[width=14cm]{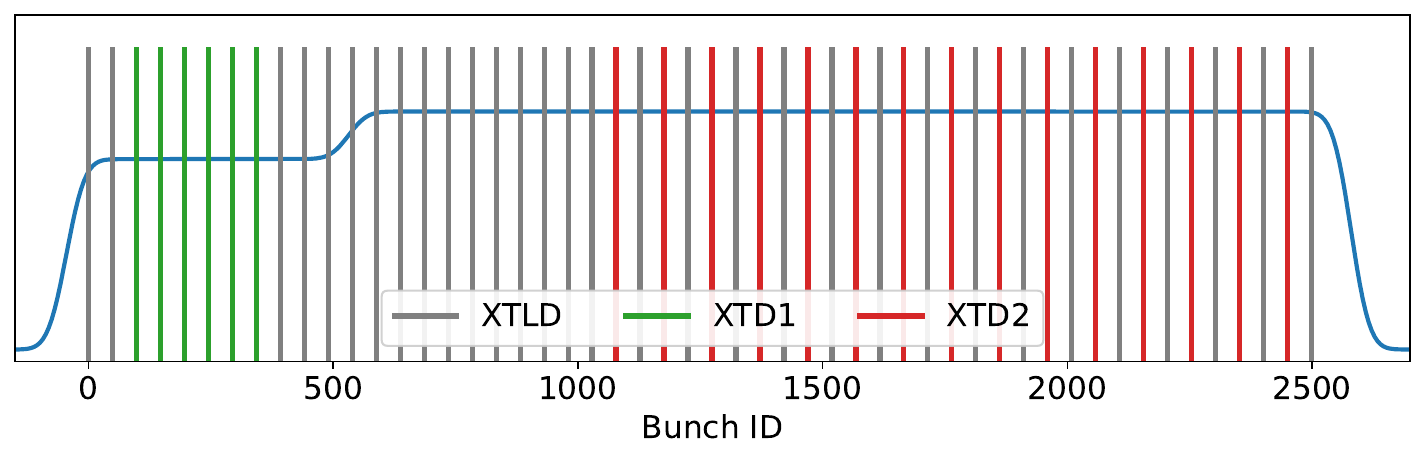}
    \caption{Bunch pattern of an example bunch train in the EuXFEL. The blue line represent one RF accelerating pulse. The step at bunch ID 600 represent a change in the bunches properties, for example a variation in bunch energy between lines implies an adjustment in the RF cavities that is not instantaneous. In red are bunches going in the XTD2 line. In green are bunches going in the XTD1 line. In gray are bunches not used for any experiment, they are sent to the switch yard dump XTLD.}
    \label{fig:euxfel_bunch_patern}
\end{figure}

In the XS1 building, the main switch yard separates electrons between the two most upstream undulator tunnels, XTD1 and XTD2. The fast kickers of the switch yard can select specific bunches in the train to go in either of the tunnels or in an electron beam dump. In Figure~\ref{fig:euxfel_bunch_patern} there is an example of the partition of the bunches inside a train.

This bunch pattern can be adapted, on a daily basis, to the needs of the experiments. 
Downstream of the switch yard, undulators are used to generate X-ray photons through self-amplified spontaneous emission (SASE)~\cite{sase}. Each SASE undulator produces photons of different properties, depending on the experiment requirements. The photon properties of the different lines are listed in Table~\ref{tab:euxfel_photon_param}.

\begin{table}[!ht]
\centering
\begin{tabular}{|c | c || c c | c c | c c |}
\hline
Value & Unit & \multicolumn{2}{c|}{SASE1} & \multicolumn{2}{c|}{SASE2} & \multicolumn{2}{c|}{SASE3} \\
\hline
\hline
Wavelength & nm & 0.05 & 0.1 & 0.1 & 0.4 & 0.4 & 1.6 \\
\hline
Photon energy & keV & 24.8 & 12.4 & 12.4 & 3.1 & 3.1 & 0.8 \\
\hline
Energy per pulse & mJ & 0.07 & 0.2-0.9\footnotemark[1]{} & 0.18-0.9\footnotemark[1]{} & 0.5-4\footnotemark[1]{} & 0.48-4\footnotemark[1]{} & 0.7-7.6\footnotemark[1]{} \\
\hline
Peak power & GW & 4.4 & 11 & 11 & 27-48\footnotemark[1]{} & 29-48\footnotemark[1]{} & 44-90\footnotemark[1]{} \\
\hline
Photon beam size & $\rm\mu m$ & 31 & 33-49\footnotemark[1]{} & 35-58\footnotemark[1]{} & 40-55\footnotemark[1]{} & 42-58\footnotemark[1]{} & 48-67\footnotemark[1]{} \\
\hline
Coherence time & fs & 0.18 & 0.19-0.24\footnotemark[1]{} & 0.19-0.25\footnotemark[1]{} & 0.39-0.41\footnotemark[1]{} & 0.36-0.39\footnotemark[1]{} & 0.9-1\footnotemark[1]{} \\
\hline
Pulse duration & fs & 17 & 17-84\footnotemark[1]{} & 17-84\footnotemark[1]{} & 17-84\footnotemark[1]{} & 17-84\footnotemark[1]{} & 17-84\footnotemark[1]{} \\
\hline
\end{tabular}
\caption{Properties of the X-ray photons for the different SASE undulators~\cite{euxfel_photons}.}
\label{tab:euxfel_photon_param} 
\end{table}

The X-ray produced by the undulators are sent to the different experiments on the right hand side in Figure~\ref{fig:euxfel_lines} and the electrons are dumped. The high energy photons sees applications for a large variety of scientific domains. The materials imaging and dynamics (MID) instrument~\cite{mid} aims to study liquid, soft-condensed and hard-condensed materials at a nanoscopic level. The small quantum systems (SQS) instrument~\cite{sqs} aims to study non-linear and multi-photon processes, time-resolved experiments on fast dynamic processes, from the atomic sizes to molecular sizes. The spectroscopy and coherent scattering (SCS) instrument~\cite{scs} is using the high repetition rate of X-ray photons for 2D imaging and holographic imaging. This is a non-exhaustive list of the current X-ray experiments at the EuXFEL and more details can be found in ~\cite{euxfel_exp}. 

\footnotetext[1]{Ranges of values are dependent on the parameters of the input electron bunch.}

\section{The ELBEX project} \label{sec:elbex}

\begin{figure}[!ht]
    \centering
    \includegraphics[width=\textwidth]{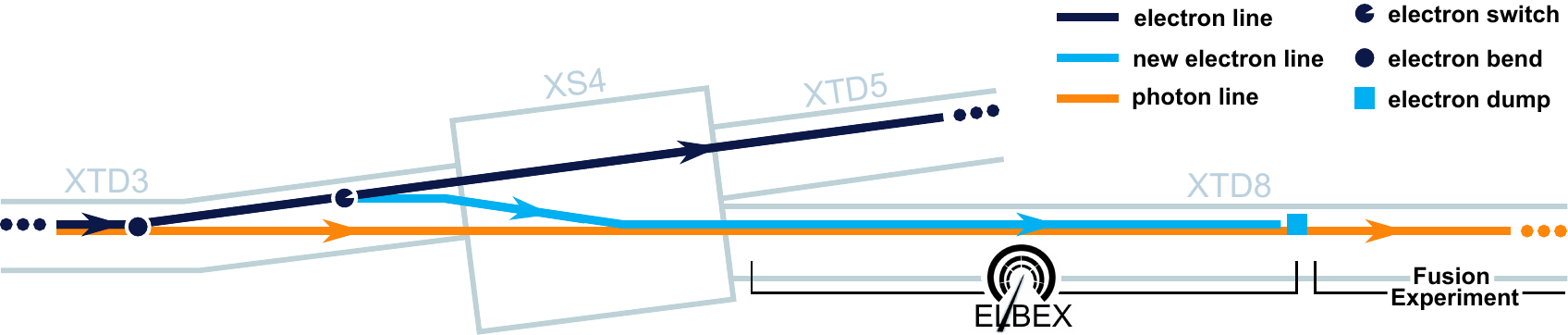}
    \caption{Schematic of the ELBEX beam line in the EuXFEL tunnels. The electrons are extracted in the XTD3 tunnel and dumped before the fusion experiment. The electrons are brought back in the photon pipe in the XS4 hall.}
    \label{fig:elbex}
\end{figure}

The electron beam line at the European XFEL (ELBEX) is an international project that aims to create a beam line for multiple experiments that directly wish to use the high energy electrons. 
This project will take place in the currently unused XTD8 tunnel of the EuXFEL. Figure~\ref{fig:elbex} shows the future location of ELBEX, the electron extraction in the XTD3 and the path into the XS4 building.

\begin{figure}[!ht]
    \centering
    \includegraphics[width=\textwidth]{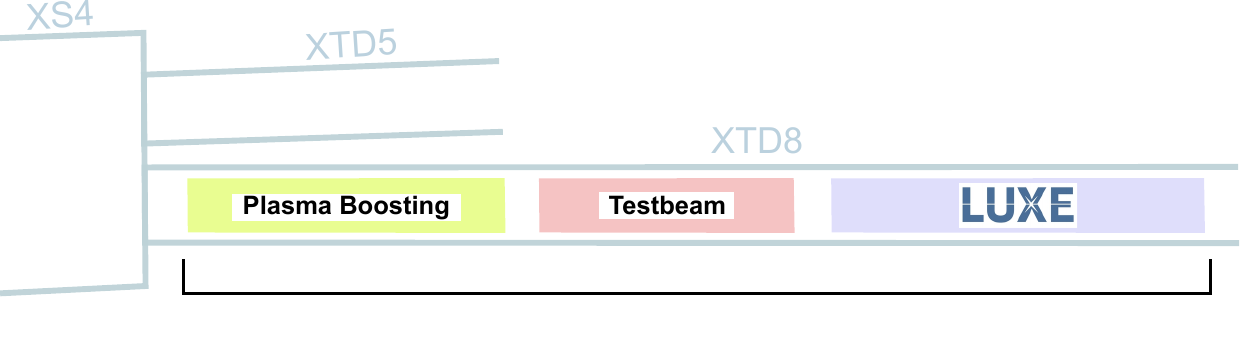}
    \caption{Plan layout of the three experiments within ELBEX in the XTD8 tunnel. The relative length of the experiments is not to scale.}
    \label{fig:elbex_experiments}
\end{figure}

In addition of the ELBEX project, this tunnel will transport photons from a SASE undulator to a proposed future fusion experiment called HAFEN~\cite{hafen}. It was decided for efficiency reasons to have a shared electron and photon beam pipe in the XTD8 tunnel instead of two separate ones. This decision does have an impact in term of radio-protection and will be accounted for in the design.
There are currently three experimental areas within the ELBEX project as shown in Figure~\ref{fig:elbex_experiments}. In beam line order there is first a plasma acceleration energy boosting experiment, then a test-beam experiment, and finally LUXE which was described in Chapter~\ref{chap:intro}.

\subsection{Plasma boosting experiment}

\begin{figure}[!ht]
    \centering
    \includegraphics[height=6cm]{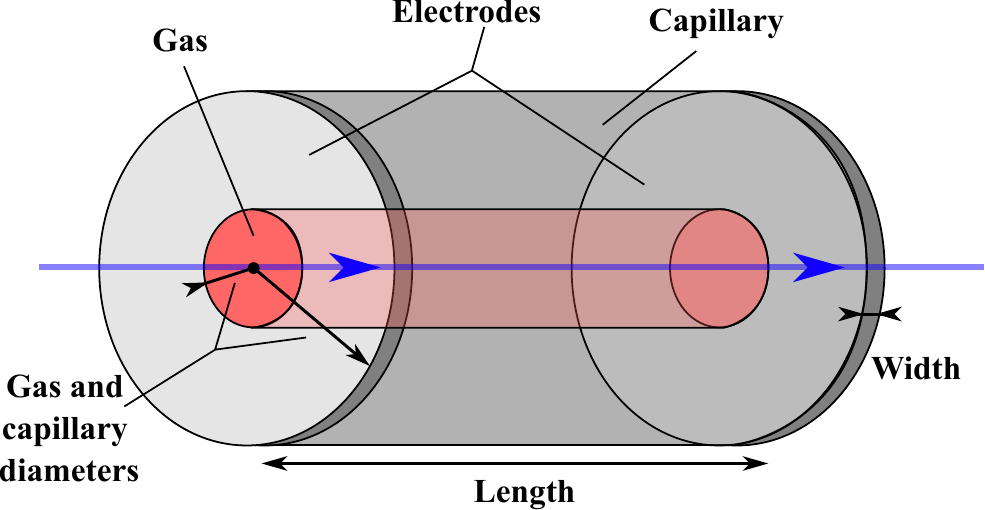}
    \caption{Schematic of a plasma discharge capillary. The cylinder of gas, centred on the beam axis, is surrounded by a capillary. On each sides there is a circular electrode with the same transverse cross section as the capillary.}
    \label{fig:gas_capillary}
\end{figure}

To accelerate a charged particle beam, the classical method is to use radio-frequency (RF) cavities~\cite{rf}. Another promising method to accelerate such a beam is called plasma wakefield acceleration~\cite{pwfa} (PWFA). Let us consider an electron bunch going through a plasma chamber as shown in Figure~\ref{fig:gas_capillary}.

As an electron bunch travels into the plasma, it will displace the mobile electrons, creating a wakefield. The wakefield is the wave of charges displacement propagating in the plasma. The induced charge separation creates a strong electric field behind the bunch. 
As a second electron bunch passes in the chamber during that time, it will experience the electric field and be accelerated. The first and second bunch are respectively called the drive bunch and the witness bunch. The wakefield usually remains in the plasma chamber for \SI{0.1}{\pico\second} to \SI{1}{\pico\second}, which is the arrival time difference needed between the two bunches.
There are three methods to generate two successive bunches at such a small distance. The first one is to send two laser pulses at the gun photo-cathode. The second method is to use a delay chicane~\cite{sb_chicane}. As two bunches are transported in a lattice, a chicane sends the leading drive bunch in a longer path than the following witness bunch. In the end the two path are merged and the time between both bunches is reduced. The final method uses a beam mask placed in a dispersive section.

\begin{figure}[!ht]
    \centering
    \includegraphics[height=6cm]{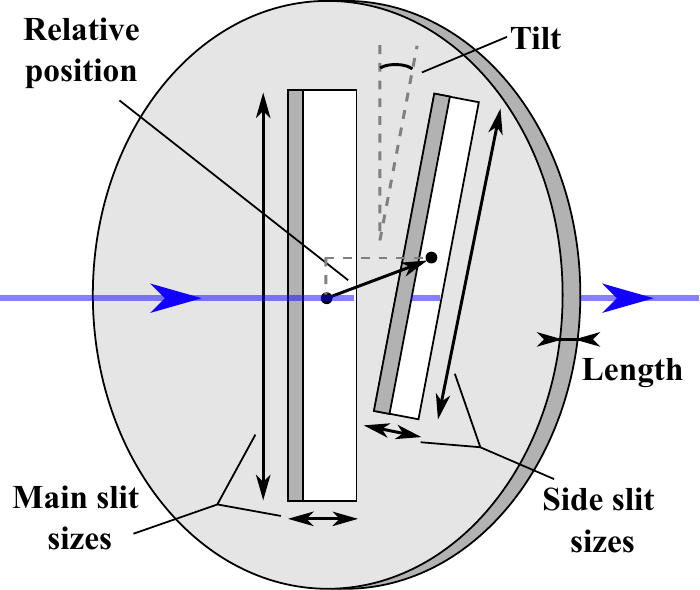}
    \caption{Schematic of a beam mask. The main slit and side slit can be moved and tilted.}
    \label{fig:beam_mask}
\end{figure}

A schematic of such mask is shown in Figure~\ref{fig:beam_mask}. The beam mask is a piece of solid material with two slits. In a dispersive section, the beam size increases as a function of the energy spread. As this large beam reaches the mask, some particles will be absorbed and two parallel bunchlets will emerge from the slits. Each bunchlet has a different transverse position, i.e. a slightly different energy. At the end of this dispersive section, the bunches will both be aligned on the reference path but with a small arrival time difference.

At DESY, the FlashForward~\cite{flash_forward} plasma boosting experiment showed promising performances, from the preservation of the energy spread and the emittance to the \SI{}{MHz} repetition rate~\cite{flash_results}.
The plasma boosting experiment in ELBEX aims to apply the successful techniques of FlashForward with the much higher energies of the EuXFEL beam. From an \SI{8.5}{\giga\electronvolt} electron beam, the plasma cell is expected to accelerate up to \SI{17}{\giga\electronvolt}. However, simulations have shown that a \SI{1.8}{\meter} plasma cell could accelerate the electron beam from \SI{17}{\giga\electronvolt} to \SI{31}{\giga\electronvolt}. If this acceleration method is proven reliable enough, it could be part of the possible continuous wave (CW) upgrade of the EuXFEL to get higher energy electron in the SASE undulators. At the moment, if the energy spread is well preserved, the emittance will experience a \SI{30}{\%} increase.

\subsection{Test-beam experiment}

The DESY II Synchrotron currently offers 3 beam lines for test beam experiments (TB21, TB22 and TB24)~\cite{test_beam}. The energy range of those lines is between \SI{1}{\giga\electronvolt} and \SI{6}{\giga\electronvolt}. The proposed test beam experiment in ELBEX should serve as a complement to the DESY II lines. It could give access to higher energy but with a lower bunch repetition rate. 

\begin{figure}[!ht]
    \centering
    \includegraphics[width=12cm]{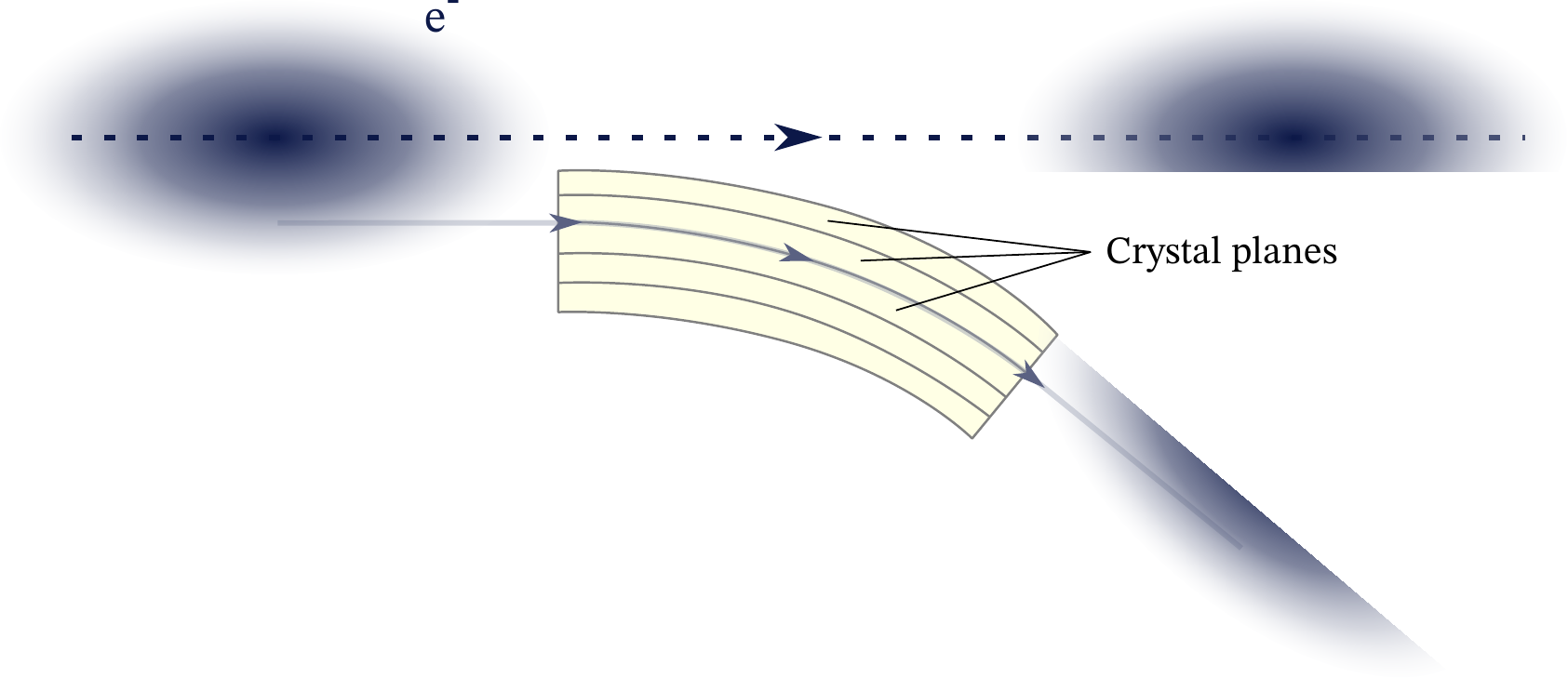}
    \caption{Schematic of an electron going through a crystal extractor.}
    \label{fig:crystal_ext}
\end{figure}

One experiment envisioned for this test beam is crystal extraction~\cite{crystal} of the beam. As an electron beam travels through a crystal, 
it is channelled by the inter-atomic electric fields in between the crystal planes.
As such, it follows a path parallel to those planes, as shown in Figure~\ref{fig:crystal_ext}.
With a bent crystal it is possible to extract the electron beam without dipole magnet. This method can also be used for beam collimation.

\chapter{Accelerator physics theory and simulation} \label{chap:theory}

\section{Accelerator physics}

As mentioned in Chapter~\ref{chap:intro}, to propagate the EuXFEL electrons towards LUXE a new beam line is needed. In this beam line, electrons behaviour will be mainly dictated by the transverse dynamic of our lattice. A full review of accelerator beam dynamics can be found in H. Wiedemann's book~\cite{accel_phys} and W. Hillert's article~\cite{tbd}. In this chapter, a brief review of transverse and longitudinal dynamics is presented as it is relevant to Chapter~\ref{chap:design} on the design of the ELBEX beam line.

\begin{figure}[!ht]
        \centering
        \includegraphics[width=8cm]{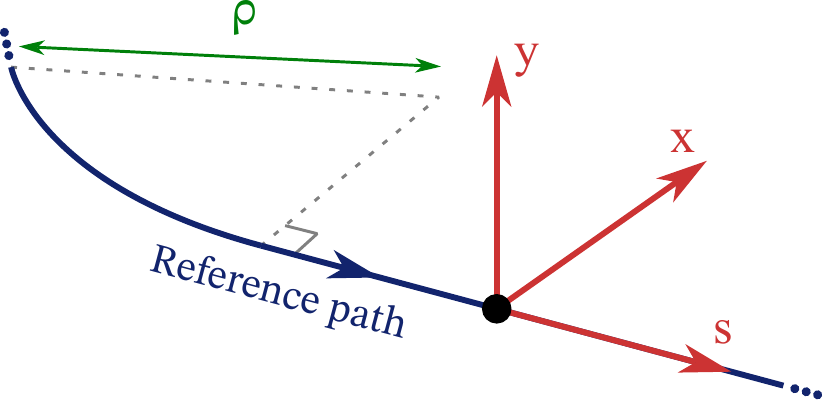}
        \caption{Coordinates chart with respect to the beam. Here $\rho$ is the radius of curvature.}
        \label{fig:repere}
\end{figure}

Let us consider a linear accelerator where particles follows a single reference path, i.e. the beam axis. Typically, the longitudinal coordinate $s$ is used to denote the particle position along said path. The set of coordinates {$X, Y, Z$} is used to describe the global position of the particle, usually with respect to the start of the reference path. At any $s$ position, the local transverse coordinates {$x, y$} are used to quantify local transverse offsets from the reference path. Similarly, the local longitudinal coordinate $z$ is used to quantify local longitudinal offsets. With an ultra-relativistic beam, $z$ is often derived from a time offset $t$ through $z=ct$, where $c$ is the speed of light.
Such coordinate system is represented in Figure~\ref{fig:repere}. For more simplicity, only the horizontal plane $x$ will be considered in the following equations but one can define a similar set of equations in the vertical plane $y$.

The beam axis in Figure~\ref{fig:repere} is also referred to as the reference path. As a charged particle of charge $q$ and velocity $\vec{v}$ travels through an EM field, it will experience a Lorentz force $\vec{F}$ expressed as

\begin{equation}
    \vec{F}=q(\vec{\mathcal{E}}+\vec{v}\times\vec{\mathcal{B}}),
\end{equation}

where $\vec{\mathcal{E}}$ and $\vec{\mathcal{B}}$ are the electric and magnetic fields respectively. The magnetic rigidity of this particle is then defined as

\begin{equation}
    \mathcal{B}\rho=\frac{p}{q},
\end{equation}

where $p$ is the particle's momentum and $\rho$ is the radius of curvature. 

\subsection{Transverse dynamics}

The transverse beam dynamics are exclusively focused on the particle movements in the $x,y$ plane. In a transport line, i.e. a line without accelerating sections, the magnets are designed for a specific reference energy $E_0$ which correspond to a reference momentum $p_0$ through

\begin{equation}
    E_0 = \sqrt{p_0^2c^2+m^2c^4},
\end{equation}

where $m$ is the particle mass. Particles with this reference momentum will be propagated along the reference path and are called ``On-momentum" particles.

\subsubsection{On-momentum particle}

Applying Lorentz's force through the fundamental principle of dynamics, one can obtain the equation of motion of a charged particle, i.e. Hill's equation~\cite{tbd}. For the horizontal position $x$ of an on-momentum particle, this equation of motion is defined as 

\begin{equation}
    x''(s) + K(s) x(s) = \frac{d^2x(s)}{ds^2}+K(s) x(s) = 0, 
    \label{eq:Hills}
\end{equation}

where $K(s)$ is a focusing strength parameter. In the following equations, the first order derivative of $x(s)$ with respect to $s$, i.e. the beam horizontal angle, will be noted as $x'(s)$, The second order derivative will be noted as $x''(s)$ . A similar equation can be established for the vertical position $y$. 
Since the focusing strength $K$ is supposed to be varying along $s$, a solution to this second order differential equation can be written as

\begin{equation}
    x(s)=\sqrt{\varepsilon_x \beta_x (s)} \cos{(\mu_x (s) + \phi)}, 
    \label{eq:beta_oscillation}
\end{equation}

where $\varepsilon_x$ is the horizontal emittance, $\beta_x$ is the horizontal beta twiss function, $\mu_x$ is the horizontal phase advance and $\phi$ is a constant. One can also get the angle $x'$ at any $s$ position by taking the derivative of equation~\ref{eq:beta_oscillation}. The angle is then expressed as

\begin{equation}
    x'(s)=\sqrt{\frac{\varepsilon_x}{\beta_x (s)}} (\alpha_x(s)\cos{(\mu_x(s) +\phi)} +\sin{(\mu_x(s) +\phi)}),
    \label{eq:beta_angle_oscillation}
\end{equation}

where $\alpha_x$ is the horizontal alpha twiss function expressed as

\begin{equation}
    \alpha_x = -\frac{1}{2}\frac{d\beta_x}{ds}.
\end{equation}

To express information not just for a single particle but for an entire bunch, the particle distribution is parametrized. As such, the horizontal emittance in $\varepsilon_x$ can be written as 

\begin{equation}
    \varepsilon_x = \gamma_x x^2 + 2\alpha_x x x' + \beta_x {x'}^2,
    \label{eq:emittance}
\end{equation}

where $\gamma_x$ is the horizontal gamma twiss function expressed as

\begin{equation}
    \gamma_x = \frac{1+\alpha_x^2}{\beta_x}.
\end{equation}

Eqn~\ref{eq:emittance} is the equation of the phase ellipse. The three twiss functions $\beta$, $\alpha$ and $\gamma$ contains all the linear transverse dynamic information of our beam in one direction. 
Propagating $x$ and $x'$ from an initial longitudinal location $s_0$ to a location $s$ is achieved by building the $2\times2$ transfer matrix $\boldsymbol{R}$, also called $R$-matrix, defined by

\begin{equation}
    \begin{pmatrix}
        x(s) \\
        x'(s)\\
    \end{pmatrix}
     = \boldsymbol{R}_{s_0\rightarrow s} \cdot
     \begin{pmatrix}
         x(s_0) \\
         x'(s_0) \\
     \end{pmatrix}
     =
     \begin{pmatrix}
        R_{11} & R_{12}\\
        R_{21} & R_{22} \\
    \end{pmatrix}\cdot
     \begin{pmatrix}
         x(s_0) \\
         x'(s_0) \\
     \end{pmatrix},
     \label{eq:rmat_x}
\end{equation}

where $R_{11}, R_{12}, R_{21}$ and $R_{22}$ are called $R$-matrix factors. Defining the transfer matrix using Eqn~\ref{eq:beta_oscillation} and Eqn~\ref{eq:beta_angle_oscillation} is not trivial. An assumption is made that the focusing strength parameter $K(s)$ is piece-wise constant, i.e. it does not vary along a small section of the reference path. Eqn~\ref{eq:Hills} then become an homogeneous second order differential equation. By setting the initial conditions $x(s_0)=x_0$ and $x'(s_0)=x'_0$, the solution takes the form

\begin{equation}
    x(s) = C(s)x_0 + S(s)x'_0,
    \label{eq:sol_cos_sin}
\end{equation}

where $C(s)$ and $S(s)$ are the principal solutions of the differential equation. They are respectively cosine-like and sin-like functions that depends on the sign of $K$. Taking the derivative of Eqn~\ref{eq:sol_cos_sin}, the angle $x'(s)$ is expressed as

\begin{equation}
    x'(s) = C'(s)x_0 + S'(s)x'_0.
\end{equation}

Identifying the factors with Eqn~\ref{eq:rmat_x}, the transfer matrix $\boldsymbol{R}$ can be expressed as

\begin{equation}
    \begin{pmatrix}
        x(s) \\
        x'(s)\\
    \end{pmatrix}
    =
    \begin{pmatrix}
        C(s) & S(s)\\
        C'(s) & S'(s) \\
    \end{pmatrix} \cdot
    \begin{pmatrix}
        x_0 \\
        x'_0\\
    \end{pmatrix}.
    \label{eq:rmat_x_cs}
\end{equation}

Assuming the particles are travelling in a lattice with linear optics and without any collimation or acceleration, the emittance is conserved between $s_0$ and $s$. As such, one can write

\begin{equation}
    \gamma_x x^2 + 2\alpha_x x x' + \beta_x {x'}^2 = 
    \gamma_{x,0} x_0^2 + 2\alpha_{x,0} x_0 x_0' + \beta_{x,0} {x_0'}^2,
    \label{eq:emitt_propag}
\end{equation}

where {$\beta_{x,0}, \alpha_{x,0}, \gamma_{x,0}$} are the twiss functions at $s_0$ and {$\beta_{x}, \alpha_{x}, \gamma_{x}$} are the twiss functions at $s$.
From Eqn~\ref{eq:rmat_x} and Eqn~\ref{eq:emitt_propag}, one can also define the transfer matrix $\boldsymbol{R}_{s_0 \rightarrow s}'$ that propagates the twiss parameters between $s_0$ and $s$. This is then expressed as


\begin{equation}
    \begin{pmatrix}
        \beta_x \\
        \alpha_x \\
        \gamma_x \\
    \end{pmatrix} = \boldsymbol{R}'_{s_0 \rightarrow s} \cdot
    \begin{pmatrix}
        \beta_{x,0} \\
        \alpha_{x,0} \\
        \gamma_{x,0} \\
    \end{pmatrix} = 
    \begin{pmatrix}
        C^2 & −2CS & S^2 \\
        −CC' & CS'+C'S & −SS' \\
        C'^2 & −2C'S' & S'^2 \\      
    \end{pmatrix}
    \cdot
    \begin{pmatrix}
        \beta_{x,0} \\
        \alpha_{x,0} \\
        \gamma_{x,0} \\
    \end{pmatrix}.
    \label{eq:twiss_mat_x}
\end{equation}

With Eqn~\ref{eq:rmat_x} and Eqn~\ref{eq:twiss_mat_x}, one can obtain the horizontal position of a particle bunch and the associated twiss functions at any $s$ using the initial conditions at $s_0$.
Taking into account the vertical position and angle, $y$ and $y'$, the transfer matrix can be expressed as

\begin{equation}
    \begin{pmatrix}
        x \\
        x' \\
        y  \\
        y' \\
    \end{pmatrix}
     =
     \begin{pmatrix}
        \begin{matrix}
            R_{11} & R_{12}\\
            R_{21} & R_{22} \\
        \end{matrix} & 0 \\ 0 &
        \begin{matrix}
            R_{33} & R_{34} \\
            R_{43} & R_{44} \\
        \end{matrix}
    \end{pmatrix}\cdot
     \begin{pmatrix}
         x_0 \\
         x'_0 \\
         y_0 \\
         y'_0 \\
     \end{pmatrix}.
     \label{eq:rmat_xy}
\end{equation}

This matrix is bloc-diagonal as there is typically no cross-correlation between the $x$ and $y$ plane. Using only the bottom right block, one can build an equation similar to Eqn~\ref{eq:twiss_mat_x} and propagate the vertical twiss functions $\beta_y$, $\alpha_y$ and $\gamma_y$. The propagations of particle beam parameters will be useful in Chapter~\ref{chap:impact_luxe}.

\subsubsection{Off-momentum particle}

The ideal case of an on-momentum particle only applies to particles with the reference momentum $p_0$. Particles with a momentum $p$ different from the reference momentum are called ``Off-momentum" particles. For an off-momentum particle, Hill's equation of motion becomes inhomogeneous and can be written as

\begin{equation}
    x''(s) + K(s) x(s) = \frac{1}{\rho} \frac{\Delta p}{p_0} = \frac{\delta}{\rho},
    \label{eq:Hills_off}
\end{equation}

where $\rho$ is the radius of curvature, $p_0$ is the reference momentum, $\Delta p = p-p_0$ is the momentum difference and $\delta = \Delta p/p_0$ is the relative energy (or momentum) deviation. A solution of Eqn~\ref{eq:Hills_off} can be written as

\begin{equation}
    x(s)=\sqrt{\varepsilon_x \beta_x(s)} \cos{(\mu_x (s) + \phi)} + \eta_x(s) \delta,
    \label{eq:position_off}
\end{equation}

where $\eta_x$ is the horizontal dispersion function. In a bending magnet of radius $\rho$, the dispersion expresses the relationship between the energy deviation $\delta$ and a transverse position deviation $\Delta x$ as $\eta_x =\Delta x / \delta$. Replacing $x$ in Eqn~\ref{eq:Hills_off}, $\eta_x$ becomes a solution of

\begin{equation}
    \eta_x''(s) + K(s) \eta_x(s) = \frac{1}{\rho}.
    \label{eq:dispersion}
\end{equation}

Since the differential equations for both position and dispersion have the same form, up to the inhomogeneous factor, both $x$ and $\eta_x$ oscillates as a function of $\beta_x$.
When propagating an off-momentum particle from a positions $s_0$ to a position $s$, Eqn~\ref{eq:rmat_x_cs} becomes

\begin{equation}
    \begin{pmatrix}
        x(s) \\
        x'(s) \\
        \delta(s) \\
    \end{pmatrix}
    =
    \begin{pmatrix}
        C & S & D\\
        C' & S' & D' \\
        0 & 0 & 1 \\
    \end{pmatrix}
    \cdot
    \begin{pmatrix}
        x_0  \\
        x'_0 \\
        \delta_0 \\
    \end{pmatrix}.
    \label{eq:rmat_x_cs_disp}
\end{equation}

From the particular solution of the inhomogeneous Hill's equation, the dispersion element $D$ is defined as

\begin{equation}
    D(s) = S(s)\int_{s_0}^s{\frac{C(\tau)}{\rho(\tau)}d\tau}\ \ - \ \ 
    C(s)\int_{s_0}^s{\frac{S(\tau)}{\rho(\tau)}d\tau} .
    \label{eq:disp_fact}
\end{equation}

Similarly, $D'$ is defined as

\begin{equation}
    D'(s) = S'(s)\int_{s_0}^s{\frac{C(\tau)}{\rho(\tau)}d\tau}\ \ - \ \ 
    C'(s)\int_{s_0}^s{\frac{S(\tau)}{\rho(\tau)}d\tau} .
    \label{eq:disp_p_fact}
\end{equation}

As $\eta_x$ follows the same betatron oscillation as $x$, an equation similar to Eqn~\ref{eq:rmat_x_cs_disp} can be built. It is expressed as

\begin{equation}
    \begin{pmatrix}
        \eta_x(s) \\
        \eta'_x(s) \\
        1
    \end{pmatrix}
    =
    \begin{pmatrix}
        C & S & D\\
        C' & S' & D' \\
        0 & 0 & 1 \\
    \end{pmatrix}
    \cdot
    \begin{pmatrix}
        \eta_{x,0} \\
        \eta'_{x,0} \\
        1
    \end{pmatrix},
    \label{eq:rmat_disp_cs_disp}
\end{equation}

where $\eta_{x,0}$ and $\eta'_{x,0}$ are the position and angle dispersion at an initial location $s_0$. With Eqn~\ref{eq:rmat_disp_cs_disp}, the dispersion function can be propagated between any two positions on the reference path just like the twiss functions.

\subsubsection{Beam sizes and Sigma matrix}

Supposing a Gaussian distribution of the particle beam in the transverse planes, the horizontal beam size $\sigma_x$ is defined as the standard deviation of the position $x$ at a given $s$. Applied to Eqn~\ref{eq:beta_oscillation}, this gives

\begin{equation}
    \sigma_x(s) = \sqrt{\langle x^2(s)\rangle} = 
    \sqrt{\varepsilon_x \beta_x (s)}.
    \label{eq:sigma_x}
\end{equation}

Similarly, the horizontal beam divergence $\sigma_{x'}$ is defined as the standard deviation of the angle $x'$ at a given $s$. Applied to Eqn~\ref{eq:beta_angle_oscillation}, this gives

\begin{equation}
    \sigma_{x'}(s) = \sqrt{\langle x'^2(s)\rangle} = \sqrt{\varepsilon_x \gamma_x (s)}.
    \label{eq:sigmap_x}
\end{equation}

Eqn~\ref{eq:sigma_x} and Eqn~\ref{eq:sigmap_x} are only valid in a lattice section where there is no horizontal dispersion $\eta_x$. In a dispersive section, an additional term dependant on the relative energy spread $\delta$ appears. The horizontal beam size then becomes

\begin{equation}
    \sigma_x(s) = \sqrt{\varepsilon_x \beta_x (s) + \eta_x^2(s)\frac{\sigma_E^2}{E_0^2}},
    \label{eq:sigma_x_disp}
\end{equation}

where $\sigma_E$ is the energy spread of the particle bunch. Similarly, the horizontal beam divergence becomes

\begin{equation}
    \sigma_{x'}(s) = \sqrt{\varepsilon_x \gamma_x (s) + \eta_x'^2(s)\frac{\sigma_E^2}{E_0^2}}.
    \label{eq:sigmap_x_disp}
\end{equation}

In the following studies, $\sigma_x$ and $\sigma_{x'}$ are exclusively calculated using Eqn~\ref{eq:sigma_x_disp} and Eqn~\ref{eq:sigmap_x_disp}. Like with the position and angle, the beam size and divergence can be propagated form an initial position $s_0$ to a position $s$ using a sigma matrix $\boldsymbol{\Sigma}_x$ expressed as

\begin{equation}
   \boldsymbol{\Sigma}_x(s)
   = 
   \begin{pmatrix}
        \sigma_{11} & \sigma_{12} \\
        \sigma_{21} & \sigma_{22} \\
    \end{pmatrix}
   =
    \begin{pmatrix}
        \langle x^2 \rangle & \langle xx' \rangle \\
        \langle x'x \rangle & \langle x'^2 \rangle \\
    \end{pmatrix}
    \stackrel{\eta_x=0}{=}
    \begin{pmatrix}
        \varepsilon\beta & -\varepsilon\alpha \\
        -\varepsilon\alpha & \varepsilon\gamma \\
    \end{pmatrix},
\end{equation}

where $\sigma_{11}, \sigma_{12}, \sigma_{21}$ and $\sigma_{22}$ are the sigma matrix factors. At each $s$, the horizontal beam size is expressed as $\sigma_x = \sqrt{\sigma_{11}}$ and the horizontal beam divergence as $\sigma_{x'} = \sqrt{\sigma_{22}}$. The two other factors, $\sigma_{12}$ and $\sigma_{21}$, are related to the correlation between $x$ and $x'$. This makes $\boldsymbol{\Sigma}_x$ the equivalent of a covariance matrix. In a non-dispersive section they are expressed as $\sigma_{12}=\sigma_{21}=-\varepsilon\alpha(s)$. The matrix $\boldsymbol{\Sigma}_x(s)$ at a location $s$ can be propagated from the matrix $\boldsymbol{\Sigma}_x(s_0)$ at a location $s_0$. First, using Eqn~\ref{eq:emittance}, one can establish the relationship

\begin{equation}
    \boldsymbol{u}_x^T \cdot \boldsymbol{\Sigma}_x
    \cdot \boldsymbol{u}_x=1,
    \label{eq:sigma_inv}
\end{equation}

where $\boldsymbol{u}_x$ is the coordinate vector defined as

\begin{equation}
    \boldsymbol{u}_x(s) =
    \begin{pmatrix}
        x(s) \\
        x'(s) \\
    \end{pmatrix}.
\end{equation}

Since Eqn~\ref{eq:sigma_inv} is verified at every $s$ location, on can write

\begin{equation}
    \boldsymbol{u}_x^T(s) \cdot 
    \boldsymbol{\Sigma}_x(s)
    \cdot \boldsymbol{u}_x(s) =
    \boldsymbol{u}_x(s_0)^T \cdot 
    \boldsymbol{\Sigma}_x(s_0)
    \cdot \boldsymbol{u}_x(s_0).
    \label{eq:sigma_inv_eq}
\end{equation}

With Eqn~\ref{eq:rmat_x} the vector $\boldsymbol{u}_x$ can be propagated between $s_0$ and $s$ as $\boldsymbol{u}_x(s) = \boldsymbol{R}_{s_0 \rightarrow s} \cdot \boldsymbol{u}_x(s_0)$. Replacing $\boldsymbol{u}_x(s_0)$ in Eqn~\ref{eq:sigma_inv_eq} finally gives

\begin{equation}
    \boldsymbol{\Sigma}_x(s) = \boldsymbol{R}_{s_0\rightarrow s}^T \cdot
    \boldsymbol{\Sigma}_x(s_0) \cdot
    \boldsymbol{R}_{s_0\rightarrow s}.
    \label{eq:sigma_propag}
\end{equation}

Using Eqn~\ref{eq:sigma_propag}, the beam size and divergence can be propagated to any position, as long as the initial conditions are known. 

\subsection{Longitudinal dynamics}

The longitudinal beam dynamics are exclusively focused on the particle variations in the $z$ axis and in relative energy deviation $\delta$. It applies of course to accelerating cavities but also in transport lines where the phase space between time $ct=z$ and energy $\delta$ still evolves.

\subsubsection{Accelerating cavities}

An accelerating cavity generates an EM field, usually through radio-frequency waves. The longitudinal electric field component $\mathcal{E}$ is used to increase the charged particle energy. On the reference path, this electric field is expressed as

\begin{equation}
    \mathcal{E}(z,t) = \mathcal{E}_0 \cos(\omega t +kz) = \mathcal{E}_0 \cos(\psi),
\end{equation}

where $\psi=\omega t +kz$ is the electric field phase, $\omega$ is the frequency of the EM wave and $k$ is the wave number. Between an initial position $s_0$ and a position $s$, this cavity increases the particle energy by $\Delta E$ which is expressed as

\begin{equation}
    \Delta E = q\int_{s_0}^{s}\mathcal{E}(z,t)dz,
\end{equation}

where $q$ is the particle charge.

\subsubsection{Propagation of time and energy}

In the longitudinal axis, the particle oscillates through a similar set of equation as in the transverse axes. One can define the longitudinal twiss functions $\beta_z$, $\alpha_z$ and $\gamma_z$ as well as the longitudinal emittance $\varepsilon_z$. In the matrix formalism, the transfer matrix $\boldsymbol{R}_{s_0\rightarrow s}$ can be completed to include both the transverse and longitudinal coordinates. The propagation of the full vector is then expressed as

\begin{equation}
    \begin{pmatrix}
        x \\
        x' \\
        y \\
        y' \\
        ct \\
        \delta \\
    \end{pmatrix} =
    \begin{pmatrix}
        \begin{matrix}
            R_{11} & R_{12}\\
            R_{21} & R_{22} \\
        \end{matrix} & 0 & 
        \begin{matrix}
            0 & D \\
            0 & D'
        \end{matrix}
        \\ 0 &
        \begin{matrix}
            R_{33} & R_{34} \\
            R_{43} & R_{44} \\
        \end{matrix} & 0 \\ 0 & 0 &
        \begin{matrix}
            R_{55} & R_{56} \\
            R_{65} & R_{66} \\
        \end{matrix}
    \end{pmatrix} \cdot
    \begin{pmatrix}
        x_0 \\
        x'_0 \\
        y_0 \\
        y'_0 \\
        ct_0 \\
        \delta_0 \\
    \end{pmatrix},
\end{equation}

where the bottom right block is dedicated to the relations between time and energy. The propagation of energy from $s_0$ to $s$ can be written as

\begin{equation}
    \delta(s) = R_{65}ct(s_0) + R_{66}\delta(s_0).
\end{equation}

A lattice without acceleration, i.e. without changes in the particle energy, implies that $\delta(s)=\delta(s_0)$. This is achieved when there is both $R_{65}=0$ and $R_{66}=1$. Similarly, the propagation of arrival time from $s_0$ to $s$ can be written as

\begin{equation}
    ct(s) = R_{55}ct(s_0)+R_{56}\delta(s_0).
\end{equation}

Assuming the energy of the particle is conserved, this implies that $R_{55}=R_{66}=1$. As such, the variations in time depends only on the relative energy $\delta$ and the $R_{56}$ factor. This last one is often referred to as the momentum compaction factor and is expressed as 

\begin{equation}
    R_{56} = \int_{s_0}^{s}\frac{\eta(\tilde{s})}{\rho(\tilde{s})}d\tilde{s},
\end{equation}

where $\eta$ is the dispersion and $\rho$ the bending radius. Adjusting this $R_{56}$ is crucial in bunch compressors and will be useful for the lattice design in Chapter~\ref{chap:design}.

\section{Simulation software} \label{sec:software}

In our studies, the aim is to design and simulate the properties of the XTD8 lattice. Another goal is to simulate electron-laser interactions happening at LUXE's IP. To perform those simulations, a variety of codes were used as no single code simulates all the features of the ELBEX beam line.

\subsection{Accelerator software}

\subsubsection{MAD-8}

MAD-8~\cite{mad8} is a Fortran based accelerator code designed in CERN. Standing for methodical accelerator design (MAD), this code was mainly used for lattice design. A model needs to define both individual elements (drifts, dipoles, quadrupoles, etc.), a sequence of those elements, and the parameters of an input beam. As a macroparticle generated from this beam is tracked, it will be propagated in the lattice with respect to the transfer matrices $\boldsymbol{R}$ for each element. Similarly, the twiss parameters $\alpha$, $\beta$ and $\gamma$ are propagated by the twiss matrix. From them, useful quantities are derived such as the phase advance $\mu$ or the transverse beam sizes $\sigma_{x}, \sigma_{y}$. MAD-8 allows for magnet strength optimisation and search of closed orbit in a circular accelerator. Those are crucial for the design of a new accelerator line. A more recent version of MAD exists, called MAD-X~\cite{madx}, but MAD-8 is still used in a lot of accelerator facilities and is still a model used in DESY for the entire EuXFEL.

\subsubsection{BDSIM}

Beam delivery simulation~\cite{bdsim} (BDSIM) is a Geant4~\cite{geant4} based accelerator tracking code which includes interactions of particles with material. Like MAD-8 and MAD-X, it allows for optics calculations and particle tracking. However, elements in BDSIM can be defined with specific materials, shapes and beam pipe apertures. One can also set the composition and density of the residual gas. A macroparticle travelling down the line can collides with the different materials and produce secondary particles. In a solid material like a magnet or a dump, the entire particle shower can be simulated. This allows the study of beam losses, energy deposition and various backgrounds. Specific physics lists can be be selected depending on what processes the user wants to observe. 

\begin{figure}[!ht]
        \centering
        \includegraphics[width=\textwidth]{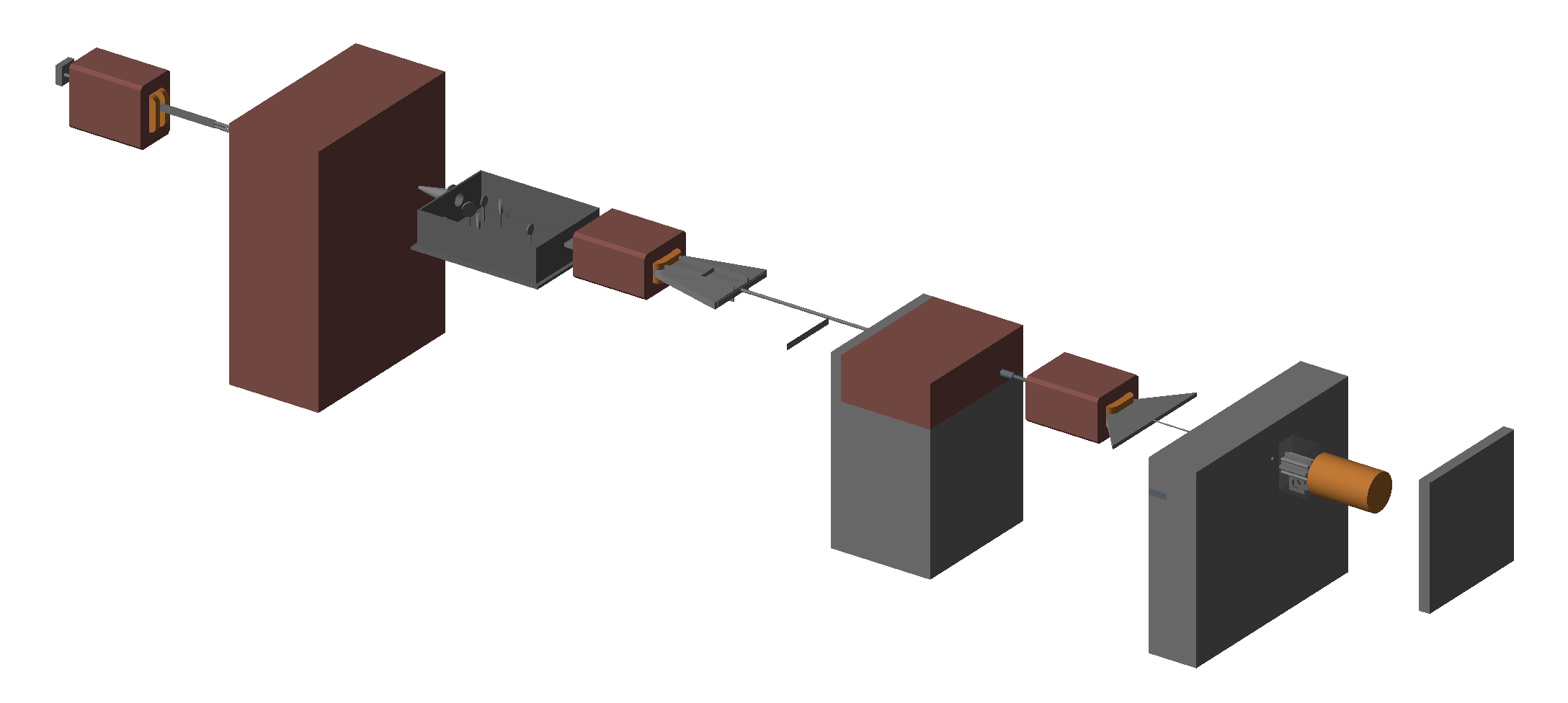}
        \caption{BDSIM visualisation of the LUXE elements loaded through GDML files.}
        \label{fig:bdsim_luxe}
\end{figure}

BDSIM offers a large variety of predefined accelerators elements to cover as many application as possible. From the most basic magnets to the more specific laser wire~\cite{laserwire} and gas capillary~\cite{capillary}. However, externally defined geometry can be loaded as a GDML~\cite{gdml} file instead of a predefined element. In Figure~\ref{fig:bdsim_luxe}, different elements of the LUXE experiment are loaded in BDSIM.

On top of the geometry, EM fields can be defined and attached to specific elements. There is a subsidiary Python package, named  \verb|pybdsim|~\cite{pybdsim}, that can be used alongside BDSIM. It provides a python interface to run BDSIM commands and some analysis tools. In this thesis it is mainly used for both way conversions between BDSIM and multiple other accelerator codes (MAD-8, MAD-X, Xsuite, Ocelot, PTC, etc.).

\subsubsection{Xsuite}

Xsuite~\cite{xsuite} is a Python based accelerator code designed in CERN. It is designed as a modern replacement for the traditional MAD codes. At its core, Xsuite is used for the design of accelerator lattice, optics calculations and particle tracking. It is also used to optimise magnet strengths in a lattice and find the closed orbit. The Python structure allows for a more accessible interface for the users. Around Xsuite, some modules can be imported for various applications : Xtrack for particle tracking, Xfield for beam-beam interaction and collective effects, Xcoll for collimation, etc.
Overall, Xsuite wants to expand the MAD code applications using a modern Python interface.

\subsubsection{Ocelot}

Ocelot~\cite{ocelot} is a Python based accelerator code designed at DESY. Like the previous codes, its main use is for the design of accelerator lines and particle tracking. Similar to Xsuite, the Python interface makes it easy to handle. However, Ocelot have a strong emphasis on free-electron lasers (FELs). It can be used for both accelerator dynamics and FEL physics. Beam matching and orbit correction are also possible with Ocelot. It can simulate the space charge effect, coherent synchrotron radiation (CSR) and wakefields.

\subsubsection{Accelerator codes comparison}

To complete the studies of the following chapters, multiple of those codes are used. Some new conversions from a model to another had to be written in the \verb|pybdsim| module. In Table~\ref{tab:codes_comparison} is a comparison of the capabilities of the different accelerator codes presented.

\begin{table}[!ht]
\centering
\begin{tabular}{|c || c c c c c |}
\hline
& MAD-8 & MAD-X & BDSIM & Xsuite & Ocelot \\
\hline
\hline
Language & Fortran & C \& Fortran & C++ & Python & Python \\
\hline
Linear optics & \checkmark & \checkmark & \checkmark & \checkmark & \checkmark \\
\hline
Nonlinear optics & Limited & \checkmark & \checkmark & \checkmark & \checkmark \\
\hline
Orbit correction & Limited & \checkmark & Limited & \checkmark & \checkmark \\
\hline
Matching & \checkmark & \checkmark & Limited & \checkmark & \checkmark \\
\hline
Optics optimisation & \checkmark & \checkmark & Limited & \checkmark & \checkmark \\
\hline
Coupling analysis & Limited & \checkmark & Limited & \checkmark & \checkmark \\
\hline
Single-particle tracking & \checkmark & \checkmark & \checkmark & \checkmark & \checkmark \\
\hline 
Multi-particle tracking & Limited & \checkmark & \checkmark & \checkmark & \checkmark \\
\hline
Space charge & $\times$ & Limited & Limited & \checkmark & \checkmark \\
\hline
Beam-beam & $\times$ & Limited & $\times$ & \checkmark & Limited \\
\hline
Wakefields & $\times$ & Limited & $\times$ & \checkmark & \checkmark \\
\hline
Particle–matter interactions & $\times$ & $\times$ & \checkmark & Limited\footnotemark{} & $\times$ \\
\hline
Free electron lasers & $\times$ & Limited & Limited & Limited & \checkmark \\
\hline
\end{tabular}
\caption{Comparison of the capabilities of the accelerator codes.}
\label{tab:codes_comparison}
\end{table}

\footnotetext{Xsuite has an interface with BDSIM and FLUKA as well as its own internal scattering routines}

\subsection{Ptarmigan}

Ptarmigan~\cite{ptarmigan} is a Monte-Carlo simulation code designed for the study of strong field QED processes. It simulates the collision of relativistic electron or photon bunches with a laser pulse. Fundamentally, Ptarmigan includes two main processes, non-linear Compton scattering and non-linear Breit-Wheeler pair creation. Individual particles can be tracked along the collision.

The QED model of Ptarmigan uses one of two approximations, the locally monochromatic approximation~\cite{ptarmigan} (LMA) or the locally constant crossed field approximation~\cite{ptarmigan} (LCFA). In the LMA, the laser pulse is divided into slices of monochromatic plane waves. Each slice has its own frequency and amplitude. An electron travelling in the laser pulse will experience different successive field states. This approximation is more suited for the transition regime between the perturbative and non-perturbative model ($a_0\sim1$). In the LCFA the laser pulse slices are defined with a constant cross field. The formation length of the QED processes is assumed to be much shorter than the laser wavelength. This approximation is valid past the transition regime, when the QED interaction are clearly non-perturbative ($a_0 \gg 1$).

The Ptarmigan simulations are controlled by a input file in the YAML format~\cite{yaml}, which is a superset of JSON. In this input file the user defines both the electron bunch properties, the laser pulse properties and various other options like the approximation used between LMA and LCFA. The output is either a text file or an HDF5 file~\cite{hdf5}. It contains for each particle type (electron, positron and photon) the position four-vector and the momentum four-vector after the collision. Each particle also has its associated weight to both account for pair-creation biasing and scale the total number of particles to the input electron bunch charge. It is also possible to store the particle trajectories. In Chapter~\ref{chap:impact_luxe} and Chapter~\ref{chap:clara}, Ptarmigan is used to reconstruct the energy spectrum for each particle type after the collision and to evaluate the number of particle produced by said collision.

\chapter{Measurements at the European XFEL} \label{chap:measurments}

\section{Beam parameters jitter}

To operate properly, the properties of the electron beam (e.g. position, size, energy, etc.) have to stay within the requirements of the different experiments. As such, it is crucial to know precisely the bunches properties as they arrive at ELBEX.
The EuXFEL bunches are created with a certain degree of error or variation in their properties. Between two consecutive bunches in a train or two consecutive trains, the properties of said bunches varies. This is referred as the beam jitter $\sigma_J$.
If the values of jitter are too high, the signal in the experiments could be impacted, e.g. if the position jitter is very large at LUXE's IP, some bunches are too far to interact with the laser. In this chapter, the aim is to evaluate the stability of the bunches produced in the EuXFEL. This is done by analysing actual measurements of the electron bunches in the main switch yard. 

\begin{table}[!ht]
\centering
\begin{tabular}{| c | c | c |}
\hline
Symbol & Unit & Central value \\
\hline
\hline
$x$ & \SI{}{\micro\meter} & 0 \\
\hline
$y$ & \SI{}{\micro\meter} & 0 \\
\hline
$\sigma_x$ & \SI{}{\micro\meter} & 30 \\
\hline
$\sigma_y$ & \SI{}{\micro\meter} & 30 \\
\hline
$E$ & GeV & 16 \\
\hline
$q$ & pC & 250 \\
\hline
$\sigma_t$ & fs & 40 \\
\hline
$t$ & fs & 0 \\
\hline
\end{tabular}
\caption{Reference values of the parameters considered for the jitter.}
\label{tab:euxfel_param_ref} 
\end{table}

In Table~\ref{tab:euxfel_param_ref} is a list of the parameters considered in the next sections for the jitter analysis. In this table $x, y$ are the transverse position of the bunch centroid, $\sigma_x, \sigma_y$ are the transverse sizes, $E$ is the energy, $q$ is the bunch charge, $\sigma_t$ is the bunch length and $t$ is the arrival time relative to a synchronous particle. 
The reference bunch considered is a \SI{16}{\giga\electronvolt} bunch of charge \SI{250}{\pico\coulomb} with a transverse beam size of \SI{30}{\micro\meter} and a longitudinal bunch length of \SI{40}{\femto\second} at the end of the linear accelerator.

\section{Diagnostics of the EuXFEL main switch yard}

\begin{figure}[!ht]
    \centering
    \includegraphics[width=12cm]{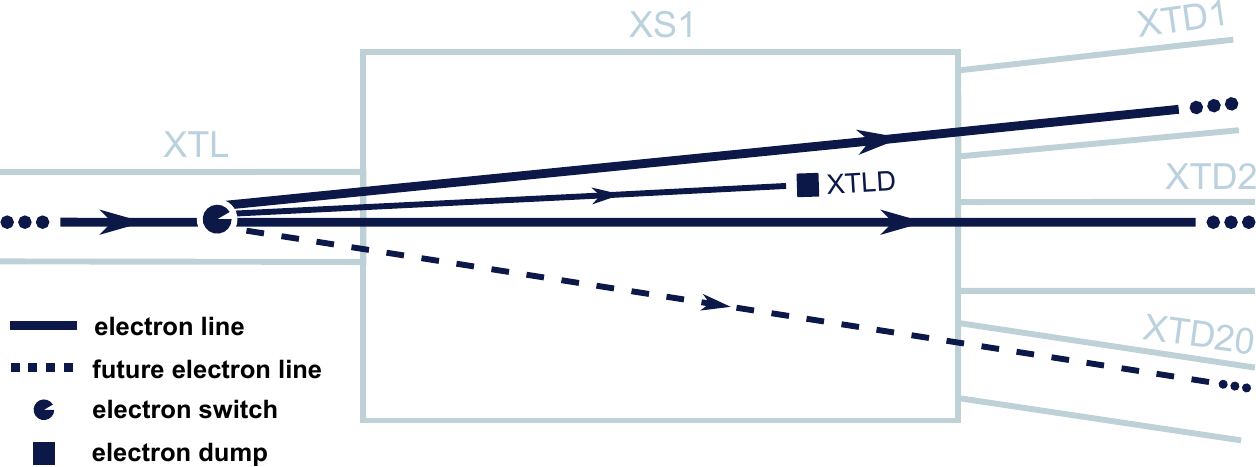}
    \caption{Schematics of the EuXFEL main switch yard around the XS1 hall.}
    \label{fig:euxfel_switchyard}
\end{figure}

The EuXFEL main switch yard distributes the electrons bunches from the transfer line (XTL) to the two main lines (XTD1 and XTD2) as shown in Figure~\ref{fig:euxfel_switchyard}.

\begin{figure}[!ht]
    \centering
    \includegraphics[width=\textwidth]{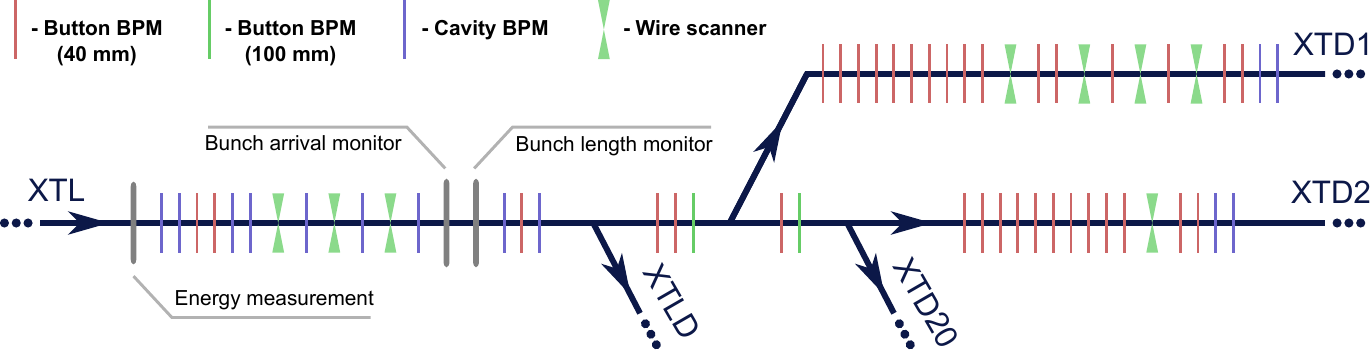}
    \caption{Plan layout of the detectors around the EuXFEL switch yard.}
    \label{fig:euxfel_switchyard_detectors}
\end{figure}

Some of the bunches are sent into the transfer line dump (XTLD) and in the future, a third main line will transport electrons in the XTD20 tunnel. Around this switch yard, different measurement devices are installed as shown in Figure~\ref{fig:euxfel_switchyard_detectors}.

The majority of the diagnostics are beam position monitors (BPMs). There are \SI{15}{} of them in the XTL, \SI{18}{} in the XTD1 and \SI{16}{} in the XTD2. But there are also 8 wire scanners, one bunch arrival monitor, one bunch length monitor and a section for energy measurements at the entrance of the XTL line.

\subsection{Beam position monitors}

A BPM~\cite{bpm} measures the transverse position of a charged particle bunch. In the EuXFEL, there are two different types of BPM used. The more frequently used one is the button BPM. For a more precise measurement of the transverse position, cavity BPMs are also used. Both of those devices can also measure the charge when a bunch passes through.

\subsubsection{Button BPMs}

\begin{figure}[!ht]
    \centering
    \includegraphics[width=6cm]{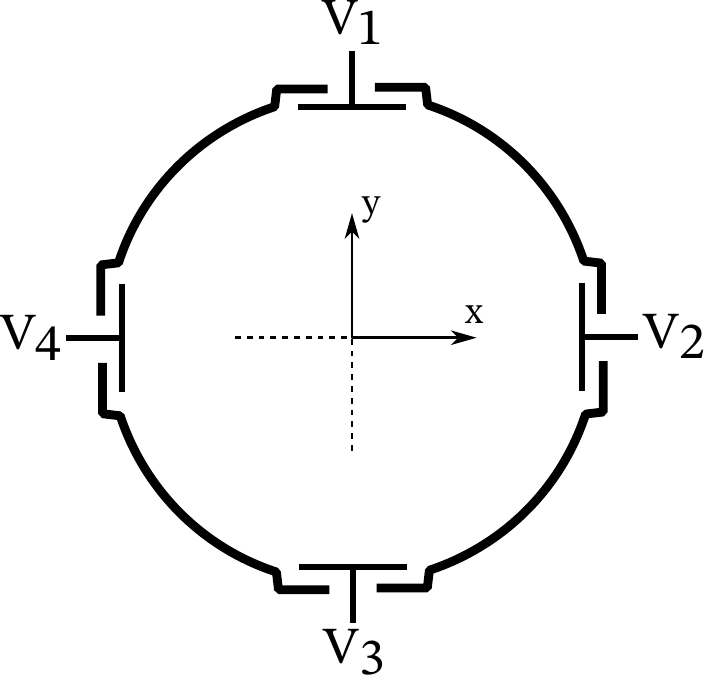}
    \caption{Schematic of the transverse profile of a button BPM. Four electrodes are orthogonally placed along the $x$ and $y$ axes. Each electrode, when a bunch is passing, produce a voltage $V_i$.}
    \label{fig:button_bpm}
\end{figure}

The button BPM~\cite{button_bpm} is a broadly used device in accelerators. Fundamentally, a charged particle bunch creates around itself an EM field. As the bunch travels in the beam pipe, image charges are created on the beam pipe. An electrode placed on any side of the beam pipe can intercept the image charges and a voltage is measured. In a typical BPM, four electrodes are arranged inside the beam pipe at every $90^{\circ}$ offsets as shown in Figure~\ref{fig:button_bpm}.


Between the two opposite electrodes on the $x$ axis, the horizontal position is expressed as

\begin{equation}
    x = k_x\frac{V_2-V_4}{V_2+V_4},
\end{equation}

where $V_2, V_4$ are the signals measured at the respective electrodes and $k_x$ is a calibration constant. Similarly, the vertical position is expressed as

\begin{equation}
    y = k_y\frac{V_1-V_3}{V_1+V_3},
\end{equation}

where $V_1, V_3$ are the signals measured at the respective electrodes and $k_y$ is another calibration constant. At the same time, the value $V=V_1+V_2+V_3+V_4$ indicates the total intensity of the signal and is used to calculate the bunch charge.


\subsubsection{Cavity BPMs}

\begin{figure}[!ht]
    \centering
    \includegraphics[width=6cm]{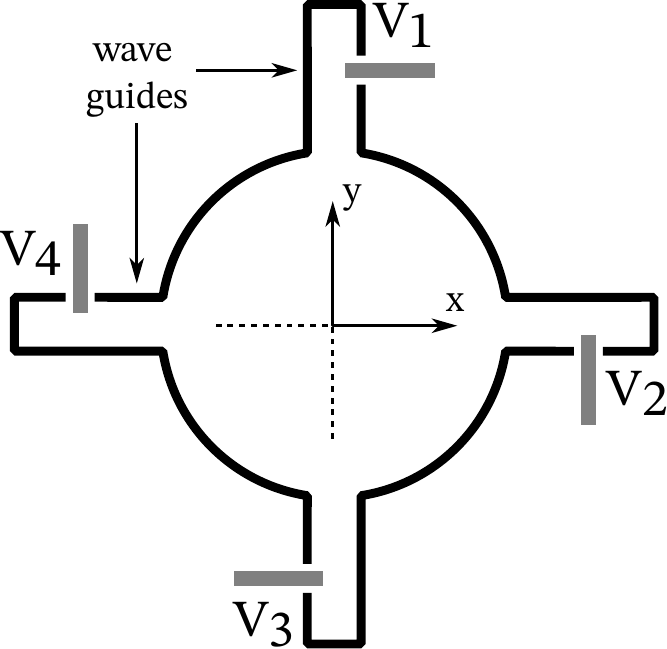}
    \caption{Schematic of the transverse profile of a cavity BPM. Four waveguides with antennas are orthogonally placed along the $x$ and $y$ axes. Each antenna, when a bunch is passing, produce a voltage $V_i$.}
    \label{fig:cavity_bpm}
\end{figure}

\begin{figure}[!ht]
    \centering
    \includegraphics[width=\textwidth]{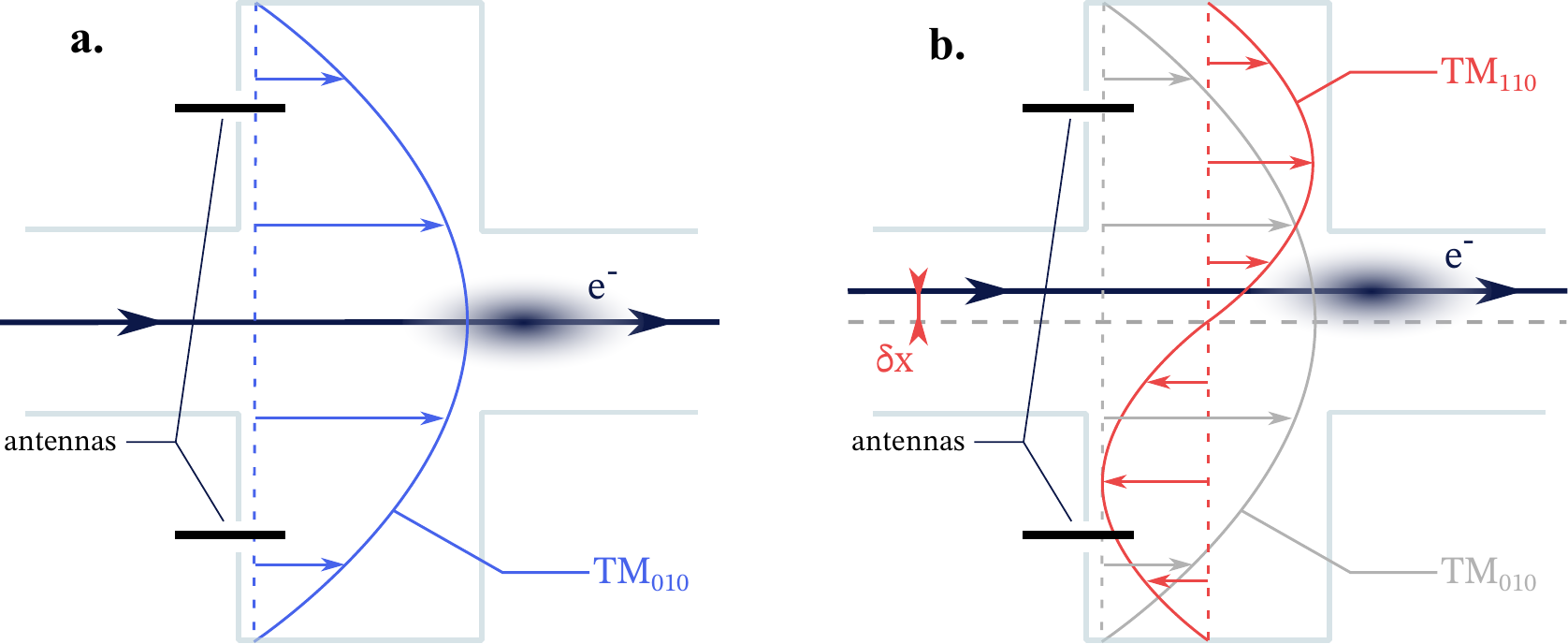}
    \caption{Schematics of the longitudinal profile of a cavity BPM. a) The electron bunch is perfectly aligned, only exiting the monopole mode $\rm TM_{010}$. b) The electron bunch arrives with a position offset $\delta x$, the dipole mode $\rm TM_{110}$ is also exited.}
    \label{fig:cavity_fields}
\end{figure}

In a cavity BPM~\cite{cavity_bpm}, the bunch traverse a chamber similar to a radio-frequency (RF) cavity. A cavity BPM transverse profile and longitudinal profile are shown in Figure~\ref{fig:cavity_bpm} and Figure~\ref{fig:cavity_fields} respectively.

As it travels inside the cavity, the bunch excites EM modes. The amplitudes of those EM modes are measured by the antennas. More specifically, the modes exited are the transverse magnetic (TM) modes~\cite{tm_modes}. It implies that the magnetic field only has transverse components and that the electric field is only longitudinal. Those modes are usually written as ${\rm TM}_{ijk}$ where $i$ is the azimuthal number , $j$ is the radial number and $k$ is the longitudinal number. The two most important modes in a cavity BPM are shown in Figure~\ref{fig:cavity_fields}. 

When perfectly in the centre of the BPM, the bunch only excites the monopole mode $\rm TM_{010}$. For this mode, $i=k=0$ so the electric field amplitude only has a radial component. The corresponding signal $V_{010}$ is proportional to the charge $q$ as

\begin{equation}
    V_{010} \propto q.
\end{equation}

With a horizontal offset $\delta x$, the bunch also excites the dipole mode $\rm TM_{110}$. In this mode, $i=j=1$, which mean that the electric field amplitude varies for both radial and azimuthal coordinates. For small values of $\delta_x$, the signal $V_{101}$ is proportional to the offset and the charge $q$ as

\begin{equation}
    V_{101} \propto q\delta x.
\end{equation}

Using the signal from both modes, the offset $\delta x$ is proportional to their ratio as

\begin{equation}
    \delta x \propto \frac{V_{010}}{V_{110}}.
\end{equation}


A typical cavity BPM is composed of two successive RF cavities like those described in Figure~\ref{fig:cavity_bpm} and Figure~\ref{fig:cavity_fields}. One to measure the monopole mode, called a ``reference" cavity and one to measure the dipole mode called the ``dipole" cavity. 

\subsubsection{BPM resolution}

For both BPM types, the variation of position $\delta x$ is expressed as a ratio of two voltages, one related to the position and the charge ($V_p$) and one only related to the charge ($V_q$). One can write

\begin{equation}
    \delta x = \frac{V_p}{V_q}.
\end{equation}

Assuming the same error on both voltages, the intrinsic error on the position measurement $\sigma_{\delta x, {\rm intrinsic}}$ is calculated through error propagation as

\begin{equation}
    \sigma_{\delta x, {\rm intrinsic}}^2=
    \left(\frac{\partial x}{\partial V_p}\right)^2
    \sigma_{V}^2 + 
    \left(\frac{\partial x}{\partial V_q}\right)^2
    \sigma_{V}^2,
    \label{eq:intr_error_bpm}
\end{equation}

with $\sigma_V$ the error on each signal measurement. Calculating the partial derivatives and inserting them into Eqn~\ref{eq:intr_error_bpm}, $\sigma_{\delta x, {\rm intrinsic}}$ is then expressed as

\begin{equation}
    \sigma_{\delta x, \rm intrinsic}^2=\left(\frac{V_q-V_V}{V_q^2}\right)\sigma_V^2.
    \label{eq:bpm_intrins_error}
\end{equation}

\begin{figure}[!ht]
    \centering
    \includegraphics[width=\textwidth]{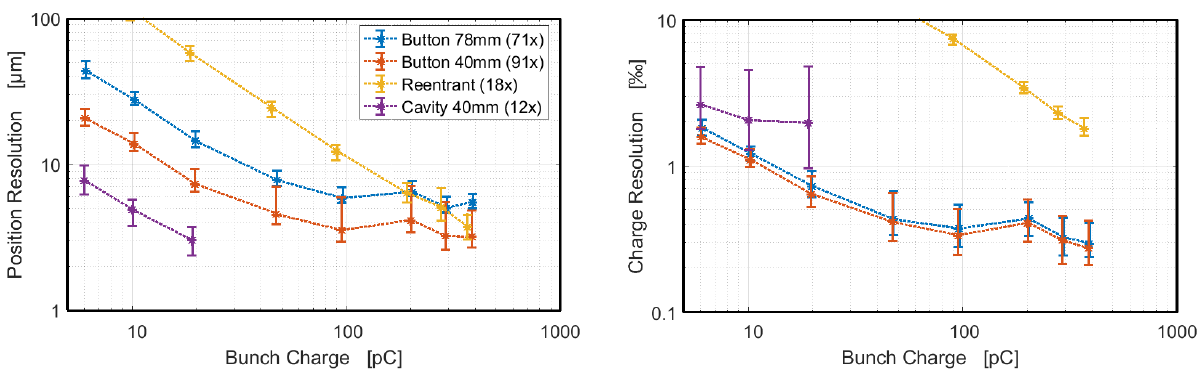}
    \caption{Resolution of the BPMs in the EuXFEL. On the left is the position resolution as a function of bunch charge. On the right is the charge resolution as a function of the bunch charge~\cite{bpm_resolution}.}
    \label{fig:bpm_resolution}
\end{figure}

The factor $\frac{V_q-V_p}{V_q^2}$ is proportional to $\frac{1}{q}$. This means the BPM resolution is better at high bunch charges. The total error on the BPM measurement is then given by

\begin{equation}
    \sigma_{\delta x}^2 = 
    \sigma_{\delta x, {\rm intrinsic}}^2 +
    \sigma_{\delta x, {\rm elec}}^2,
\end{equation}

where $\sigma_{\delta x, {\rm elec}}$ is the error produced by the electric circuit measuring the voltage. When the charge $q$ of the beam increases, $\sigma_{\delta x, {\rm intrinsic}}$ follow a $\frac{1}{q}$ dependence. At high charges, only the electrical component $\sigma_{\delta x, {\rm elec}}$ remains. This behaviour can be seen in Figure~\ref{fig:bpm_resolution} where resolution measurements for EuXFEL BPMs~\cite{bpm_resolution} were made using a virtual measurement method which is explained in Section~\ref{sec:analysis}. In addition to the button BPMs and cavity BPMs, measurements were made for re-entrant BPMs~\cite{reentrant_bpm}. These are special cavity BPMs, easier to clean and which can be located near the superconducting structures. 

\begin{figure}[!ht]
    \centering
    \includegraphics[width=8cm]{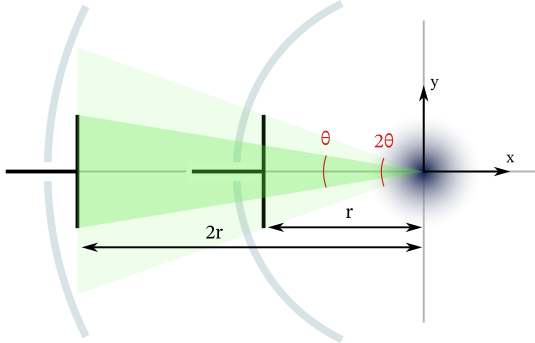}
    \caption{Schematics of a BPM electrode placed at different values of radius $r$.}
    \label{fig:bpm_radius}
\end{figure}

At \SI{250}{\pico\coulomb}, measurements of individual button BPMs have shown a resolution of around \SI{3}{\micro\meter}-\SI{5}{\micro\meter}. For cavity BPMs, the resolution is down to \SI{0.3}{\micro\meter} for as low as \SI{20}{\pico\coulomb}. In the EuXFEL, button BPMs of different apertures are used. At ultra-relativistic speed, the electric field of a charged particle is almost entirely radial. As shown in Figure~\ref{fig:bpm_radius}, an electrode twice as far from the centre of the beam will receive half the amount of electric field. The voltage of an electrode is then proportional to the inverse of the pipe radius. This can be expressed as

\begin{equation}
    V \propto \frac{1}{r}.
\end{equation}

The intrinsic error defined in Eqn~\ref{eq:bpm_intrins_error} then becomes

\begin{equation}
    \sigma_{\delta x, \rm intrinsic}^2 \propto \left(\frac{1}{\frac{1}{r^2}}\right)\sigma_V^2 = r^2\sigma_V^2.
    \label{eq:linear_proportional_error_bpm}
\end{equation}

So the intrinsic error on the position $\sigma_{\delta x, \rm intrinsic}$ is linearly proportional to $r$.

\subsection{Wire scanners}

\begin{figure}[!ht]
    \centering
    \includegraphics[width=12cm]{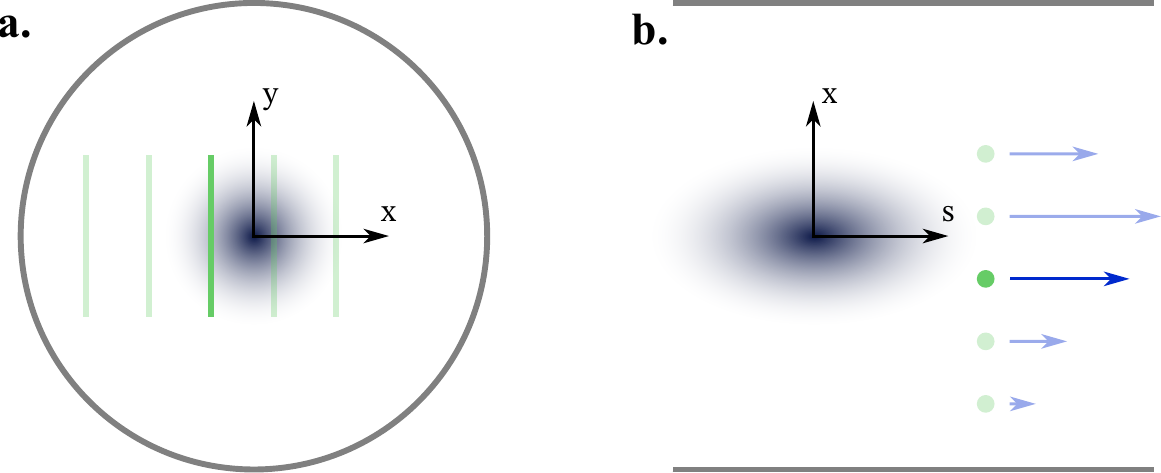}
    \caption{Schematic of a wire scanner (green line or green circles) passing through the bunch (shaded dark blue ellipse). a.) Transverse profile of the beam pipe with different positions of the wire. b.) Longitudinal profile of the beam pipe with different positions of the wire. Blue arrows indicates the direction and intensity of the signal for each position of the wire.}
    \label{fig:wire_scanner}
\end{figure}

A wire scanner~\cite{wire} is a device used to measure the transverse beam sizes $\sigma_{x, y}$ and, by extension, the transverse emittances $\varepsilon_{x, y}$. The method consist on driving progressively a thin wire (usually made of tungsten) into the beam. At each position of the wire, a bunch is fired and interacts with the wire, mainly through scattering processes and bremsstrahlung. The scattered particles are then detected downstream by beam loss monitors (BLM)~\cite{blm}. The number of detected particles is related to the distance between the wire and the centre of the bunch. By taking multiple measurements, the transverse profile of the bunch can be obtained along the wire's axis of motion, as shown in Figure~\ref{fig:wire_scanner}. 

\begin{figure}[!ht]
    \centering
    \includegraphics[width=10cm]{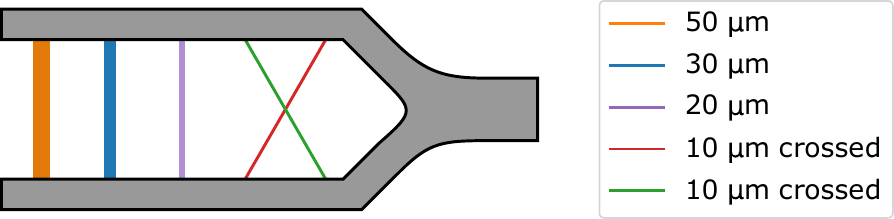}
    \caption{Layout (as seen from the beam direction) of the five wires on the mount with the corresponding thicknesses. The legend indicates the wire diameters.}
    \label{fig:wire_scanner_mount}
\end{figure}

This is a relatively slow process that uses multiple bunches to be completed. Using this method, it is not possible to get the beam size of each of the bunches in a train.
The EuXFEL wire scanner stations are composed of two arms with a \SI{90}{\degree} offset to measure the beam size in both transverse axes~\cite{wire_euxfel}. Each arm carries a set of five wires with varying thicknesses and angles as shown in Figure~\ref{fig:wire_scanner_mount}. 

A wire with a circular cross-section inserted along the $x$ plane have a thickness along $z$ expressed as 

\begin{equation}
    dz = 2\sqrt{R^2-x^2},
\end{equation}

where $dz$ is the wire thickness and $R$ the wire radius. The normalised distribution function $w(x)$ can then be expressed as

\begin{equation}
    w(x) = \frac{1}{\pi R^2}dz(x) = \frac{2}{\pi R^2}\sqrt{R^2-x^2}.
\end{equation}

The distribution of produced particles is expressed as the convolution between this wire thickness and the beam distribution which is assumed to be Gaussian. The goal is then to find the standard deviation of the wire distribution $\sigma_{\rm wire}$. It is expressed as 

\begin{equation}
    \sigma_{\rm wire}^2 = \langle w(x)^2 \rangle - \langle w(x) \rangle^2.
    \label{eq:std_wire}
\end{equation}

Assuming the wire is centred on $x=0$, its symmetric distribution implies that the mean $\langle w(x) \rangle$ is equal to 0. As such, $\sigma_{\rm wire}$ is only expressed with the second moment of the distribution $dz$ and Eqn~\ref{eq:std_wire} becomes 

\begin{equation}
    \sigma_{\rm wire}^2 = \langle w(x)^2 \rangle = \frac{2}{\pi R^2} \int^{+R}_{-R} \sqrt{R^2-x^2}x^2dx.
\end{equation}

Using the substitution $x=R\sin(\theta)$, one can solve this integral and prove that $\sigma_{\rm wire}^2=\frac{R^2}{4}$. At the measurement screen, the distribution of the signal received has a standard deviation $\sigma_{\rm meas}$ expressed by

\begin{equation}
    \sigma_{\rm meas}^2 = \sigma_{\rm beam}^2+\sigma_{\rm wire}^2,
\end{equation}

where $\sigma_{\rm beam}$ is the beam size. If $\sigma_{\rm beam} \ll \sigma_{\rm wire}$, the signal distribution will be dominated by the wire and it will be difficult to measure the beam size. Reducing the wire radius $R$ also reduces $\sigma_{\rm wire}$ and allows for a better resolution on the beam size measurement. It can be estimated that the smallest $\sigma_{\rm beam}$ measurable is of the order of $\sigma_{\rm wire}$. Figure~\ref{fig:wire_scanner_mount} showed that in the EuXFEL wire scanner mounts, the smallest wires have a diameter of \SI{10}{\micro\meter}. For such a wire, $\sigma_{\rm wire}=\frac{5\ {\rm \mu m}}{2}=2.5\ {\rm \mu m}$, so it could measure beam size as small as \SI{2.5}{\micro\meter}. However, a smaller wire produces less particles so the signal received at the screen has a lower intensity.
By combining the results of all ten wires, the transverse emittance and transverse beam size can be obtained in both planes.


\subsection{Bunch arrival monitor}

\begin{figure}[!ht]
    \centering
    \includegraphics[width=\textwidth]{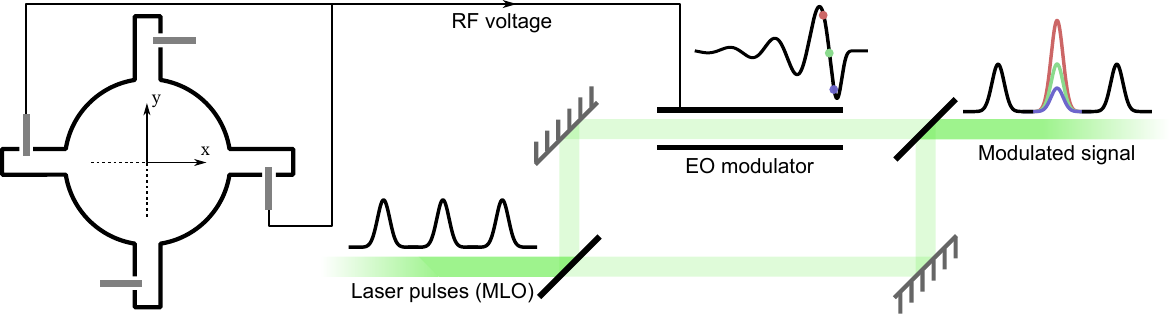}
    \caption{Schematic of the bunch arrival monitor. On the left is the RF cavity measuring the electron bunch signal. On the right is the EOM where laser pulses are modulated with respect to the RF signal timing. The dots on the RF signal indicates the timing of the laser pulse in different configuration. The green dot and curve correspond to and electron bunch in time. In red an early bunch and in blue a late bunch.}
    \label{fig:bam}
\end{figure}

A bunch arrival time monitor (BAM)~\cite{bam} measures the arrival time $t$ of a bunch with respect to the central clock of the accelerator. Similar to a cavity BPM, the device is composed of an RF cavity in which a passing bunch generates EM fields. Four antennas are orthogonally placed around the cavity to measure those fields. Signals from opposite antennas are combined to reduce the transverse position dependency. This combined signal, as well as a laser pulse from the master laser oscillator~\cite{mlo} (MLO), are sent to an electo-optic modulator~\cite{eom} (EOM). A schematic of the BAM is shown in Figure~\ref{fig:bam}.

As a laser pulse arrives at the EOM, if there is no RF signal or if the signal is synchronised with the laser pulse, the amplitude of the laser pulse is unchanged. If the RF signal is not synchronised, i.e. if it arrive early or late, this will impact the laser pulse amplitude.
In the RF cavity, both pair of antennas are calibrated for different bunch charges. One combined signal is used to measure low bunch charges, in the orders of \SI{10}{\pico\coulomb}. This channel has a theoretical resolution on the arrival time between \SI{5}{\femto\second} and \SI{10}{\femto\second} . The other combined signal is used to measure high bunch charges, in the orders of \SI{100}{\pico\coulomb}. This channel has a theoretical resolution between \SI{10}{\femto\second} and \SI{20}{\femto\second}. Measurements on multiple BAM in the EuXFEL~\cite{bam_meas} have shown values of noise below \SI{5}{\femto\second} at each BAM.

\subsection{Bunch length monitor}

\begin{figure}[!ht]
    \centering
    \includegraphics[width=10cm]{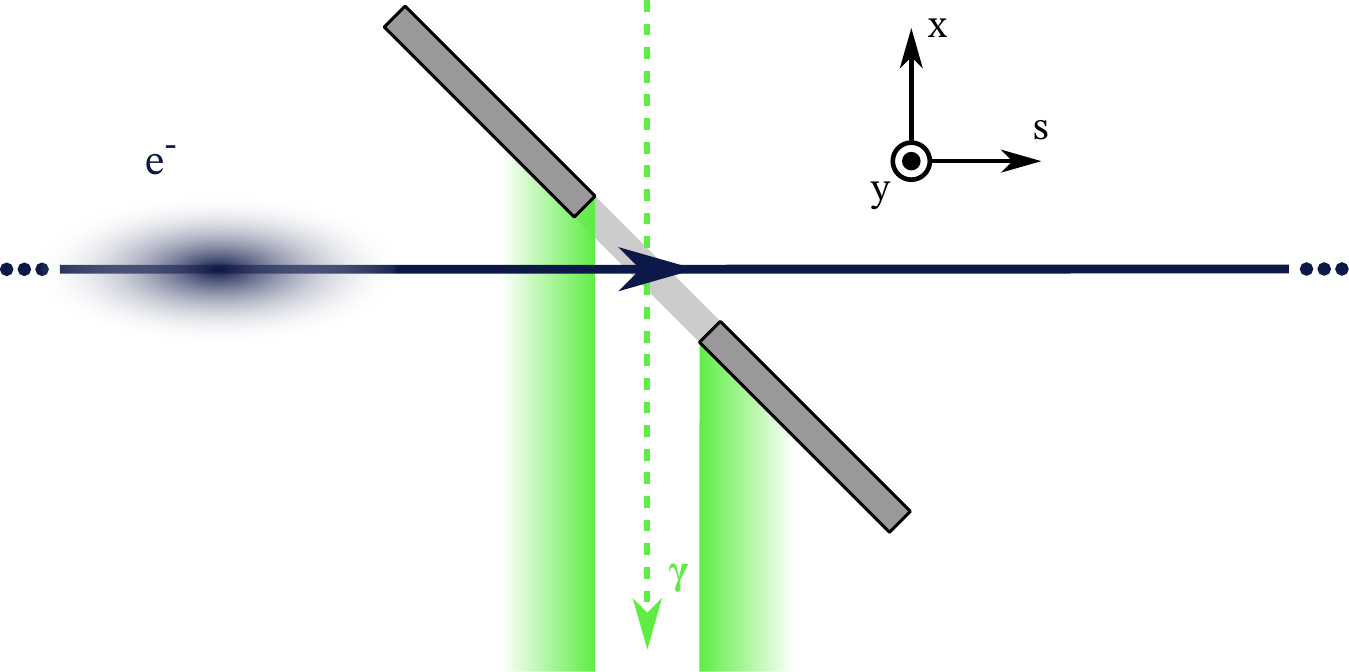}
    \caption{Plan schematic of the coherent diffraction radiation (CoDR) at the aluminium screen. The screen is placed at a \SI{45}{\degree} to have a perpendicular CoDR. The shaded green region represents the halo photon beam produced by CoDR}
    \label{fig:diffraction_rad}
\end{figure}

In the EuXFEL, the bunch length $\sigma_t$ is measured through a coherent radiation intensity spectrometer~\cite{crisp} (CRISP). This device aims to measure the longitudinal profile of the beam from which the bunch length can be computed.
At the detector, the electron bunch passes near a dielectric material to produce coherent diffraction radiation~\cite{cdr} (CoDR). In the switch yard, this material is an aluminium screen of \SI{1}{\milli\meter} thickness. On this screen the electron beam passes through one of two holes (\SI{5}{\milli\meter} or \SI{7}{\milli\meter}).
A schematic of the CoDR at the screen can be seen in Figure~\ref{fig:diffraction_rad}. 

Considering $\frac{dU_1}{df}$ the intensity per unit frequency emitted by a single electron, then for a bunch of $N_e$ electrons the intensity is expressed as

\begin{equation}
    \frac{dU}{df} = \frac{dU_1}{df}N_e^2|F(f)|^2,
\end{equation}

where $F(f)$ is both the longitudinal form factor of the bunch and the Fourier transform of the normalised bunch charge density $\rho(t)$. $F(f)$ is epxressed as

\begin{equation}
    F(f) = \int \rho(t)\exp(-2\pi ift)dt.
\end{equation}

The photons radiated by the screen are then transported to the spectrometer which then reconstruct the charge profile $\rho(t)$ through an inverse Fourier transform. From this profile the bunch length can be calculated. In an ideal Gaussian profile it is just the standard deviation of the distribution. More often the distribution will be non-Gaussian and the bunch length is calculated with a \SI{68}{\%} cumulative sum.
This CRISP method have the advantage of being non-destructive for the electron bunch. In theory this measurement can be applied to all buncesh in a single train. However, at MHz repetition rates, the CoDR induces mechanical oscillations and heat which causes a distortion of the subsequent bunches measurements. At the EuXFEL the detector is limited to the measurement of a specific bunch ID in the train. In the end the bunch length can be evaluated with a resolution of the order of \SI{8}{\femto\second}.


\subsection{Energy measurement}

\begin{figure}[!ht]
    \centering
    \includegraphics[width=\textwidth]{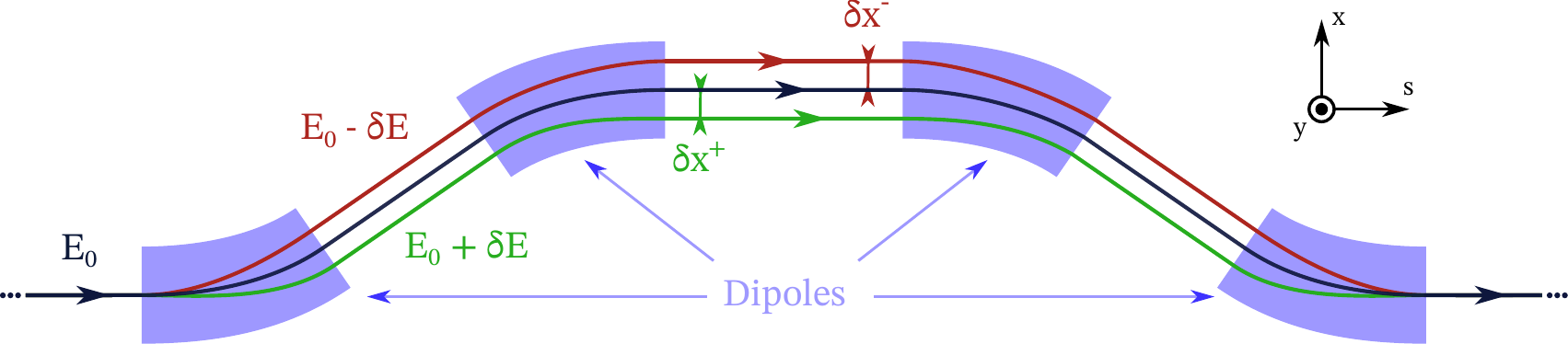}
    \caption{Plan diagram of a four bend energy spectrometer. Three different energy particles are shown, nominal $E_0$ in dark blue, lower $E_0-\delta E$ in red and higher $E_0+\delta E$ in green.}
    \label{fig:ebpm}
\end{figure}

In the transfer line of the EuXFEL, the energy is measured using energy beam position monitors (EBPMs)~\cite{ebpm}. Those EBPMs are placed in a dispersive section of the lattice, i.e. where a variation in energy $\delta E$ will be converted to a transverse deviation $\delta x$ as shown in Figure~\ref{fig:ebpm}.

The relative variation of energy can be expressed as

\begin{equation}
    \frac{\delta E}{E_0} = \frac{\delta x}{\eta_x},
\end{equation}

where $E_0$ is the reference energy and $\eta_x$ is the horizontal dispersion. The error on the energy measurement is then

\begin{equation}
    \sigma_{\delta E}^2 = 
    \left(\frac{\partial \delta E}{\partial \delta x}\right)^2
    \sigma_{\delta x}^2 + 
    \left(\frac{\partial \delta E}{\partial \eta_x}\right)^2
    \sigma_{\eta_x}^2.
\end{equation}

Assuming the dispersion $\eta_x$ is measured with a position measurement in the dispersive section, the error on $\eta_x$ is equivalent to the error on $\delta x$ ($\sigma_{\delta x}\sim\sigma_{\eta_x}$). The error on energy then becomes

\begin{equation}
    \sigma_{\delta E}^2 = E_0^2\left(\frac{\eta_x-\delta x}{\eta_x^2}\right)\sigma_{\delta x}^2,
\end{equation}

and the relative energy error can be expressed as

\begin{equation}
    \frac{\sigma_{\delta E}}{E_0} = \sqrt{\frac{\eta_x-\delta x}{\eta_x^2}}\sigma_{\delta x}.
\end{equation}

At a typical value of dispersion $\eta_x=400\ \rm mm$~\cite{ebpm} the relative error is approximately \SI{1.58e-3}{\%}.


\section{Analysis of the measurements} \label{sec:analysis}

Measurements from all the devices listed above were taken from the EuXFEL control room. For all the parameters listed in Table~\ref{tab:euxfel_param_ref} the associated jitter $\sigma_J$ must be calculated.

\subsection{Transverse position}

Using the 49 BPMs shown in Figure~\ref{fig:euxfel_switchyard_detectors}, data was took for over 3000 trains each containing up to 2700 bunches. 

\subsubsection{Single BPM measurements}

\begin{figure}[!ht]
    \centering
    \includegraphics[width=\textwidth]{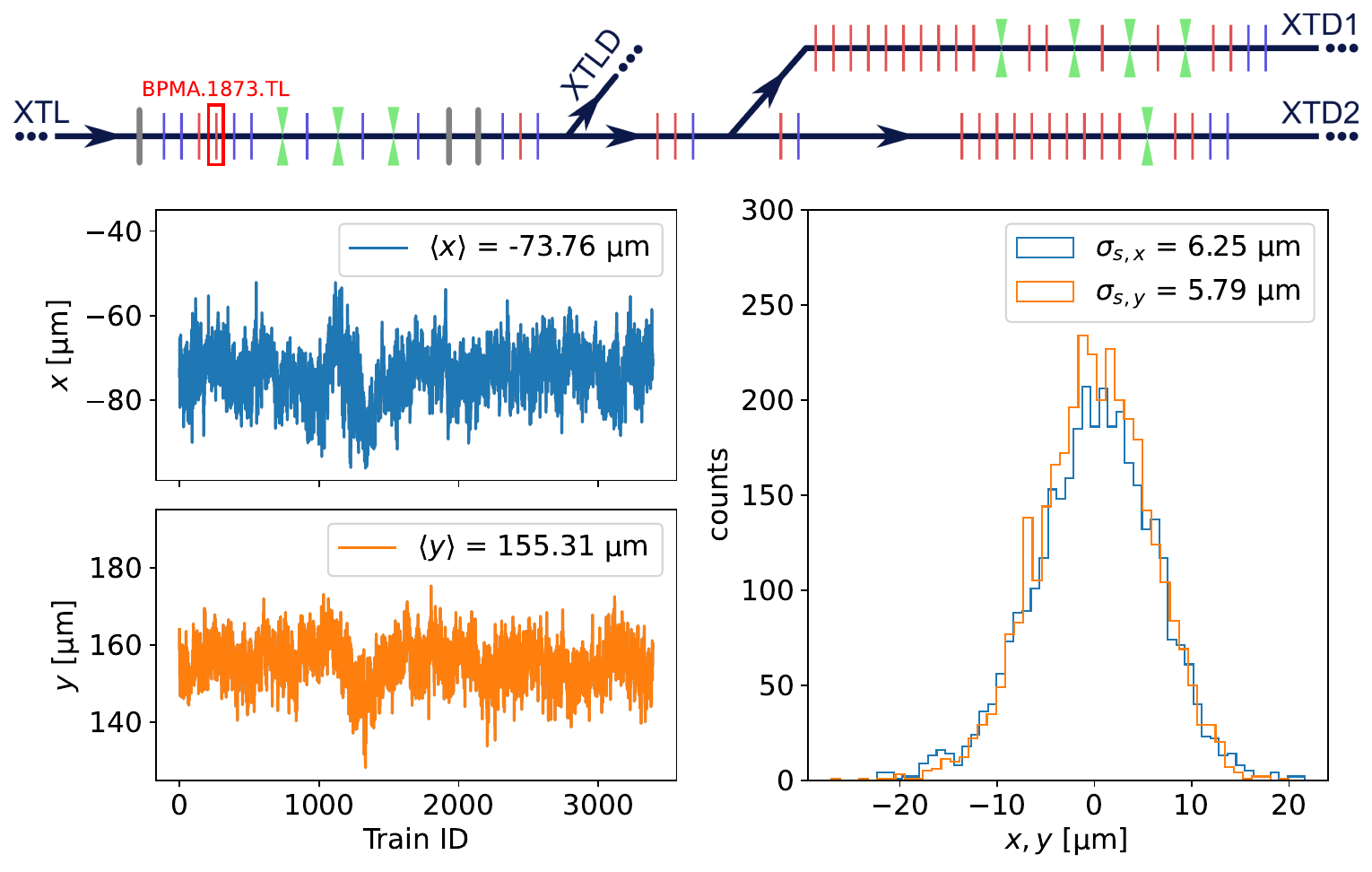}
    \caption{Variation of the positions $x, y$ in the BPMA.1873.TL button BPM between each train. On the top is the diagnostics layout with a red rectangle indicating the BPM considered. On the left are the positions $x, y$ for each bunch. On the right are the corresponding histograms.}
    \label{fig:bpm_tl_raw_train}
\end{figure}

At a specific BPM, the variation of the transverse positions $x, y$ can be observed between different trains. 
Figure~\ref{fig:bpm_tl_raw_train} shows the measurement for a single BPM, named \verb|BPMA.1873.TL|, in the XTL line. Among the bunches that are directed towards the XTLD dump, the transverse position a single arbitrary bunch ID inside the train is observed.
For this bunch ID in the train, the $x$ position varies between \SI{-96}{\micro\meter} and \SI{-52}{\micro\meter} and the $y$ position varies between \SI{128}{\micro\meter} and \SI{175}{\micro\meter}. At this BPM, the standard deviation of the position signal $\sigma_S$ is around \SI{6}{\micro\meter} for each plane. This signal variation is a combination of multiple correlated and uncorrelated sources. Slight changes at the gun, the RF-cavities or the magnet strengths are independent from the measuring device and have an impact on all BPMs. This correlated variation of the beam is the jitter $\sigma_J$. In contrast, variations due to the electronics or the misalignments are different from one BPM to another, they are uncorrelated. This uncorrelated variation is referred as the noise $\sigma_N$, i.e. the resolution of the BPM. 

The signal variation $\sigma_S$ can be expressed as a combination of the jitter and the BPM noise, 

\begin{equation}
    \sigma_S^2 = \sigma_J^2+\sigma_N^2.
    \label{eq:signal_decomposition}
\end{equation}

Using the data from a single BPM, separating $\sigma_J$ from $\sigma_N$ is not easily possible.

\subsubsection{Multiple BPM measurements}

\begin{figure}[!ht]
    \centering
    \includegraphics[width=\textwidth]{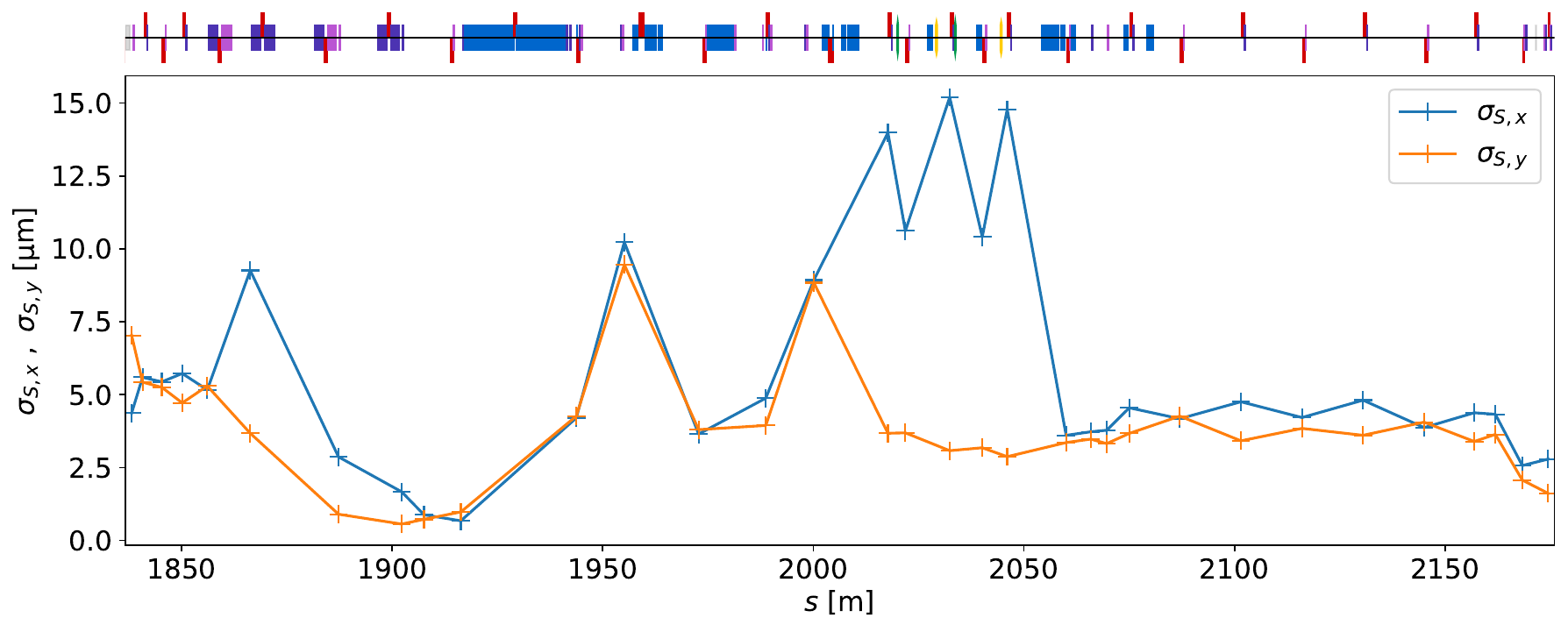}
    \caption{Variation of the standard deviation of the signal $\sigma_s$ along the XTD1 path. On top is the magnets layout, following BDSIM colour convention.}
    \label{fig:xtd1_raw_position}
\end{figure}

To consider the variation of the signal along the lattice, the value of $\sigma_{S,x},\sigma_{S,y}$ is plotted in Figure~\ref{fig:xtd1_raw_position} for each BPM in the XTD1 path and for a single bunch ID.
Since there are multiple BPMs along the beam lines it is possible to perform a virtual measurement~\cite{vm} (VM). Let us consider a beam line with $N$ BPMs in which the positions of $M$ individual bunches are measured. The position values are expressed as $x_{ij}$ where $i$ is the bunch number and $j$ is the BPM number. 
The virtual measurement method states that for a reference BPM $k$ among the BPMs in the line, there is a set of coefficient $c_j$ so that the position $x_{ik}$ is expressed as

\begin{equation}
    x_{ik} = \sum_{j\neq k}^{N} c_j x_{ij}.
    \label{eq:correl_sum}
\end{equation}

In essence, the position measured at a reference BPM $k$ is a linear combination of all the positions measured at the other BPMs. The coefficient $c_j$ represents how much the measurement at the BPM $j$ contributes to $x_{ik}$. Those coefficients are independent of the reference BPM choice.
Building Eqn~\ref{eq:correl_sum}, the assumption is made that the lattice is mainly formed of linear magnetic elements (like drifts, dipoles and quadrupoles). In the matrix formalism, Eqn~\ref{eq:correl_sum} can be re-written as

\begin{equation}
    \boldsymbol{x}_{k} = 
    \begin{pmatrix}
        x_1 \\ \vdots\\ x_i \\ \vdots \\ x_M
    \end{pmatrix} = 
    \begin{pmatrix}
        x_{11} & \dots & x_{i,1} & \dots & x_{M1} \\
        \vdots \\
         x_{1,j\neq k} & & \ddots & & \vdots \\
       \vdots  \\
        x_{1N} & & \dots & & x_{MN}
    \end{pmatrix} \cdot
    \begin{pmatrix}
        c_1 \\ \vdots \\ c_{j\neq k} \\ \vdots \\ c_N
    \end{pmatrix}
    = \boldsymbol{X}_k \cdot \boldsymbol{c}_k,
    \label{eq:correl_matrix}
\end{equation}

where $\boldsymbol{x}_k$ is a vector of all the horizontal positions at the reference BPM, $\boldsymbol{c}_k$ is a vector of the coefficients for each BPM except the reference one and $\boldsymbol{X}_k$ is the matrix containing all horizontal positions for each bunch and each BPM except the reference one. The coefficient vector $\boldsymbol{c}_k$ can then be expressed as

\begin{equation}
    \boldsymbol{c}_k = \boldsymbol{X}_k^{-1} \cdot \boldsymbol{x}_k.
    \label{eq:coeff}
\end{equation}

Inverting the non-diagonal and non-square matrix $\boldsymbol{X}_k$ is not trivial. The chosen solution is a Singular Value Decomposition~\cite{svd} (SVD) that allows to re-write the matrix as

\begin{equation}
    \boldsymbol{X}_k = \boldsymbol{U} \cdot \boldsymbol{D} \cdot \boldsymbol{V}^T,
    \label{eq:SVD}
\end{equation}

where $\boldsymbol{D}$ is a diagonal matrix and $\boldsymbol{U}$ and $\boldsymbol{V}$ are two unitary matrices. These three matrices can be easily inverted and Eqn~\ref{eq:coeff} becomes

\begin{equation}
    \boldsymbol{c_k} = \boldsymbol{V} \cdot \boldsymbol{D^{-1}} \cdot \boldsymbol{U^T} \cdot \boldsymbol{x_k}.
    \label{eq:correlation_coeff_vector_SVD}
\end{equation}

With this set $\boldsymbol{c_k}$ of coefficients, the positions calculated with the virtual measurement can be expressed as 

\begin{equation}
    \boldsymbol{x}_{k,\rm VM} = \boldsymbol{X}_k \cdot \boldsymbol{c}_k.
    \label{eq:position_measured}
\end{equation}

The virtual measurement vector $\boldsymbol{x}_{k,\rm VM}$ can be related with the real positions $\boldsymbol{x}_k$ by

\begin{equation}
    \boldsymbol{x}_k = 
    \boldsymbol{x}_{k,\rm VM} +
    \boldsymbol{\delta x}_k,
    \label{eq:position_with_error}
\end{equation}

where $\boldsymbol{\delta x}_k$ is the vector of errors of our virtual measurement. The correlated variation in position, i.e. the position jitter, is the standard deviation of the virtual measurement vector. It can be expressed as

\begin{equation}
    \sigma_{J, k} = \langle 
    \boldsymbol{x}_{k,\rm VM}^2 \rangle.
    \label{eq:jitter}
\end{equation}

Similarly, the uncorrelated variation in position, i.e. the BPM noise, can be calculated with

\begin{equation}
    \sigma_{N, k} = \langle 
    \boldsymbol{\delta x}_k^2 \rangle.
    \label{eq:noise}
\end{equation}

\begin{figure}[!ht]
    \centering
    \includegraphics[width=15cm]{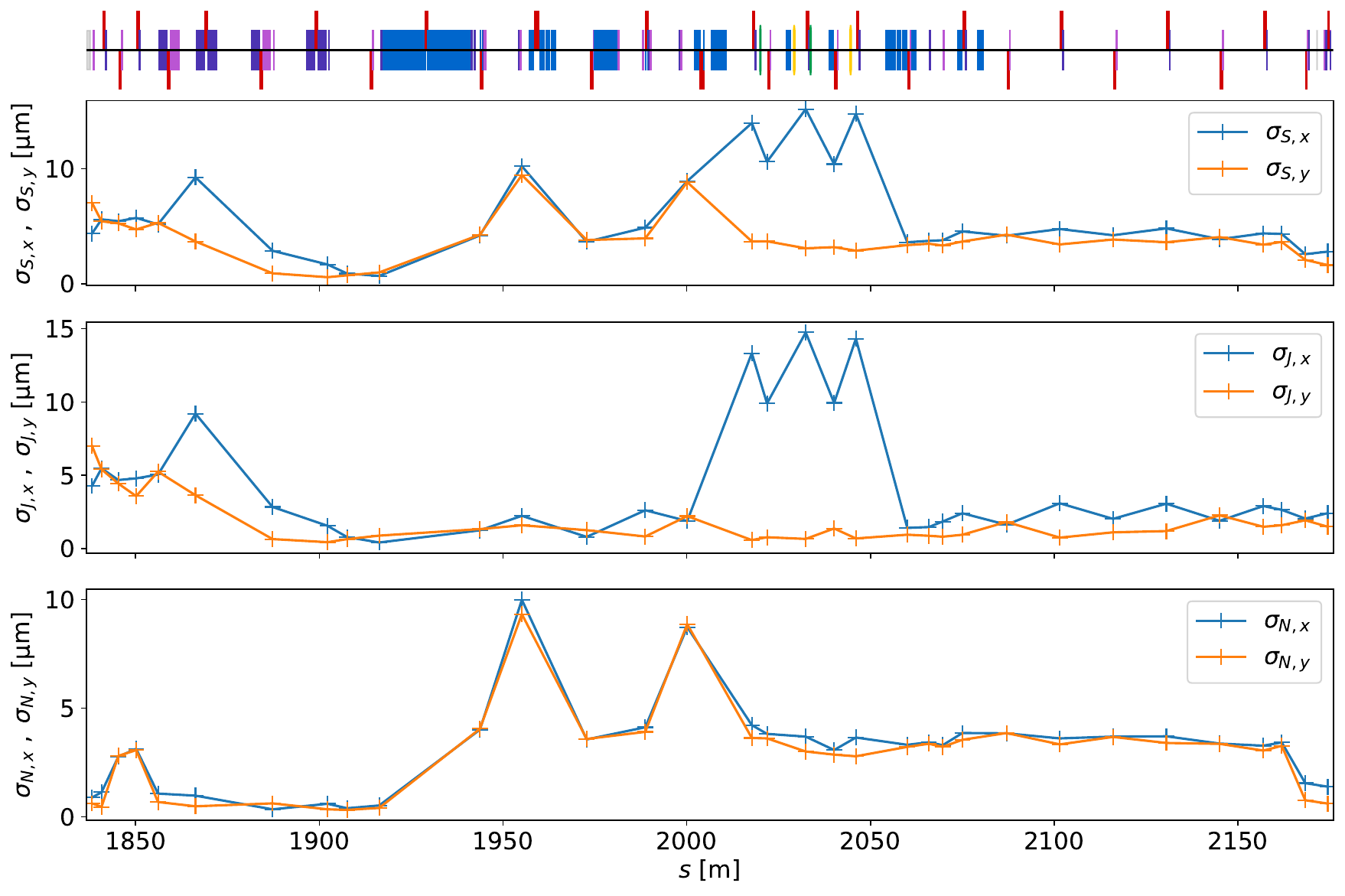}
    \caption{Plots of the position signal decomposition using virtual measurement in both transverse planes. The top plot is the raw signal variation $\sigma_S$. The middle plot is the position jitter $\sigma_J$. The bottom plot is the BPMs noise $\sigma_N$.}
    \label{fig:xtd1_svd_position}
\end{figure}

\begin{figure}[!ht]
    \centering
    \includegraphics[width=15cm]{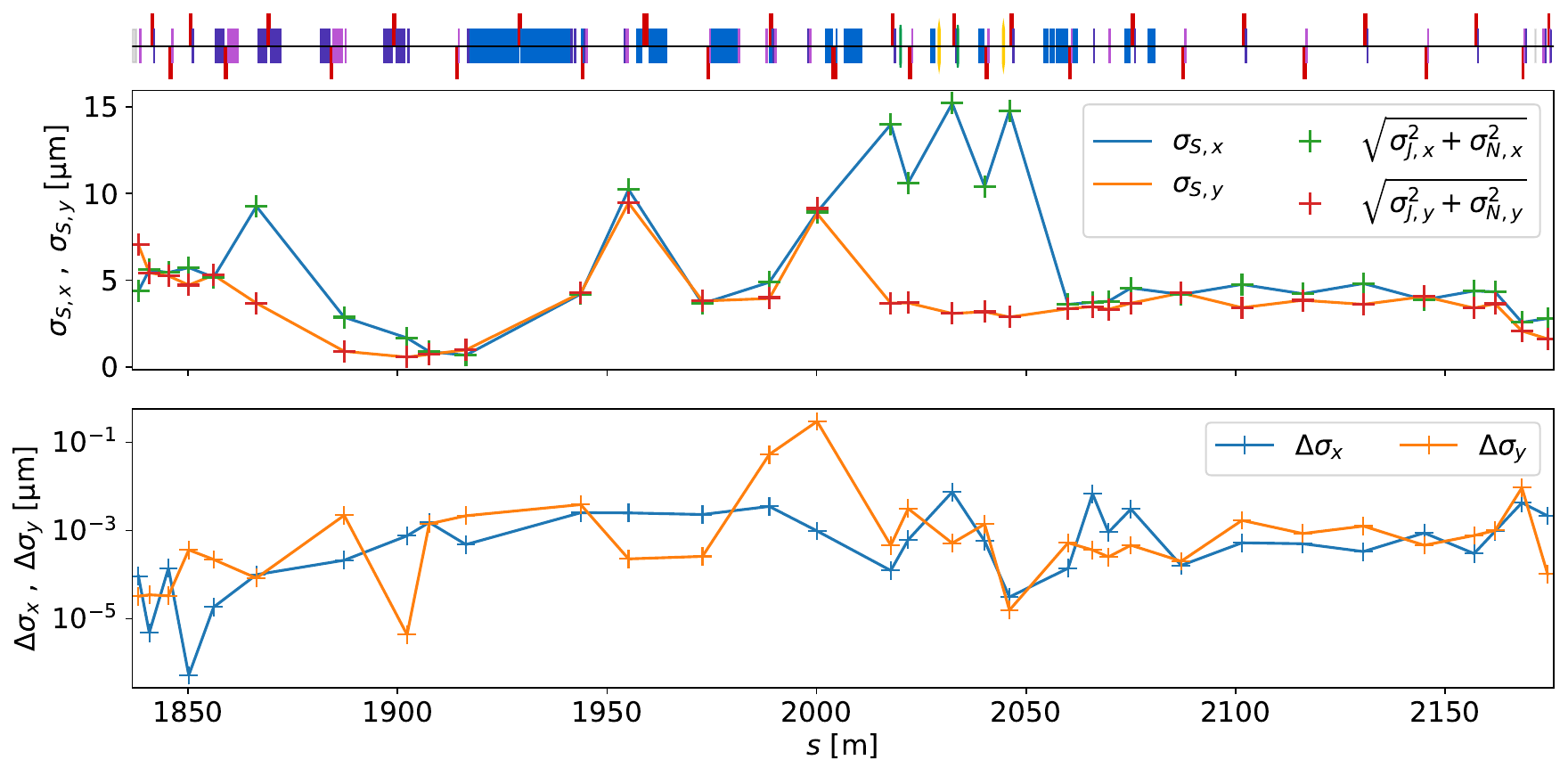}
    \caption{Reconstruction of the signal from virtual measurement and comparison with the raw signal $\sigma_S$. The top plot is the original signal $\sigma_S$ and the reconstructed one. The bottom plot show the absolute difference between both.}
    \label{fig:xtd1_svd_reconstruction}
\end{figure}

Using the data from the switch yard BPMs, a virtual measurement can be performed to calculate both $\sigma_J$ and $\sigma_N$ at each BPM. The raw signal variation $\sigma_S = \langle \boldsymbol{x}_k^2 \rangle$ is also calculated and the results are shown in Figure~\ref{fig:xtd1_svd_position}.
It is now possible to identify the contributions from jitter and noise on the signal variation. For example, the peaks in the horizontal plane between \SI{2000}{\meter} and \SI{2050}{\meter} are coming entirely from the jitter, most probably because of the kick into XTD1. 
To confirm the validity of the virtual measurement, the raw signal is recalculated using Eqn~\ref{eq:signal_decomposition}, as shown in Figure~\ref{fig:xtd1_svd_reconstruction}.

\begin{figure}[!ht]
    \centering
    \includegraphics[width=15cm]{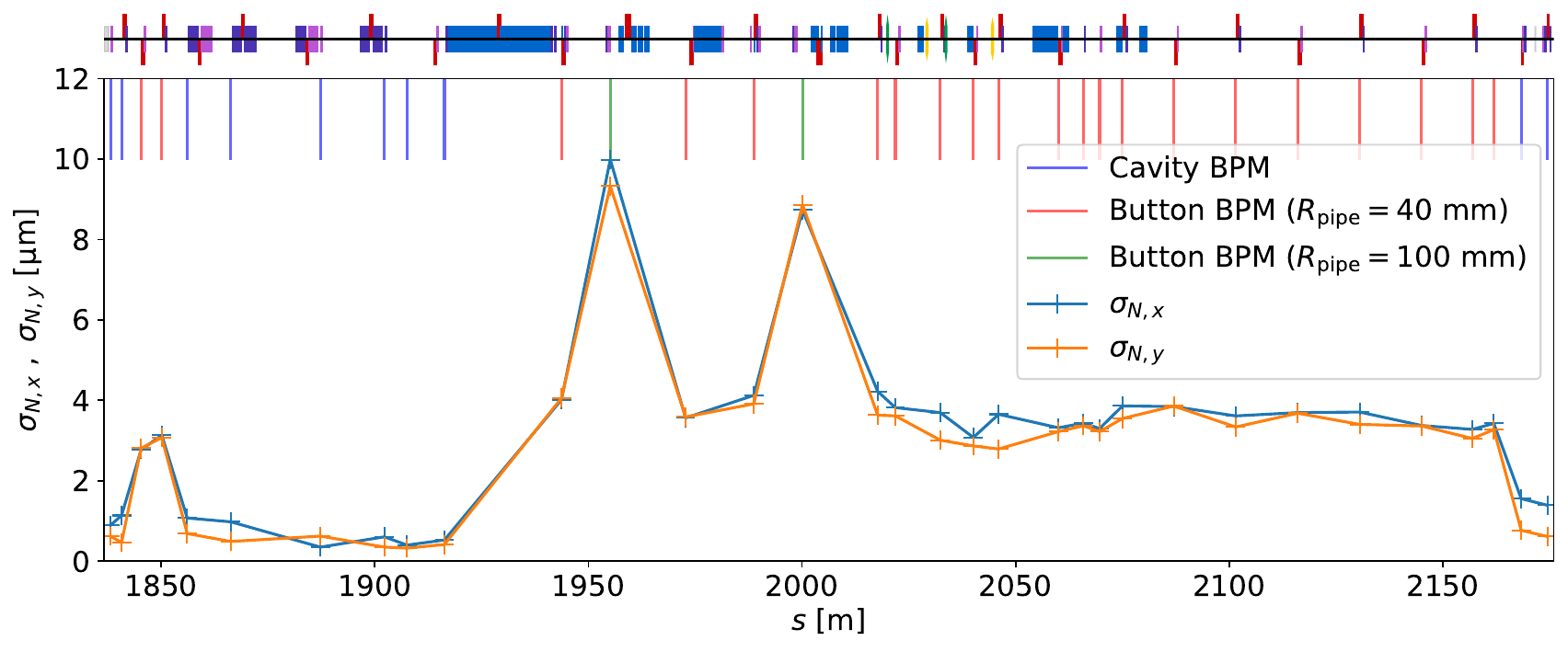}
    \caption{Plot of the BPM noise $\sigma_N$ as a function of the BPM location $s$. Highlighted in red and green are the button BPMs and in blue are the cavity BPMs.}
    \label{fig:xtd1_noise_bpm}
\end{figure}

Apart from one BPM at around \SI{2000}{\meter} which shows a \SI{0.1}{\micro\meter} difference in $y$ from the raw signal, the virtual measurement accurately separated the correlated part $\sigma_J$ of the signal and the uncorrelated part $\sigma_N$ which is specific for each BPM.
Using only the uncorrelated part $\sigma_N$ the resolution of each BPM can be evaluated, as shown in Figure~\ref{fig:xtd1_noise_bpm}.

\begin{figure}[!ht]
    \centering
    \includegraphics[width=15cm]{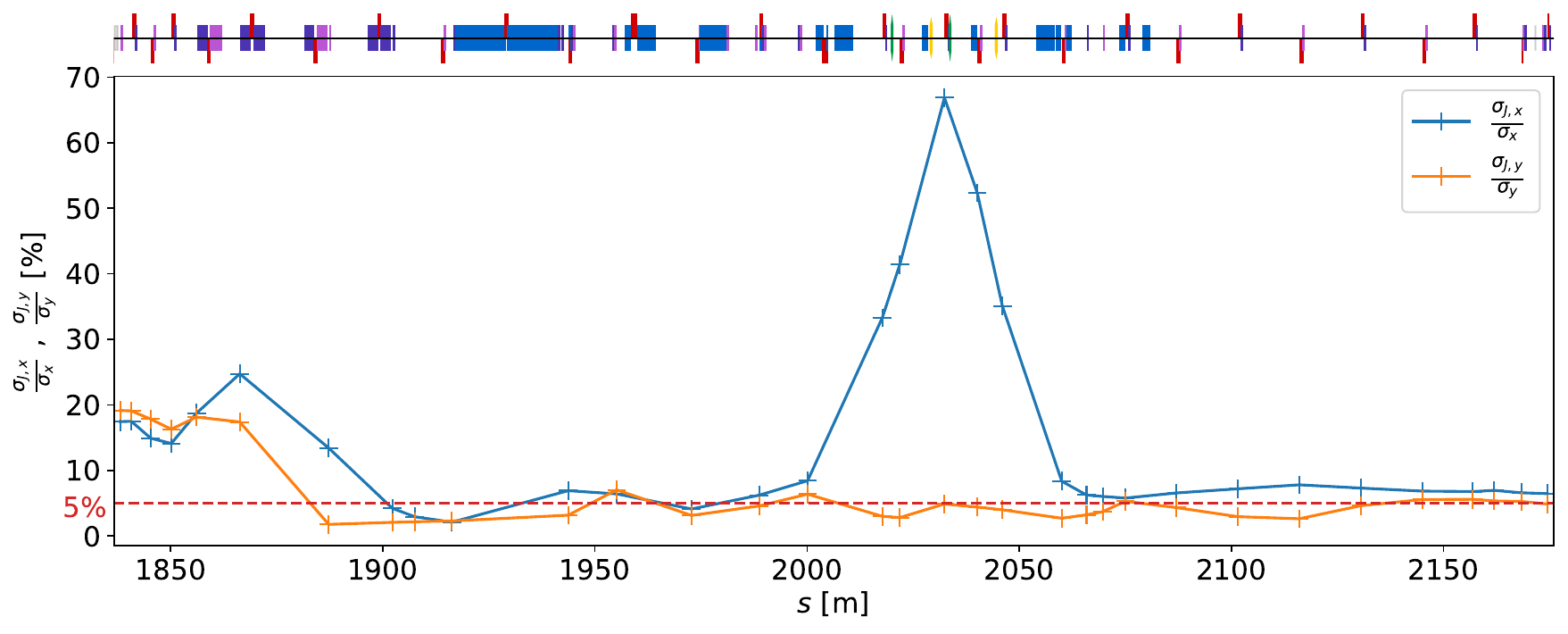}
    \caption{Ratio of the jitter $\sigma_J$ and the beam size $\sigma$ plotted against the BPM positions in the XTD1 path.}
    \label{fig:xtd1_jitter_ratio_bpm}
\end{figure}

The regular \SI{40}{\milli\meter} aperture button BPMs have a noise $\sigma_N$ between \SI{3}{\micro\meter} and \SI{4}{\micro\meter}. The two large aperture (\SI{100}{\milli\meter}) button BPMs have a higher noise, between \SI{8}{\micro\meter} and \SI{10}{\micro\meter}. In both button BPMs $\sigma_N/R_{\rm pipe} \approx 8.8$ so this confirms the linear dependence between the pipe radius and the BPM resolution established in Eqn~\ref{eq:linear_proportional_error_bpm}. For the cavity BPMs, $\sigma_N$ is of the order of \SI{1}{\micro\meter} and even below at around \SI{1900}{\meter}. Those values confirm the resolutions shown in Figure~\ref{fig:bpm_resolution}. Since both the position jitter and the beam size are dictated by the $\beta$ function, it is expected for this jitter to be linearly proportional to the beam size~\cite{jitter_size_proport}. In Figure~\ref{fig:xtd1_jitter_ratio_bpm} the ratio between $\sigma_J$ and the beam size $\sigma$ is plotted.

Apart from the the high peak at \SI{2030}{\meter}, which relates to the dispersion of the XTD1 bend, and the high values before \SI{1900}{\meter}, the beam jitter $\sigma_J$ is rather stable in both transverse planes, especially in the XTD1, at \SI{5}{\%} of the beam size. In the following studies, both the absolute jitter of around \SI{1.5}{\micro\meter} and the relative jitter of around \SI{5}{\%} of the beam size will be considered for LUXE's collision.

\subsubsection{Variations inside a bunch train}

\begin{figure}[!ht]
    \centering
    \includegraphics[width=\textwidth]{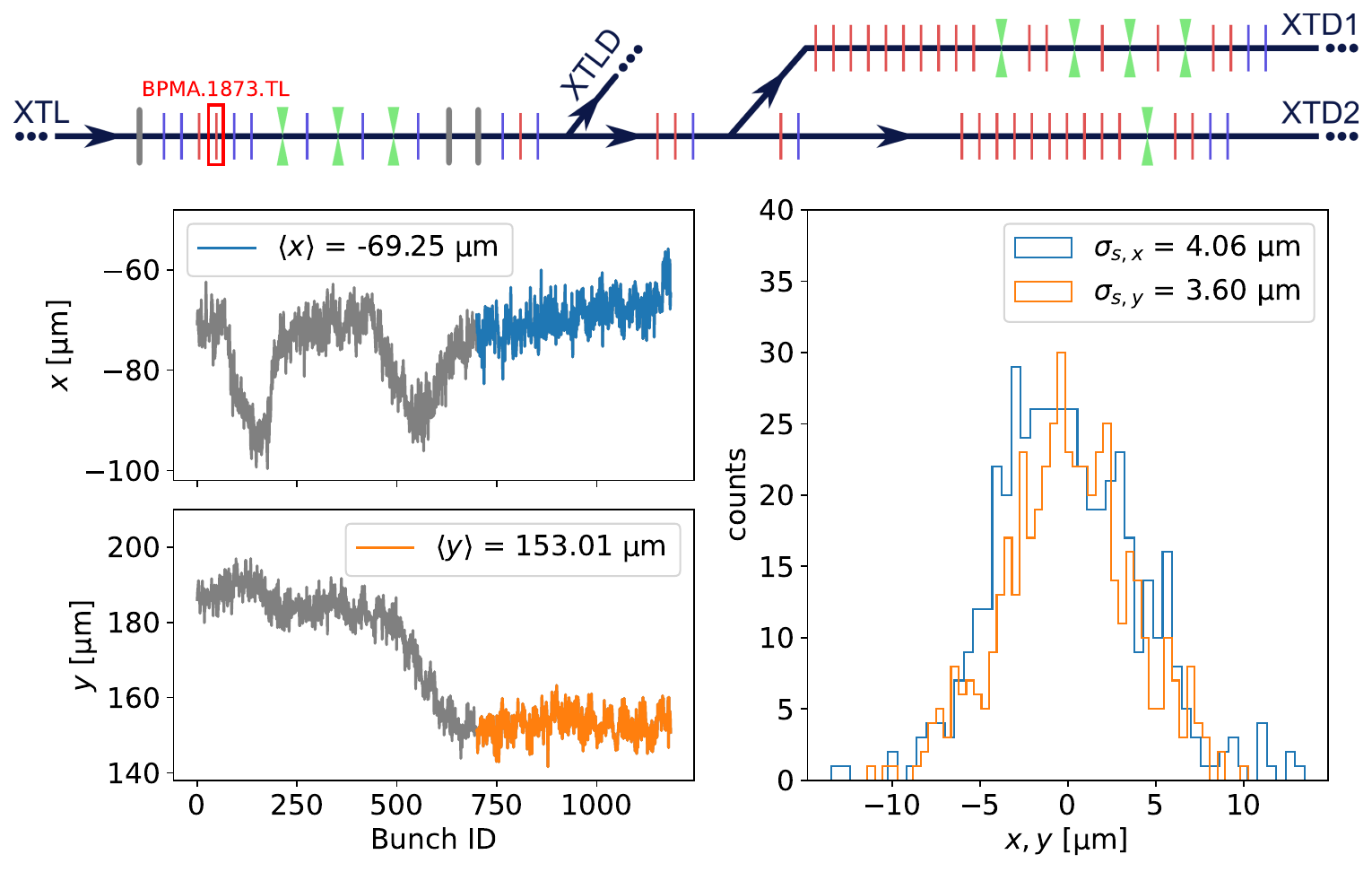}
    \caption{Variation of the positions $x,y$ in the BPMA.1873.TL button BPM between each bunch in the train. On the top is the diagnostics layout with a red rectangle indicating the BPM considered. On the left are the positions $x,y$ as function of bunch ID. On the right are the corresponding histogram. The greyed part of each signal is not used to measure the means $\langle x\rangle, \langle y\rangle$ and for the histograms.}
    \label{fig:bpm_tl_raw_bunch}
\end{figure}

The analysis presented in the previous section is for a single bunch ID in the train. But there is potentially some position variation along the \SI{2700}{} bunches of the same train. 
In the XTL line, all bunches created are present as they have not yet been distributed between the lines and the dump. Only the BPMs in this XTL line can measure the transverse positions $x, y$ for every bunches in a train. Figure~\ref{fig:bpm_tl_raw_bunch} shows the measurement in the same BPM \verb|BPMA.1873.TL| of every bunches in an arbitrary train.

\begin{figure}[!ht]
    \centering
    \includegraphics[width=\textwidth]{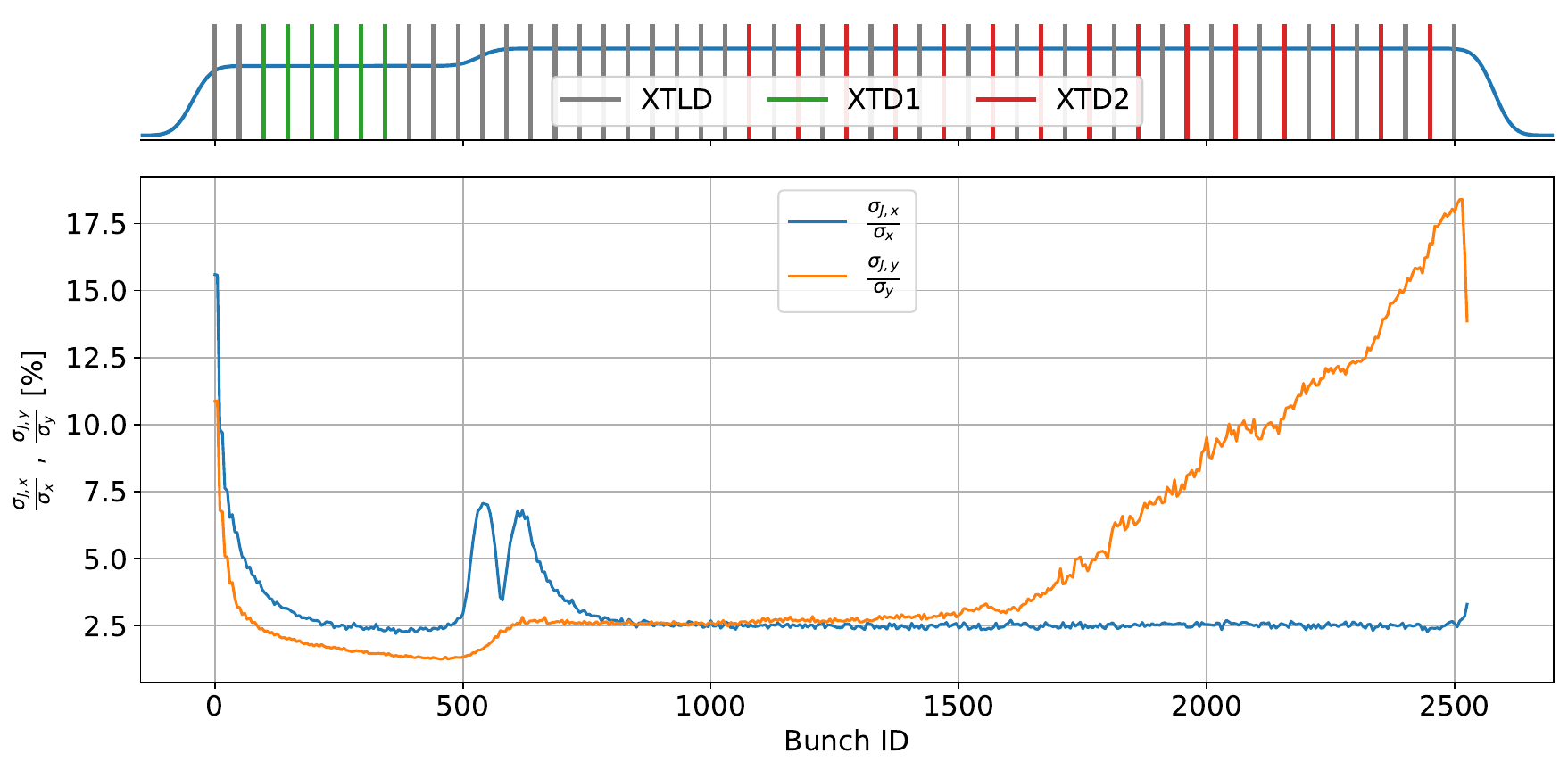}
    \caption{Plot of the ratio between the jitter $\sigma_J$ and the beam size $\sigma$ as a function of the bunch ID. On top is the bunch pattern, in red are the bunches going in the XTD1 and in green are the bunches going in the XTD2.}
    \label{fig:bunch_jitter_ratio}
\end{figure}

For the signal itself, the position mean varies by around \SI{30}{\micro\meter} in $x$ and around \SI{40}{\micro\meter} in $y$. Looking only at the stable end section of this bunch train, the standard deviation of the signal $\sigma_S$ is around \SI{4}{\micro\meter} in both planes. Using all the bunch trains and all BPMs in the XTL line, the same virtual measurement can be performed. The position jitter to beam size ratio $\frac{\sigma_J}{\sigma}$ is calculated for each bunch ID and shown in Figure~\ref{fig:bunch_jitter_ratio}.
In this specific bunch pattern, there is a high relative jitter on the early bunches. This can probably be attributed to the RF ramping time, i.e. the time needed for the RF cavity to stabilise, which induces more variations on the bunch position. The values stabilizes at around \SI{2.5}{\%}. Towards the end of the train there is a progressive increase of $\sigma_{J, y}$. Between the bunch IDs 500 and 750 there are two spikes in the $x$ plane and a progressive step in the $y$ plane. As those changes are located between the bunches going in the XTD1 and those going in the XTD2, they may be caused by some modification of the bunch properties. For example if the experiments downstream of XTD1 requires a different energy from the experiments downstream of XTD2.

\subsection{Bunch charge}

\begin{figure}[!ht]
    \centering
    \includegraphics[width=\textwidth]{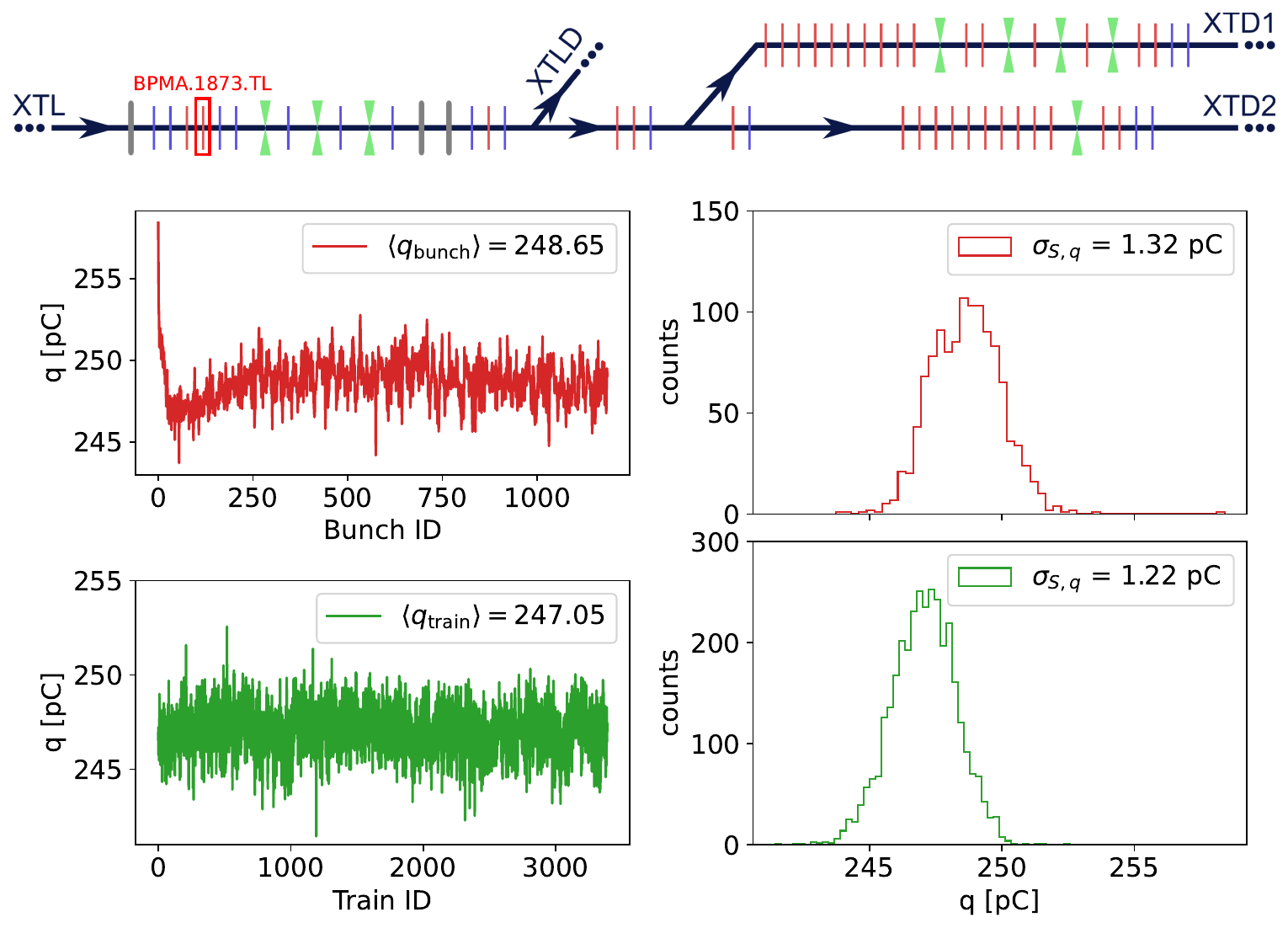}
    \caption{Variations of the charge in a button BPM of the XTL line for a specific bunch ID between each train. On the top is the diagnostics layout with a red rectangle indicating the BPM considered. On the left are the charge signals, per bunch and per train. On the right are the corresponding histograms.}
    \label{fig:bpm_tl_raw_charge}
\end{figure}

As explained previously, the bunch charge $q$ is also measured with both the button BPMs and cavity BPMs. Similar to the position measurements, the charge can vary for a specific bunch ID between trains and between every bunches in a single train. Both variations can be measured in the button BPM \verb|BPMA.1873.TL| and the results are shown in Figure~\ref{fig:bpm_tl_raw_charge} for an arbitrary bunch ID and an arbitrary train ID.
For both ``between bunches" and ``between trains" variations, the average charge is between \SI{247}{\pico\coulomb} and \SI{249}{\pico\coulomb}. The signal is quite stable apart from a peak for early bunch IDs. Similar to Figure~\ref{fig:bunch_jitter_ratio}, this peak can probably be attributed to the RF ramping time.
Between the bunches of the same train, the signal variation $\sigma_S$ is \SI{1.32}{\pico\coulomb}, which is around \SI{0.5}{\%} of the bunch charge. Between the same bunch ID in different trains, there is a similar variation of \SI{1.22}{\pico\coulomb}.

\begin{figure}[!ht]
    \centering
    \includegraphics[width=\textwidth]{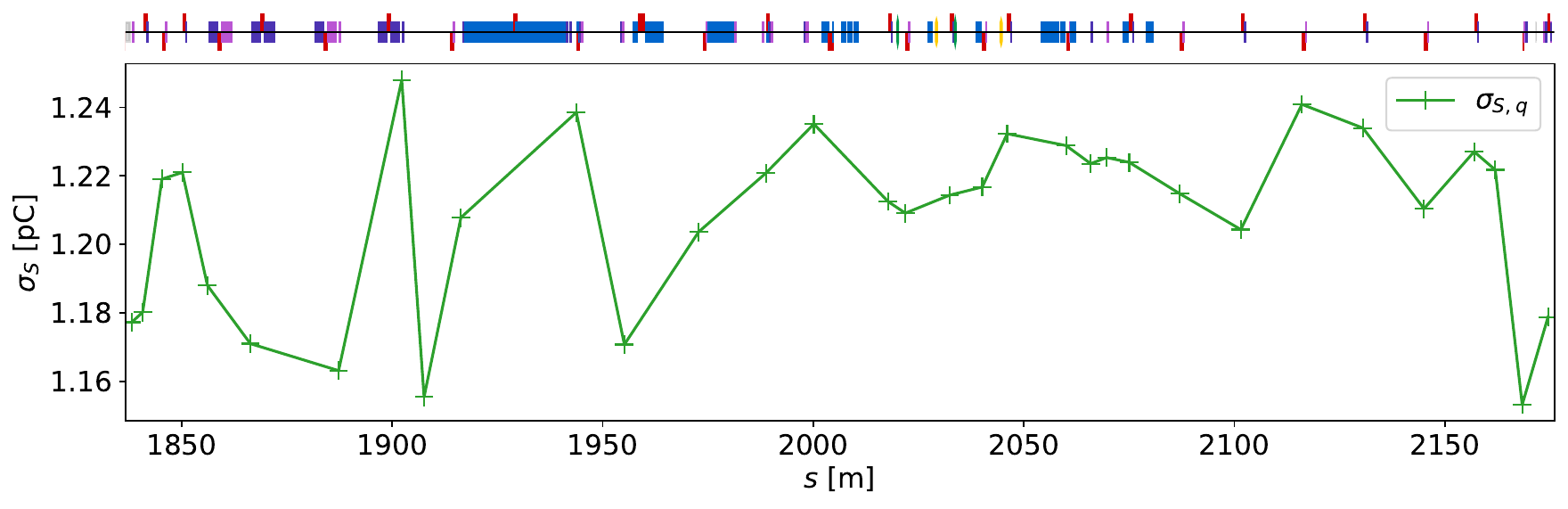}
    \caption{Variations of the standard deviation of the charge signal $\sigma_s$ along the XTD1 path.}
    \label{fig:xtd1_raw_charge}
\end{figure}

\begin{figure}[!ht]
    \centering
    \includegraphics[width=\textwidth]{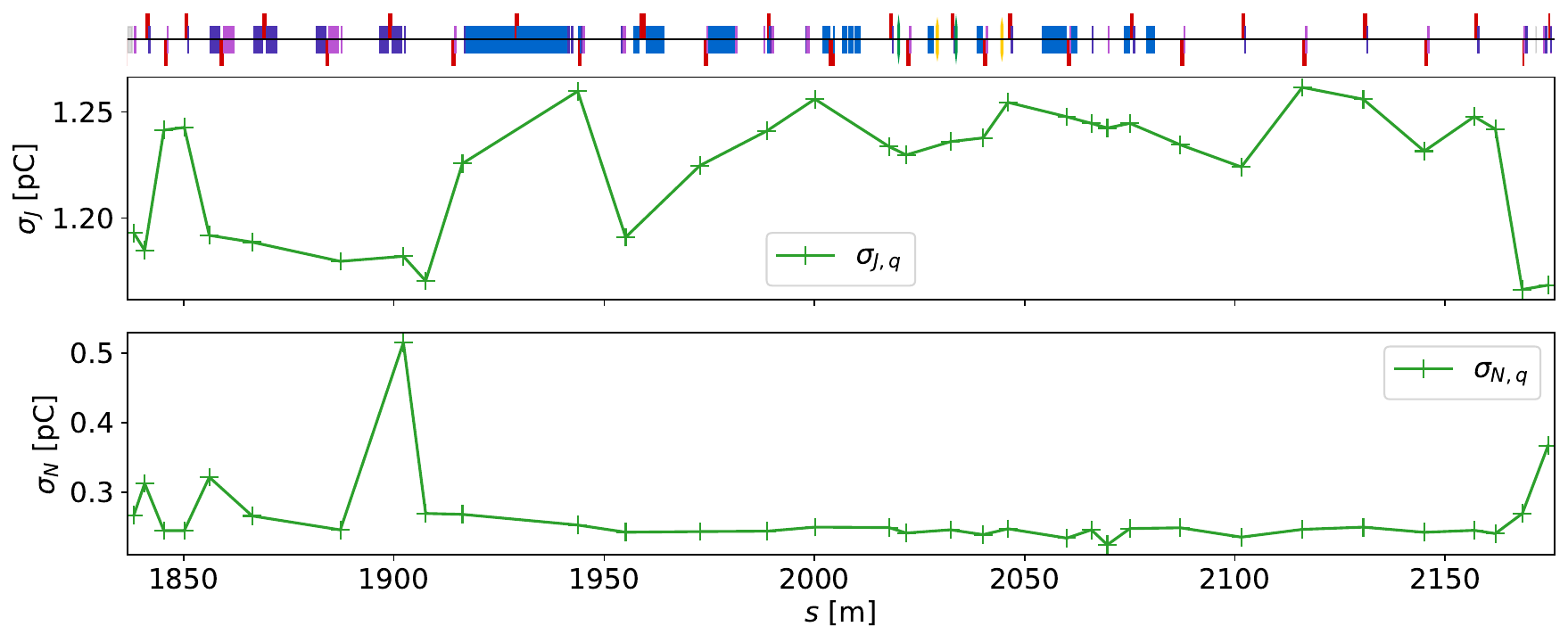}
    \caption{Decomposition of the charge signal variation $\sigma_{S,q}$ using the virtual measurement. The top plot is the charge jitter $\sigma_{J, q}$. The bottom plot is the BPMs noise $\sigma_{N, q}$.}
    \label{fig:xtd1_svd_charge}
\end{figure}

The value of $\sigma_S$ for an arbitrary bunch going in the XTD1 tunnel is shown in Figure~\ref{fig:xtd1_raw_charge}.
The signal is rather stable along the lattice between \SI{1.15}{\pico\coulomb} and \SI{1.25}{\pico\coulomb}. Using the same virtual measurement method shown previously, the signal can be decomposed between the actual jitter $\sigma_J$ and the BPM noise $\sigma_N$, as shown in Figure~\ref{fig:xtd1_svd_charge}.

\newpage

\begin{figure}[!ht]
    \centering
    \includegraphics[width=\textwidth]{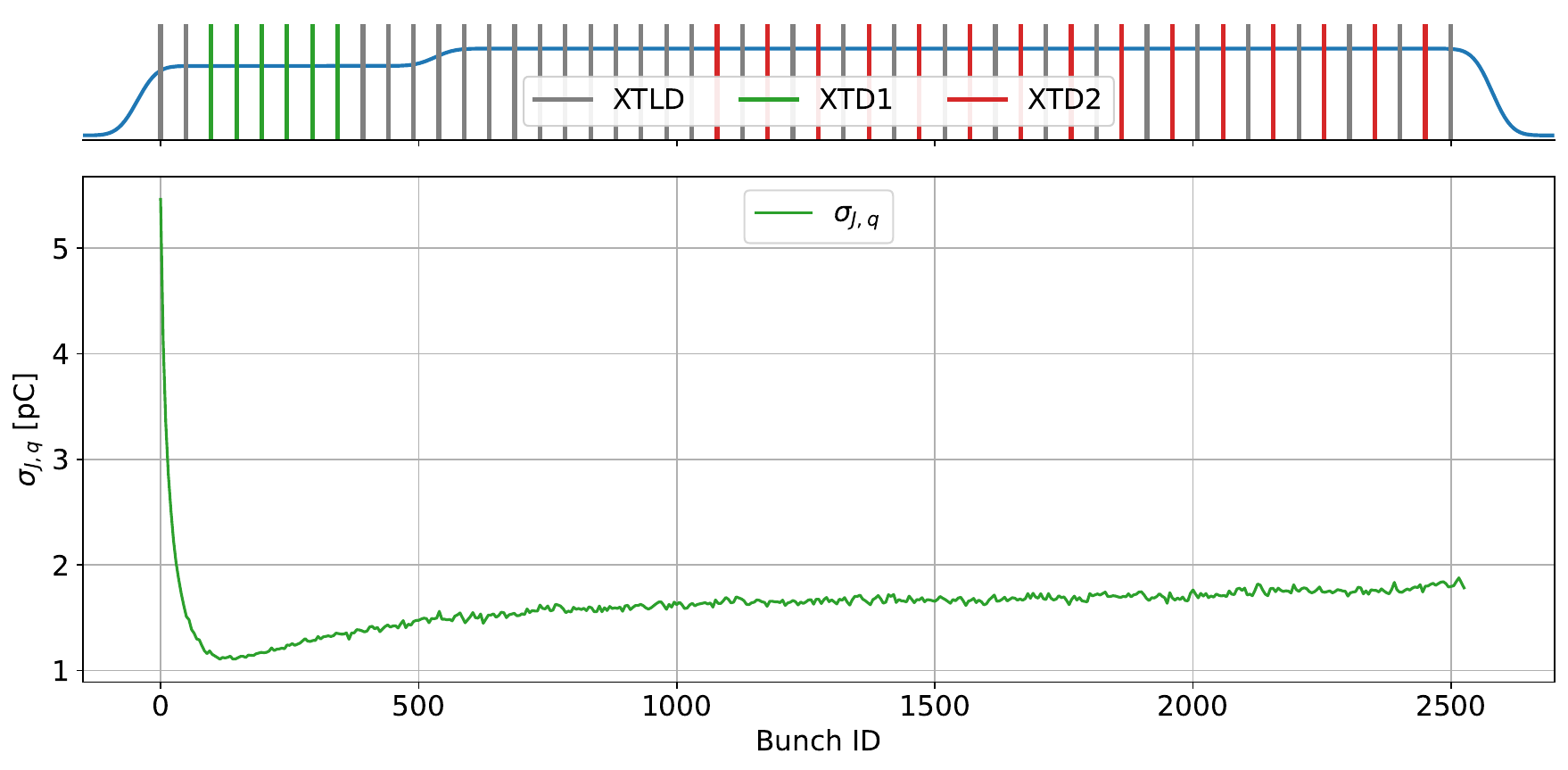}
    \caption{Variation of the charge jitter $\sigma_{J, q}$ for each bunch in the train. On top is the bunch pattern, in red are the bunches going in the XTD1 and in green are the bunches going in the XTD2.}
    \label{fig:bunch_jitter_charge}
\end{figure}

It appears that the majority of charge signal variation $\sigma_{S,q}$ is contained in the jitter $\sigma_{J,q}$. This jitter could be attributed to the accelerator gun photo-cathode interaction or variations in the laser itself. The BPM noise on the charge shows a peak of \SI{0.5}{\pico\coulomb} at around \SI{1900}{\meter} but is otherwise quite stable at around \SI{0.2}{\pico\coulomb}. The variation of the jitter $\sigma_{J, q}$ along the bunch train, can be  observed in Figure~\ref{fig:bunch_jitter_charge}.
Similar to Figure~\ref{fig:bunch_jitter_ratio}, there is a higher jitter at the beginning of the bunch train. This can also be attributed to the RF ramping time. Then it slowly stabilises to around \SI{1.7}{\pico\coulomb}. For every bunch ID beyond 100, the charge jitter $\sigma_{J,q}$ stays smaller than \SI{2}{\pico\coulomb}, hence smaller than \SI{1}{\%} of the bunch charge.

\subsection{Beam size}

\begin{figure}[!ht]
    \centering
    \includegraphics[width=\textwidth]{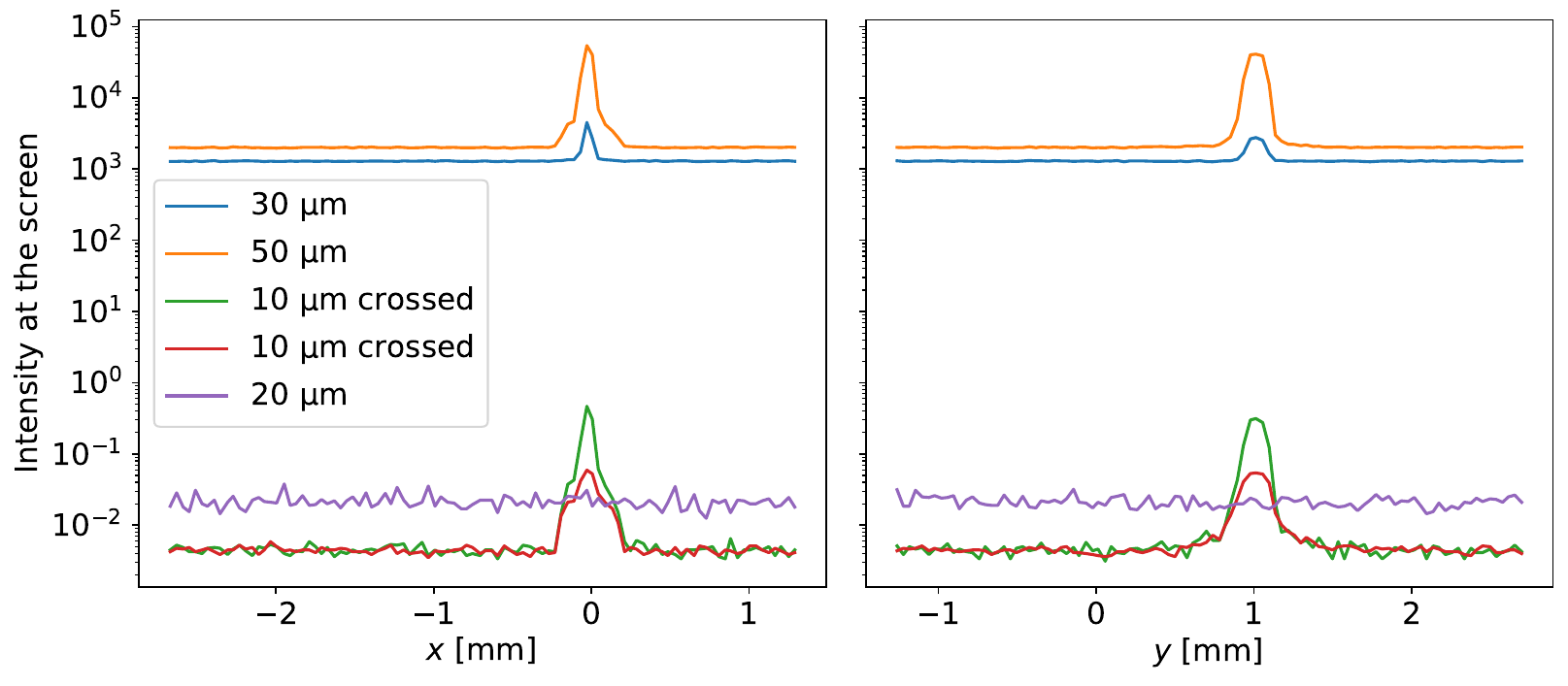}
    \caption{Example of a wire scan at the first wire and first quadrupole configuration.}
    \label{fig:wire_scan}
\end{figure}

As previously described, the beam size measurement is a slow process that can neither be applied to every bunch in a train nor to a single bunch ID for every train. So getting the jitter $\sigma_J$ for the beam size with this method is not possible.
But with the data from the wire scanners, the resolution $\sigma_N$ of the beam size measurement can be evaluated. 
In the XDT1 tunnel there are four wire scanners as shown in Figure~\ref{fig:euxfel_switchyard_detectors}. The measurements are done with different quadrupole configurations. An example of all configurations for a single measurement can be found in Table~\ref{tab:wire_setup}.
For each line of this table, the two mounts (one horizontal and one vertical) of five wires are inserted in the beam pipe. At each position of a wire, the signal is measured at a downstream BLM. An example of the measured signals is shown in Figure~\ref{fig:wire_scan}.

\begin{table}[!ht]
\centering
\begin{tabular}{|c || c | c | c | c |}
\hline
Wire name & $\beta_x$ / m &	$\beta_y$ / m & $\psi_x$ & $\psi_y$ \\
\hline
\hline
OTRBW.2117.T1 & 33.15 & 37.76 & 45.93 & 37.44 \\
\hline
OTRBW.2146.T1 & 34.64 & 26.90 & 93.68 & 92.91 \\
\hline
OTRBW.2161.T1 & 32.23 & 26.37 & 113.34 & 134.84 \\
\hline
OTRBW.2174.T1 & 28.19 & 31.03 & 145.64 & 156.00 \\
\hline
OTRBW.2117.T1 & 66.74 & 48.08 & 27.34 & 41.06 \\
\hline
OTRBW.2146.T1 & 41.36 & 100.06 & 57.45 & 62.08 \\
\hline
OTRBW.2161.T1 & 31.51 & 36.10 & 76.34 & 78.85 \\
\hline
OTRBW.2174.T1 & 15.53 & 24.40 & 115.56 & 103.5 \\
\hline
\end{tabular}
\caption{Example of all the $\beta$ and phase advance $\psi$ for a beam size measurement. There are four wire scanners and two optic configurations per wire.}
\label{tab:wire_setup}
\end{table}

By combining the measurements from all wires in both planes, the value of the beam sizes can be calculated. The results of this scan are shown in Table~\ref{tab:wire_results}. 
Using the measured beam size values of Table~\ref{tab:wire_results} and the $\beta$ function, the emittance can then be computed as 

\begin{equation}
    \varepsilon = \frac{\sigma^2}{\beta}.
\end{equation}

For all measurements, the error on the beam size $\delta\sigma$ is estimated at \SI{5}{\%} of the beam size.


\begin{table}[!ht]
\centering
\begin{tabular}{|c || c | c | c | c |}
\hline
Wire name & $\sigma_x$ / $\rm \mu m$ & $\sigma_y$ / $\rm \mu m$ \\
\hline
\hline
OTRBW.2117.T1 & 22.75 & 50.11 \\
\hline
OTRBW.2146.T1 & 27.67 & 39.71 \\
\hline
OTRBW.2161.T1 & 32.52 & 67.11 \\
\hline
OTRBW.2174.T1 & 34.93 & 89.27 \\
\hline
OTRBW.2117.T1 & 40.06 & 48.00 \\
\hline
OTRBW.2146.T1 & 35.76 & 48.75 \\
\hline
OTRBW.2161.T1 & 24.36 & 28.18 \\
\hline
OTRBW.2174.T1 & 21.95 & 48.40 \\
\hline
\end{tabular}
\caption{Results of the scan for each configuration.}
\label{tab:wire_results}
\end{table}

\subsection{Bunch arrival time}

\begin{figure}[!ht]
    \centering
    \includegraphics[width=\textwidth]{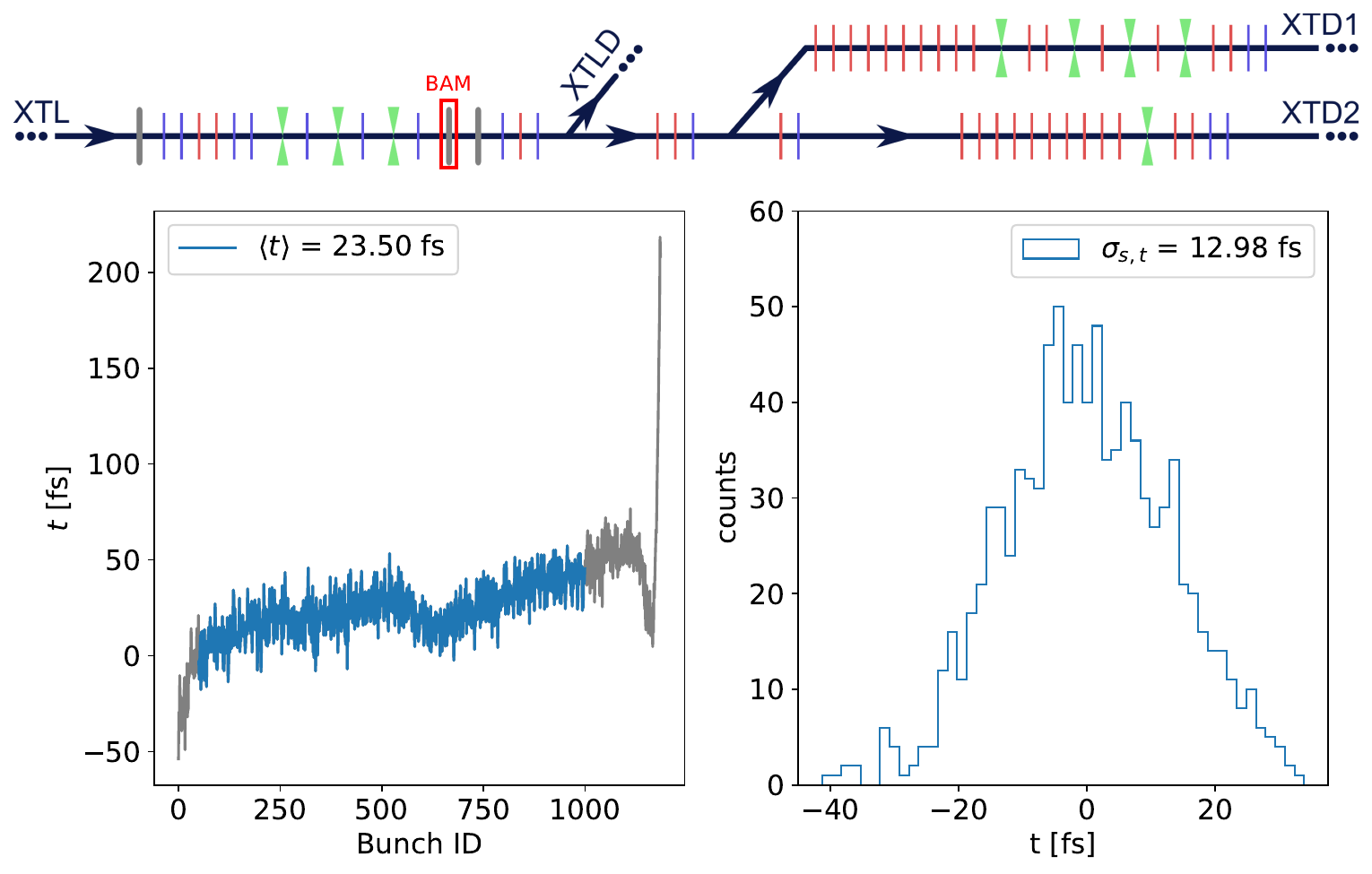}
    \caption{Variations of the arrival time $t$ inside the bunch train. On the left is the arrival time $t$ for each bunch. On the top is the diagnostics layout with a red rectangle indicating the bunch arrival monitor. On the right is the corresponding histogram. The greyed part of the signal is not used to measure the mean $\langle t\rangle$ and for the histogram.}
    \label{fig:time_per_bunch}
\end{figure}

The arrival time $t$ of the electron bunches is measured in the XTL line. Similar to the BPM measurements, the variation of $t$ can be measured both between bunches of the same train or for the same bunch ID between different trains.
Figure~\ref{fig:time_per_bunch} shows the arrival time of each bunch along an arbitrary train.
Inside a bunch train, the signal variation $\sigma_{S,t}$ is around \SI{12.98}{\femto\second}.

\begin{figure}[!ht]
    \centering
    \includegraphics[width=\textwidth]{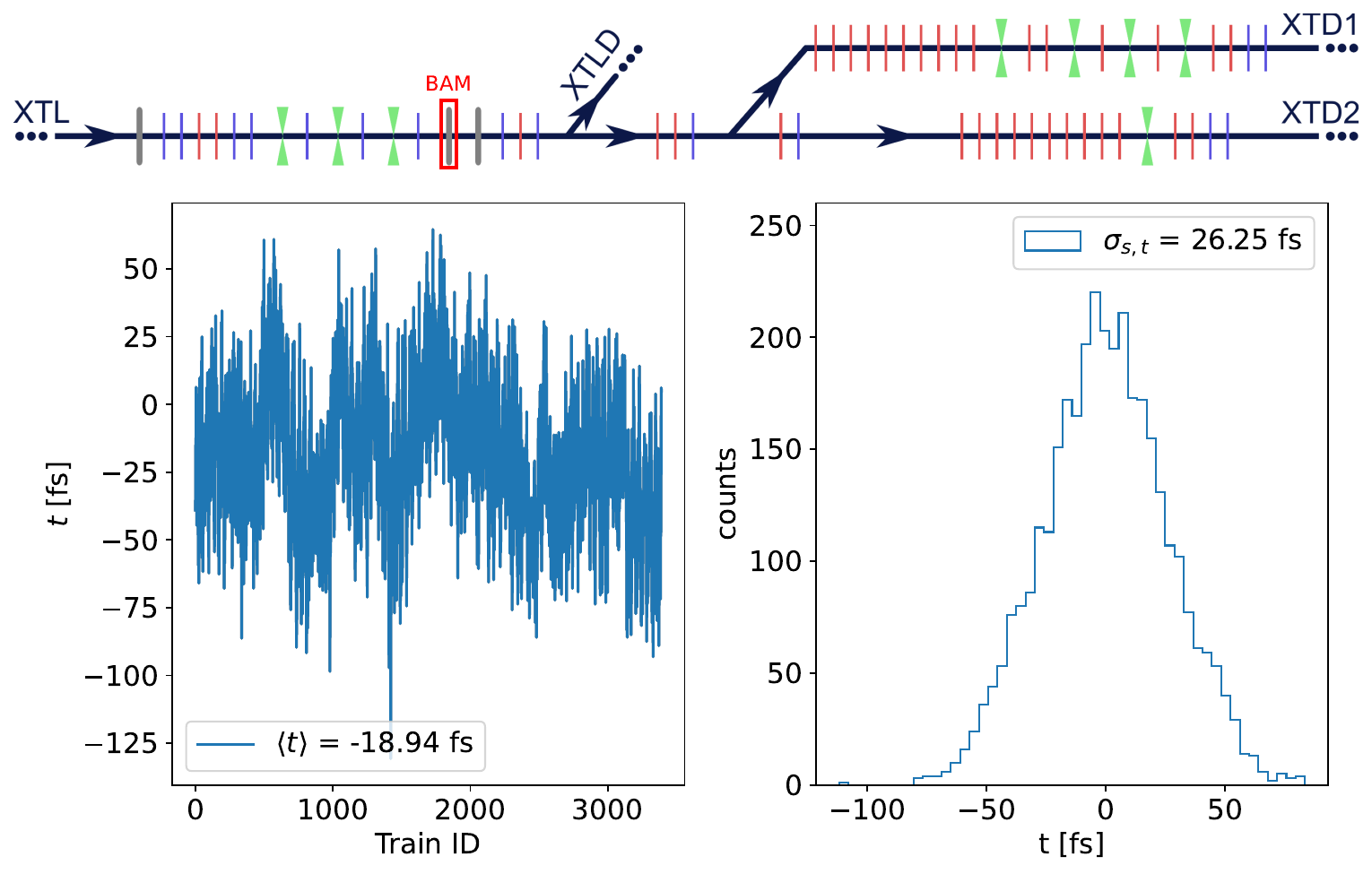}
    \caption{Variations of the arrival time $t$ between trains. On the top is the diagnostics layout with a red rectangle indicating the bunch arrival monitor. On the left is the arrival time $t$ for each train. On the right is the corresponding histogram.}
    \label{fig:time_per_train}
\end{figure}

From the literature on BAM~\cite{bam}, the resolution for a single BAM was established at a maximum value of \SI{5}{\femto\second}. Assuming that $\sigma_{N,t}=\ 5\ \rm fs$ the jitter can be expressed as

\begin{equation}
    \sigma_{J,t} = \sqrt{\sigma_{S,t}^2 - \sigma_{N,t}^2},
    \label{eq:jitter_from_noise}
\end{equation}

which gives an arrival time jitter of \SI{11.99}{\femto\second} between bunches of the same train. In Figure~\ref{fig:time_per_train} are the variations of the arrival time between bunch trains.
Between train, the signal variation $\sigma_{S, t}$ is higher, at around \SI{26.25}{\femto\second}. 
Since the BAM noise $\sigma_{N,t}$ is known, at around \SI{5}{\femto\second}, Eqn~\ref{eq:jitter_from_noise} can be used to calculate the jitter $\sigma_{J,t}$. 
This gives an estimated arrival time jitter of around \SI{25.78}{\femto\second}.

\subsection{Bunch length}

\begin{figure}[!b]
    \centering
    \includegraphics[width=\textwidth]{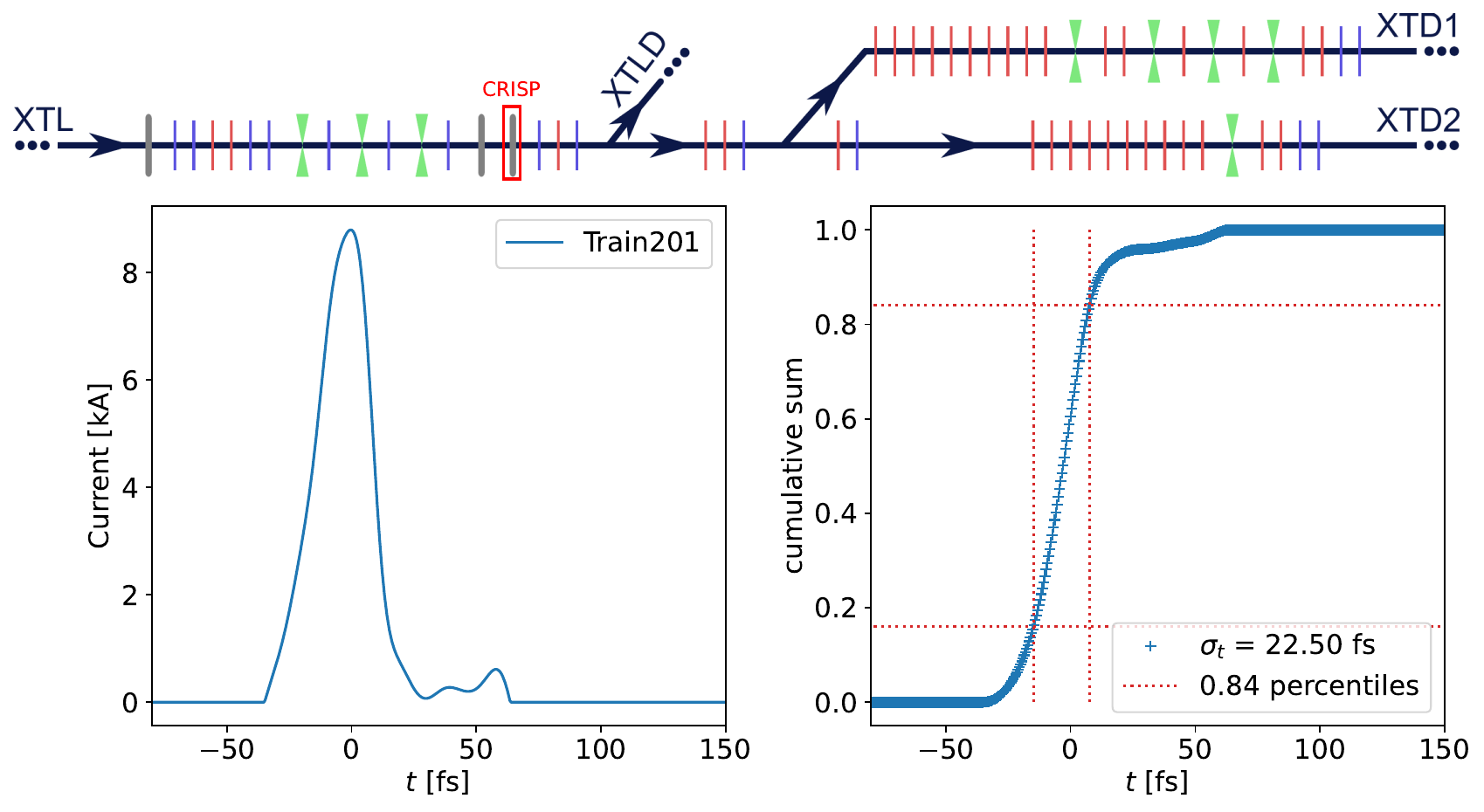}
    \caption{Example of the longitudinal profile of one bunch. On the top is the diagnostics layout with a red rectangle indicating the bunch length monitor. On the left is the signal as a function of time. On the right is the cumulated sum of the current used to measure the bunch length $\sigma_t$.}
    \label{fig:length_cumsum}
\end{figure}

As previously explained, the CRISP detector at the EuXFEL can only measure the length of one specific bunch ID at a time. For this measurement, a bunch ID sent to the XTD1 line was chosen.

An example of a bunch profile measured this way is shown in Figure~\ref{fig:length_cumsum}.
Because the profile shape is often non-Gaussian, the length $\sigma_t$ is calculated using a cumulative sum on the measured current. To be analogue with the standard deviation, the signal is summed between the \SI{0.16}{} and \SI{0.84}{} percentiles, i.e. the length that contains \SI{68}{\%} of the signal. From the example in Figure~\ref{fig:length_cumsum}, the bunch length is around \SI{22.5}{\femto\second}. 

\begin{figure}[!ht]
    \centering
    \includegraphics[width=\textwidth]{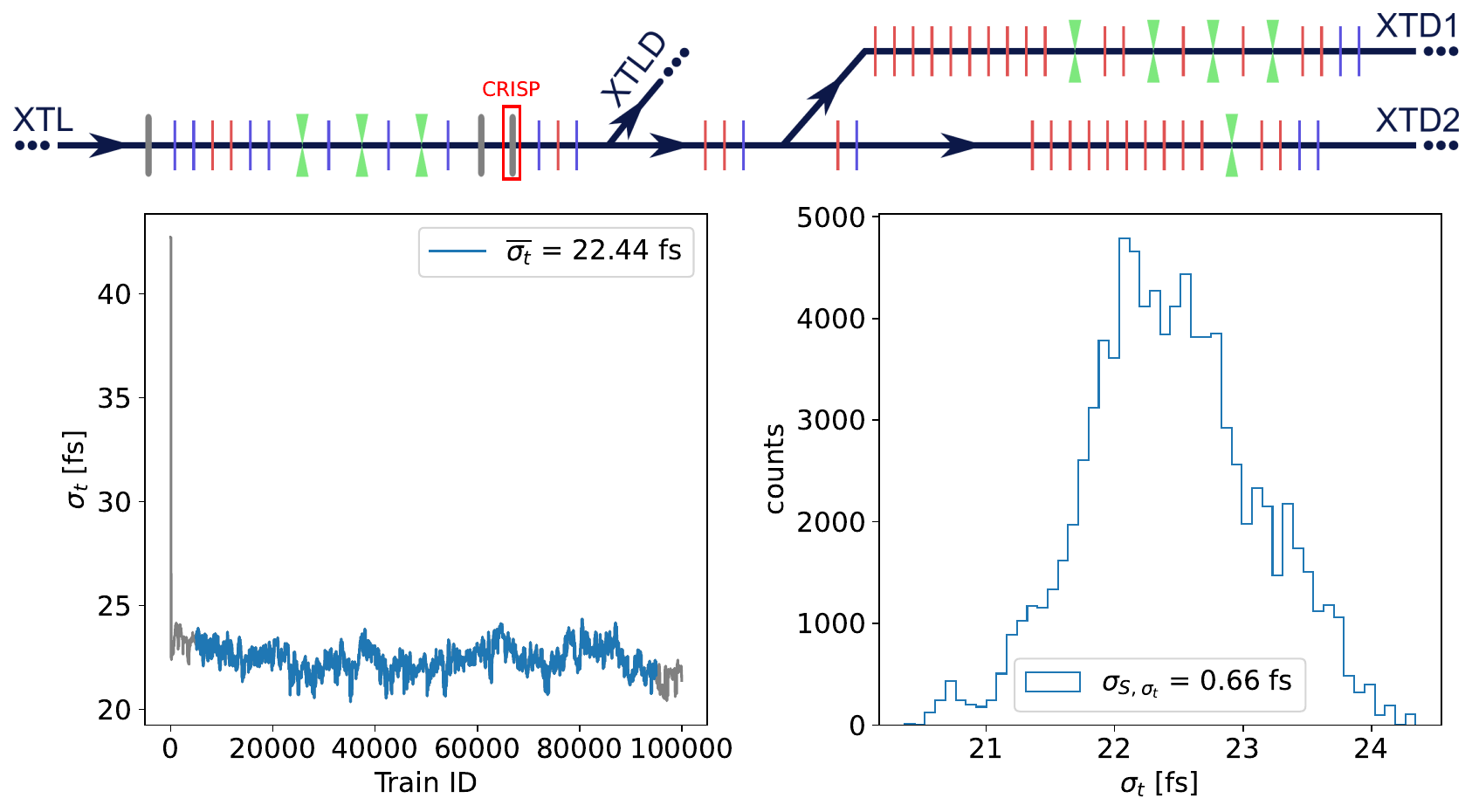}
    \caption{Variations of the bunch length $\sigma_t$ between trains. On the top is the diagnostics layout with a red rectangle indicating the bunch length monitor. On the left is the bunch length $\sigma_t$ for each train. On the right is the corresponding histogram. The greyed part of each signal is not used to measure the mean $\langle \sigma_t \rangle$ and for the histogram.}
    \label{fig:length_per_train}
\end{figure}

Measurements for $10^{5}$ consecutive bunch trains were performed and the resulting signal is shown in Figure~\ref{fig:length_per_train}.
Apart from the first bunch with lengths above \SI{40}{\femto\second}, the average bunch length value is around \SI{22.44}{\femto\second}. The total signal variation $\sigma_{S, \sigma_t}$ is around \SI{0.66}{\femto\second}. However, in multiple papers describing the CRISP diagnostic~\cite{crisp}, the bunch length resolution is estimated around \SI{8}{\femto\second}. This noise is way higher than the signal variation $\sigma_{S, \sigma_t}=0.66\ \rm fs$ actually measured in the EuXFEL switch yard. For our studies, the bunch length jitter $\sigma_{J, \sigma_t}$ will be estimated around \SI{1}{\femto\second}.


\subsection{Bunch energy}

\begin{figure}[!ht]
    \centering
    \includegraphics[width=\textwidth]{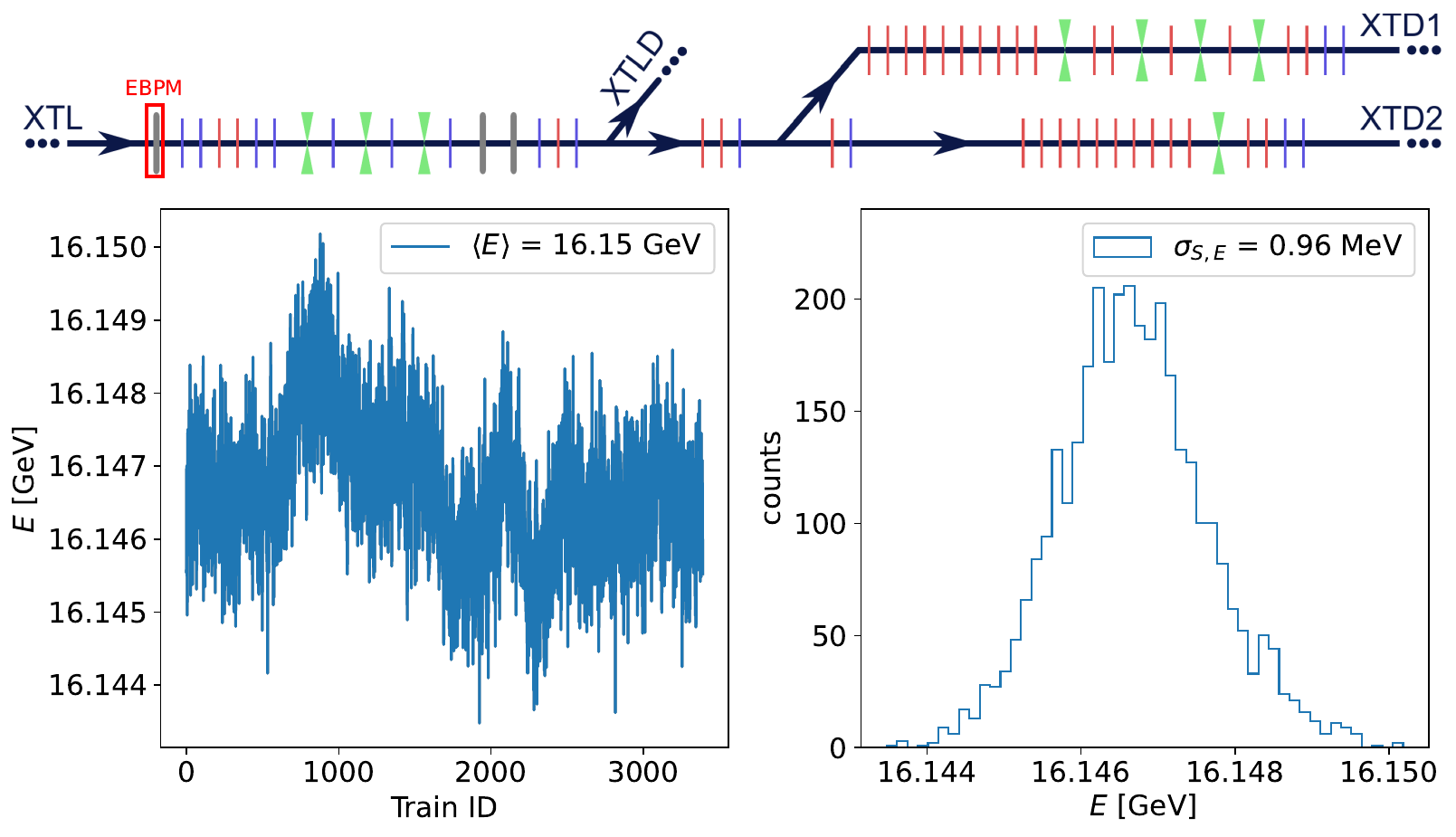}
    \caption{Variations of the bunch energy the XTL line between trains. On the top is the diagnostics layout with a red rectangle indicating the energy monitor. On the left is the bunch energy for each train. On the right is the corresponding histogram.}
    \label{fig:energy_per_train}
\end{figure}

As with the other measurements, the energy of the electron beam can be measured for all bunches in the train or for a single bunch ID between trains. Figure~\ref{fig:energy_per_train} shows the energy variation measurement at the EBPM for an arbitrary bunch ID between trains.

\newpage

\begin{figure}[!ht]
    \centering
    \includegraphics[width=\textwidth]{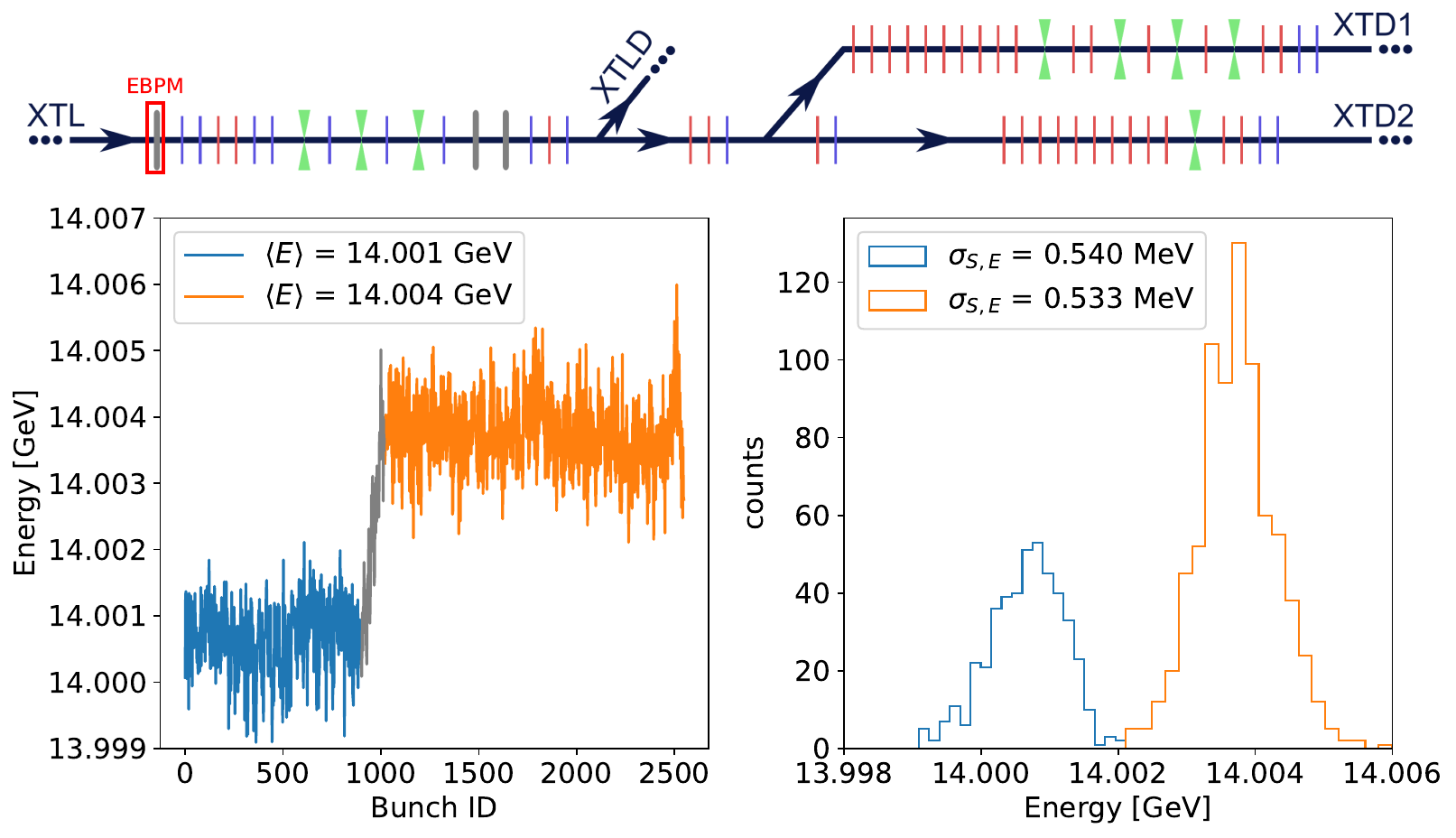}
    \caption{Variations of the bunch energy the XTL line inside a bunch train. On the top is the diagnostics layout with a red rectangle indicating the energy monitor. On the left is the bunch energy for each bunch ID. On the right is the corresponding histogram.}
    \label{fig:energy_per_bunch}
\end{figure}

The signal variation $\sigma_{S, E}$ is of the order of \SI{1}{\mega\electronvolt} which corresponds to a relative error $\frac{\sigma_{S,E}}{E_0} = 6.21\times10^{-3}\%$. In Figure~\ref{fig:energy_per_bunch}, a measurement of the energy variation inside a bunch train is also performed. This was done in a different configuration of the EuXFEL and with a reference energy $E_0 = 14\ \rm GeV$. 

Around the bunch ID 1000, there is a jump of the average energy $\langle E \rangle$ from \SI{14.001}{\giga\electronvolt} to \SI{14.004}{\giga\electronvolt}. Similar to the spikes seen in Figure~\ref{fig:bunch_jitter_ratio}, this might be attributed to a change in RF settings between the two undulator lines. The signal variation $\sigma_{S, E}$ can be calculated on both stable regions of the bunch pattern. In both cases, $\sigma_{S, E}$ stays around \SI{0.5}{\mega\electronvolt}. This correspond to a relative error of \SI{3.11e-3}{\%} of the central value.

\subsection{Summary table}

All the jitter measurements performed in this section are summarized in Table~\ref{tab:euxfel_param_jitter}. In this table, only the variation between trains are considered. This is because LUXE was expected to run at a rate of one bunch per train.

\begin{table}[!ht]
\centering
\begin{tabular}{| c | c || c | c | c |}
\hline
Parameter & Unit & Central value & $\sigma_J$ & $\sigma_N$ \\
\hline
\hline
$x$ & \SI{}{\micro\meter} & 0 & 1.5 & 1-4\footnotemark[1]{} \\
\hline
$y$ & \SI{}{\micro\meter} & 0 & 1.5 & 1-4\footnotemark[1]{} \\
\hline
$\sigma_x$ & \SI{}{\micro\meter} & 30 & - & 1.5 \\
\hline
$\sigma_y$ & \SI{}{\micro\meter} & 30 & - & 1.5 \\
\hline
$E$ & GeV & 16 & \SI{1e-3}{} & \SI{2.53e-4}{} \\
\hline
$q$ & pC & 250 & 1.25 & 0.5 \\
\hline
$\sigma_t$ & fs & 40 & $\sim1$ & $<1$ \\
\hline
$t$ & fs & 0 & 25.78 & 5 \\
\hline
\end{tabular}
\caption{Values of jitter $\sigma_J$ and noise $\sigma_N$ for all the measured beam parameters.}
\label{tab:euxfel_param_jitter} 
\end{table}

\footnotetext[1]{The noise for $x$ and $y$ is dependant of the BPM considered.}

The jitter $\sigma_J$ for the beam sizes can not be directly measured with the wires but if the beam size increases by a value greater than the noise $\sigma_N$ it could be identified with a wire scan. So for our studies, a maximum jitter equal to $\sigma_N$ will be assumed. This is now a good overview of the electron beam jitter at the end on the switch yard section, i.e. before the SASE undulators. 

To simulate accurate electron-laser interactions at LUXE's IP, those jitter must be propagated from the switch yard to the hypothetical location of the IP. The transport lattices should not impact the jitter and noise significantly. However, electron bunches will pass through the SASE2 undulator before reaching ELBEX and LUXE. As X-ray photons are emitted, the the electron bunch energy decreases whereas the emittance and the energy spread increases~\cite{undulator_impact}. This could be another source of jitter that is not considered in this study.

\chapter{ELBEX beam line design} \label{chap:design}

\section{Design requirements for the lattice}

In order to send electrons from the existing beam line to the ELBEX experiments, a new lattice needs to be designed. Among the requirements for this lattice, the first one is the ability to select and kick specific bunches in a bunch train towards the XTD8 tunnel. The lattice must then transport those bunches to the different experiments and satisfy the transverse size requirements for each focus point. 
This lattice must also be achromatic~\cite{achromatic}, i.e. it must produce an electron beam with no dispersion at the experimental locations. Looking at the total transfer matrix $\boldsymbol{R}$, this means that the dispersion factors $R_{16}$ and $R_{26}$ should be as small as possible. A non-zero dispersion would increase the beam size, particularly at the focus points of each experiments.
For the longitudinal part of the beam, to first order the lattice must conserve the upstream bunch length, i.e it must be an isochronous~\cite{isochronous} lattice. Looking at the transfer matrix $\boldsymbol{R}$, this means that the time-energy factor $R_{56}$ should be as small as possible. A non-zero value of $R_{56}$ would mean a compression or expansion of the beam longitudinally. If the design allows for a variable value of $R_{56}$, this could be useful for the uses to adapt the electron bunch length to the experiments needs. 
In addition, the lattice must minimise the impact of coherent synchrotron radiation (CSR) as much as possible. With a bending radius $\rho$ too low, the dipole could produce CSR photons and increases the beam emittance. Quadrupoles are used to optimise the experimental focus and can correct a variation of the beam size. However, such a correction is complicated if the CSR is not minimised.
%
Finally, the beam pipe in the XTD8 tunnel is shared between the ELBEX electrons and the X-ray photons going towards the downstream fusion experiment. The dipole angles must be constrained so that the electron path is collinear with the photon path after the last extraction dipole.

\section{Lattice layout}

\begin{figure}[!ht]
    \centering
    \includegraphics[width=\textwidth]{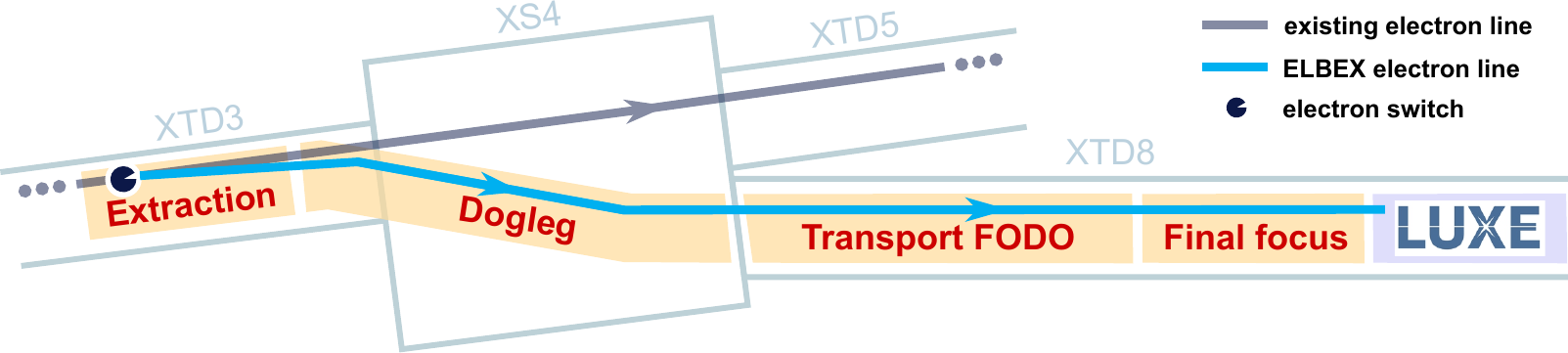}
    \caption{Plan layout of the four sections of the new beam line in the XTD8 tunnel.}
    \label{fig:lattice_sections}
\end{figure}

The XTD8 lattice is composed of four different sections. First the extraction to kick specific electron bunches from the train. Then a dogleg section to bend them back to the XTD8 tunnel. Next a FODO transport line to bring the beam to the experiments. Lastly the final focuses to get the required transverse size for each experiment. All those sections are shown in Figure~\ref{fig:lattice_sections}.
In the following lattice design, the electrons are transported from the existing line to LUXE's IP. Additional focus points for the other experiments will be potentially inserted in the FODO line in the future. 

\subsection{Extraction from the XTD5 lattice}

To provide ELBEX with electrons without impacting the beam usage 
downstream of the XTD5 line, a fast kicker system is needed to select specific bunches and send them towards the XTD8. Such a system is similar to what already exists at the EuXFEL switch yard.

\begin{figure}[!h]
    \centering
    \includegraphics[width=\textwidth]{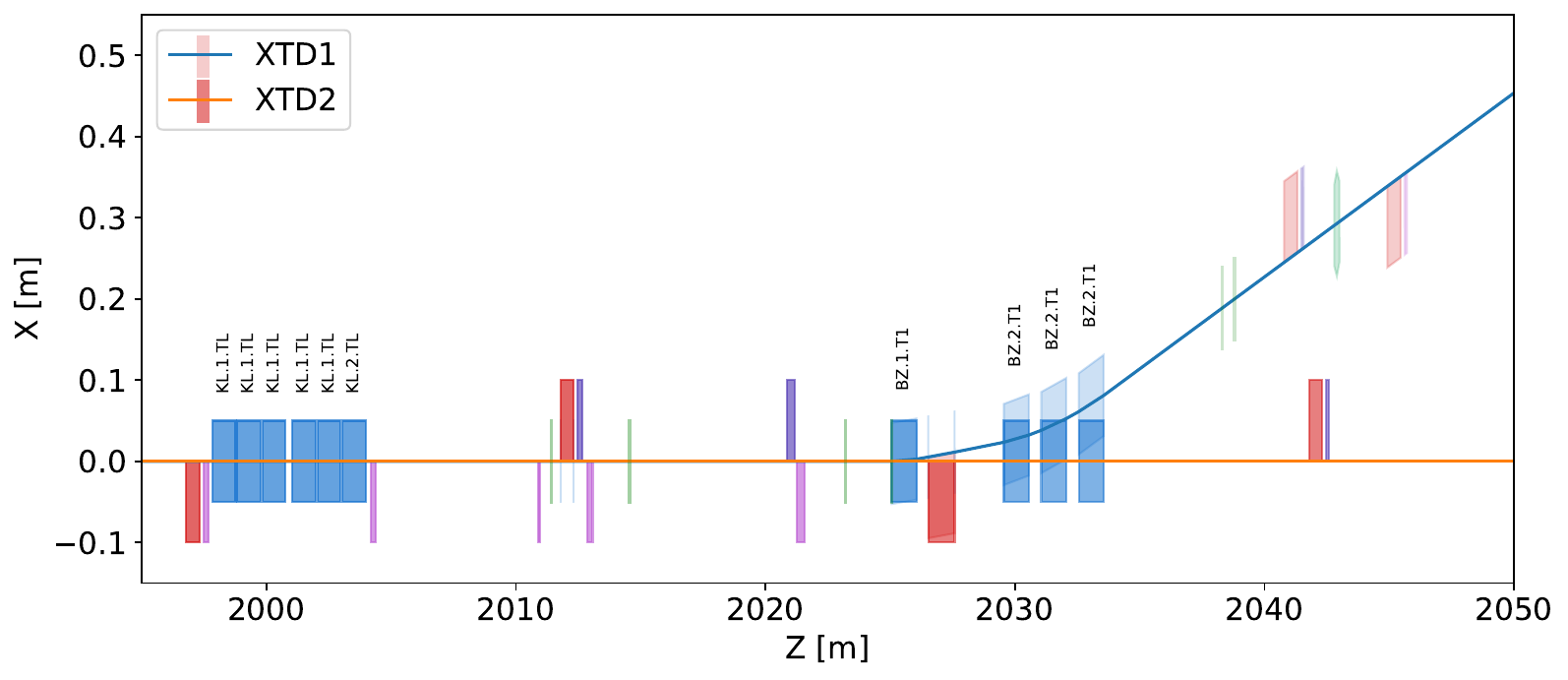}
    \caption{Plan layout of the EuXFEL's main switch yard. The blue line and light shaded elements represent the XTD1 path. The orange line and dark shaded elements represent the XTD2 path. Blue elements are the dipoles, red are quadrupoles, yellow are sextupoles, green are octupoles, purple and magenta are horizontal and vertical correctors.}
    \label{fig:euxfel_switchyard_magnets}
\end{figure}

In Figure~\ref{fig:euxfel_switchyard_magnets} there is a set of six 'KL' kicker dipoles about \SI{30}{\meter} upstream of the fork. Those kickers have a small bending angle $\alpha$ but can be switched on and off at a fast rate in order to affect only specific bunches in the train. On the \SI{30}{\meter} distance, those bunches will progressively deviate in $x$ from the XTD2 reference trajectory. Arriving at the fork, a set of four 'BZ' septa dipoles will affect only the bunches that are off-centred and properly kick them in the XTD1 beam line.

\begin{figure}[!h]
    \centering
    \includegraphics[width=\textwidth]{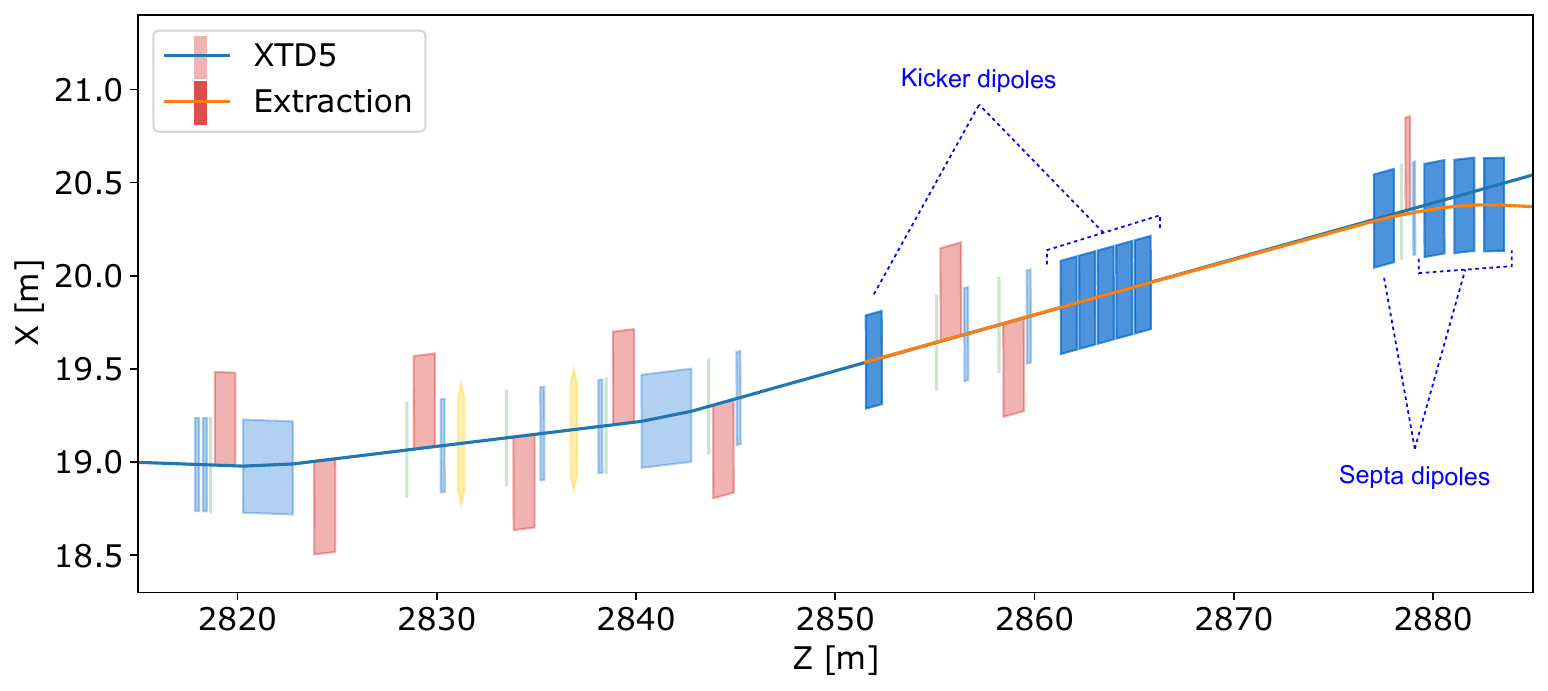}
    \caption{Plan layout of the XTD8 extraction. The blue line and light shaded elements represent the existing XTD5 path. The orange line and dark shaded elements represent the proposed XTD8 extraction.}
    \label{fig:XTD8_extraction}
\end{figure}

A similar design has been proposed for the XTD8 extraction in Figure~\ref{fig:XTD8_extraction}. There are again six horizontal kickers with the first one around \SI{30}{\meter} upstream of the beam line separation. The other 5 are separated to accommodate for quadrupoles already present in the XTD5 lattice. In the same way, four septa dipoles are used to do the proper line separation. There is enough space between the two first septas to fit another XTD5 quadrupole.
The exact placement of those \SI{10}{} magnets along $s$ is not a major constraint as long as there is enough distance between the kickers and the septas to separate the bunches.

\begin{figure}[!ht]
    \centering
    \includegraphics[width=11cm]{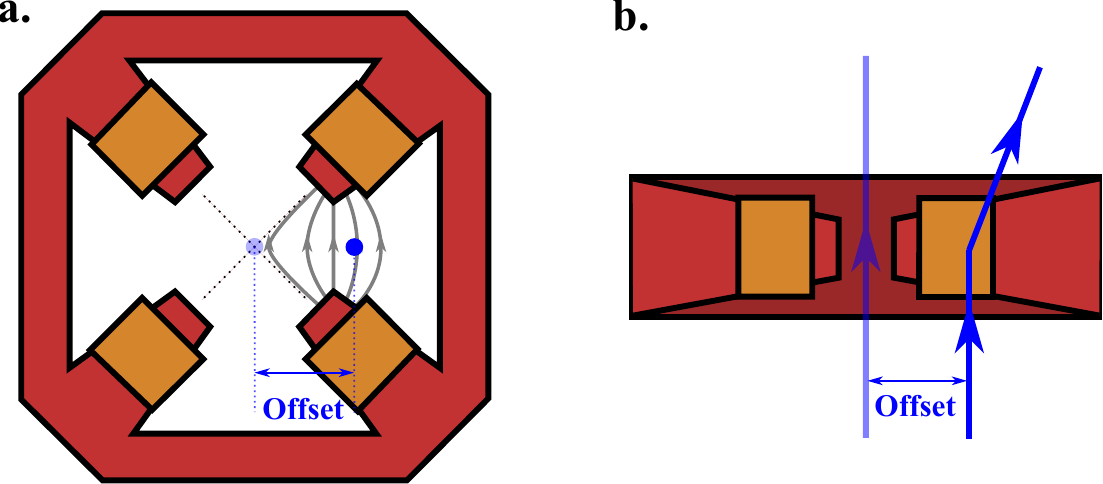}
    \caption{Schematics of an off-centred beam going through a quadrupole. a.) Front side with the field lines. b.) Top side with the beam path.}
    \label{fig:quad_off_centered}
\end{figure}

Even if moving the existing magnets is a possibility, this design aims to maintain the optics of the downstream XTD5 without changing the strength of the mentioned quadrupoles.
As a trade-off, the impact of those quadrupoles on the XTD8 reference path must be considered. When an off-centred electron bunch crosses a quadrupole it will experience a horizontal dipole kick as shown in Figure~\ref{fig:quad_off_centered}. 
The transfer matrix for a quadrupole in the thin-lens approximation propagates as

\begin{equation}
    \begin{pmatrix}
        x \\
        x' \\
    \end{pmatrix}
    =
    \begin{pmatrix}
        1 & 0 \\
        \pm k_1L_Q & 1\\
    \end{pmatrix}
    \cdot
    \begin{pmatrix}
        x_0 \\
        x'_0 \\
    \end{pmatrix},
\end{equation}

where $(x_0, x'_0)$ are the input position and angle, $(x, x')$ are the output position and angle, $L_Q$ is the length of the quadrupole and $k_1$ is the quadrupole strength. The bottom left term's sign indicates if the quadrupole is focusing or defocusing in the $x$ plane. Considering an input beam with a position offset $\delta x$, the output angle becomes

\begin{equation}
    x'=\pm k_1L_Q\delta x + x'_0.
\end{equation}

However, it is not trivial to model this off-centred quadrupole in a tracking code like Xsuite. The reference orbit is expected to go through the centre of each lattice element. Changing the reference orbit can not be done with a quadrupole, a dipole must be used for that. But the quadrupolar field needs to be conserved to correctly affect particles with a $y$ offset.
A workaround method is to define the off-centred quadrupole as a dipole of bending angle $\alpha=\pm k_1L_Q\delta x$ with an additional quadrupolar field of strength $k_1$. 

\subsection{Dogleg to XTD8}

\begin{figure}[!ht]
    \centering
    \includegraphics[width=\textwidth]{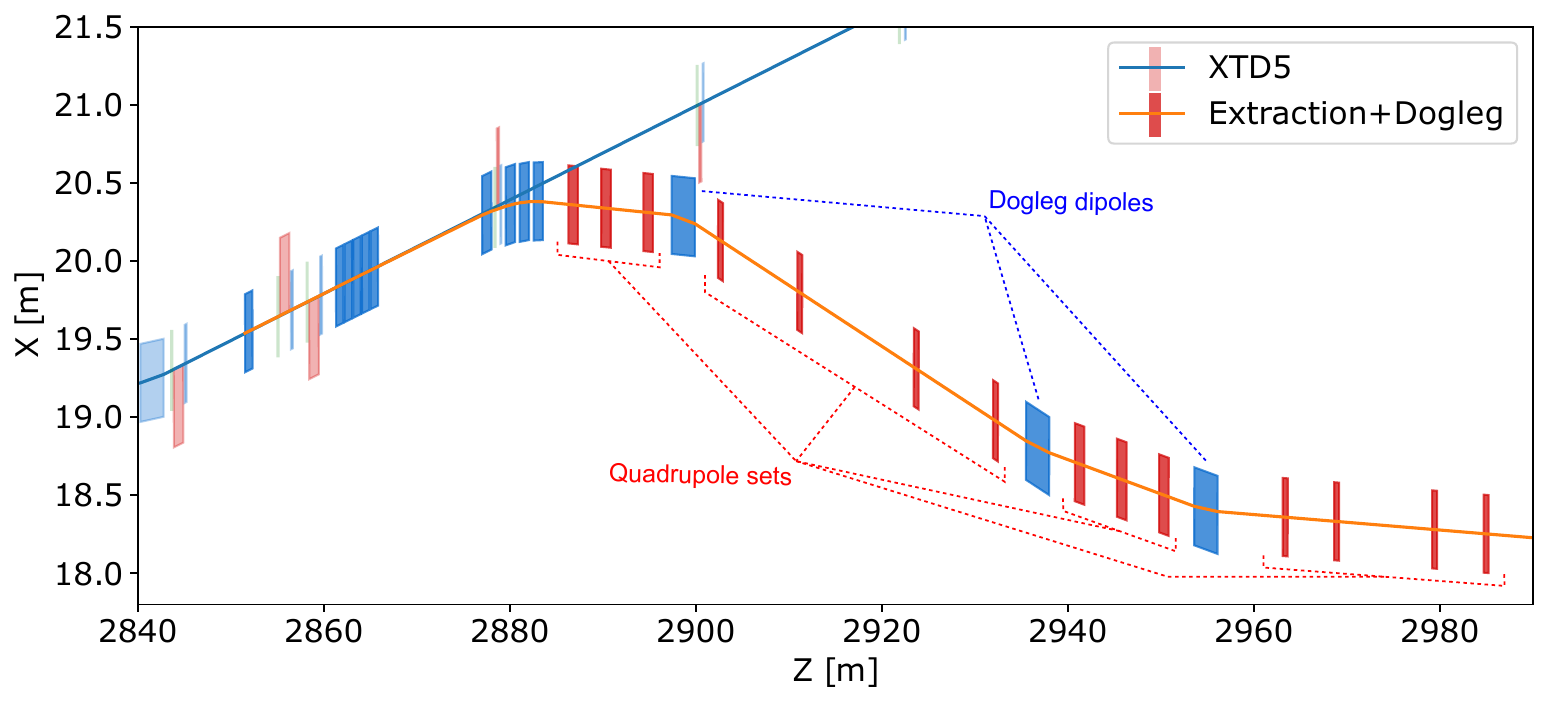}
    \caption{Plan layout of the XTD8 extraction and dogleg. The blue line and light shaded elements represent the existing XTD5 path. The orange line and solid shaded elements represent the XTD8 extraction and dogleg. The three dipoles are indicated with the blue dash lines, and the fourteen quadrupoles with the red dash lines.}
    \label{fig:XTD8_extraction_dogleg}
\end{figure}

After the extraction, a dogleg lattice is used to bring the electrons back onto the photon path. Figure~\ref{fig:XTD8_extraction_dogleg} shows that the dogleg is composed of three dipoles in between quadrupole triplets and quadruplets. There are a total of fourteen quadrupoles in the dogleg.
Beyond the geometric constraints, quadrupoles of this dogleg are initially placed at reasonable locations between the dipoles. 
Optimizing the location and the strength of those quadrupoles using Xsuite will allow for both the achromatic and isochronous properties of the lattice.
Similarly, the bending angles of the dipoles were initially set to coincide approximately with the photon path. In the complete lattice design they are adjusted using Xsuite to be properly aligned.

\subsection{FODO transport} \label{sec:FODO}

\begin{figure}[!ht]
    \centering
    \includegraphics[width=12cm]{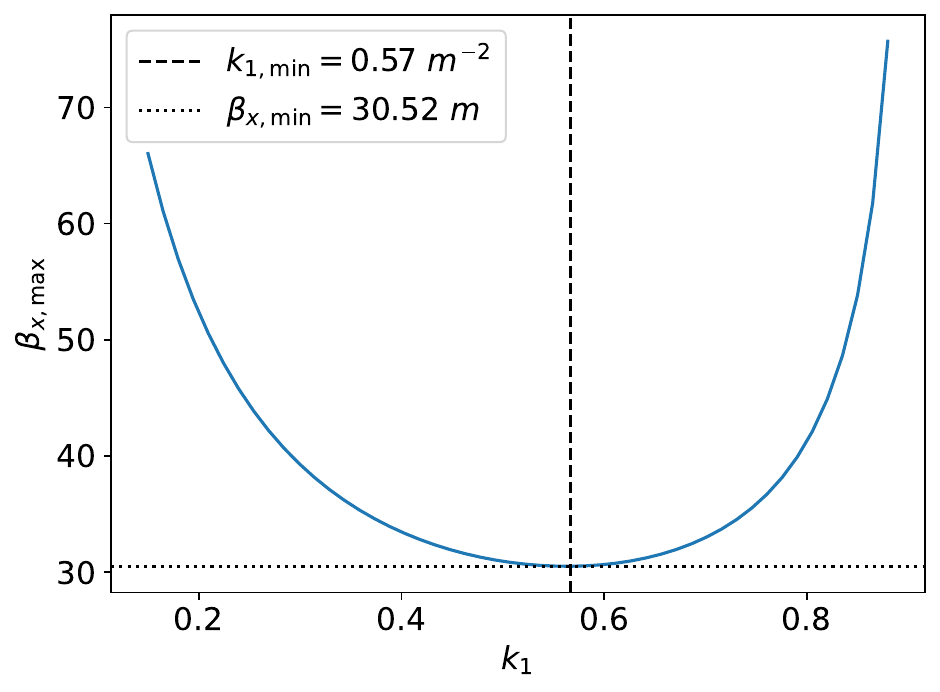}
    \caption{Scan of the maximum value of horizontal beta $\beta_{x, \rm max}$ with respect to the quadrupole strength $k_1$.}
    \label{fig:opt_beta}
\end{figure}

In the XTD8 tunnel the goal of the lattice is to transport the electron beam to the different experiments. A basic FODO cell is a sufficient choice for this purpose. The quadrupoles used are QE~\cite{euxfel_quads} quadrupoles which are standard FODO quadrupoles in the EuXFEL.
For a given drift length and quadrupole length, a single FODO cell can be modelled. By scanning for different values of $k_1$, one can find the maximum value of $\beta_x$ and $\beta_y$. The range of possible $k_1$ for the FODO cell is bound by the stability condition~\cite{tbd} which is expressed as

\begin{equation}
    |{\rm Tr}(R_{\rm FODO})| > 2,
\end{equation}

where $R_{\rm FODO}$ is the transfer matrix of one FODO cell. For a thin quadrupole FODO, this condition is reduced to 

\begin{equation}
    \frac{L_D}{f} > 4,
\end{equation}

where $L_D$ is the drift length and $f$ the quadrupole focal length. To get from the end of the dogleg to LUXE's final focus, a total length of \SI{74.1}{\meter} is estimated. The length $L_Q$ of the QE quadrupole is \SI{0.24}{\meter}. With four FODO cells $L_D$ is then around \SI{9.023}{\meter}. 
In Figure~\ref{fig:opt_beta} the single FODO cell is modelled for different values of $k_1$ with $L_D=9.023\ \rm m$ and $L_Q=0.24\ \rm m$. The minimal value of $\beta_{x, {\rm max}}$ is found at $k_{1, {\rm min}} = 0.57\ \rm m ^{-2}$.

\begin{figure}[!ht]
    \centering
    \includegraphics[width=\textwidth]{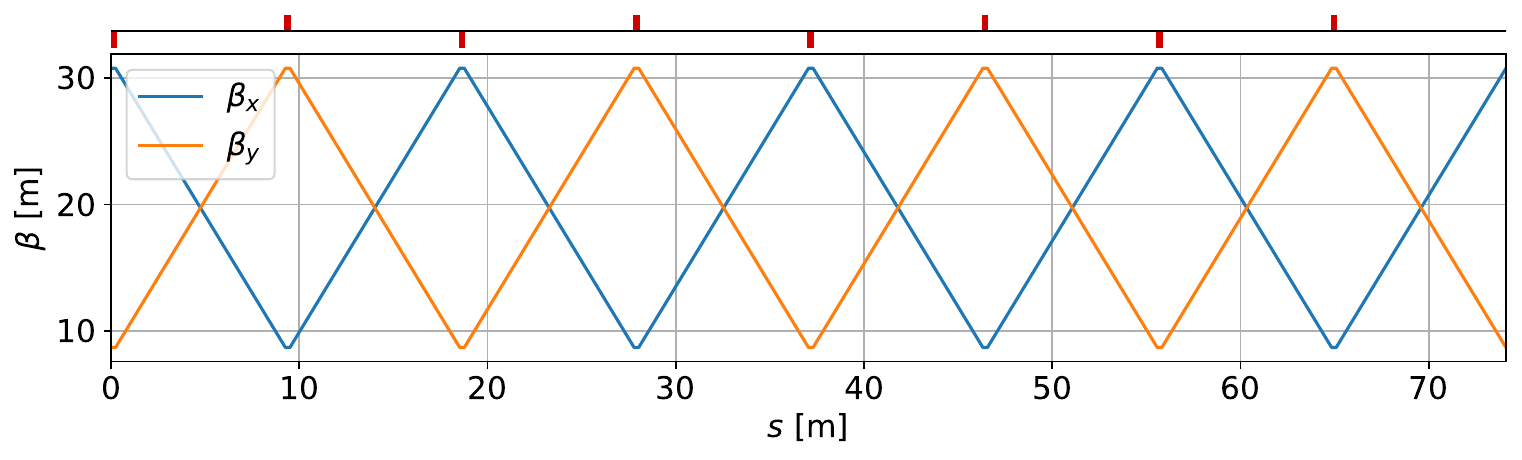}
    \caption{Beta functions of the FODO transport line. It is composed of 4 cells.}
    \label{fig:FODO_beta}
\end{figure}

With this value of $k_{1,{\rm min}}$ found in Figure~\ref{fig:opt_beta}, the four cells lattice in Figure~\ref{fig:FODO_beta} can be built.
The total length of this lattice, hence the length of the drifts $L_D$, are arbitrarily set to fill the distance between the dogleg and LUXE's final focus. It can be easily adjusted afterwards to the precise location of LUXE or to accommodate for the other experiments.

\subsection{Final focus for LUXE} \label{sec:FF}

The role of a final focus is to reduce the transverse size of the beam to a required value $\sigma^*$ at a distance $L^*$ of the last magnet. The most basic method to focus a beam in both planes is with a focusing doublet~\cite{doublet}.
A single quadrupole is focusing in one transverse plane and defocusing in the other. The transfer matrix $\boldsymbol{R}_{\rm QF}$ for a focusing quadrupole in the $x$ plane propagates the positions and angles in both transverse planes as

\begin{equation}
    \begin{pmatrix}
        x \\ x' \\ y \\ y' \\
    \end{pmatrix} = \boldsymbol{R}_{\rm QF} \cdot
    \begin{pmatrix}
        x_0 \\ x'_0 \\ y_0 \\ y'_0 \\
    \end{pmatrix} =
    \begin{pmatrix}
        1 & 0 & & \\
        -k_1L_Q & 1 & & \\
        & & 1 & 0 \\
        & & k_1L_Q & 1
    \end{pmatrix} \cdot
    \begin{pmatrix}
        x_0 \\ x'_0 \\ y_0 \\ y'_0 \\
    \end{pmatrix}.
\end{equation}

A focusing doublet uses a focusing and a defocusing quadrupole separated by a drift of length $L_D$. The resulting transfer matrix is

\begin{equation}
    \boldsymbol{R}_{\rm doublet} = 
    \boldsymbol{R}_{\rm QF}
    \boldsymbol{R}_{\rm drift}
    \boldsymbol{R}_{\rm QD} =
    \begin{pmatrix}
        1+k_1L_QL_D & L_D & & \\
        -L_D(k_1L_Q)^2 & 1-k_1L_QL_D & & \\
        & & 1-k_1L_QL_D & L_D \\
        & & -L_D(k_1L_Q)^2 & 1+k_1L_QL_D \\
    \end{pmatrix}.
\end{equation}

Assuming that $L_D >> k_1L_Q$, the propagation of the transverse positions and angles is expressed by

\begin{equation}
    \begin{pmatrix}
        x \\ x' \\ y \\ y' \\
    \end{pmatrix} = 
    \begin{pmatrix}
        1 & L_D &  & \\
        -L_D(k_1L_Q)^2 & 1 & & \\
        & & 1 & L_D \\
        & & -L_D(k_1L_Q)^2 & 1 \\
    \end{pmatrix} \cdot
    \begin{pmatrix}
        x_0 \\ x'_0 \\ y_0 \\ y'_0 \\
    \end{pmatrix}.
\end{equation}

\begin{figure}[!ht]
    \centering
    \includegraphics[width=12cm]{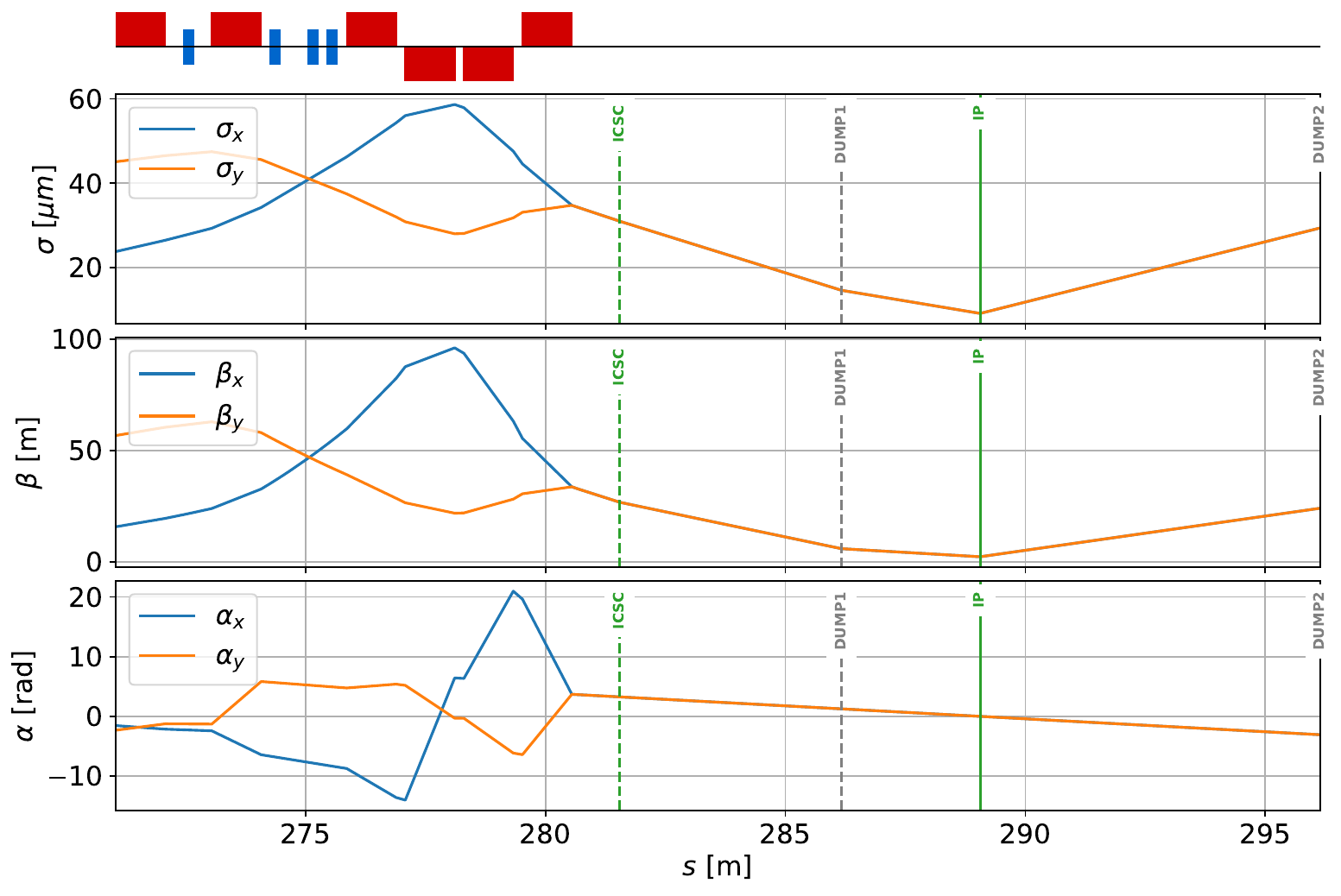}
    \caption{Twiss parameters as function of $s$ for the final focus proposed in the XTD20 lattice. On top is the beam size, in the middle the beta and on the bottom the alpha.}
    \label{fig:XTD20_FF_IP_optics}
\end{figure}

In both planes, the matrix is equivalent to a focusing quadrupole of length $L_D$ and quadrupole strength $(k_1L_Q)^2$. As such a quadrupole doublet focuses the beam on both planes. In a final focus, additional quadrupoles are used to match with the upstream twiss functions.
For both of LUXE's focus points, $\sigma^*$ must be at most \SI{10}{\micro\meter}. At first order, the final focus used for LUXE is the same proposed in Figure~\ref{fig:XTL_XTD20} for the XTD20 lattice. The optics of this final focus are shown in  Figure~\ref{fig:XTD20_FF_IP_optics}

At LUXE's IP the beam is focused to a transverse size of around \SI{9.1}{\micro\meter}. In this configuration, the highest value of $\beta$ is \SI{100}{\meter} which correspond to a beam size below \SI{60}{\micro\meter}. This final focus was matched with a specific set of input twiss parameters, different from the transport FODO output twiss parameters shown in Figure~\ref{fig:FODO_beta}. To adjust this final focus for the XTD8 lattice, the matching functions of Xsuite are used. To perform the matching of a lattice, Xsuite requires the user to define both variables and targets: 
\begin{itemize}
    \item The variables can be a magnet strength, an element angle or a numerical combination like the sum of all drifts length in the lattice. Each variable is given a range of value and a step size. 
    \item The targets are usually twiss parameters ($\beta$, $\alpha$, $\eta$, etc.) but can be define as a custom function by the user. Each target has location on lattice, a value to target and a tolerance.
\end{itemize}

\begin{figure}[!b]
    \centering
    \includegraphics[width=12cm]{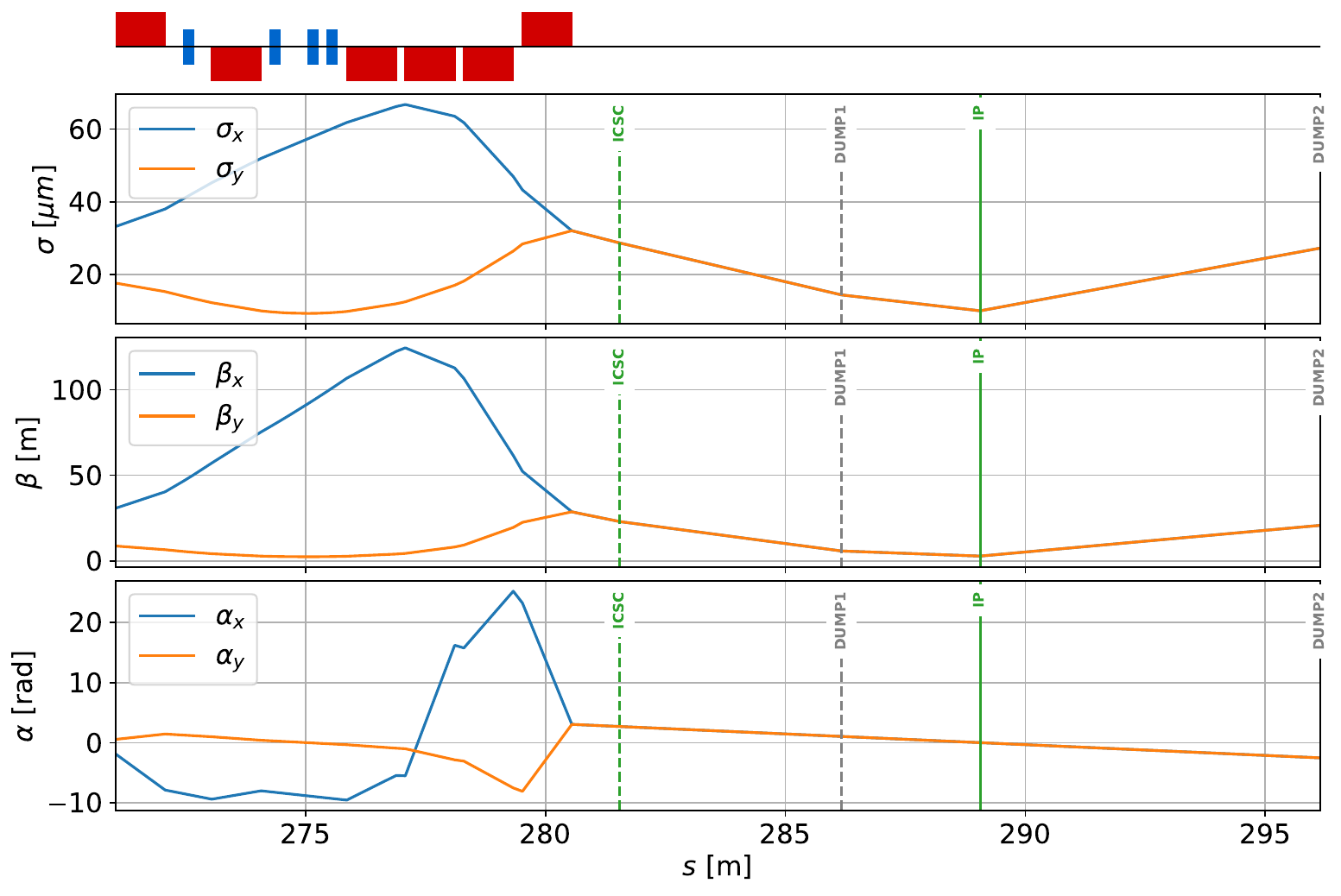}
    \caption{Twiss parameters as function of $s$ for the final focus rematched to LUXE's IP in XTD8. On top is the beam size, in the middle the beta and on the bottom the alpha.}
    \label{fig:XTD8_FF_IP_optics}
\end{figure}

\begin{figure}[!ht]
    \centering
    \includegraphics[width=12cm]{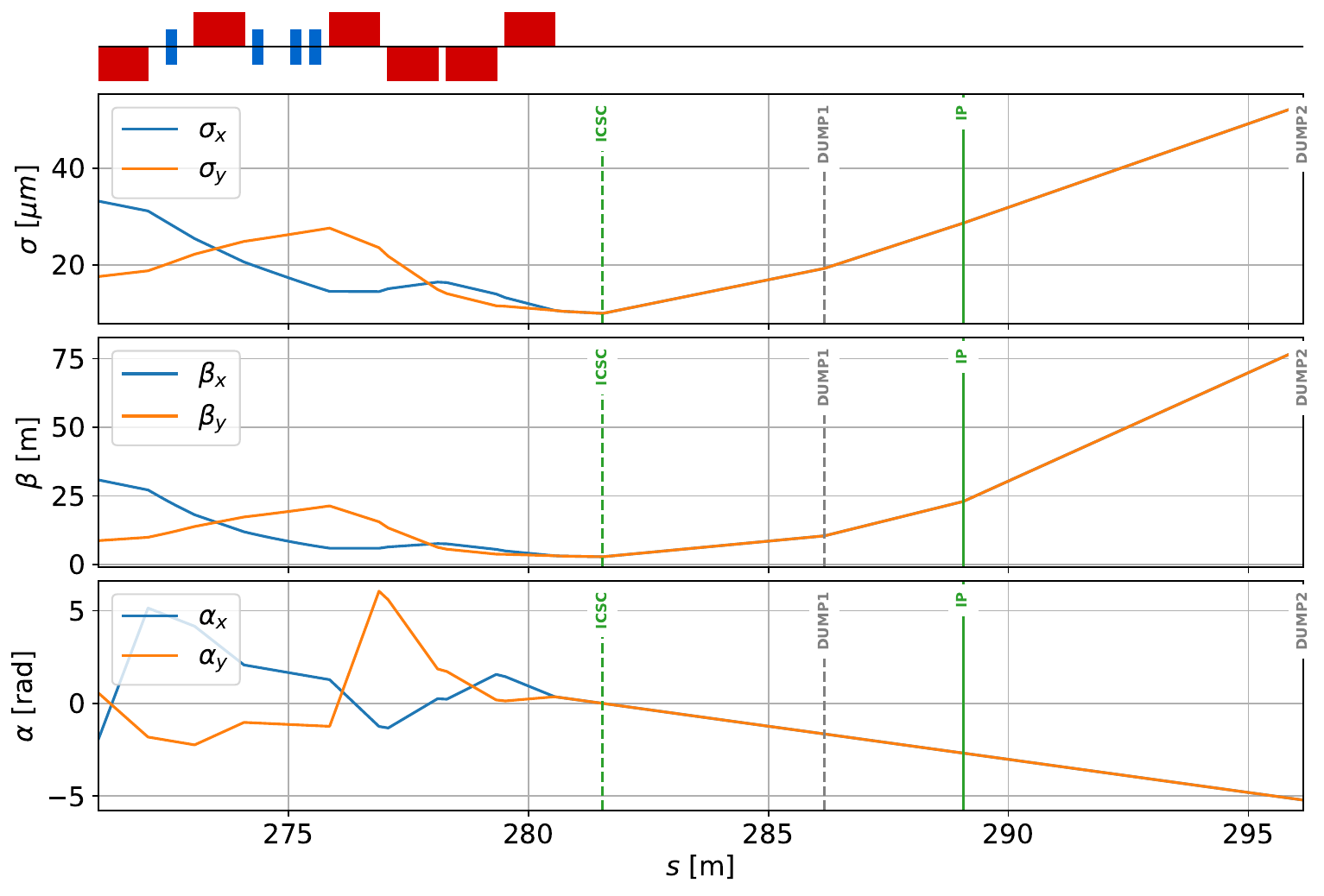}
    \caption{Twiss parameters as function of $s$ for the final focus rematch to LUXE's ICS chamber in XTD8. On top is the beam size, in the middle the beta and on the bottom the alpha.}
    \label{fig:XTD8_FF_ICSC_optics}
\end{figure}

 When running the match function, Xsuite uses an optimisation algorithm to change every variable as independent knobs and get all targets to the desired values within the tolerances. In the context of LUXE's final focus there are two sets of targets. First, the transverse beam size at focus must be at most \SI{10}{\micro\meter} in both planes. Then the twiss function $\beta$ and $\alpha$ at the beginning of the final focus must match the end of the transport FODO. The strengths $k_1$ of the six quadrupoles are the variables. The result functions in the IP focus configuration are shown in Figure~\ref{fig:XTD8_FF_IP_optics}.
Here the maximum value of $\beta$ is a bit higher than \SI{100}{\meter} and the beam size is just above \SI{60}{\micro\meter}. This slight increase stays reasonable for what is expected in the final focus.
In the same way, the lattice is matched to get the focus point in the ICS chamber. The results are shown in Figure~\ref{fig:XTD8_FF_ICSC_optics}. The \SI{10}{\micro\meter} beam size is also achieved in this configuration. Due to the small $L^*$ of the ICS chamber, the values of $\beta$ are kept rather low along the final focus. 

\begin{figure}[!ht]
    \centering
    \includegraphics[width=\textwidth]{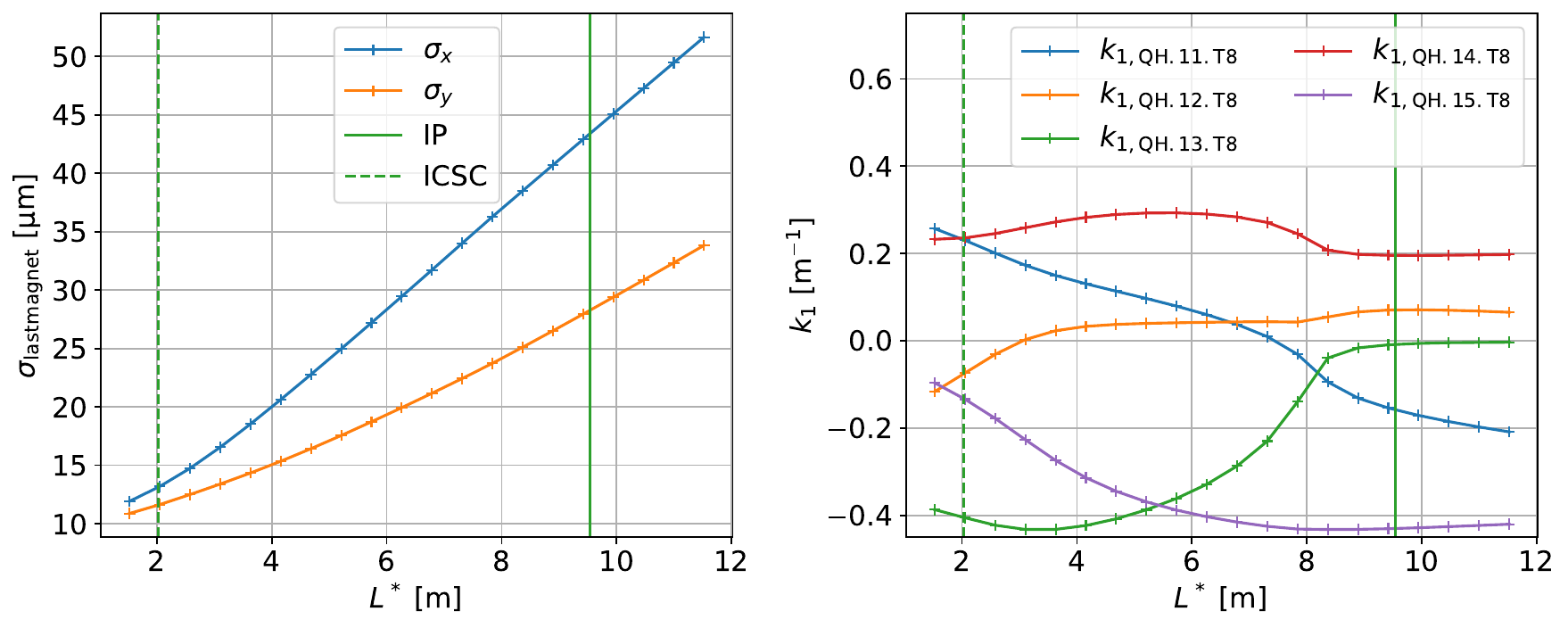}
    \caption{Scan for different focusing distance $L^*$. All points are focused at $\sigma^*=10\mu m$. On the left is the impact on the beam size in the last final focus magnet $\sigma_{\rm lastmagnet}$. On the right are the values of $k_1$ for the different values of $L^*$.}
    \label{fig:knob_LStar}
\end{figure}

\begin{figure}[!ht]
    \centering
    \includegraphics[width=\textwidth]{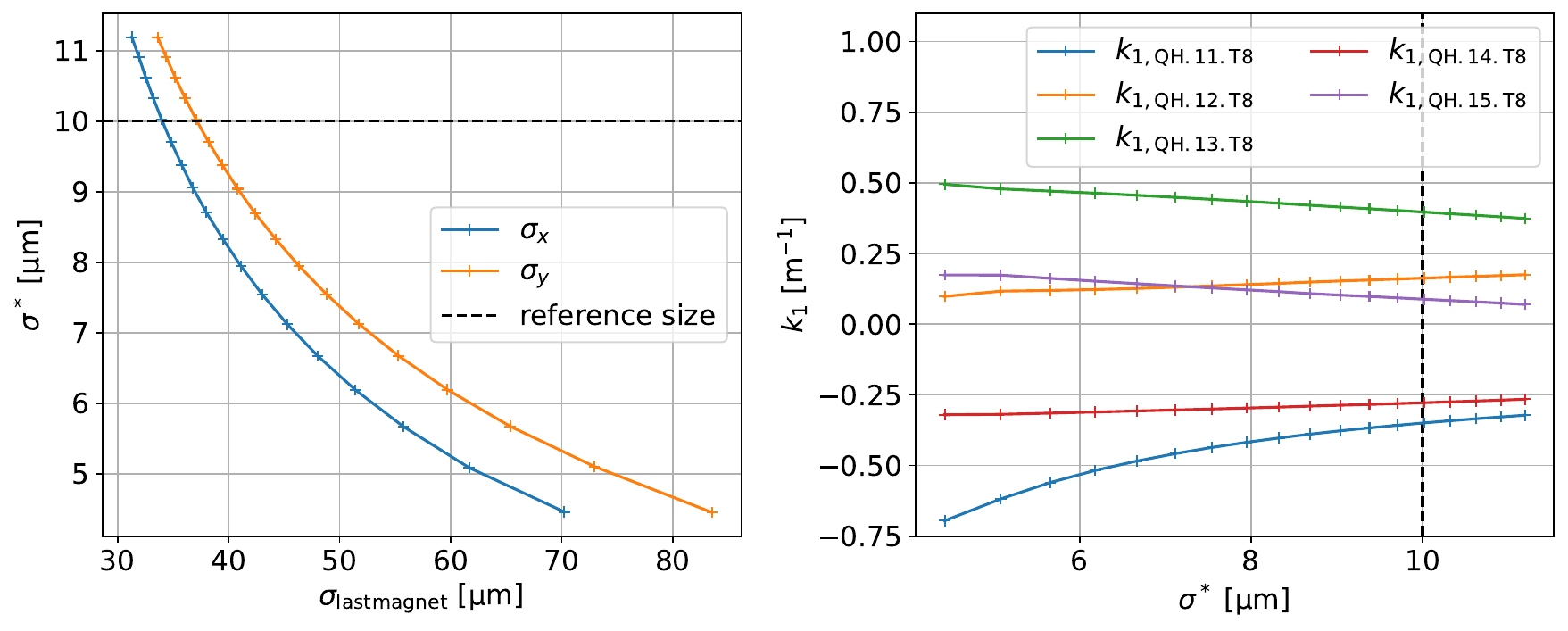}
    \caption{Scan for different beam sizes $\sigma^*$ at LUXE's IP. On the left is the impact on the beam size in the last final focus magnet. On the right are the values of $k_1$ for the different values of $\sigma^*$.}
    \label{fig:knob_sigma_IP}
\end{figure}

The IP and the ICS chamber are currently the two main points where the electron beam will be focused for LUXE. In the future, adjustments on the experiment could require a focus point at a different $L^*$. To offer some flexibility, it is useful to establish a $L^*$ knob, i.e. to associate a set of quadrupole strength $k_1$ to each value of $L^*$. It is done by changing the focus position and running the Xsuite match again. In Figure~\ref{fig:knob_LStar} a range of $L^*$ between \SI{1.5}{\meter} and \SI{11.5}{\meter} is explored. For all points, $\sigma^*$ is set to $10\ \rm \mu m$.
Changing the distance between the quadrupoles and the focus point has an impact on the beam size at the last quadrupole $\sigma_{\rm lastmagnet}$. The higher the $L^*$, the larger beam size is needed in the quadrupoles to focus at \SI{10}{\micro\meter}. Apart for the low values of $L^*$ the relationship between $L^*$ and $\sigma_{\rm lastmagnet}$ is linear. At LUXE's IP, the intensity of the produced signal is dependant on the electron beam transverse size. A beam smaller than the required \SI{10}{\micro\meter} will produce more strong-field QED particles but will requires stronger quadrupoles. Similar to the $L^*$ knob, creating a $\sigma^*$ knob will offer a set of quadrupole strengths for each desired beam size. In Figure~\ref{fig:knob_sigma_IP} the final focus has been rematched for a range of $\sigma^*$ between \SI{11}{\micro\meter} and \SI{4.5}{\micro\meter} at the IP.

As expected, a lower $\sigma^*$ requires both higher values of $k_1$ and a larger beam size in the quadrupoles. In this configuration, the lowest beam size achieved is below \SI{5}{\micro\meter}. Attempts at even lower $\sigma^*$ were run but the Xsuite optimizer did not converge on a solution. Even if a $\sigma^*$ below \SI{4.5}{\micro\meter} is possible, 
it might require a redesign of the final focus.
Below $\sigma^* = 5\ \rm \mu m$ the values of $k_1$ are going as low as \SI{-0.7}{\meter^{-2}}. The QH quadrupoles in the EuXFEL can produce values of $|k_1|$ up to \SI{0.55}{\meter^{-2}}, so a different type of quadrupole must be use to reach \SI{-0.7}{\meter^{-2}}. 

\begin{figure}[!ht]
    \centering
    \includegraphics[width=\textwidth]{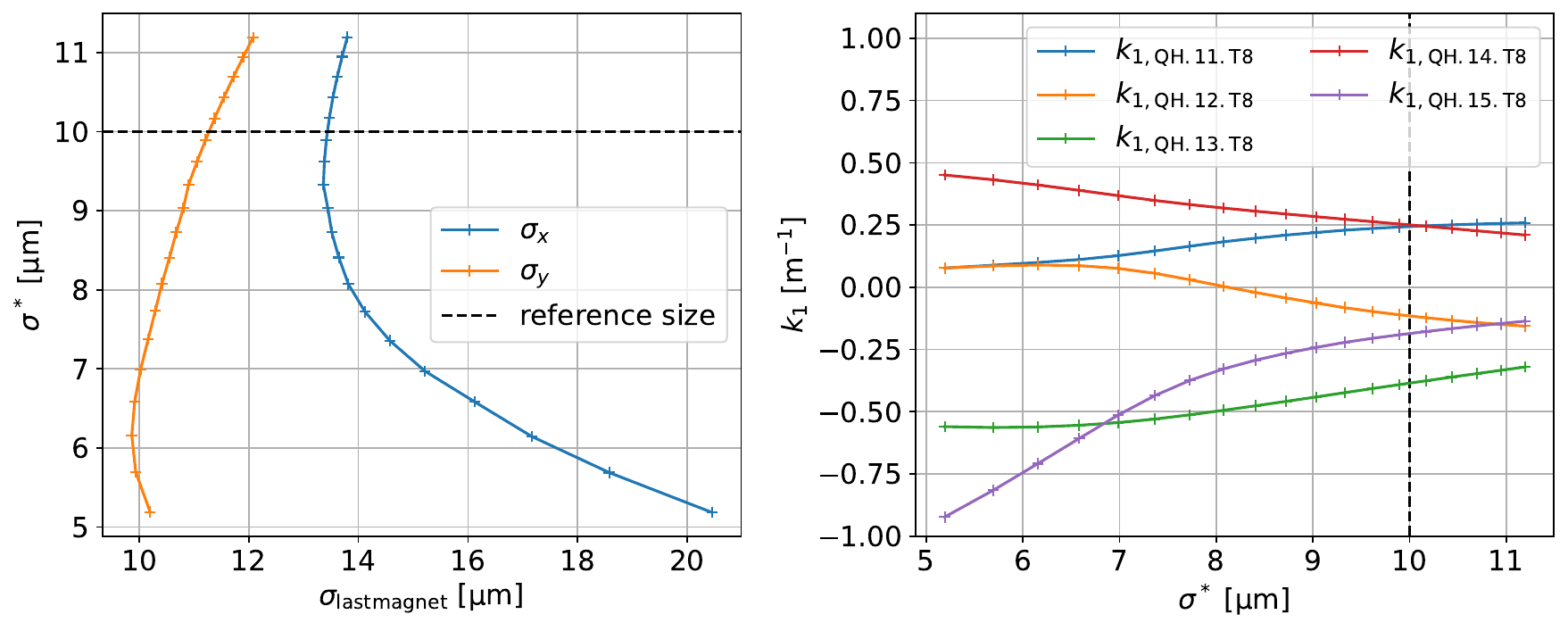}
    \caption{Scan for different beam sizes $\sigma^*$ at LUXE's ICS chamber. On the left is the impact on the beam size in the last final focus magnet. On the right are the values of $k_1$ for the different values of $\sigma^*$.}
    \label{fig:knob_sigma_ICSC}
\end{figure}

Similarly in Figure~\ref{fig:knob_sigma_ICSC}, a $\sigma^*$ knob has been made for the ICS target focus.
The Xsuite optimization got to a beam size of almost \SI{5}{\micro\meter}. The amplitude in $\sigma_{\rm lastmagnet}$ is lower than for the previous knob, especially in the $y$ plane. This could be explained by the low $L^*$ value for the ICS chamber. In Figure~\ref{fig:XTD8_FF_ICSC_optics}, the values of $\beta$ (thus the values of $\sigma$) did not increase in the quadrupoles to get the required \SI{10}{\micro\meter}. At the lowest beam size, the last quadrupole $k_1$ goes down to \SI{-0.9}{\meter^{-2}} which is even lower than for the IP knob.

\section{Matching the complete lattice}

\begin{figure}[!ht]
    \centering
    \includegraphics[width=\textwidth]{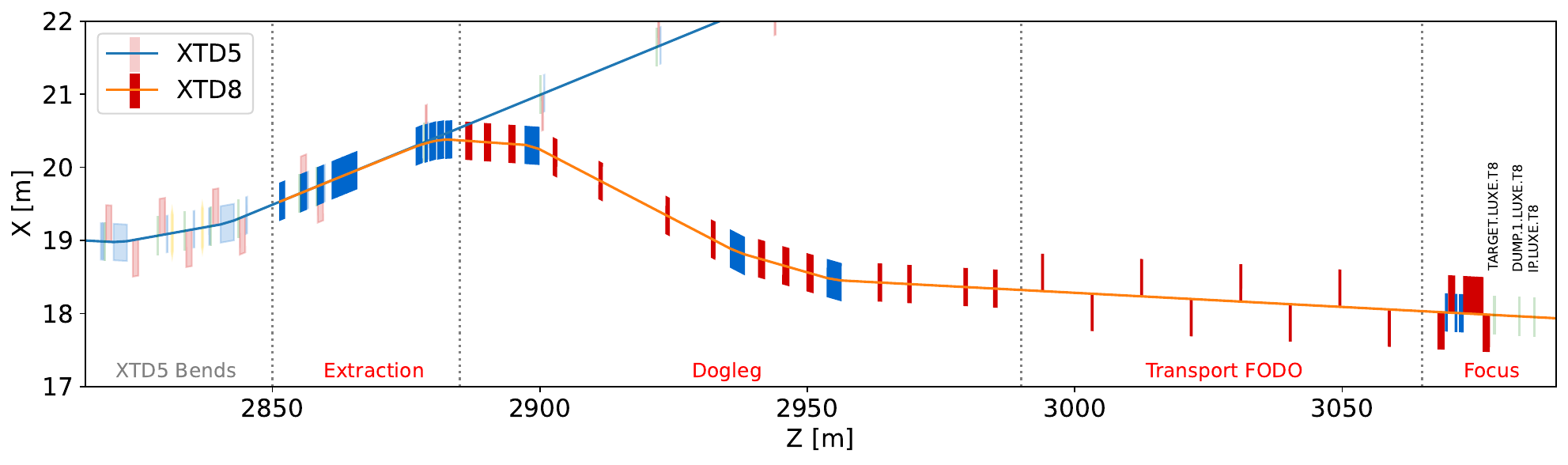}
    \caption{Plan layout of the entire XDT8 lattice. The blue line and light shaded elements represent the existing XTD5 path. The orange line and solid shaded elements represent the proposed XTD8 lattice.}
    \label{fig:XTD8_top_view_details}
\end{figure}

All the previous lattices combined form a beam line of around \SI{250}{\meter} for which the layout can be seen in Figure~\ref{fig:XTD8_top_view_details}. A summary of all magnets used in each section can be found in Table~\ref{tab:nb_magnets}.
The position and angle of the kickers, septas and sector bends are already defined because of geometric constraints. Moreover, the quadrupoles in the FODO line and the final focus where also adjusted in Section~\ref{sec:FODO} and Section~\ref{sec:FF}. Only the quadrupoles in the dogleg are left without a defined strength and with currently arbitrary positions. The goal is now to optimize the strength and position of those quadrupoles to match the rest of the line.

\begin{table}[!ht]
\centering
\begin{tabular}{| c || c | c | c | c || c |}
\hline
 & Extraction & Dogleg & FODO & Focus & Total \\
\hline
\hline
Fast kicker & 6 & - & - & - & 6 \\
\hline
Septa & 4 & - & - & - & 4 \\
\hline
Sector bend & - & 3 & - & - & 3 \\
\hline
Correctors & - & - & - & 4 & 4 \\
\hline
Shared quadrupoles & 3 & - & - & - & 3 \\
\hline
New quadrupoles & - & 14 & 8 & 6 & 28 \\
\hline
\hline
Total & 13 & 17 & 8 & 10 & 48 \\
\hline
\end{tabular}
\caption{Number of magnets for each section of the new line.}
\label{tab:nb_magnets} 
\end{table}

\subsection{Twiss functions matching}

\begin{figure}[!ht]
    \centering
    \includegraphics[width=\textwidth]{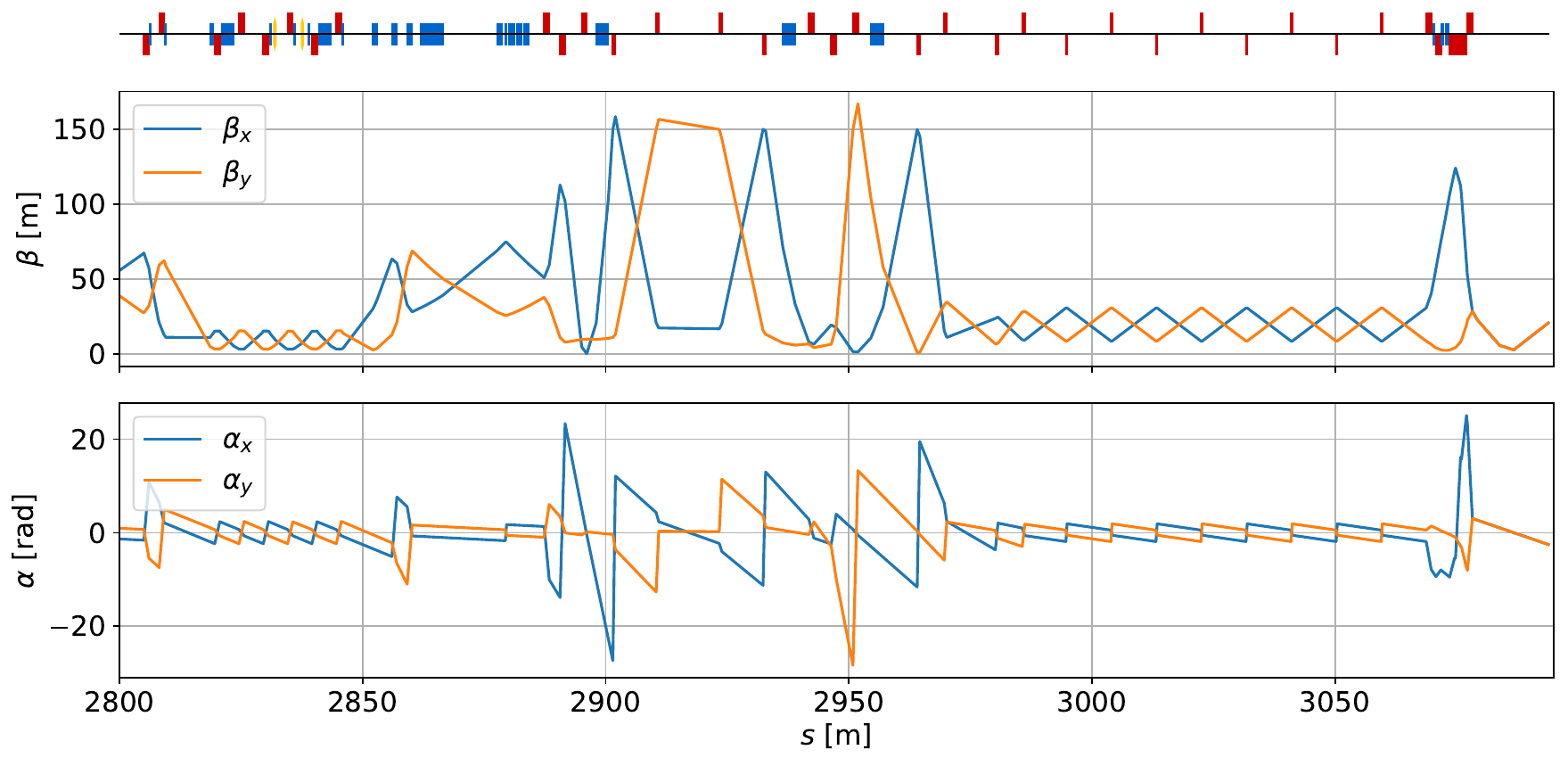}
    \caption{Plots of the twiss functions $\beta$ and $\alpha$ as a function of the longitudinal position $s$. This configuration focuses the electron beam at LUXE's IP. On top is the magnets sequence of the XTD8 lattice.}
    \label{fig:XTD8_beta_alpha}
\end{figure}

At first order, the lattice design must ensure a \SI{10}{\micro\meter} transverse beam size at the focus point. This is done using again the matching functionality of Xsuite. Since the final focus is already matched to the exit of the transport FODO, the only targets required are the values of $\beta$ and $\alpha$ at the entrance of the FODO. The variables are then all 14 quadrupole strengths $k_1$ in the dogleg. 
For each quadrupole, an additional target was made to limit the maximum values of $\beta$, thus keeping the beam sizes within reasonable values. This upper limit for $\beta$ was progressively reduced from \SI{350}{\meter} to \SI{180}{\meter}.
As those quadrupoles currently had no $k_1$ value in the design, they are given an initial alternating values pattern of $k_1=\pm0.2\ \rm m^{-2}$. The resulting $\beta$ and $\alpha$ function after the Xsuite match are shown in Figure~\ref{fig:XTD8_beta_alpha}.

Both $\beta$ and $\alpha$ functions are recovered in the FODO transport and the final focus. The maximum value of $\beta$ is below \SI{170}{\meter} which fairly reasonable in an extraction. However, the beam size depends on both the $\beta$ function and the dispersion $\eta$. The FODO and final focus lattice were designed assuming no dispersion. The new target $\eta_{x,y}=0$ must be added in the Xsuite matching. This corresponds to an achromatic lattice.

\subsection{Achromaticity}

In an arbitrary beam line, the position $x$ and angle $x'$ of an off-momentum particle are propagated from $s_0$ to $s$ with respect to Eqn~\ref{eq:rmat_x_cs_disp}. Similarly, the dispersions $\eta_x$ and $\eta'_x$ are propagated as in Eqn~\ref{eq:rmat_disp_cs_disp}.
In an element with no bending angle, like a drift or a quadrupole, the bending radius $\rho$ tends to infinity. As such, the dispersive factors expressed in Eqn~\ref{eq:disp_fact} and Eqn~\ref{eq:disp_p_fact} becomes 0. For those elements, $D=0$ and $D'=0$ in their transfer matrices. This means that, no matter the combination of drifts and quadrupole, it is not possible to introduce or correct the dispersion. However for a sector dipole the transfer matrix can be expressed as

\begin{equation}
    \begin{pmatrix}
        \eta_x(s) \\
        \eta'_x(s) \\
        1
    \end{pmatrix}
    =
    \begin{pmatrix}
        \cos\frac{L}{\rho} & \rho\sin\frac{L}{\rho} & \rho\left[1-\cos\frac{L}{\rho}\right]\\
        -\frac{1}{\rho}\sin\frac{L}{\rho} & \cos\frac{L}{\rho} & \sin\frac{L}{\rho} \\
        0 & 0 & 1 \\
    \end{pmatrix}
    \cdot
    \begin{pmatrix}
        \eta_x(s_0) \\
        \eta'_x(s_0) \\
        1
    \end{pmatrix}.
\end{equation}

\begin{figure}[!b]
    \centering
    \includegraphics[width=\textwidth]{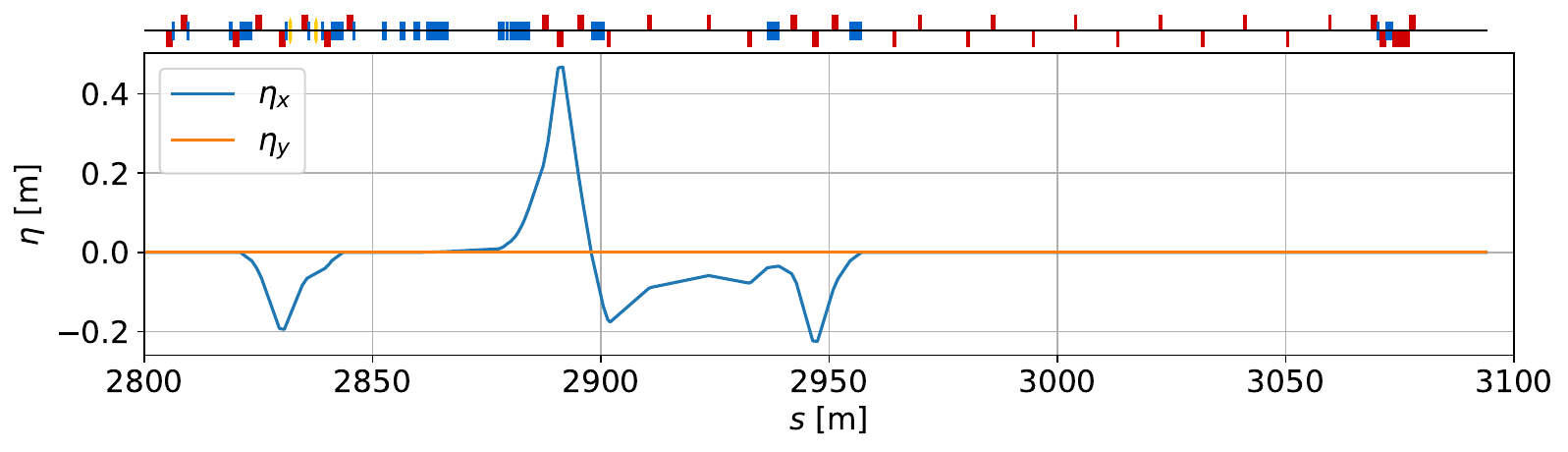}
    \caption{Values of the dispersion function $\eta$ along $s$ in both planes. On top is the sequence of magnets.}
    \label{fig:XTD8_disp}
\end{figure}

In order to have an achromatic lattice, $\eta_{x,y}$ and $\eta'_{x,y}$ must both be equal to $0$ after the dogleg. According to the previous equations, this is only possible if the lattice has dipoles. A classic example of an achromatic lattice is the double bend achromat~\cite{dba} (DBA). It is composed of two dipole magnets and 3 sets of quadrupoles, one in between the dipoles and the others on both sides (the number of quadrupoles used can vary between designs). This layout, as well as the achromats of higher order, is often used in storage rings to bend the beam between two sections without introducing dispersion.
In the XTD8 lattice design, a dispersion in the horizontal plane $\eta_x$ is introduced by the extraction dipoles. Together with the dogleg, this portion of the lattice can be treated as a multi-bend achromat~\cite{mba} (MBA) of order four. The dispersion must be closed, i.e. brought back to zero, at the last dipole of the dogleg. The Xsuite match is ran again with the additional targets $\eta_x=0$ and $\eta'_x=0$ at the end of the dogleg and the resulting dispersion function is shown in Figure~\ref{fig:XTD8_disp}

Between \SI{2820}{\meter} and \SI{2850}{\meter} the existing magnets for the XTD5 line already follows an independent DBA pattern. The dispersion from those magnets is closed before the beginning of XTD8. Between \SI{2850}{\meter} and \SI{2890}{\meter}, $\eta_x$ increases through the kickers and the septa dipoles, reaching a maximum of $\eta_x \approx 0.47\ \rm m$ in the first quadrupole triplet of the dogleg. This dispersion is then closed at the last dogleg dipole (around \SI{2960}{\meter}). Similarly, the extraction magnets creates a non-zero angular dispersion $\eta_{x'}$. Both $\eta_x$ and $\eta_{x'}$ must equal zero after the last dipole of the dogleg. The target $\eta_{x'}=0$ was also added to the Xsuite match function.






\subsection{Isochronicity}

In an arbitrary lattice, the time and energy components are propagated through the transfer matrix as

\begin{equation}
    \begin{pmatrix}
        ct(s) \\
        \delta(s) \\
    \end{pmatrix}
    =
    \boldsymbol{R} \cdot
    \begin{pmatrix}
        ct(s_0) \\
        \delta(s_0) \\
    \end{pmatrix}.
    =
    \begin{pmatrix}
        R_{55} & R_{56} \\
        R_{65} & R_{66}\\
    \end{pmatrix}
    \cdot
    \begin{pmatrix}
        ct(s_0) \\
        \delta(s_0) \\
    \end{pmatrix},
\end{equation}

where $t$ is the relative time deviation and $\boldsymbol{R}$ is a reduced part of the $6\times6$ transfer matrix. Outside of RF cavities, and if the losses are ignored, the energy of the beam is conserved. This mean that $R_{65}=0$ and $R_{55}=R_{66}=1$. The previous equation then becomes

\begin{equation}
    \begin{pmatrix}
        ct(s) \\
        \delta(s) \\
    \end{pmatrix}
    =
    \begin{pmatrix}
        1 & R_{56} \\
        0 & 1\\
    \end{pmatrix}
    \cdot
    \begin{pmatrix}
        ct(s_0) \\
        \delta(s_0) \\
    \end{pmatrix}.
\end{equation}

At first order, the relative time propagation can then be expressed as 

\begin{equation}
    ct(s) = ct(s_0) + R_{56} \delta(s_0).
\end{equation}

The linear factor $R_{56}$ is often calculated as~\cite{r56}

\begin{equation}
    R_{56}=\int_{s_0}^s\frac{\eta(\tilde{s})}{\rho(\tilde{s})}d\tilde{s},
    \label{eq:r56}
\end{equation}

where $\eta$ is the dispersion and $\rho$ is the curvature radius along $s$. The energy deviation $\delta$ can be expressed as a combination of a correlated and uncorrelated part $\delta = \delta_{\rm c}+\delta_{\rm u}$. A variation of the longitudinal position $\tau=ct$ is then expressed as~\cite{bunch_compressor}

\begin{equation}
    d\tau(s) = d\tau(s_0) + R_{56}(d\delta_{\rm c} + d\delta_{\rm u}) = d\tau(s_0)\left(1+R_{56}\frac{d\delta_{\rm c}}{d\tau(s_0)}\right)+R_{56}d\delta_{\rm u}.
\end{equation}

The correlated energy deviation $\delta_{\rm c}$ relates to $\tau$ through a linear energy chirp $h$. The previous equation then becomes

\begin{equation}
    d\tau(s) = d\tau(s_0)(1+h(s_0)R_{56})+R_{56}d\delta_{\rm u},
\end{equation}

where $h(s_0)=\frac{d\delta_c}{d\tau(s_0)}$ is the initial chirp. The compression factor $C$ of the lattice is defined as

\begin{equation}
    C = (1+h(s_0)R_{56})^{-1}.
    \label{eq:compression_factor}
\end{equation}

This factor tends to infinity when $R_{56} = \frac{-1}{h(s_0)}$. This is the value of $R_{56}$ where the bunch compression is maximal. Assuming $t$ and $\delta$ have a Gaussian distribution, the bunch length $\sigma_\tau$ can be expressed as

\begin{equation}
    \sigma_\tau(s) = \sqrt{[1+h(s_0)R_{56}]^2\sigma_\tau(s_0)^2 + [R_{56} \sigma_\delta(s_0)]^2},
    \label{eq:bunch_length_compress}
\end{equation}

where $\sigma_\delta(s_0)$ is the energy spread at $s_0$. At maximal compression, the bunch length reaches a minimal value $\sigma_{\tau, \rm min}=R_{56}\sigma_\delta(s_0)$. For a beam line to be isochronous, a particle should have the same relative time $t$ before and after the line. This means that $t(s)=t(s_0)$ and by extension $\sigma_\tau(s)=\sigma_\tau(s_0)$. From the previous equations, this is only possible if $R_{56}=0$. A lattice with a non-zero value of $R_{56}$ is either compressing or stretching $\sigma_\tau$. This is the main principle used to create bunch compressors~\cite{bunch_compressor}. A variation in the bunch length could have an impact on the LUXE experiment signal. At first order XTD8 should produce no compression of the bunch length, i.e. it should have a total value of $R_{56}$ as close to zero as possible. 

\begin{figure}[!ht]
    \centering
    \includegraphics[width=\textwidth]{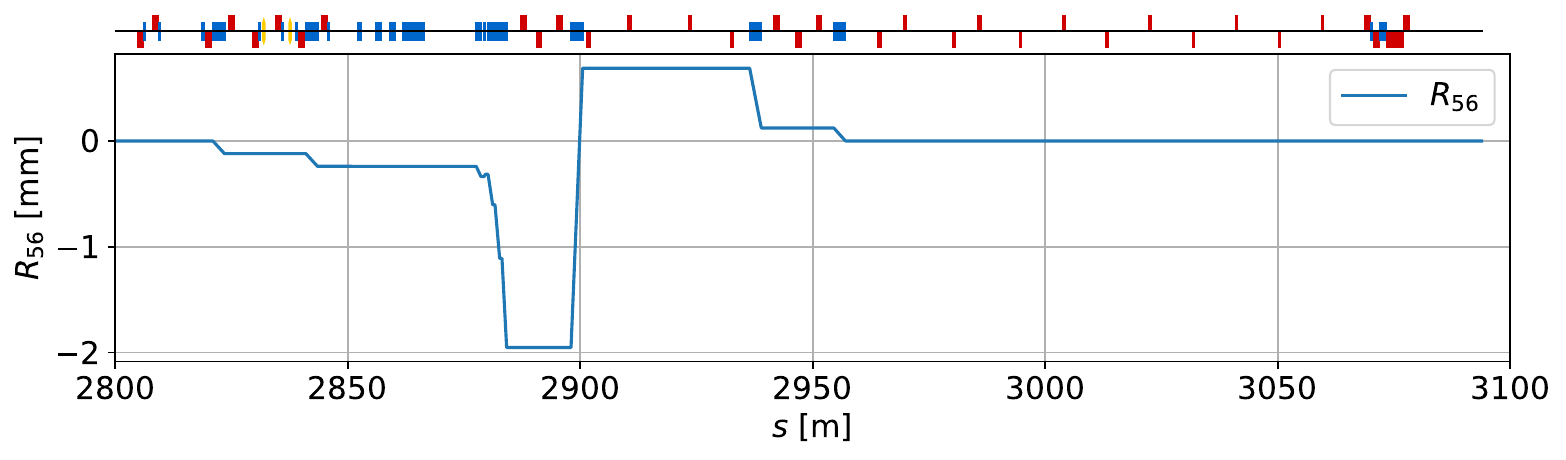}
    \caption{Values of the matrix element $R_{56}$ along $s$. On top is the sequence of magnets. At each point $R_{56}$ is calculated with respect to the starting element of the lattice.}
    \label{fig:XTD8_R56}
\end{figure}

Using Xsuite, it is possible to track the value of $R_{56}$ between the start of the lattice and any given $s$. The target $R_{56}=0$ at the end of the lattice was added to the Xsuite matching function. The result $R_{56}$ evolution is shown in Figure~\ref{fig:XTD8_R56}.
The changes of $R_{56}$ occurs exclusively in the dipoles as they are the only elements introducing a curvature radius $\rho(s)$. Before \SI{2850}{\meter} the XTD5 bends already produces a $R_{56}$ value of around \SI{0.25}{\milli\meter}. The most important contributions in $R_{56}$ comes from the septas (around \SI{2880}{\meter}) and the first dogleg dipole (around \SI{2900}{\meter}). As seen in Eqn~\ref{eq:r56}, the amplitude and sign of $R_{56}$ variations depends on the dispersion and the curvature radius. If $\eta_x(s)>0$ and $\rho(s)>0$ in a dipole, $R_{56}$ increases. Looking at Figure~\ref{fig:XTD8_disp} for the dispersion sign and Figure~\ref{fig:XTD8_top_view_details} for the dipole angle sign, the results of Figure~\ref{fig:XTD8_R56} are consistent. The residual $R_{56}$ is finally closed in the last two dipoles of the dogleg. In this configuration, the XTD8 lattice will not affect the bunch length.

\subsection{Matching results and beam size}

To obtain a beam line which is achromatic, isochronous, and which focuses to a \SI{10}{\micro\meter} beam size at LUXE's IP, all the targets mentioned previously must be added to the Xsuite match. In addition to the global constraint $\beta_{x,y}<180\ \rm m$, a summary of the targets at the FODO entrance can be found in Table~\ref{tab:matching_targets}.

\begin{table}[!ht]
    \centering
    \begin{tabular}{| c | c || c | c |}
        \hline
        Target & Unit & Value & Tolerance \\
        \hline
        \hline
        $\beta_x$ & \SI{}{\meter} & 30.78 & $10^{-3}$ \\
        \hline
        $\beta_y$ & \SI{}{\meter} & 8.68 & $10^{-3}$ \\
        \hline
        $\alpha_x$ & \SI{}{\radian} & 0 & $10^{-3}$ \\
        \hline
        $\alpha_y$ & \SI{}{\radian} & 0 & $10^{-3}$ \\
        \hline
        $\eta_x$ & \SI{}{\meter} & 0 & $10^{-10}$ \\
        \hline
        $\eta_{x'}$ & \SI{}{\radian} & 0 & $10^{-10}$ \\
        \hline
        $R_{56}$ & \SI{}{\meter} & 0 & $10^{-10}$ \\
        \hline
    \end{tabular}
    \caption{Target values and respective tolerances at the entrance of the FODO transport line. Those are inputs to the Xsuite matching function.}
    \label{tab:matching_targets} 
\end{table}

\begin{figure}[!ht]
    \centering
    \includegraphics[width=\textwidth]{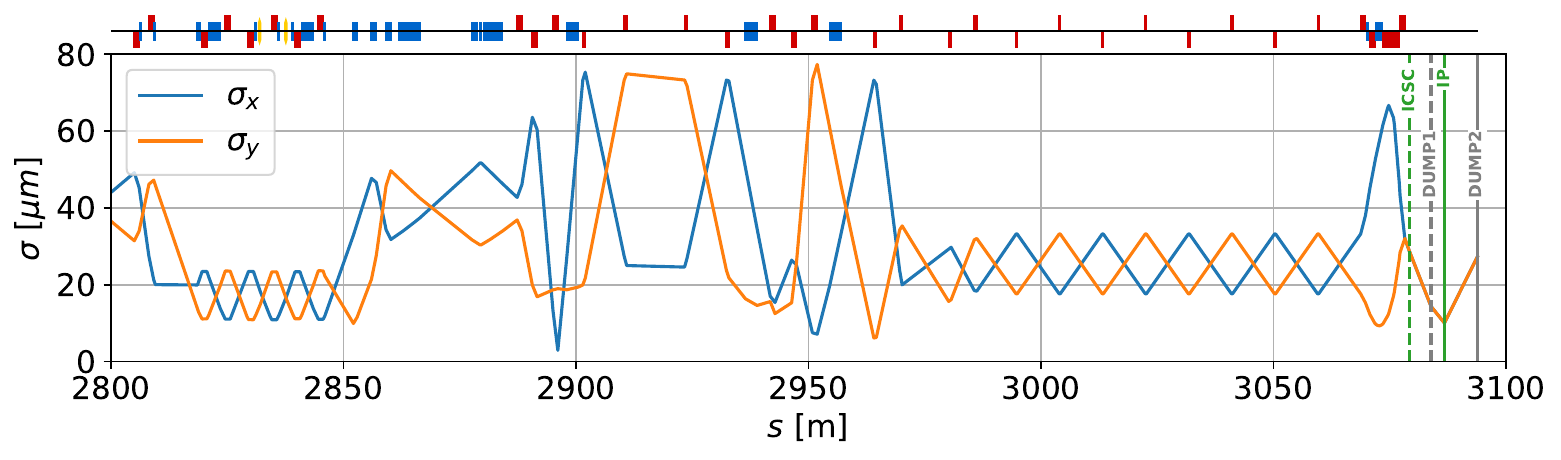}
    \caption{Beam size $\sigma$ in both planes along $s$. On top is the sequence of magnets.}
    \label{fig:XTD8_sigma}
\end{figure}

As the Xsuite optimiser is running, it varies the strengths and position of all 14 dogleg quadrupoles to find a configuration where all the targets are fulfilled. A solution was found and the resulting magnet strengths and positions can be found in Table~\ref{tab:matched_quad}
The maximum absolute value of $k_1$ is \SI{-0.527}{\meter^{-2}} for the quadrupole QH.2941.T8 in the last triplet. This is an achievable strength but is at the limit of what a QH quadrupole can produce. The transverse beam sizes along the lattice are shown in Figure~\ref{fig:XTD8_sigma}.

\begin{table}[!ht]
    \centering
    \begin{tabular}{| c || c | c |}
        \hline
        Name & $s$ / m & $k_1$ / $\rm m^{-2}$ \\
        \hline
        \hline
        QH.2887.T8 & 2887.34 & -0.198 \\
        QH.2890.T8 & 2890.61 & 0.330 \\
        QH.2892.T8 & 2895.06 & -0.074 \\
        QF.2906.T8 & 2901.46 & 0.479 \\
        QF.2911.T8 & 2910.36 & -0.161 \\
        QF.2920.T8 & 2923.34 & -0.145 \\
        QF.2925.T8 & 2932.35 & 0.303 \\
        QH.2941.T8 & 2941.76 & -0.527 \\
        QH.2944.T8 & 2946.37 & 0.331 \\
        QH.2946.T8 & 2950.84 & -0.253 \\
        QF.2955.T8 & 2964.05 & 0.394 \\
        QF.2961.T8 & 2969.61 & -0.454 \\
        QF.2981.T8 & 2980.13 & 0.441 \\
        QF.2987.T8 & 2985.65 & -0.315 \\
        \hline
    \end{tabular}
    \caption{Strengths and positions of the 14 dogleg quadrupoles after the Xsuite matching.}
    \label{tab:matched_quad}
\end{table}

In the dogleg, the maximum values of $\sigma_{x,y}$ are below \SI{80}{\micro\meter}. This is a reasonable transverse size when compared to the 0.02413 m beam pipe radius. At LUXE's IP, the \SI{10}{\micro\meter} beam size is achieved. A summary of the electron beam parameters at the IP can be found in Table~\ref{tab:ip_params}.
In the end, the beam line is achromatic and isochronous with maximum beam size less than \SI{80}{\micro\meter} and with a \SI{10}{\micro\meter} beam size at the focus point.

\begin{table}[!ht]
\centering
\begin{tabular}{| c | c | c |}
\hline
Parameter & Unit & Value \\
\hline
\hline
$\sigma_x$ & $\rm \mu m$ & 10.01  \\
\hline
$\sigma_y$ & $\rm \mu m$ & 10.01 \\
\hline
E & GeV & 17.5 \\
\hline
$L^*$ & m & 9.54 \\
\hline
$\beta_x^*$ & m & 2.8 \\
\hline
$\varepsilon_x$ & m.rad & \SI{3.58e-11}{} \\ 
\hline
$\varepsilon_y$ & m.rad & \SI{3.58e-11}{} \\ 
\hline
$R_{56}$ & m & \SI{4.09e-9}{} \\
\hline
\end{tabular}
\caption{Parameters of the electron beam when focusing at LUXE's IP.}
\label{tab:ip_params} 
\end{table}

\section{Impact of coherent synchrotron radiation}

\begin{figure}[!ht]
    \centering
    \includegraphics[width=8cm]{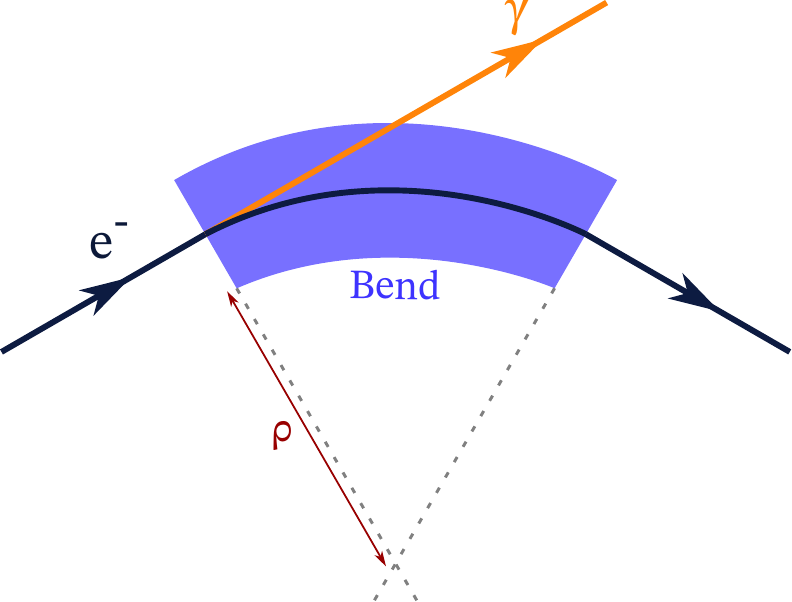}
    \caption{Simple case of SR production by a single bend.}
    \label{fig:csr_bend_diagram}
\end{figure}

When charged particles are going through a dipole, they can produce photons in a path tangent to the electron path (Figure~\ref{fig:csr_bend_diagram}). The smaller the bending radius $\rho=l/|\theta|$ of the magnet, the higher number of photons are produced by this bend. This effect is called synchrotron radiation~\cite{sr} (SR).

Moreover, if the photons wavelength is close to the electron bunch length, the photons are emitted in phase and this becomes coherent synchrotron radiation~\cite{csr} (CSR). This property is used in undulators to produce the photon beams in FELs. However, the CSR photons can interact with the electron beam and disrupt the beam parameters. The dipoles in the rest of the accelerator are designed to avoid CSR as it degrades the emittance and reduces the beam energy. In the XTD8 there are seven sector dipoles, the four septas from the extraction and the three dogleg dipoles. Their lengths, angles and bending radii are listed in Table~\ref{tab:sbend} .

\begin{table}[!ht]
    \centering
    \begin{tabular}{| c || c | c | c |}
        \hline
         Name & $l$ / m & $\theta$ / rad & $\rho$ / m \\
         \hline
         \hline
         BZ.2876.T8 & 1.0 & 0.0089 & 112.66 \\
         BZ.2878.T8 & 1.0 & 0.0089 & 112.66 \\
         BZ.2880.T8 & 1.0 & 0.0089 & 112.66 \\
         BZ.2881.T8 & 1.0 & 0.0089 & 112.66 \\
         \hline
         BE.2899.T8 & 2.5 & 0.0330 & 75.83 \\
         BE.2935.T8 & 2.5 & -0.0171 & 146.63 \\
         BE.2953.T8 & 2.5 & -0.0172 & 146.62 \\
         \hline
    \end{tabular}
    \caption{List of the lengths, angles and bending radii of XTD8 sector dipoles.}
    \label{tab:sbend}
\end{table}

Those lengths $l$ are standards in the EuXFEL but the angles $\theta$ were only set to obey the geometric constraints. The dipoles with high $\theta$, i.e. small $\rho$, are more likely to create CSR radiation. To evaluate the sources of CSR in the XTD8 lattice and the impact on the beam properties, calculating the optics function with Xsuite is not sufficient and a tracking simulation that includes CSR production must be ran. 
To do this, the Xsuite model is converted to Ocelot as it is more suited to run CSR calculations. In addition of the input twiss parameters ($\beta$, $\alpha$, $\varepsilon$, $\eta$, etc.), the particle tracking requires the definition of a bunch charge $q$, a bunch length $\sigma_\tau$, an momentum spread $\sigma_p$ and an energy chirp $h$. 
Simulations of an electron bunch propagated along the entire EuXFEL have been done at DESY. A list of the bunch distributions at different locations along the beam lines can be found on DESY's website~\cite{desy_bunch}. 
For the purpose of our Ocelot simulations, the electron bunch parameters are extracted from the bunch distributions after the SASE2 undulator. The transport beam line between SASE2 and the ELBEX location should not have a significant impact on the beam parameters. Table~\ref{tab:ocelot_input} shows the additional bunch parameters extracted from the bunch distribution after SASE2.


\begin{table}[!ht]
    \centering
    \begin{tabular}{| c | c | c |}
        \hline
         Parameter & Unit & Value \\
         \hline
         \hline
         $E_0$ & GeV & 17.5 \\
         \hline
         $q$ & pC & 250 \\
         \hline
         $\sigma_\tau$ & $\rm \mu m$ & 6.02 \\
         \hline
         $\sigma_p$ & 1 & \SI{6e-4}{} \\
         \hline
         $h$ & $\rm m^{-1}$ & 8.33 \\
         \hline
    \end{tabular}
    \caption{Additional input parameters extracted from the bunch distribution.}
    \label{tab:ocelot_input}
\end{table}

\begin{figure}[!ht]
    \centering
    \includegraphics[width=\textwidth]{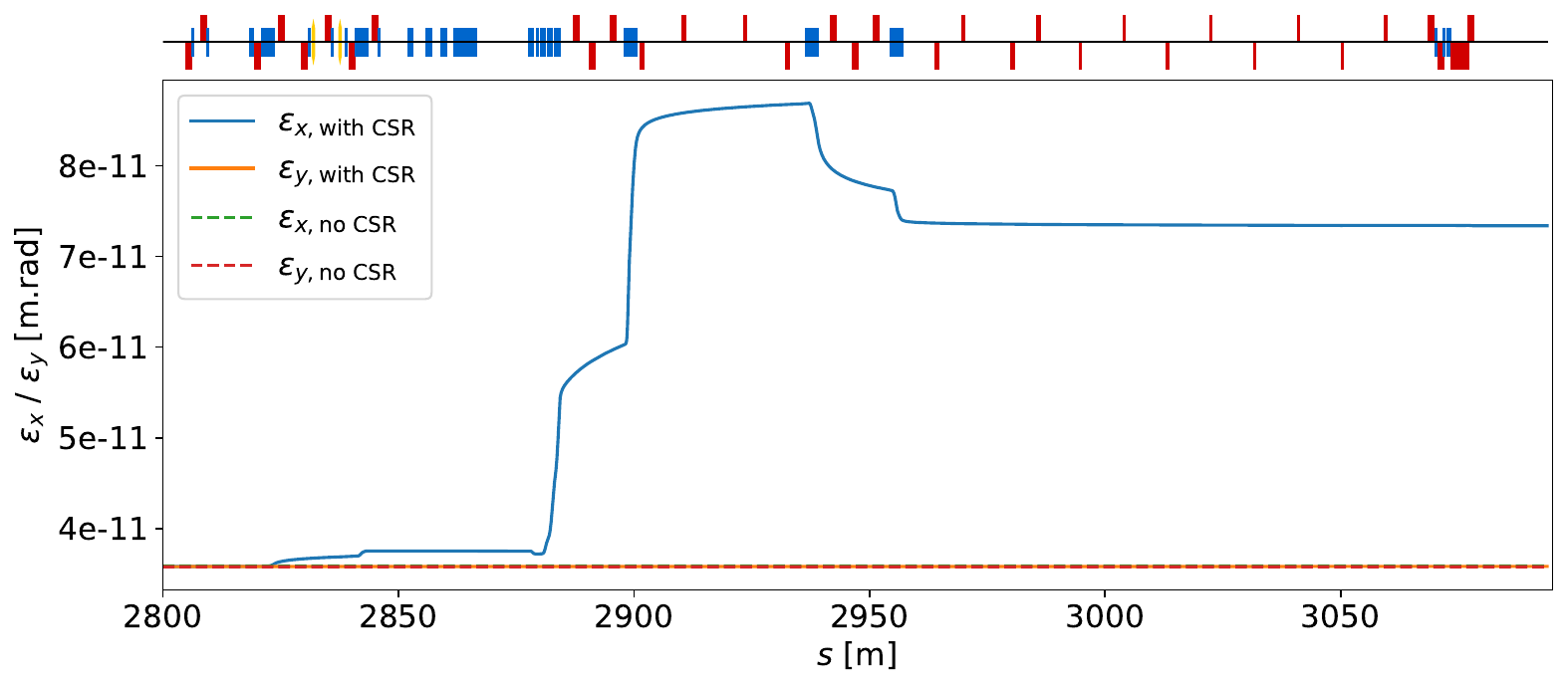}
    \caption{Comparison of the emittance $\varepsilon$ in both planes between tracking with and without CSR.}
    \label{fig:tracking_csr_emitt}
\end{figure}

An Ocelot tracking simulation with those parameters was run twice, with and without CSR calculation enabled. For each run, $10^6$ primary electron macroparticles were simulated. The resulting variations of transverse emittance $\varepsilon_{x,y}$ are shown in Figure~\ref{fig:tracking_csr_emitt}.

\begin{figure}[!ht]
    \centering
    \includegraphics[width=\textwidth]{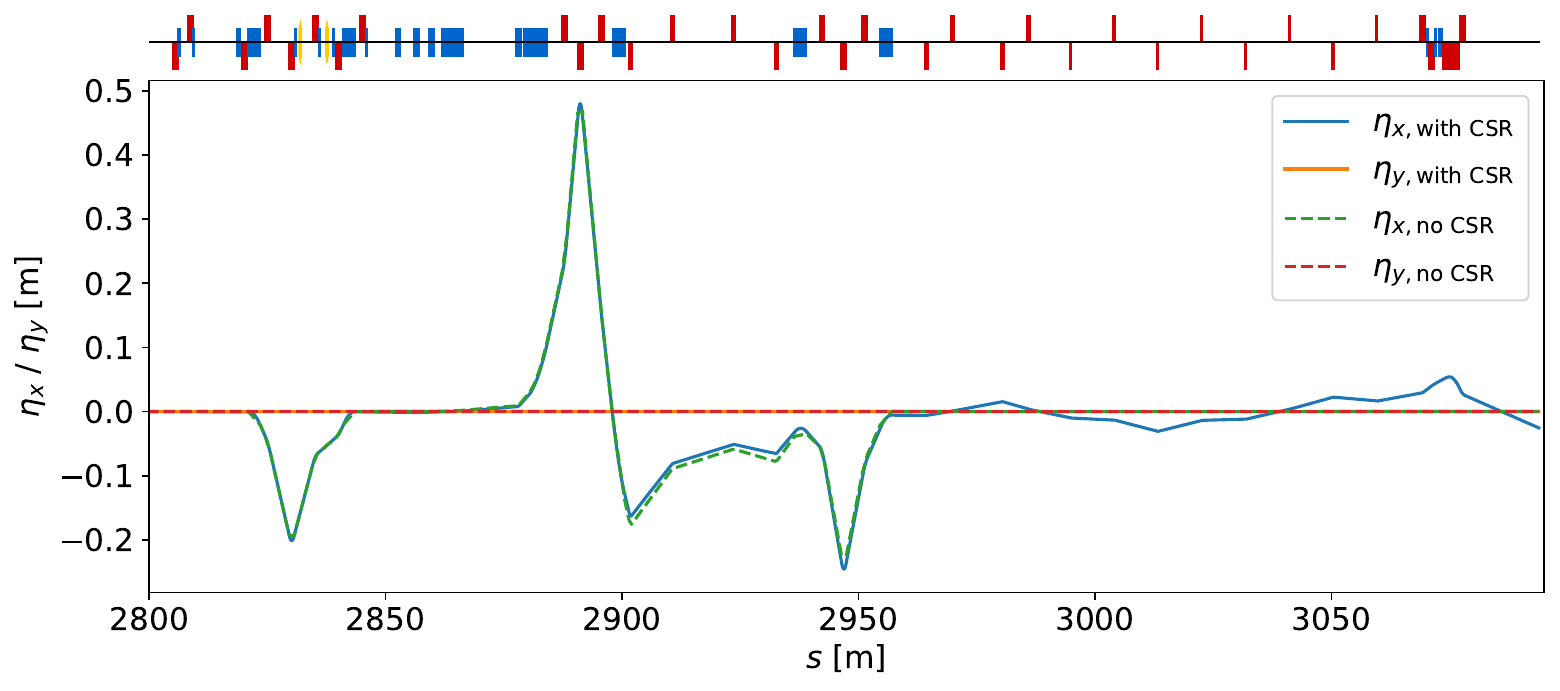}
    \caption{Comparison of the dispersion $\eta$ in both planes between tracking with and without CSR.}
    \label{fig:tracking_csr_disp}
\end{figure}

As there is no dipole bending in the vertical plane $y$, there is no impact of the CSR on $\varepsilon_y$. In the horizontal plane $x$, there is an increase from the initial emittance \SI{3.58e-11}{\meter.rad} to \SI{7.34e-11}{\meter.rad} at the end of the lattice. Below \SI{2850}{\meter}, the XTD5 dipoles, already designed to avoid CSR, produces very little changes in the horizontal emittance. The highest contributions for the increase are coming from the septas (at around \SI{2880}{\meter}) and the first dogleg dipole (at around \SI{2900}{\meter}). CSR also has an impact on the dispersion, as shown in Figure~\ref{fig:tracking_csr_disp}.

\begin{figure}[!ht]
    \centering
    \includegraphics[width=\textwidth]{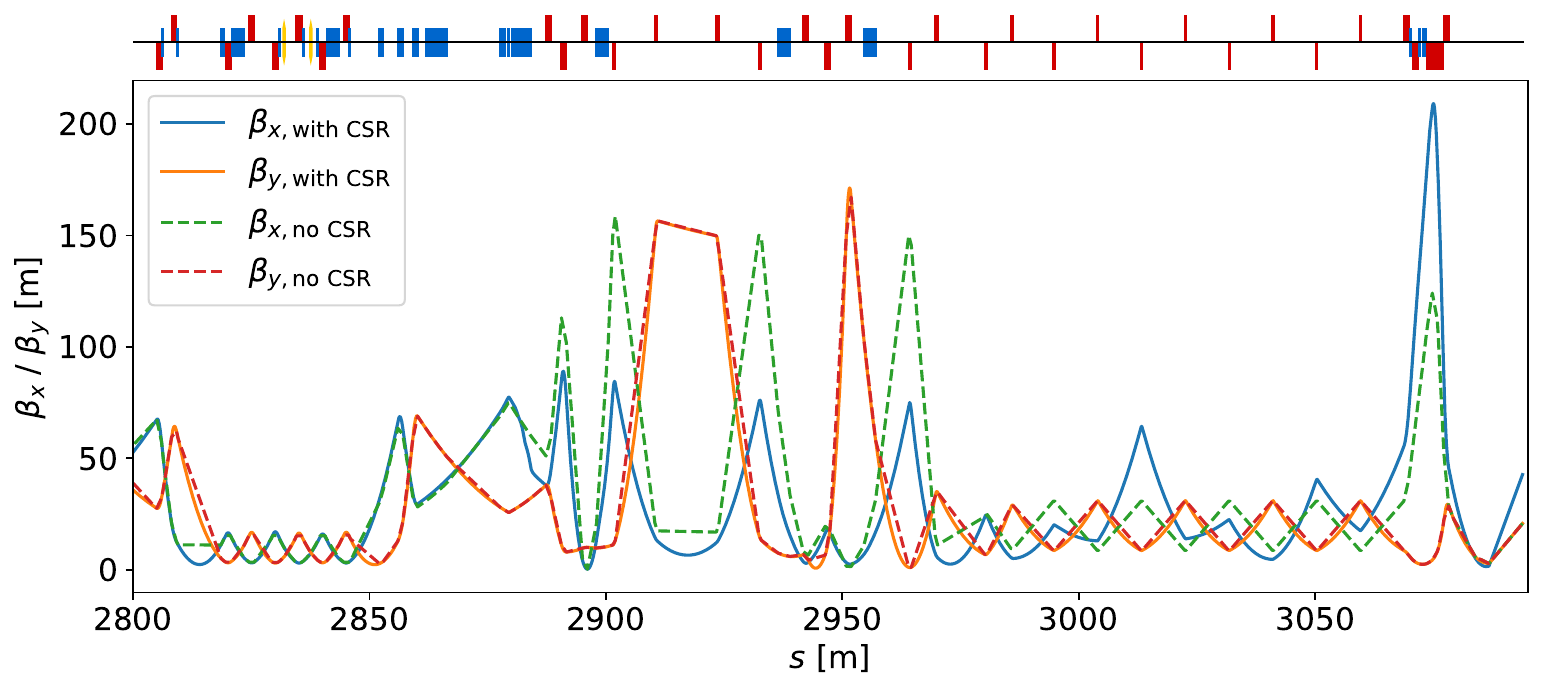}
    \caption{Comparison of the beta functions $\beta$ in both planes between tracking with and without CSR.}
    \label{fig:tracking_csr_beta}
\end{figure}

Up until the dogleg, $\eta_x$ is unchanged by the CSR. From $s=2900\ \rm m$ a slight variation of around \SI{0.01}{\meter} appears and is not fully closed at the end of the dogleg. It is referred as a dispersion leak. With the impact of CSR, the lattice is no longer fully achromatic. Some residual dispersion gets propagated to the focus point and will impact the beam size. Finally, there is also an impact of CSR on the $\beta$ function, as shown in Figure~\ref{fig:tracking_csr_beta}.

\begin{figure}[!ht]
    \centering
    \includegraphics[width=\textwidth]{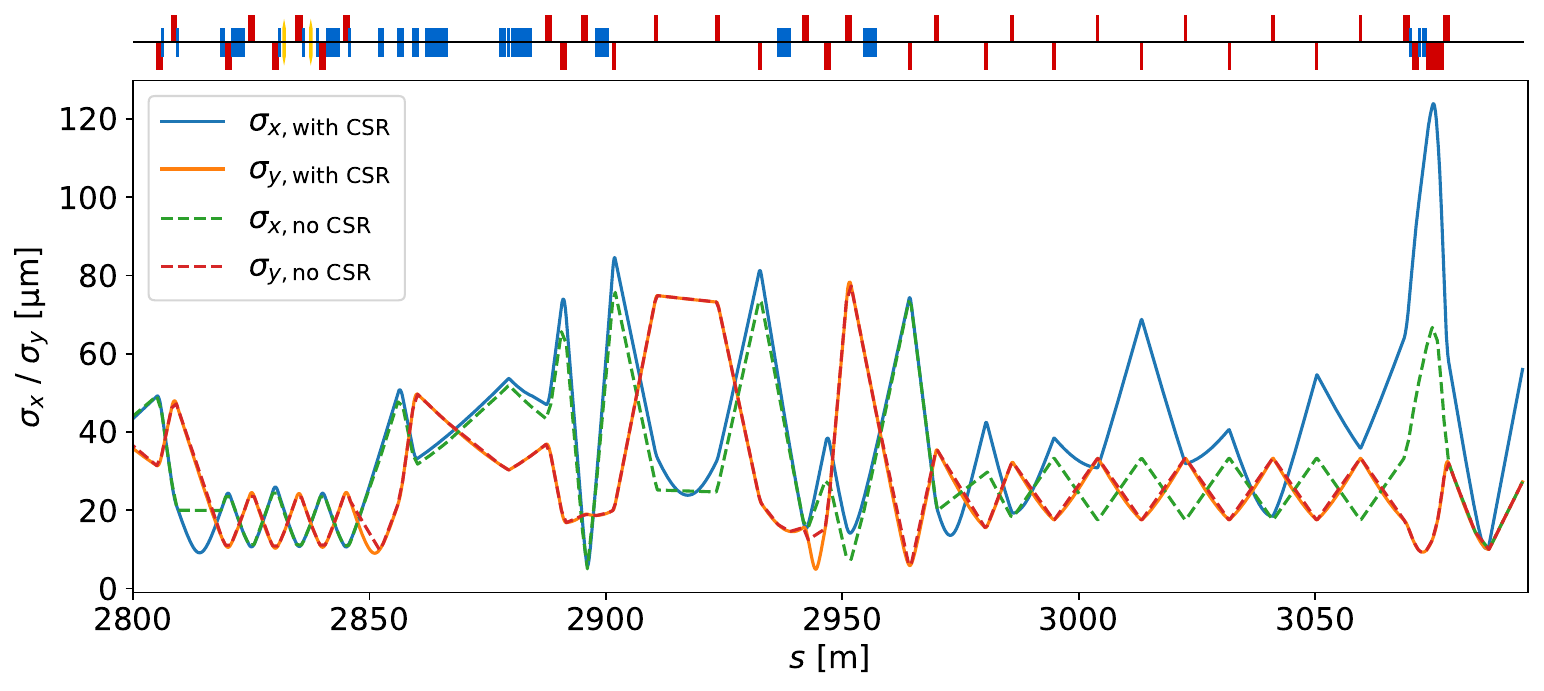}
    \caption{Comparison of the beam sizes $\sigma$ in both planes between tracking with and without CSR.}
    \label{fig:tracking_csr_sigma}
\end{figure}

There is a divergence in $\beta_x$ starting at the septa dipoles (around \SI{2880}{\meter}). The peaks in the dogleg are significantly reduced with the CSR effects. After the dogleg, a beta-beating~\cite{beta_beating} pattern can be observed in the transport FODO. In the final focus, $\beta_x$ spike above \SI{200}{\meter}, almost twice its value without CSR. However, the value of $\beta_x$ at the focus point is really close to its default value.
Following Eqn~\ref{eq:sigma_x_disp}, the beam sizes $\sigma_{x,y}$ can be computed from the previous results and is shown in Figure~\ref{fig:tracking_csr_sigma}.

The impact on $\sigma_x$ is minimal up until the end of the dogleg. In the FODO there is again the beta-beating and the beam goes up to \SI{120}{\micro\meter} in the final focus. At the focus point, the horizontal beam size $\sigma_x$ increased from \SI{10.01}{\micro\meter} to \SI{11.04}{\micro\meter}. This \SI{10}{\%} increase in beam size could impact the intensity of strong-field QED products at LUXE's detectors. A comparison of the electron beam parameters at the IP with and without CSR can be found in Table~\ref{tab:ip_params_csr}

\begin{table}[!ht]
    \centering
    \begin{tabular}{| c | c || c | c |}
        \hline
        Parameter & Unit & Without CSR & With CSR \\
        \hline
        \hline
        $\sigma_x$ & $\rm \mu m$& 10.01 & 11.04 \\
        \hline
        $\sigma_y$ & $\rm \mu m$ & \multicolumn{2}{c|}{10.01} \\
        \hline
        $L^*$ & m & \multicolumn{2}{c|}{9.54} \\
        \hline
        $\beta_x^*$ & m & 2.8 & 1.66 \\
        \hline
        $\eta_x^*$ & m & \SI{2.56e-5}{} & \SI{2.49e-3}{} \\
        \hline
        $\varepsilon_x$ & m.rad & \SI{3.58e-11}{} & \SI{7.34e-11}{} \\ 
        \hline
        $\varepsilon_y$ & m.rad & \multicolumn{2}{c|}{\SI{3.58e-11}{}} \\ 
        \hline
        $R_{56}$ & mm & \multicolumn{2}{c|}{\SI{4.09e-9}{}} \\
        \hline
    \end{tabular}
    \caption{Parameters of the electron beam at the IP.}
    \label{tab:ip_params_csr} 
\end{table}


\subsection{CSR scans for charge and length}

\begin{figure}[!ht]
    \centering
    \includegraphics[width=\textwidth]{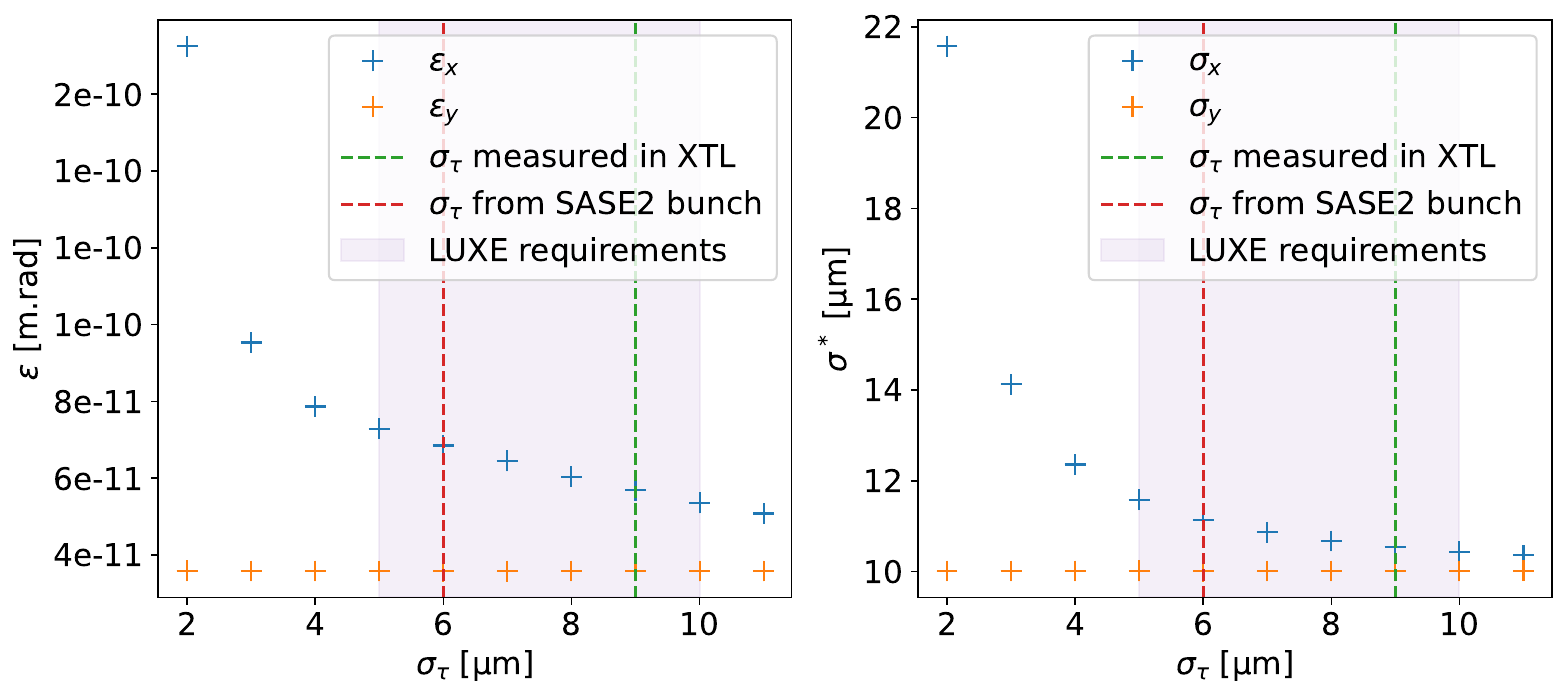}
    \caption{Plots of the impact of the bunch length $\sigma_\tau$ with CSR. On the left the transverse emittances $\varepsilon_{x,y}$ as a function of $\sigma_\tau$. On the right the transverse beam sizes at focus $\sigma^*_{x,y}$ as a function of $\sigma_\tau$.}
    \label{fig:scan_length}
\end{figure}

The power $P_{\rm CSR}$ radiated by CSR from a Gaussian bunch in a bending magnet depends on the bunch charge $q$ and the bunch length $\sigma_\tau$. It can be expressed as~\cite{csr}

\begin{equation}
    P_{\rm CSR} \propto \frac{q^2}{{\rho}^{2/3}{\sigma_\tau}^{4/3}},
    \label{eq:csr_power}
\end{equation}

where $\rho$ is the bending radius of the magnet. With an electron beam of higher charge $q$ and lower $\sigma_\tau$, the effect of CSR should be increased. In the last section, the Ocelot simulation showed a \SI{10}{\%} increase with an electron beam of charge \SI{250}{\pico\coulomb} and length \SI{6.02}{\micro\meter}. By running multiple Ocelot simulations and varying those parameters, the impact on the final horizontal emittance $\varepsilon_x$ and the focus transverse beam size $\sigma^*$ can be determined. The impact of $\sigma_\tau$ is shown in Figure~\ref{fig:scan_length}

\begin{figure}[!ht]
    \centering
    \includegraphics[width=\textwidth]{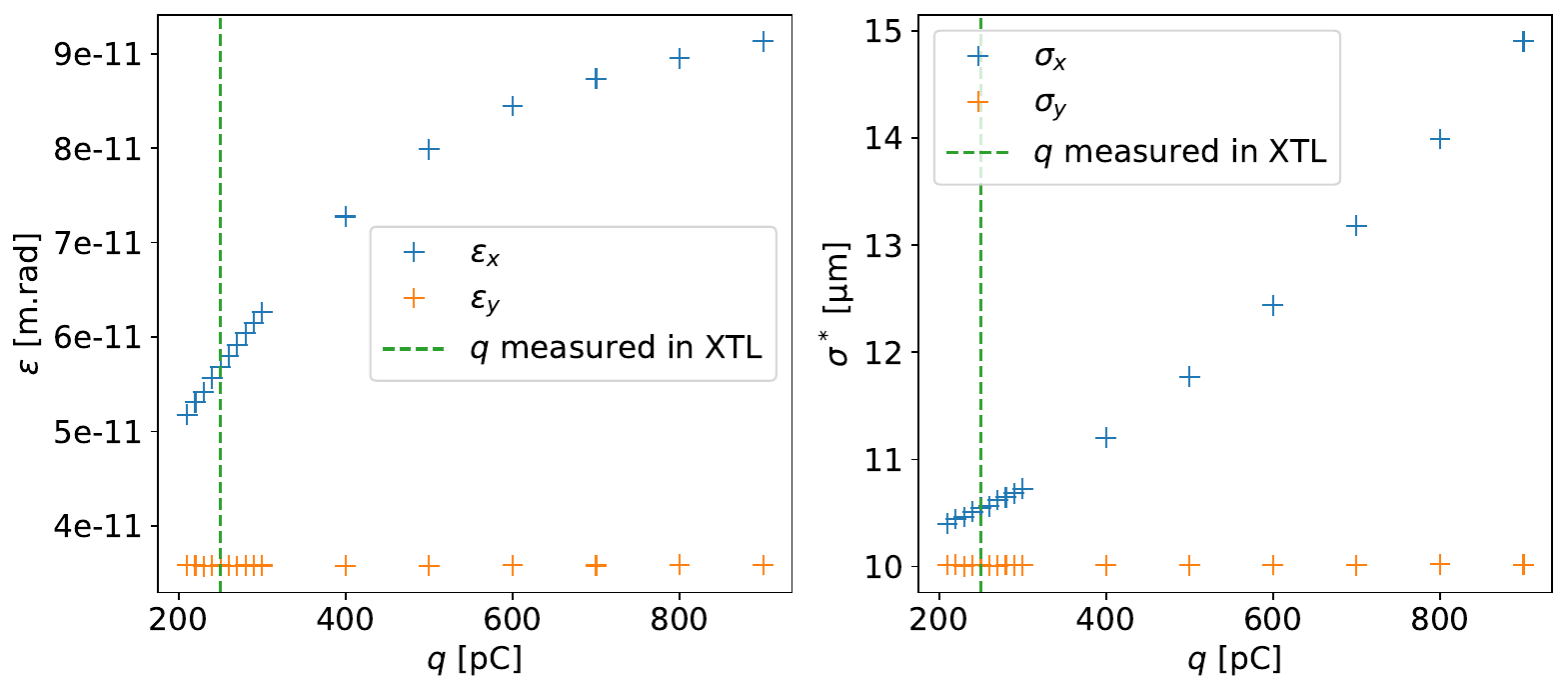}
    \caption{Plots of the impact of the bunch charge $q$ with CSR. On the left the transverse emittances $\varepsilon_{x,y}$ as a function of $q$. On the right the transverse beam sizes at focus $\sigma^*_{x,y}$ as a function of $q$.}
    \label{fig:scan_charge}
\end{figure}

As expected from Eqn~\ref{eq:csr_power}, the impact of CSR on the emittance and the beam size is higher for shorter electron bunches. In LUXE's TDR~\cite{luxe_tdr}, the bunch length at the IP is expected to be between \SI{5}{\micro\meter} and \SI{10}{\micro\meter}. At $\sigma_\tau=5\ \rm \mu m$, the simulations shows an increase in transverse emittance $\varepsilon_x$ by a factor 2. In the same configuration, the horizontal beam size is increased by \SI{15}{\%}. Similarly, the impact of $q$ is shown in Figure~\ref{fig:scan_charge}

As $q$ increases, the impact on the emittance and the beam size is higher. Between \SI{200}{\pico\coulomb} and \SI{300}{\pico\coulomb}, the emittance increases by \SI{20}{\%} and the beam size by \SI{3}{\%}. This shows that the CSR is less sensitive to variations of $q$ than for variations of $\sigma_\tau$, in agreement with Eqn~\ref{eq:csr_power}. In addition, the bunch charge jitter $\sigma_{J,q}$ was measured in the switch yard in Chapter~\ref{chap:measurments} at around \SI{1.25}{\pico\coulomb}. Variations of that amplitude should have little to no impact on the CSR.

\subsection{Correction of the CSR effects}

The effects of CSR does significantly impact the quality and precision of the electron beam at LUXE's IP. To reduce those, there are two different approaches possible. The first one is to reduce the sources of synchrotron radiation. From Eqn~\ref{eq:csr_power} it can be done in three ways, reducing the bunch charge $q$, increasing the bunch length $\sigma_\tau$ or increasing the bending radius $\rho$. The Ocelot simulation showed a low sensitivity from the CSR on $q$, so reducing the $\sigma_\tau$ seems a better approach. However, both the charge and length of the bunch will impact the LUXE's signal. This will be detailed in Chapter~\ref{chap:impact_luxe}. The other solution is to increase $\rho=l/\theta$, either by increasing the length $l$ of the most contributing dipoles or by splitting them into dipoles with smaller angle $\theta$. In terms off emittance growth, the highest contributors are the septa dipoles and the first dipole of the dogleg. A solution could be considered were those five dipoles are either elongated or split into dipoles with lower $\theta$.

The second approach is to match the lattice with the CSR included. At each step of our optimization, Xsuite calculates the new twiss functions, especially $\beta$ and $\eta$, until the targets are achieved at the transport FODO entrance. Instead of using the internal twiss calculation, every Xsuite step could call an Ocelot tracking with CSR enabled. The targets $\beta_{x,y}$ and $\eta_x$ at the entrance of the FODO could then be achieved no matter the amount of CSR. After the dogleg, no significant modification of the optics due to CSR should be observed.
Some attempts were made to use the calculations of Ocelot in Xsuite optimisation, but integrating these two codes is left for future work. More details about this can be found in Chapter~\ref{chap:next}.

\section{$R_{56}$ tunability}

The XTD8 was designed in the previous sections to avoid any variation of the bunch length, i.e. to obey $R_{56}=0$. However, for the purpose of the experiments, having a control on the bunch length can be quite useful. Without moving the quadrupoles, the Xsuite optimizer can be rerun to get specific values of $R_{56}$ at the end of the lattice. The extraction and dogleg will then act as a bunch compressor that can be adjusted for the experiments needs. 
This is done while keeping the \SI{10}{\micro\meter} beam size target at LUXE's IP and having no dispersion after the last dipole.
With the current XTD8 design it was easily possible to get values of $R_{56}$ from \SI{-20}{\milli\meter} to \SI{15}{\milli\meter}. 
The variations of quadrupole strengths in the dogleg are small but this have an impact on the beam optics.

\begin{figure}[!ht]
    \centering
    \includegraphics[width=\textwidth]{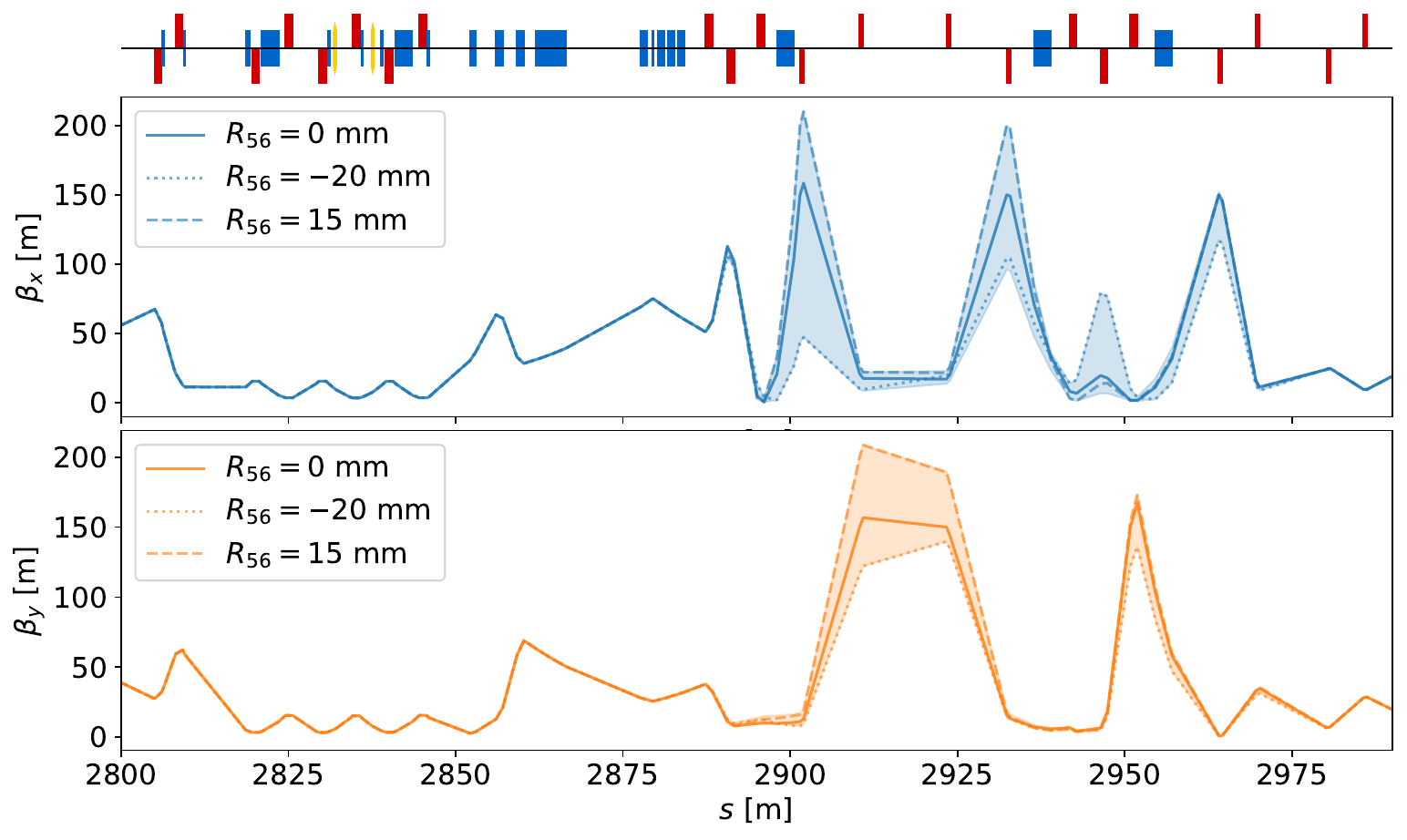}
    \caption{Plots of the beta functions $\beta_{x,y}$ as a function of $s$ for different values of $R_{56}$. On top is $\beta_x$ and on the bottom is $\beta_y$. Both envelopes contain 10 curves ranging from $R_{56}=-20\ \rm mm$ and $R_{56}=15\ \rm mm$. The solid lines correspond to the isochronous optics established previously.}
    \label{fig:r56_beta_env}
\end{figure}

\begin{figure}[!ht]
    \centering
    \includegraphics[width=\textwidth]{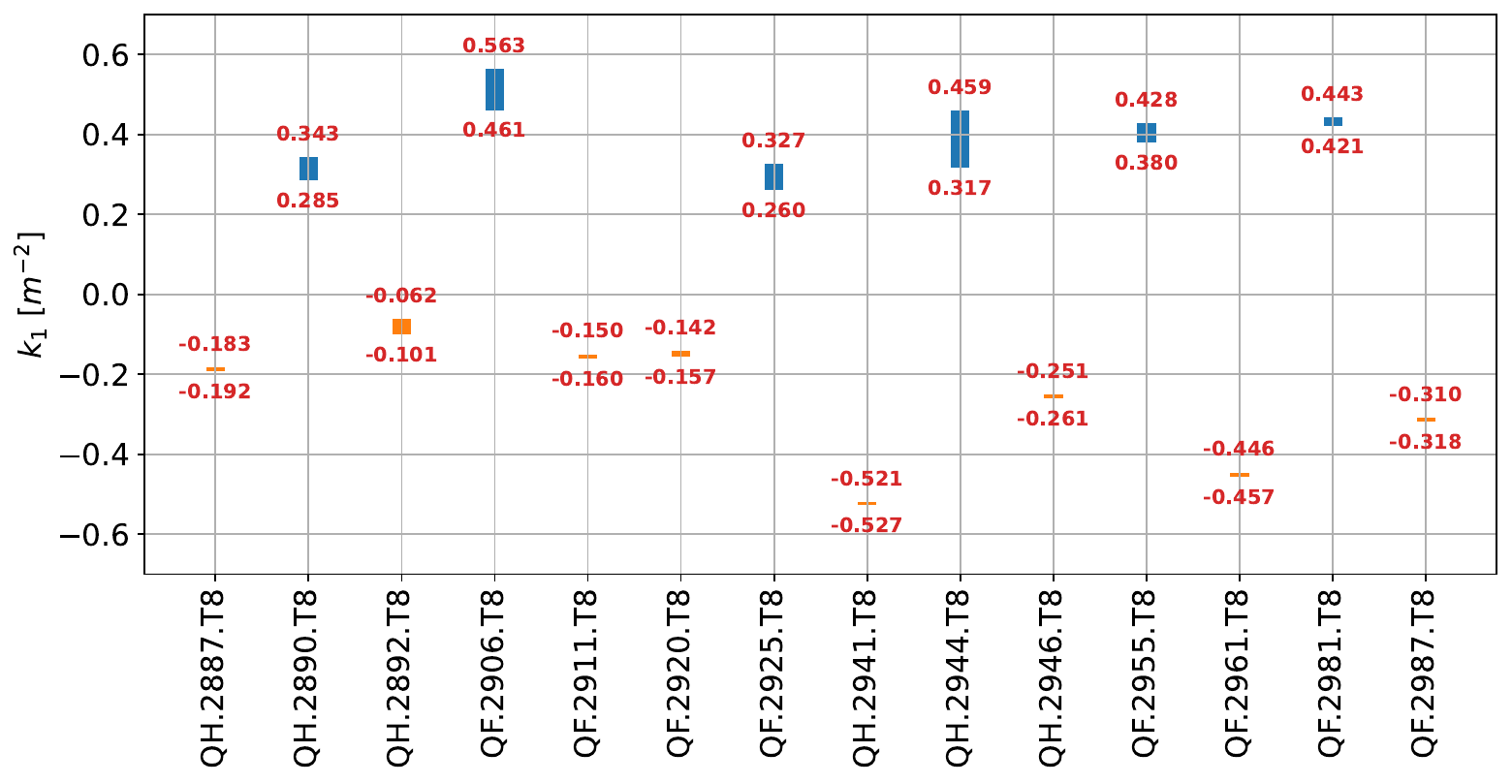}
    \caption{Plot of the ranges of quadrupole strength $k_1$ for each quadrupoles in the dogleg and for different values of $R_{56}$. Red numbers are the maximum and minimum value for each quadrupole.}
    \label{fig:k1_bars_wrt_R56}
\end{figure}

\newpage

A range of the beta function values can be found in Figure~\ref{fig:r56_beta_env}. In addition, Figure~\ref{fig:k1_bars_wrt_R56} shows the ranges of quadrupole strength $k_1$ required for each quadrupole in the dogleg.
Inside the dogleg, the maximum values of $\beta$ increases from \SI{150}{\meter} to \SI{200}{\meter} for the configuration where $R_{56}=15\ \rm mm$. Even this \SI{33}{\%} increase is reasonable for the XTD8 optics. The Eqn~\ref{eq:compression_factor} showed that the compression is maximal when $R_{56}=-1/h$. With a range of $R_{56}$ from \SI{-20}{\milli\meter} to \SI{15}{\milli\meter}, the XTD8 lattice could in theory recompress electron beams with chirp above \SI{50}{\meter^{-1}} and below \SI{-66}{\meter^{-1}}. 

\begin{figure}[!ht]
    \centering
    \includegraphics[width=10cm]{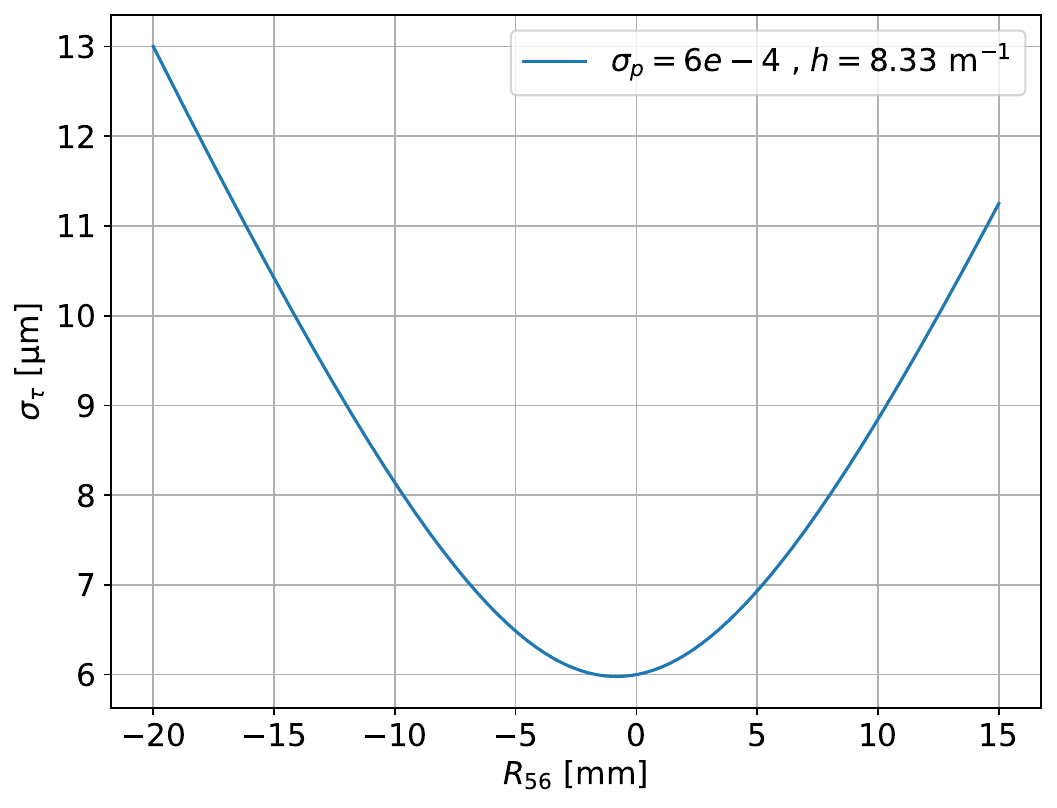}
    \caption{Plot of the bunch length $\sigma_\tau$ as a function of $R_{56}$.}
    \label{fig:bunch_length_wrt_R56}
\end{figure}

An electron beam with the properties listed in Table~\ref{tab:ocelot_input} can be compressed to a bunch length $\sigma_\tau$ with respect to Eqn~\ref{eq:bunch_length_compress}. A range of the achievable bunch lengths is shown in Figure~\ref{fig:bunch_length_wrt_R56}.
The electron bunch simulated after SASE2 is almost fully compressed and has a low chirp $h=8.33\ \rm m^{-1}$. In the $R_{56}$ range accessible with XTD8, $\sigma_\tau$ can take values from \SI{6}{\micro\meter} to \SI{13}{\micro\meter}.

\section{Diagnostics and corrector placement}

As the accelerator is running, one must insure that despite variations in the beam optics, the electron beam arrives with the right properties. To achieve this, the XTD8 lattice must contain diagnostic instruments for various parameters. In placing the beam instrumentation, the ability to monitor LUXE's IP or other focus points is most important. Both the transverse positions $x,y$ and angles $x',y'$ can be measured using button BPMs. In addition, the lattice must contain corrector magnets to adjust those parameters. Similarly, the transverse beam sizes $\sigma_x$ and $\sigma_y$ can be measured using wire scanners. This section aims to find suitable positions for those diagnostics in the XTD8.

\subsection{BPM placement}

The electron beam transverse position is driven by Eq~\ref{eq:beta_oscillation}. Between two longitudinal position $s_0$ and $s$, there is a phase advance difference $\delta\mu=\mu(s)-\mu(s_0)$. A position variation at the first location $\delta x(s_0)$ is linearly proportional to the induced variation at the second location $\delta x(s)$ if the phase advance respects $\sin(\delta\mu)=0$. This corresponds to a phase difference of

\begin{equation}
    \delta\mu=n\pi,
    \label{eq:correl_mu_position}
\end{equation}

where $n$ is a integer. A BPM placed at a position $s$ will be maximally correlated in $x$ to LUXE's IP if

\begin{equation}
    \delta\mu_{x, \rm IP}(s) = \mu_x(s)-\mu_{x, \rm IP} = n\pi,
\end{equation}

where $\mu_{x, \rm IP}$ is the horizontal phase advance at the IP. This phase difference also indicates a linear correlation between an angle variation $\delta x'(s_0)$ and the induced angle variation at the IP $\delta x'_{\rm IP}$. Similarly, with Eqn~\ref{eq:beta_angle_oscillation} a correlation can be established between position and angle. A position variation $\delta x(s_0)$ is correlated to the induced angle variation $\delta x'(s)$ if the phase advance difference is

\begin{figure}[!b]
    \centering
    \includegraphics[width=15cm]{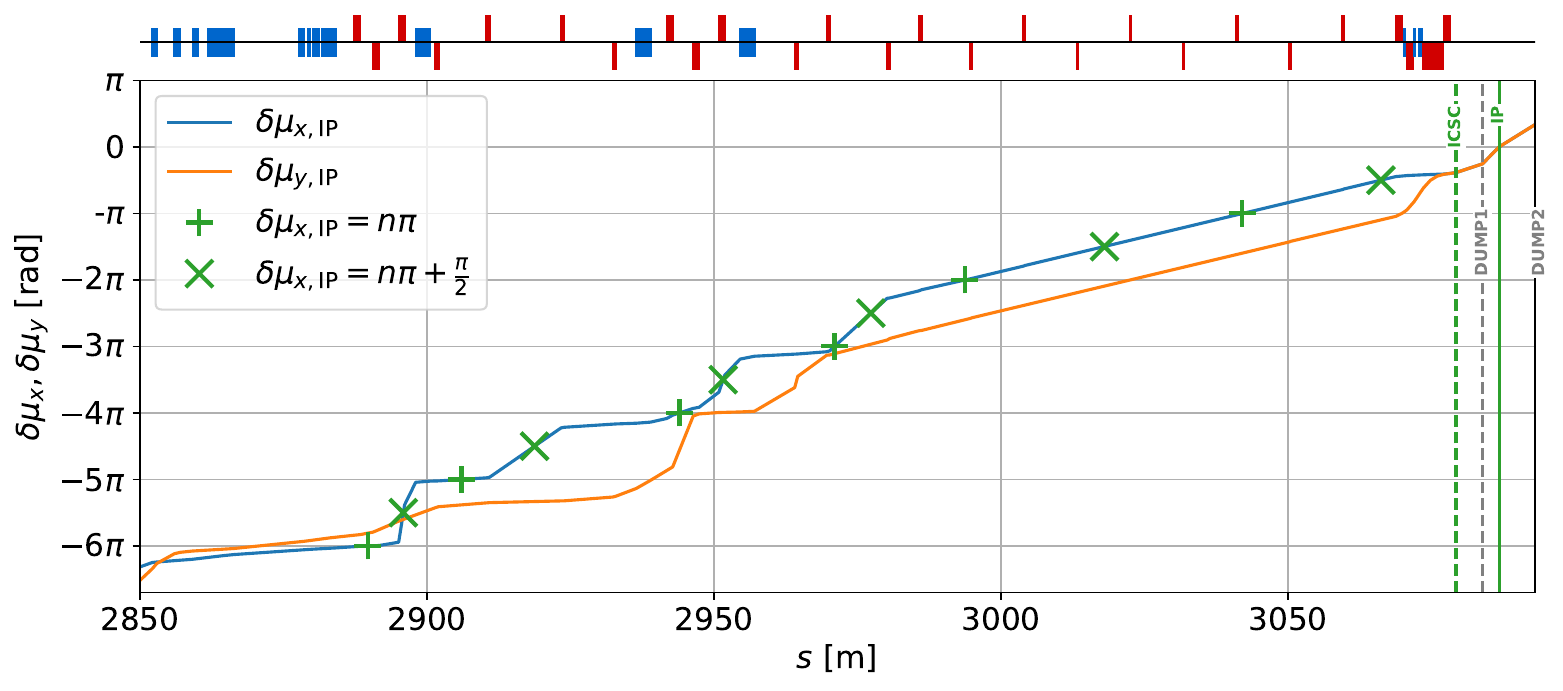}
    \caption{Plot of the phase advance difference to LUXE's IP in both planes $\delta\mu_{x, \rm IP}$ and $\delta\mu_{y, \rm IP}$ as a function of $s$. The crosses indicates the locations where Eqn~\ref{eq:correl_mu_position} is satisfied in the $x$ plane. The diagonal crosses indicates the locations where Eqn~\ref{eq:correl_mu_angle} is satisfied in the $x$ plane.}
    \label{fig:correl_mux}
\end{figure}

\begin{equation}
    \delta\mu=n\pi + \frac{\pi}{2}.
    \label{eq:correl_mu_angle}
\end{equation}

This is also true for an angle variation $\delta x'(s_0)$ and the induced position variation $\delta x(s)$. Along the XTD8 lattice there are multiple positions where those correlation locations to the IP can be found, as shown in Figure~\ref{fig:correl_mux}.
There are six locations with a horizontal position to position (or angle to angle) correlation to the IP. As well, there are six locations with high horizontal correlation between position and angle. 

\begin{figure}[!ht]
    \centering
    \includegraphics[width=15cm]{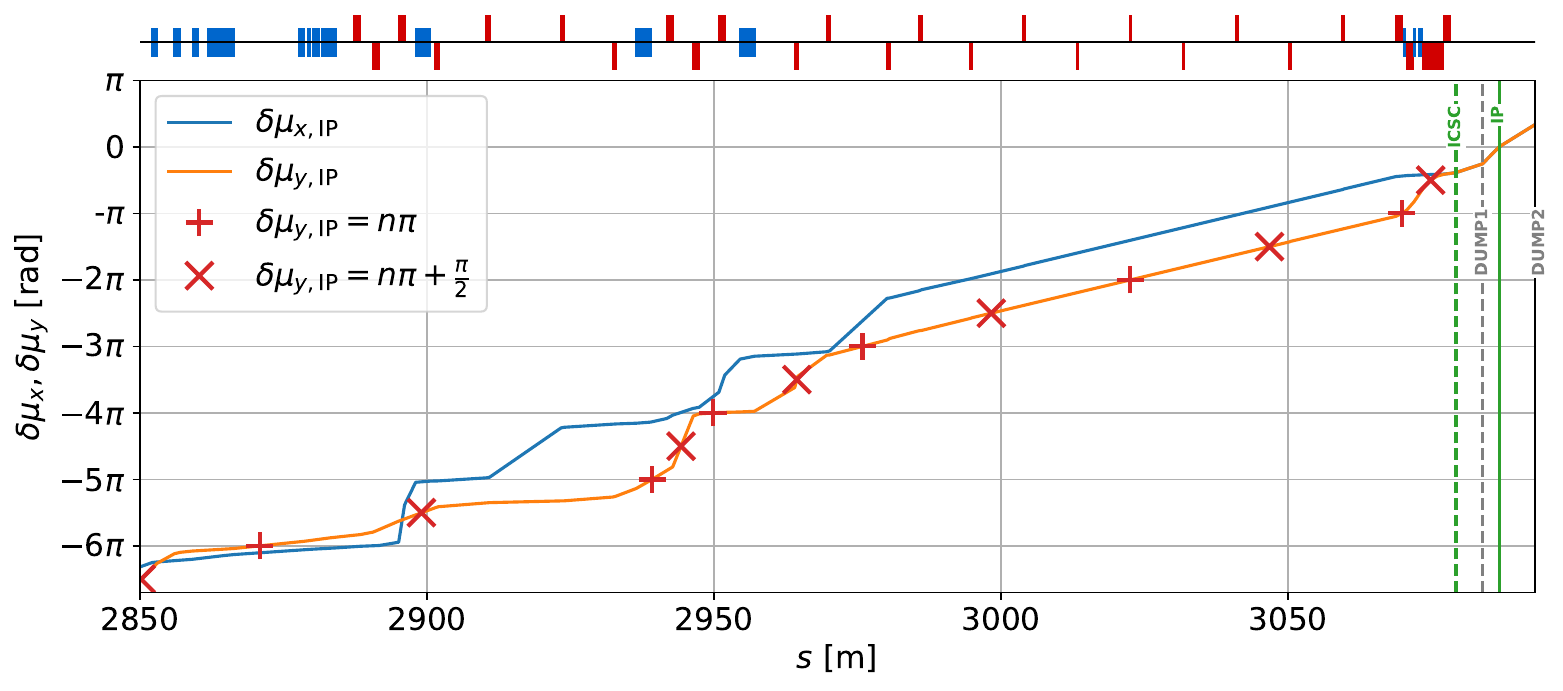}
    \caption{Plot of the phase advance difference to LUXE's IP in both planes $\delta\mu_{x, \rm IP}$ and $\delta\mu_{y, \rm IP}$ as a function of $s$. The crosses indicates the locations where Eqn~\ref{eq:correl_mu_position} is satisfied in the $y$ plane. The diagonal crosses indicates the locations where Eqn~\ref{eq:correl_mu_angle} is satisfied in the $y$ plane.}
    \label{fig:correl_muy}
\end{figure}

Similarly, the locations correlated in the $y$ plane can be found in Figure~\ref{fig:correl_muy}.
In the vertical plane there are six locations with high position to position correlation and seven locations with high correlation between position and angle.
In theory, to measure the transverse position and angle at the IP, the lattice needs only four BPMs, one for each transverse angle and position. Among the different locations available from Figure~\ref{fig:correl_mux} and Figure~\ref{fig:correl_muy}, one can select BPM locations with the highest sensitivity and sufficient longitudinal space. 

\begin{figure}[!ht]
    \centering
    \includegraphics[width=15cm]{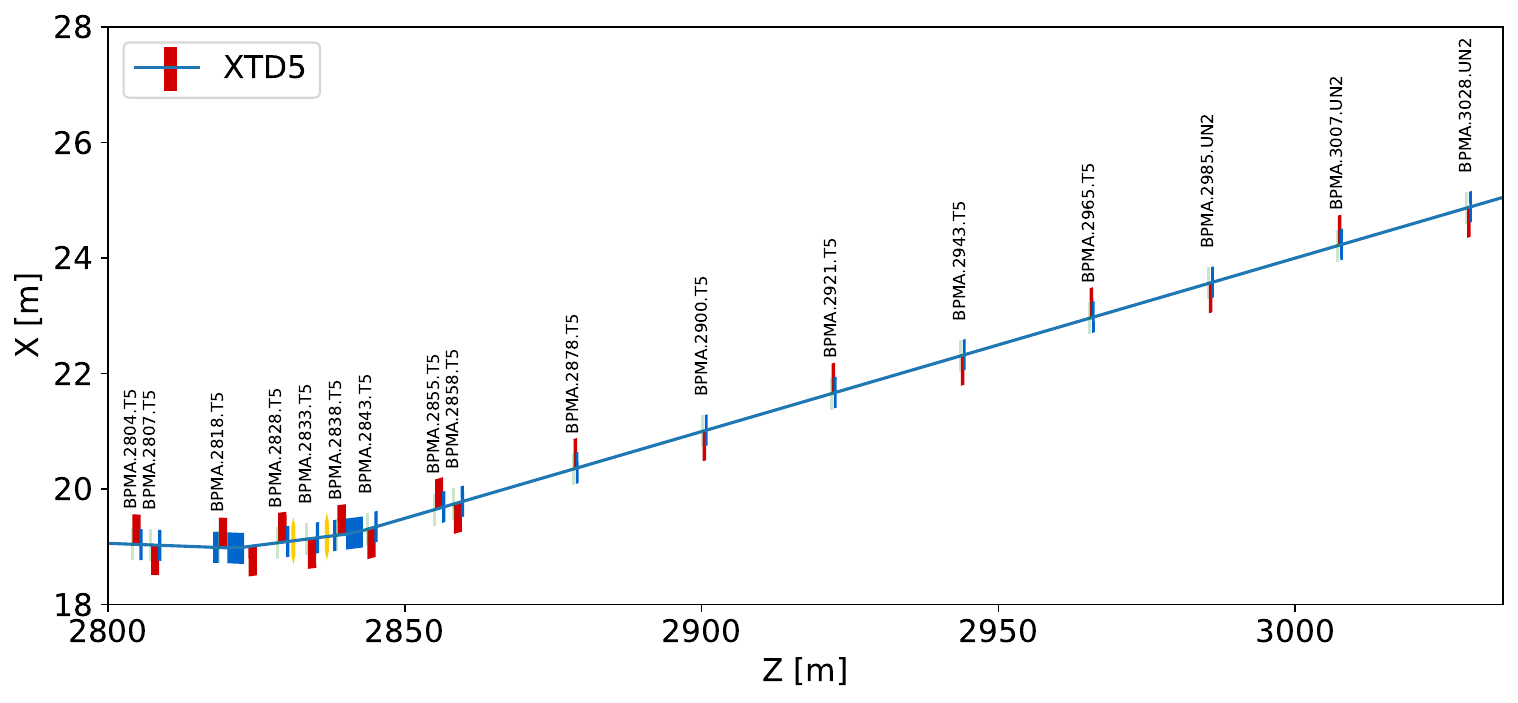}
    \caption{Schematic of the XTD5 bending section and transport line. All BPM name are indicated at their respective positions.}
    \label{fig:xtd5_bpms}
\end{figure}

However, in the EuXFEL transport lines there is usually one BPM associated to every quadrupole. An example of the XTD5 lattice can be found in Figure~\ref{fig:xtd5_bpms}.
If the XTD8 lattice features a BPM at every quadrupole, the majority of them will not be perfectly correlated to the IP phase advance. But a high number of position diagnostics allow the use of the virtual measurement method described in Chapter~\ref{chap:measurments}, to infer the position and angle at the LUXE IP.

\subsection{Wire scanner placement}

\begin{figure}[!ht]
    \centering
    \includegraphics[width=\textwidth]{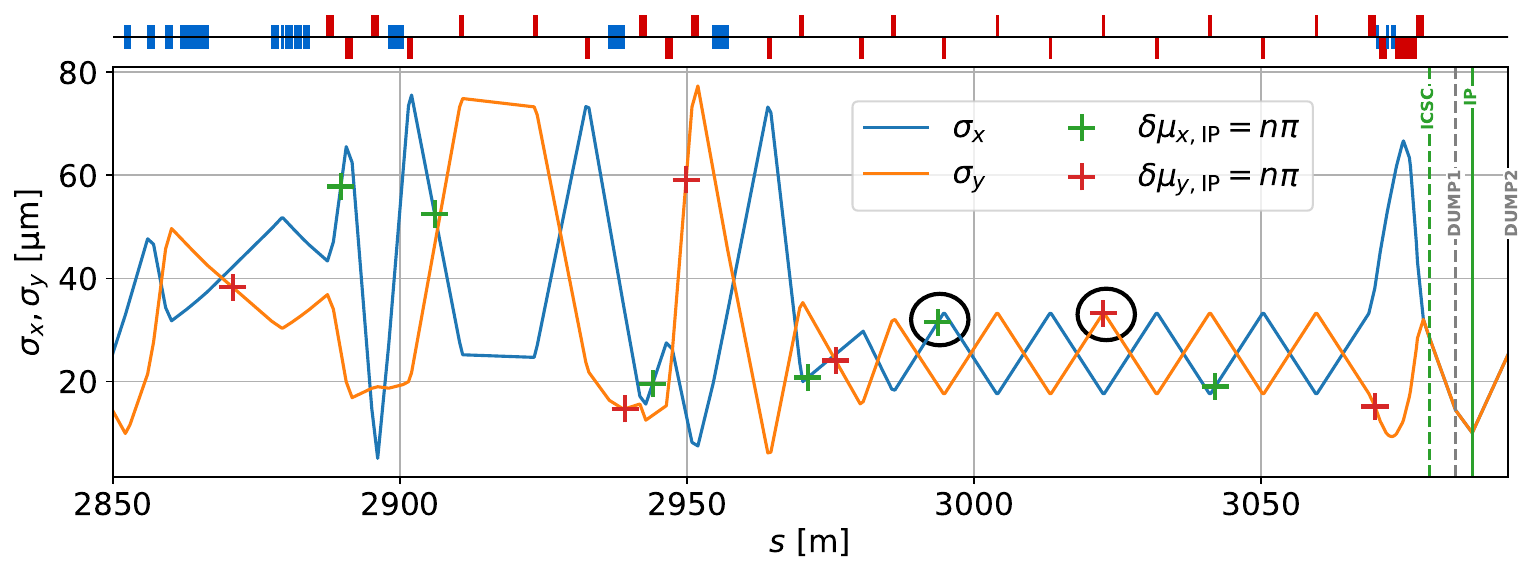}
    \caption{Plot of the transverse beam sizes $\sigma_{x,y}$ as a function of $s$. The crosses indicates the location and beam sizes where $\delta\mu = n\pi$ for both planes. The two black circles indicates the best locations in $x$ and $y$ for the wire scanners.}
    \label{fig:xtd8_sigma_correl}
\end{figure}

The horizontal beam size is driven by Eqn~\ref{eq:sigma_x_disp}. Like the $\beta$ function, it oscillates as a function of the phase advance $\mu_x$. As such, at the position to position correlated locations found in Figure~\ref{fig:correl_mux}, the beam size $\sigma_x$ will be linearly correlated to $\sigma_{x, \rm IP}$. Placed at any location where $\delta\mu_{\rm IP} = n\pi$, a wire scanner could directly relate to the IP transverse spot size. The main argument to select a location for the scanners is the resolution achievable for a variations at the IP. This can be easily evaluated in Figure~\ref{fig:xtd8_sigma_correl} with the evolution of the default beam size.

A measurement of the horizontal beam size can be done in any of the six locations in Figure~\ref{fig:xtd8_sigma_correl}. Suppose $\sigma_x$ is measured at the first one ($s\approx2890\ \rm m$) where the beam size is almost \SI{60}{\micro\meter} and at the last one ($s\approx3042\ \rm m$) where the beam size is below \SI{20}{\micro\meter}. If there is a \SI{5}{\%} increase for $\sigma_{x, \rm IP}$, the beam size will increase by \SI{3}{\micro\meter} at the first wire scanner and only by \SI{1}{\micro\meter} at the second one. Given the wire scanner resolution is fixed, a location with a larger beam size will be more sensitive to a variation at the IP. However, in the dispersive sections, a variation of the beam size could be attributed to a change in the energy spread. As such, the correlation points in the extraction and dogleg, i.e. below $s=2960\ \rm m $, must be avoided. In the end the two best locations are around \SI{2994}{\meter} for a horizontal wire scanner and around \SI{3022}{\meter} for a vertical one.

\subsection{Corrector placement}

When a transverse displacement (compared to the laser focus), either in position or angle, is detected at the IP, corrector magnets~\cite{correctors} are used to realign the beam. A corrector is a small dipole with adjustable strength used to change the angle of the electron beam at a given location. Depending on its orientation, it is referred as a horizontal or vertical corrector, i.e. bending on the $x$ plane or $y$ respectively. As with the BPMs, the placement of correctors is driven by the phase advance. In this situation, it requires positions where an angle variation $\delta x'(s_0)$ in a corrector is correlated to a position variation $\delta x(s)$ or an angle variation $\delta x'(s)$. The correlated location are identical to the ones found in Figure~\ref{fig:correl_mux} and Figure~\ref{fig:correl_muy}. In the same way, only four corrector are needed in theory to adjust both $x, y, x'$ and $y'$. 

However, some phase advance favoured locations are located quite far upstream in the extraction or dogleg.
When a corrector is on, the downstream beam deviates from the reference trajectory, i.e. the corrector introduces orbit. Such a beam will not cross magnets and corrector in their transverse centre. Going through and off-centred quadrupole will introduce some dispersion, with respect to the energy spread of the beam. In a BPM, a bunch passing significantly closer to an electrode could introduce non-linearities in the voltage response of said electrode. As such, it is preferred to place the correctors as close as possible to the IP.
The resolution of the transverse position at the IP is then limited by the electrical current resolution of the corrector's digital-to-analogue converter (DAC). For example, if the DAC is controlled with 10 bits, the total number of settings will be $2^{10} = 1024$. Assuming the alimentation's range is between $\pm$\SI{10}{\ampere}, the minimum current variation possible is around \SI{9.8}{\milli\ampere}.
Having correctors closer to the IP should allow a more precise tunnability of the position and angle.

\begin{figure}[!ht]
    \centering
    \includegraphics[width=15cm]{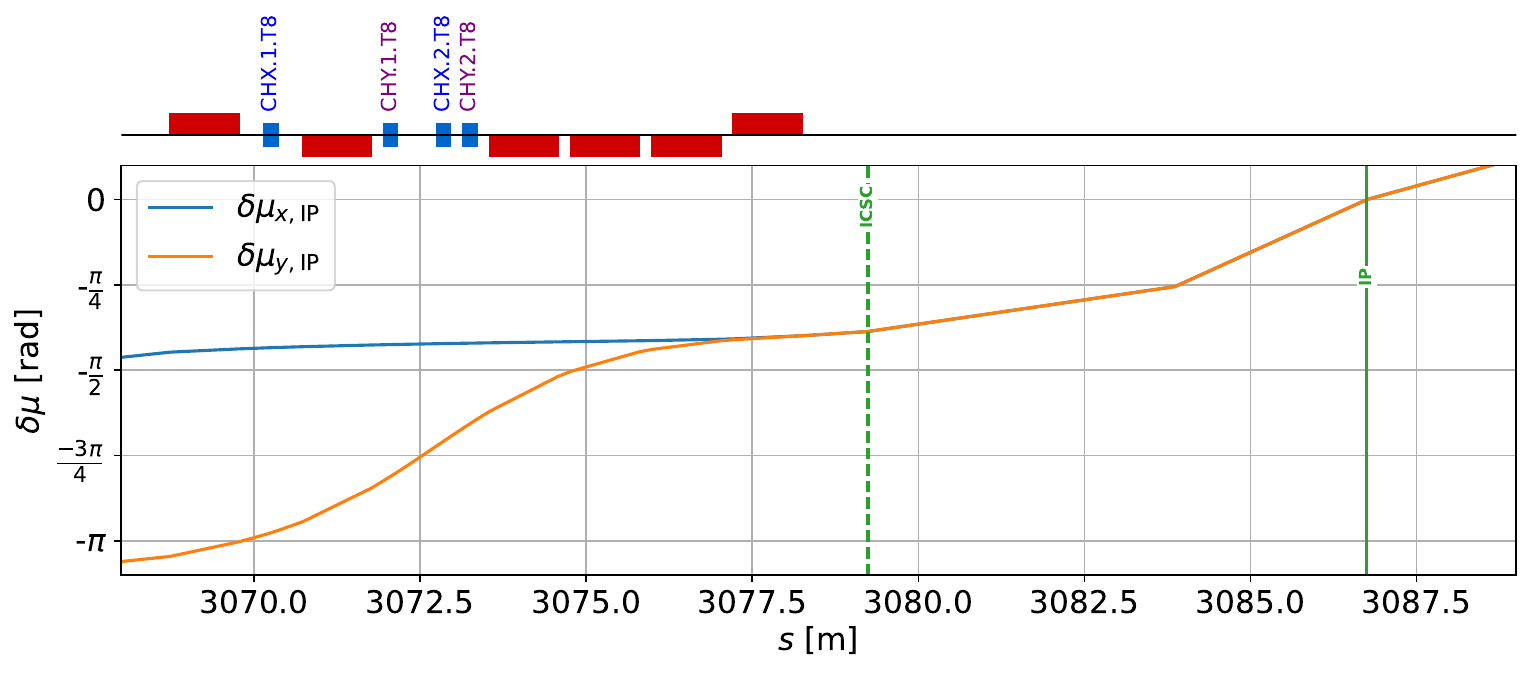}
    \caption{Plot of the phase advance difference $\delta \mu_{\rm IP}$ in both planes as a function of $s$. The four corrector names are indicated on the machine diagram.}
    \label{fig:xtd8_ff_mu}
\end{figure}

\begin{figure}[!ht]
    \centering
    \includegraphics[width=\textwidth]{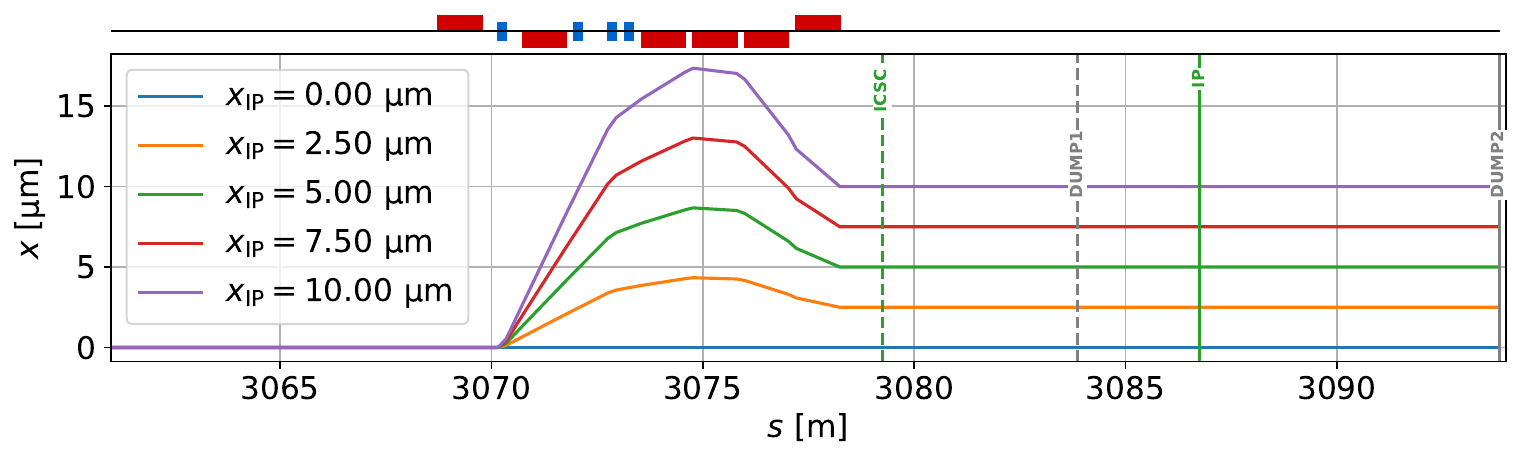}
    \caption{Plot of the local horizontal position $x$ as a function of $s$. Each curve corresponds to a different position at the IP $x_{\rm IP}$.}
    \label{fig:xtd8_ff_scan_x}
\end{figure}

Instead of using one corrector for $x$ and one for $x'$, it is possible to combine two correctors, placed at different phase advances, to adjust simultaneously $x$ and $x'$ at the IP. In the current final focus design, four correctors (two horizontal correctors CHX and two vertical correctors CHY) are already placed, as shown in Figure~\ref{fig:xtd8_ff_mu}.
On the vertical plane there is almost a $\frac{\pi}{4}$ phase advance difference between the two CHY correctors. On the horizontal plane, the phase advance difference is quite small between the two CHX correctors. To demonstrate a correction in the horizontal plane, the Xsuite optimizer was used. The variables are both correctors strength $k_0$ and the target is the position and angle at LUXE's IP $x_{\rm IP}$ and $x'_{\rm IP}$. In Figure~\ref{fig:xtd8_ff_scan_x}, the horizontal trajectory are shown for $x'_{\rm IP}=0$ and multiple values of $x_{\rm IP}$ between 0 and \SI{10}{\micro\meter}.

\begin{figure}[!ht]
    \centering
    \includegraphics[width=\textwidth]{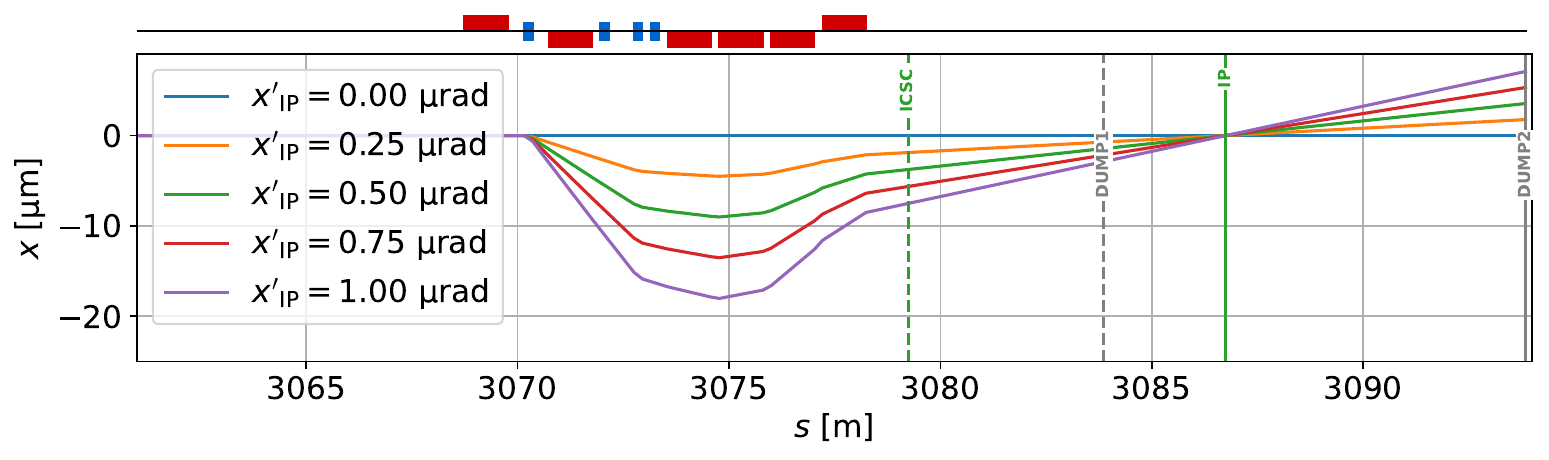}
    \caption{Plot of the local horizontal position $x$ as a function of $s$. Each curve corresponds to a different angle at the IP $x'_{\rm IP}$.}
    \label{fig:xtd8_ff_scan_px}
\end{figure}

A correction of \SI{10}{\micro\meter} in horizontal position implies a maximum displacement below \SI{20}{\micro\meter} in the final focus which is reasonable compared to the beam pipe radius. In Figure~\ref{fig:xtd8_ff_scan_px} the same correctors are optimized for $x_{\rm IP}=0$ and multiple values of $x'_{\rm IP}$ between 0 and \SI{1}{\micro\radian}

\begin{figure}[!ht]
    \centering
    \includegraphics[width=\textwidth]{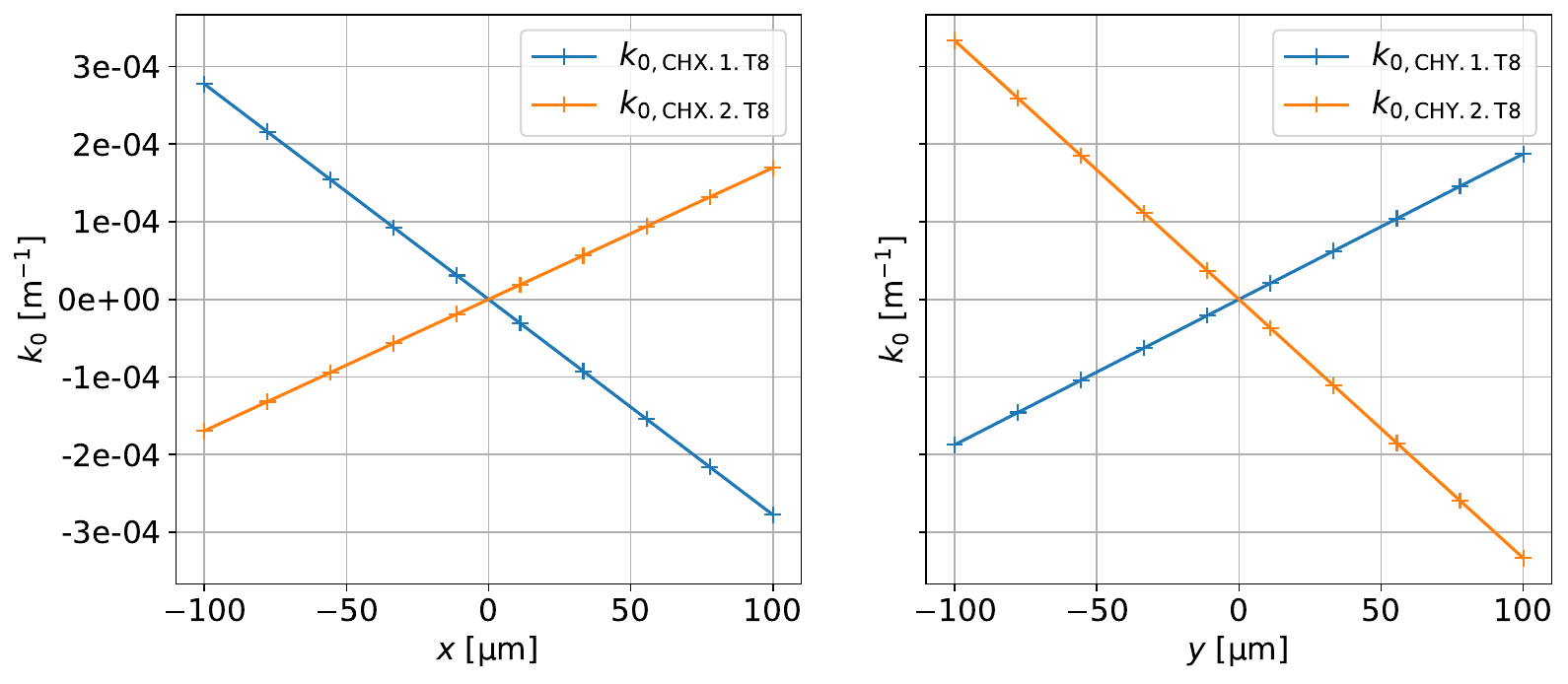}
    \caption{Plots of the horizontal and vertical position knobs. On the left are the horizontal corrector strengths as a function of $x$. On the right are the vertical corrector strengths as a function of $y$.}
    \label{fig:xtd8_ff_knob_pos}
\end{figure}

A correction of \SI{1}{\micro\radian} in horizontal position implies a maximum displacement below \SI{20}{\micro\meter} in the final focus which is reasonable considering the beam pipe radius. The required corrector strengths for both transverse positions are shown in Figure~\ref{fig:xtd8_ff_knob_pos}.

\begin{figure}[!ht]
    \centering
    \includegraphics[width=\textwidth]{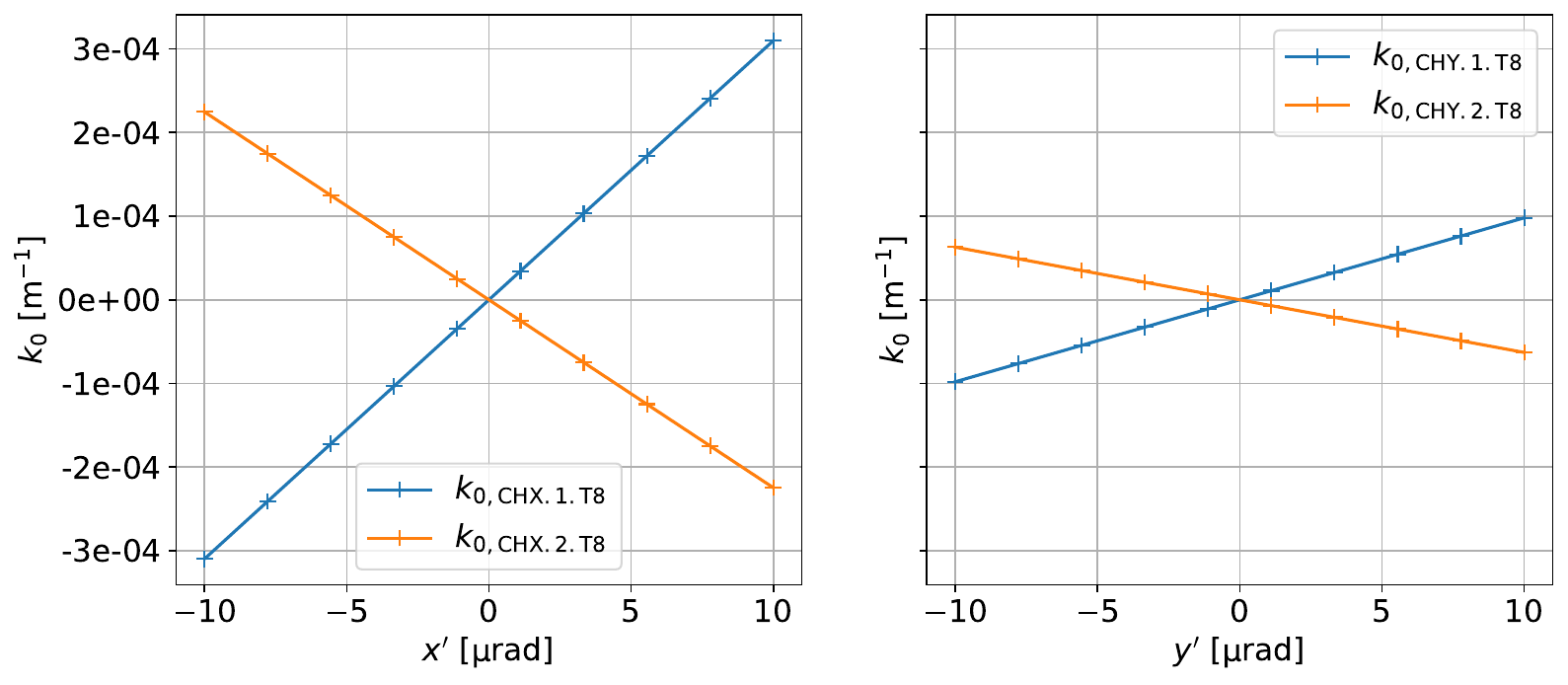}
    \caption{Plots of the horizontal and vertical angle knobs. On the left are the horizontal corrector strengths as a function of $x'$. On the right are the vertical corrector strengths as a function of $y'$.}
    \label{fig:xtd8_ff_knob_ang}
\end{figure}

In both configurations there is a linear relationship between the corrector strengths $k_0$ and the position. For a position variation between \SI{-100}{\micro\meter} and \SI{100}{\micro\meter} the corrector strength $k_0$ are at most $\pm$\SI{3.3e-4}{\meter^{-1}}. Similar scans and knobs for angle correction are shown in Figure~\ref{fig:xtd8_ff_knob_ang}.

For an angle variation between \SI{-10}{\micro\radian} and \SI{10}{\micro\radian} the corrector strength $k_0$ needed are up to \SI{3e-4}{\meter^{-1}} in $x'$ but below \SI{1e-4}{\meter^{-1}} in $y'$. This might be due to the higher phase difference between the vertical correctors.

\newpage

\section{Vertical bend for radioprotection} \label{sec:vbend}

\begin{figure}[!ht]
    \centering
    \includegraphics[width=\textwidth]{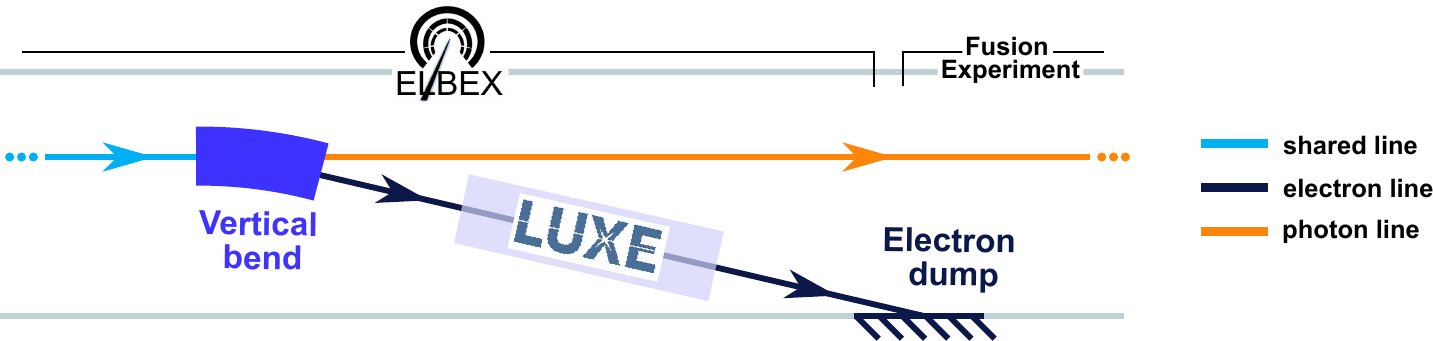}
    \caption{Longitudinal schematic of the vertical bend placement in the XTD8 tunnel.}
    \label{fig:vbend}
\end{figure}

As Figure~\ref{fig:elbex} showed, the ELBEX project is located in the XTD8 tunnel upstream of a fusion experiment that will use undulator produced photons. Up to this point, the lattice design kept a shared beam pipe between electrons for LUXE and photons for the fusion experiment. 
To avoid any direct electrons propagating to the fusion experiment and any background coming from the EuXFEL beam or LUXE's products, it was decided to bend the electron beam line vertically, starting at the final focus. A separated photon pipe will then continue straight downstream. This way, the \SI{17.5}{\giga\electronvolt} electrons and the QED products will be dumped towards the ground before the beginning of the fusion experiment. In the previous sections, the dogleg was optimized to get an achromatic XTD8 beam line. Bending vertically the beam line while conserving an achromatic design would require another double-bend achromat. This will add two dipoles, at least six quadrupoles and will take quite some longitudinal space. To circumvent that, another solution is possible with only a single dipole, as shown in Figure~\ref{fig:vbend}.

\begin{figure}[!ht]
    \centering
    \includegraphics[width=\textwidth]{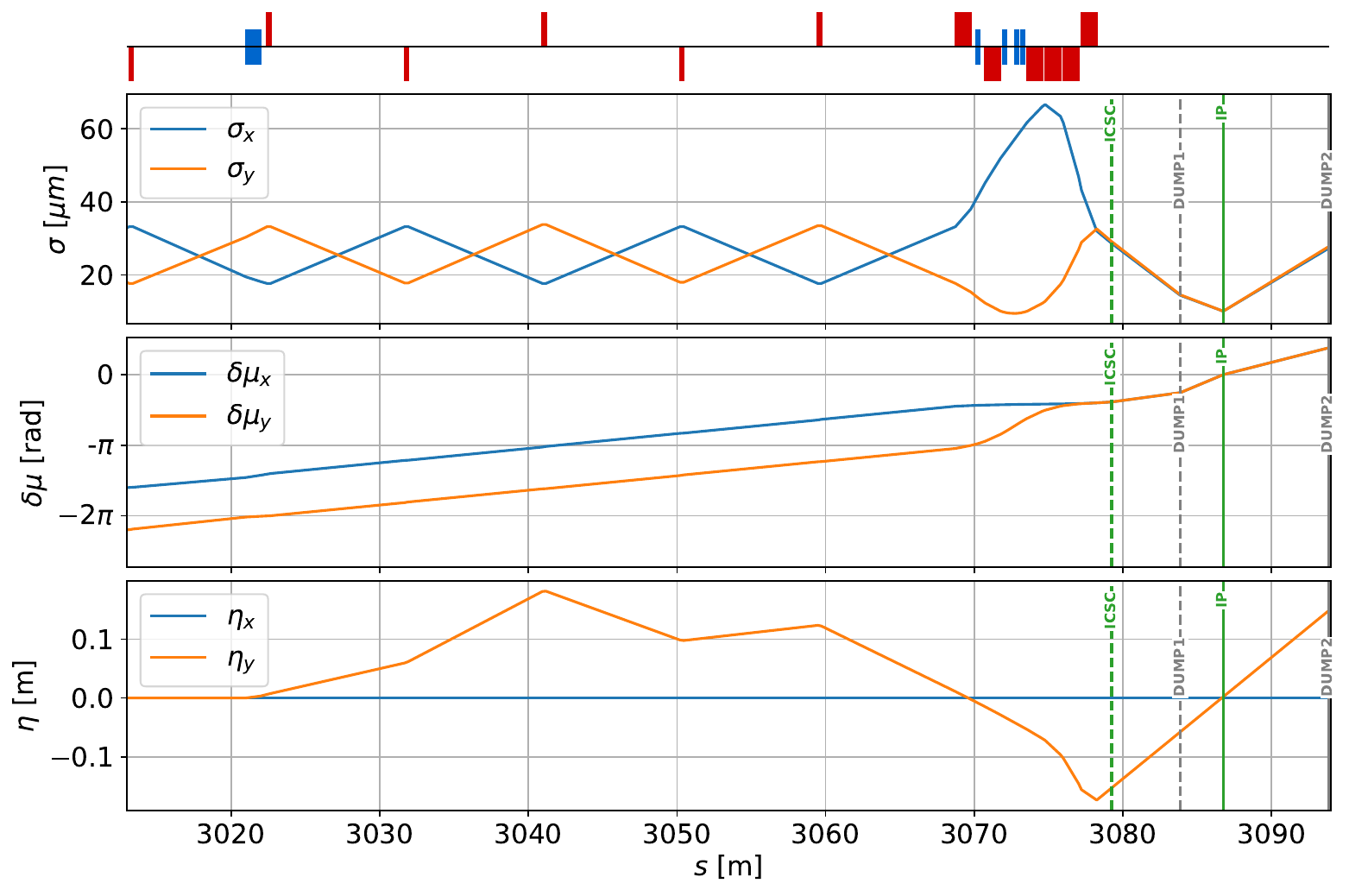}
    \caption{Plots of beam size, phase difference and dispersion for the XTD8 transport FODO and final focus along $s$ with a vertical bend at $\delta \mu_y = -2\pi$. On top is the beam sizes $\sigma_{x,y}$, On the middle is the phase advance differences $\delta \mu_{x,y}$. On the bottom is the dispersions $\eta_{x,y}$.}
    \label{fig:xtd8_vbend_2pi}
\end{figure}

With one dipole added in the lattice, it will generate some vertical dispersion $\eta_y$. However, like the betatron oscillation, the dispersion will oscillate as a function of the phase advance $\mu_y$. If the dipole is placed at a $n\pi$ phase advance difference from the IP then the optics will show $\eta_y=0$ at the IP. This is shown in Figure~\ref{fig:xtd8_vbend_2pi}.

\begin{figure}[!ht]
    \centering
    \includegraphics[width=10cm]{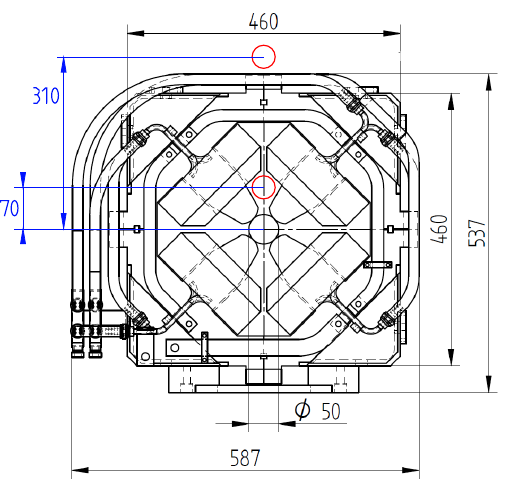}
    \caption{Transverse schematic of the XTD8 transport and final focus quadrupole. The red circles indicate the potential position of a separate photon pipe going through the quadrupole pocket or clearing the quadrupole completely. All measurements are indicated in millimetres.}
    \label{fig:QE_quad}
\end{figure}

With a vertical bend placed at a $2\pi$ phase advance difference from the IP, i.e. $s\approx 3021\ \rm m$, the beam size is barely changing. Without any re-matching of the final focus quadrupole strengths, the \SI{10}{\micro\meter} spot size at the IP is recovered as the vertical dispersion $\eta_y$ crosses 0 at that point. This is true no matter the vertical bend angle. Such a solution works regarding the optics but does not consider the geometric constraints of the photon beam pipe. After the separation with the electron beam pipe, the photon pipe has to avoid any overlap with the quadrupoles geometry. As the electron pipe stays in the centre of the magnets, the photon pipe must either already be above the entire geometry or go through what is called the 'pocket' of the quadrupole. These two pipe position are shown in Figure~\ref{fig:QE_quad}.

In the quadrupoles used in the XTD8, the pocket is around \SI{70}{\milli\meter} above the magnet centre. To clear the quadrupole completely, the pipe must be above \SI{310}{\milli\meter} of the magnet centre. With the current location of our vertical bend, its distance from the next quadrupole is around \SI{0.42}{\meter}. The angle needed to clear this quadrupole is then around \ang{36.4}, this is way too high. For the photon pipe to go through the quadrupole pocket, the angle goes down to around \ang{9.5}. Even then, the end of LUXE which is around \SI{80}{\meter} downstream will be \SI{13}{\meter} below the photon pipe, i.e. quite deep in the ground. The beginning of the fusion experiment is estimated at $s\approx 3123\ \rm m$. At this location, the vertical displacement of the electron line must be around \SI{1}{\meter}.

\begin{figure}[!ht]
    \centering
    \includegraphics[width=\textwidth]{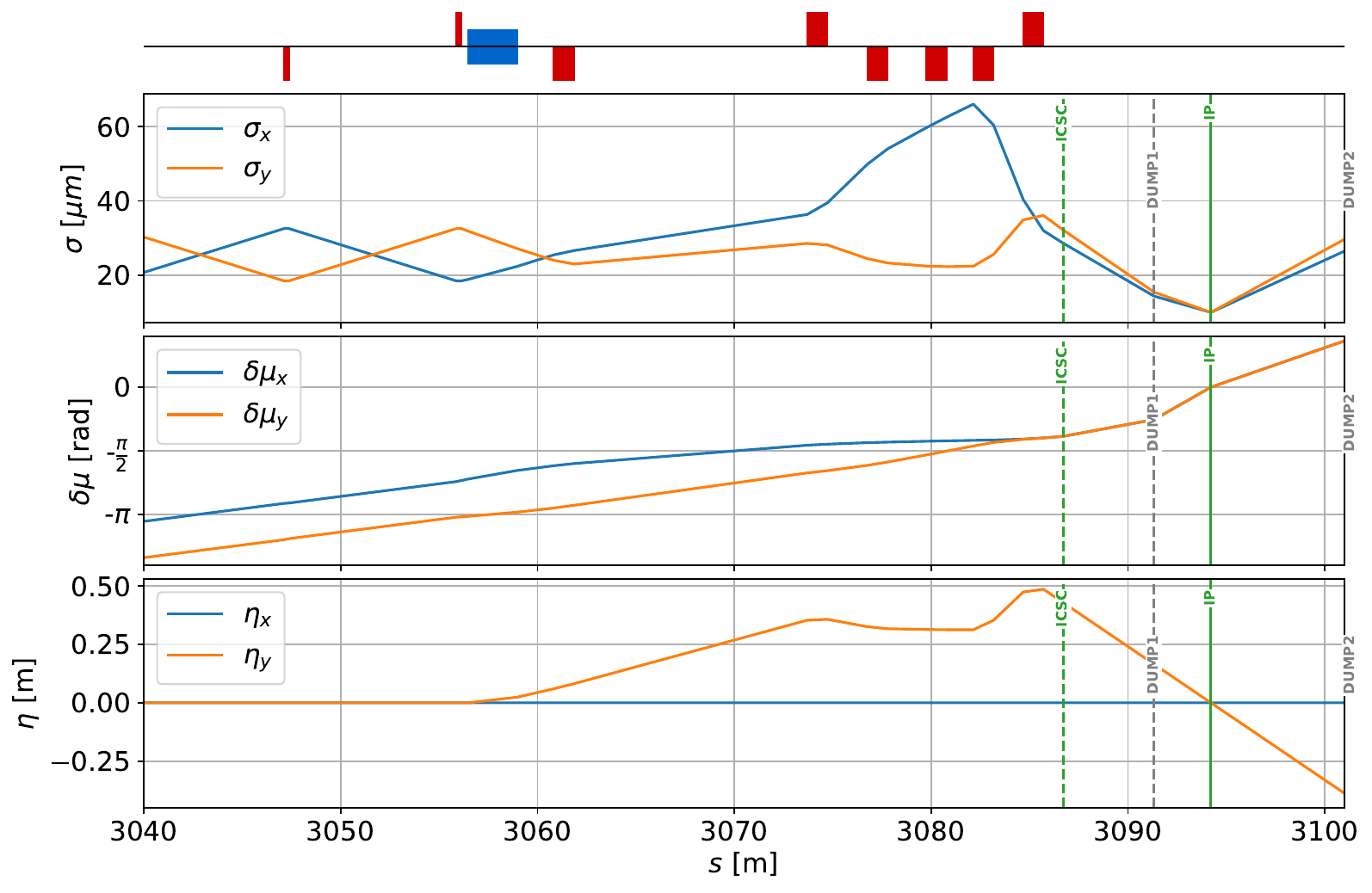}
    \caption{Plots of beam size, phase difference and dispersion for the XTD8 transport FODO and final focus along $s$ with a vertical bend after the FODO. On top is the beam sizes $\sigma_{x,y}$, On the middle is the phase advance differences $\delta \mu_{x,y}$. On the bottom is the dispersions $\eta_{x,y}$.}
    \label{fig:xtd8_vbend_pocket}
\end{figure}

To fulfil both the phase advance and geometric constraints, the easiest way is to redesign the final focus. First, the vertical bend is placed closer to the final focus ($s\approx3030\ \rm m$) as the phase advance will be readjusted by matching in Xsuite. The variables for Xsuite are the position $s$ and strengths $k_1$ of the six final focus quadrupoles as well as the vertical bend angle. The targets are $\eta_x=0\ \rm m$ and $\sigma_{x,y}=10\ \rm \mu m$ at the IP. Additionally, a constraint is put on the first two quadrupoles so that the photon beam pipe will pass through the pocket of the first one and above the second one. The results of this match are shown in Figure~\ref{fig:xtd8_vbend_pocket}.

With the rematched final focus, the vertical bend is at a phase advance $\delta\mu_y=-\pi$. As such, the vertical dispersion $\eta_y$ is equal to zero at the IP. Compared to Figure~\ref{fig:xtd8_vbend_2pi}, the maximum dispersion increased to almost \SI{0.5}{\meter}. The vertical bend angle is around \ang{1.11} which corresponds to an offset in $y$ of around \SI{1.29}{\meter} at $s=3123\ \rm m$. Now that the quadrupoles are placed, the Xsuite match can be rerun to get a focus at the ICS chamber or at any $L^*$.

\section{Permanent magnets}

In typical accelerator facility, the majority of the magnets used in the lattice are electromagnets. For a long time it was the favoured method to reach the required magnetic fields between 1 and \SI{2}{\tesla}. Superconducting (SC)~\cite{sc_mag} magnets can even reach as high as \SI{15}{\tesla}. Those electromagnets requires a power and water cooling supply and even cryogenic systems for SC magnets. This adds constraints on the magnets installation and is a non-negligible cost when the facility is running. An alternate method is to use permanent magnets~\cite{perm_mag_1} as they do not require a power supply to generate a magnetic field. 
Even if the magnetic field produced by a single permanent magnet is constant, it is possible to make a magnetic element with varying field with permanent magnets. For a quadrupole magnet, there are two sets of permanent magnets. Four static magnets closer to the beam pipe that produces the major part of the magnetic field and four magnets around the first set that can be rotated to adjust the field strength~\cite{perm_mag_2}. The adjustment range of such quadrupole is of the order of \SI{100}{} to \SI{200}{\tesla/\meter}. However, a complete change of quadrupole polarity, which is doable on an electromagnet, is not currently possible for the permanent magnets. In the quadrupole example, this would require moving the central set of magnets.
Applications of permanent magnets in accelerator facilities started in the early 1980's and the technology continues to see improvements~\cite{perm_mag_2}. In the context of the future XTD8 beam line, having some permanent magnets among the fifty new magnets could reduce the required power supplies and reduce the total cost of the line. This section aims to evaluate which magnets can be permanent in the current design.

\subsection{Extraction and transport}

The XTD8 extraction is composed of six fast kickers and 4 septa dipoles. The kickers must be able to select a precise electron bunch in the train and send it down the XTD8 line. Bunches in the EuXFEL trains have a minimal spacing of \SI{220}{\nano\second}. Permanent magnets can not operate at such speeds and such are not suited for a fast kicker application. The septas are dipoles that only kick bunches with a sufficient offset. Their magnetic field does not require a fast variation so permanent magnets are, in theory, a viable option here.
However, due to the unique geometry of a septa, there is currently no proven design for a septa with permanent magnets. As such, an electromagnetic design is still the most suitable option for a septa magnet.
In the transport sections of the XDT8, including both the dogleg and the FODO, the magnets strengths should see little to no variations. Even with a adjustment of the $R_{56}$ factor, Figure~\ref{fig:k1_bars_wrt_R56} showed that the impact of the quadrupole strengths $k_1$ is minimal. As such, permanent magnet are a viable option for quadrupoles in the dogleg and the FODO.

\subsection{Final focus}

\begin{figure}[!ht]
    \centering
    \includegraphics[width=14cm]{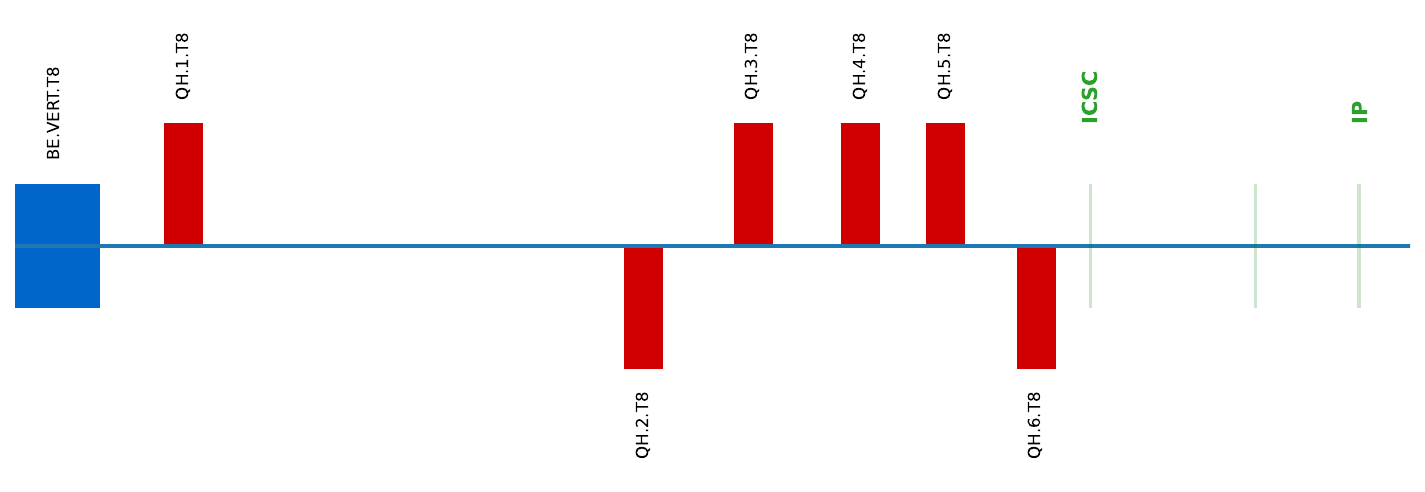}
    \caption{Schematic of the final focus with the element names.}
    \label{fig:FF_vbend_lattice}
\end{figure}

\begin{figure}[!ht]
    \centering
    \includegraphics[width=\textwidth]{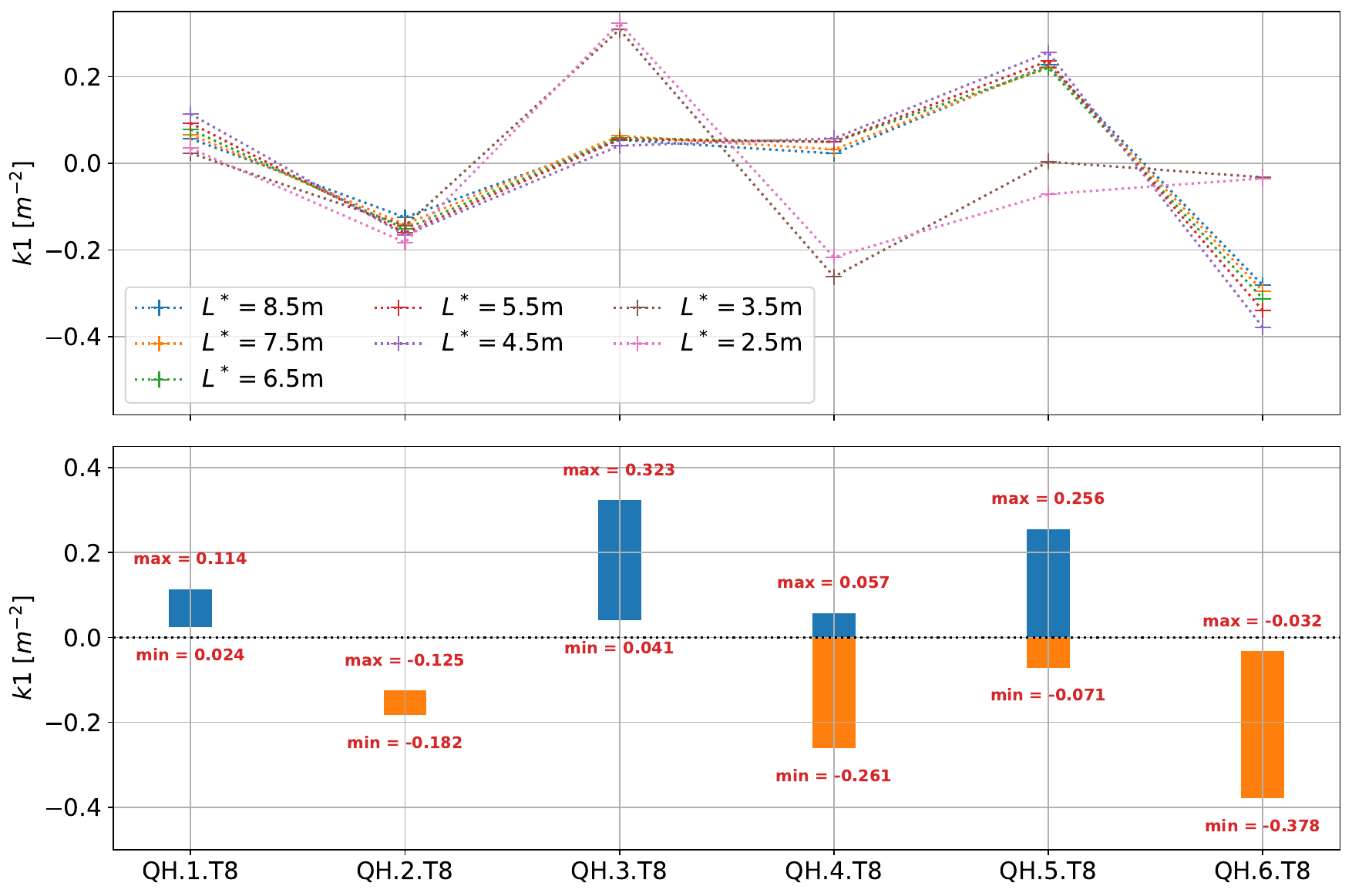}
    \caption{Plots of the quadrupole strengths $k_1$ for the six quadrupoles of the final focus. On the top are the curves corresponding to each $L^*$. On the bottom are the bars containing all values of $k_1$ taken by each quadrupole.}
    \label{fig:FF_K1_curves_bars}
\end{figure}

The XTD8 final focus was designed such that the spot size $\sigma^*$ and the distance from last quadrupole $L^*$ can be adjusted to the experiment needs. As it is possible to modify the strength $k_1$ of the permanent magnets, changing the polarity is not possible. 
Using the final focus layout shown in Figure~\ref{fig:FF_vbend_lattice}, multiple Xsuite optimisations were ran to focus the beam at different $L^*$ between \SI{2.5}{\meter} and \SI{8.5}{\meter}. The resulting $k_1$ for each quadrupole in all configuration are shown in Figure~\ref{fig:FF_K1_curves_bars}.

\newpage

For the values of $L^*$ above \SI{4}{\meter}, the quadrupoles strengths variations are quite small and there is no polarity change. However, for $L^*=3.5\ \rm m$ and $L^*=2.5\ \rm m$, the last four quadrupoles sees more significant $k_1$ changes. Especially the fourth and fifth quadrupoles that do change polarity. The ICS chamber for LUXE, which is one of the focus locations, is close to $L^*=2\ \rm m$ so it is currently not possible to design the final focus with permanent magnets. 
A similar set of simulation can be established for different values of the focus size $\sigma^*$. Looking a the quadrupoles strengths shown in Figure~\ref{fig:knob_sigma_ICSC}, one of the quadrupole did change polarity at $\sigma^*=8\ \rm \mu m$. This further confirms that permanent magnets can not be used in LUXE's final focus.

\chapter{Beam impact on LUXE} \label{chap:impact_luxe}

For the LUXE experiment, the aim is to collide an electron beam with either a laser or a converter target. Preliminary simulations have shown that strong field QED processes can be obtained with the high energy electrons of the EuXFEL. However, the quality of signal produced in LUXE's interaction chamber depends on the quality of the electron beam. This chapter aims to evaluate the properties of the electron beam as it enters LUXE and study the impact of the beam variations to the experiment signal. The first two sections are studying the beam interaction with the residual gas and the beam pipe. In the third section a virtual measurement method is used to measure the transverse beam position at LUXE's IP. The fourth section propagates the beam jitter established in Table~\ref{tab:euxfel_param_jitter} up to LUXE's IP. Finally, the last section applies those jitter to the electron-laser collision at LUXE and evaluates the impact of each parameters on the QED signal.

\section{Background from beam gas interactions}

Inside the EuXFEL electron beam pipes, there is an average residual gas pressure of \SI{1e-7}{} mbar~\cite{vacuum}. In specific sections of the accelerator, it is even lower (\SI{1e-10}{} mbar in the superconducting RF cavities and \SI{1e-9}{} mbar in the undulators). The typical composition of the residual gas is \SI{80}{\%} of $\rm H_2$ and \SI{20}{\%} of $\rm CO$~\cite{euxfel_tdr}. 
As the electron beam travels along the \SI{3.4}{\kilo\meter} of beam pipe it will interact with the residual gas and produce a background spectrum of particles. Those secondary particles could impact LUXE measurements if they are produced in high enough numbers. The goal is to simulate the interaction between the electron beam and the residual gas to insure that this induce background will not perturb the experiment. To perform those simulations, a BSDIM model was build from the beam line designed in Chapter~\ref{chap:design}. However, this BDSIM model does not include the vertical bend proposed for radioprotection.

\subsection{Physics processes}

\begin{figure}[!ht]
        \centering
        \includegraphics[width=12cm]{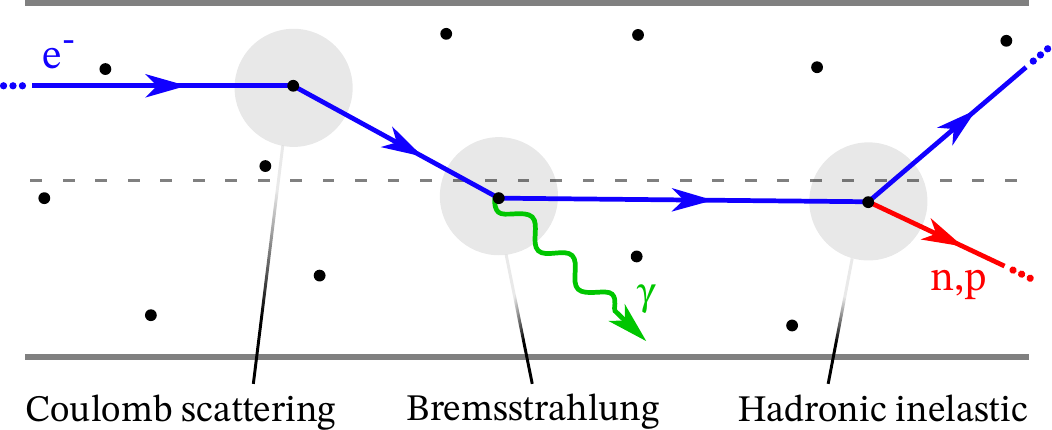}
        \caption{Schematic of a beam pipe longitudinal profile with residual gas. An electron successively interacts via Coulomb scattering, bremsstrahlung and hadronic inelastic collision.}
        \label{fig:pipe_sketch}
\end{figure}

Along the \SI{300}{\meter} of the XTD8 lattice, the electron interact with the residual gas through various processes. Among all those processes, three of them will produce most of the secondary particles : Coulomb scattering~\cite{coulomb}, bremsstrahlung~\cite{bremss} and hadronic inelastic collisions~\cite{hadronic}.
Figure~\ref{fig:pipe_sketch} shows these processes. 

\subsubsection{Coulomb scattering}

\begin{figure}[!ht]
        \centering
        \includegraphics[width=6cm]{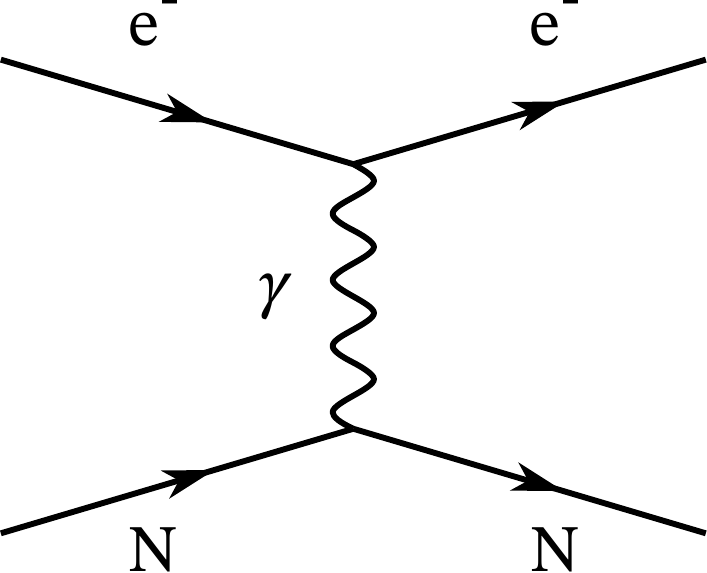}
        \caption{Feynman diagram of the Coulomb scattering interaction of an electron on a nucleus $N$.}
        \label{fig:coulomb}
\end{figure}

The Coulomb scattering interaction is an elastic collision between two particles. As an electron approaches a nucleus of the gas it is deflected by the electromagnetic force of the nucleus as shown in Figure~\ref{fig:coulomb}. 

For this elastic collision, the conservation of energy can be written as

\begin{equation}
    E_{e,\rm in} + E_{N,\rm in} = 
    E_{e,\rm out} + E_{N,\rm out},
\end{equation}

where $E_{e,\rm in}$ and $E_{e,\rm out}$ are respectively the input and output energy of the electron and where $E_{N,\rm in}$ and $E_{N,\rm out}$ are respectively the input and output energy of the nucleus.
The electron can usually transfer some of its kinetic energy to the nucleus.
The energy transferred to the nucleus $E_T$ can be expressed as~\cite{geant4_manual}

\begin{equation}
    E_T = \frac{2m_Nc^2(pc)^2}{E^2_{\rm cm}}\sin\frac{\theta}{2},
\end{equation}

where $m_e$ and $m_N$ are respectively the masses of the electron and the nucleus, $\theta$ is the scattered angle, $p$ is the input electron momentum and $E_{\rm cm}$ is the energy of the centre of mass. The output electron energy is the expressed as

\begin{equation}
    E_{e,\rm out} = E_{e,\rm in} - E_T.
\end{equation}

The Coulomb scattering process does not produce secondary particles but it degrades the electrons energy and induce a scatter of the individual electron trajectories. 


\subsubsection{Bremsstrahlung}

\begin{figure}[!ht]
        \centering
        \includegraphics[width=6cm]{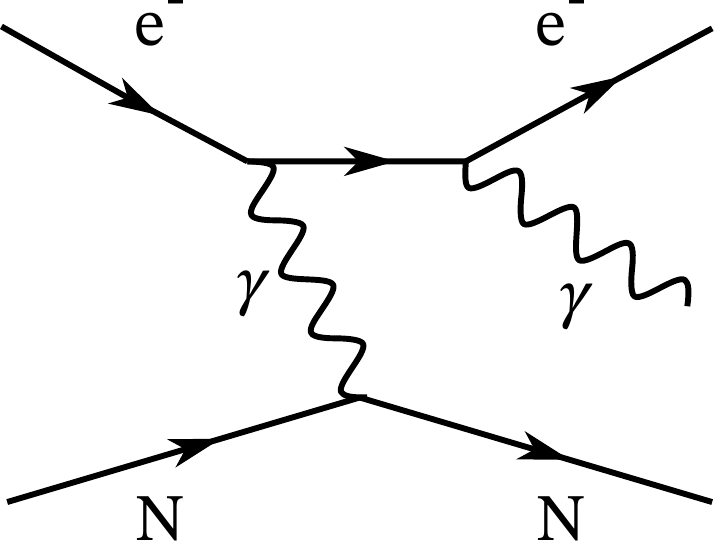}
        \caption{Feynman diagram of the bremsstrahlung interaction of an electron on a nucleus $N$.}
        \label{fig:bremsstrahlung}
\end{figure}

As an electron approaches a nucleus, there is a probability for the electron to radiate a photon, as shown in Figure~\ref{fig:bremsstrahlung}

For this elastic collision, the conservation of energy can be written as

\begin{equation}
    E_{e,\rm in} + E_{N,\rm in} = 
    E_{e,\rm out} + E_{N,\rm out} + E_\gamma,
\end{equation}

where $E_{e,\rm in}$ and $E_{e,\rm out}$ are respectively the input and output energy of the electron, $E_{N,\rm in}$ and $E_{N,\rm out}$ are respectively the input and output energy of the nucleus and $E_\gamma$ is the energy of the produced photon.
As with Coulomb scattering, this interaction decreases the electron initial energy $E_{e, \rm in}$. It also produce a secondary photon that could interact with the residual gas and create other secondary particles. 
The cross section for the bremsstrahlung interaction is derived from the Bethe-Heitler equation~\cite{bremss}. 




\subsubsection{Hadronic inelastic collision}

\begin{figure}[!ht]
        \centering
        \includegraphics[width=6cm]{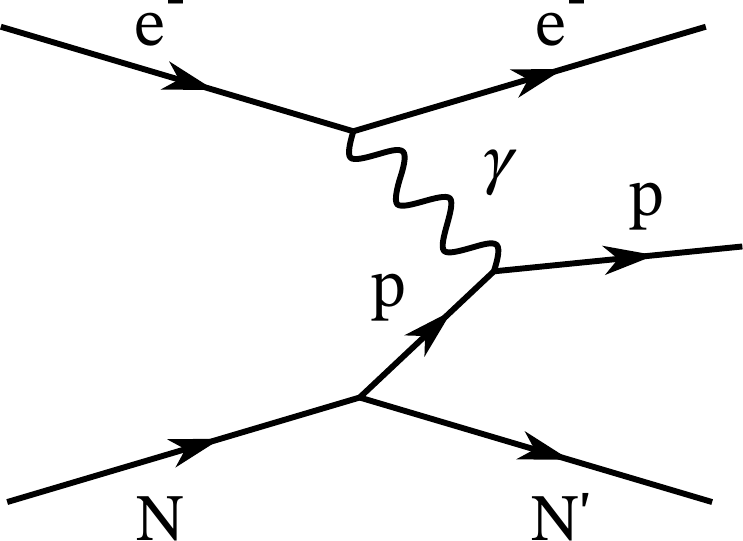}
        \caption{Feynman diagram of the hadronic inelastic interaction of an electron on a nucleus $N$. In this example a secondary proton p is ejected from the nucleus.}
        \label{fig:hadronic}
\end{figure}

If an electron collides with a nucleus with sufficient energy and at a close enough distance, there is a probability of an inelastic collision~\cite{hadronic} that ejects a proton or a neutron as shown in Figure~\ref{fig:hadronic}.

In this process, the incoming electron interacts with a specific nucleon. It receives enough energy to be ejected from the nucleus. Like with the bremsstrahlung interaction, this secondary nucleon can then collides with other nuclei of the residual gas.

\subsection{Impact on the primary electrons}

The probability each process occurs at is driven by a cross section $\sigma$. The mean free path of an electron for a process is then defined by

\begin{equation}
    \lambda = \frac{1}{\sigma n},
\end{equation}

where $n$ is the number density of the gas. The number of electrons $N_e$ will vary along $s$ by following the differential equation 

\begin{equation}
    \frac{dN_e}{ds} + \frac{1}{\lambda}N_e = 0.
\end{equation}

Solving this equation, the number of electrons at a given $s$ can be expressed as an exponential attenuation law

\begin{equation}
    N_e(s) = N_{e,0}\cdot e^{-\frac{s}{\lambda}},
    \label{eq:attenuation}
\end{equation}

where $N_{e,0}$ it the initial number of electrons. Assuming that those are independent processes, the total cross section $\sigma_{\rm total}$ is simply expressed as

\begin{equation}
    \sigma_{\rm total} = \sigma_{\rm Coulomb} + \sigma_{\rm brem} + \sigma_{\rm hadronic},
\end{equation}

where $\sigma_{\rm Coulomb}$, $\sigma_{\rm brem}$ and $\sigma_{\rm hadronic}$ are the cross sections for each process. As such, Eqn~\ref{eq:attenuation} can be written as

\begin{equation}
    N_e(s) = N_{e,0}\cdot e^{-\frac{s}{\lambda_{\rm total}}}
    = N_{e,0}\cdot e^{-\sigma_{\rm total}ns}.
\end{equation}

In the EuXFEL there is an average residual gas pressure $P$ of \SI{1e-7}{} mbar. It is assumed that it follows the ideal gas law

\begin{equation}
    P=\frac{n}{\mathcal{N}_A}RT,
\end{equation}

where is $T$ the temperature, $R$ is the the ideal gas constant and $\mathcal{N}_A$ is the Avogadro number. With a typical temperature of \SI{300}{\kelvin}, the number density becomes $n=2.41\times10^{15}\ \rm m^{-3}$. With the typical composition at the EuXFEL this is equivalent to a mass density $n_{\rm mass} = 2.89\times10^{-14}\ \rm g.cm^{-3}$. This last number is more relevant as it is the unit used in BDSIM.
Let us consider the example of Coulomb scattering at this density. In the Geant4 physics manual~\cite{geant4_coulomb}, the Coulomb cross section is calculated from Mott differential cross section (MDCS)~\cite{mdcs} approximations. In this model, the scatters once on a stationary, point-like nucleus and without any screening considered.
With a \SI{17.5}{GeV} electron beam, values of $\sigma_{\rm Coulomb}$ are of the order $10^{-23}\ \rm m^2$. If the beam travels along a distance $D$ of the beam pipe the percentage of particle lost is expressed as

\begin{equation}
    \frac{N_{e,0}-N_e(D)}{N_{e,0}} = 1-e^{-\sigma_{\rm Coulomb} nD}.
\end{equation}

When running simulations for our \SI{300}{\meter} lattice, one can expect from this equation than only \SI{0.001}{\%} of the primary electrons will interact via Coulomb scattering.

\subsection{Cross section biasing}

In the context of BDSIM simulations, the low rate of interaction would require to generate multiple bunch crossing in order to evaluate the particle spectrum for LUXE. To circumvent this, the cross section can be artificially increased, hence the mean free path $\lambda$ is decreased. This is called biasing. There are two ways to bias $\lambda$, either increasing the gas density $n$ or increasing the cross section $\sigma$ for each processes~\cite{bias}. Let us consider a process of initial cross section $\sigma$ for which a biasing factor $\chi$ is applied. The number of electrons differential equation becomes

\begin{equation}
    \frac{dN_{e, \rm bias}}{ds} + \frac{1}{\lambda_{\rm bias}}N_{e, \rm bias} = 0,
\end{equation}

where $\lambda_{\rm bias} = \frac{1}{\chi\sigma n}$ is the biased mean free path. For any interaction $i$, the number of particles produced with this biasing 
can be related to the real number of particle produced 
via a weight $\omega_i$.
%
%
This weight is also applied to the electrons that interacts through the biased process. The real number of electrons $N_e(s)$ can be expressed as

\begin{equation}
    N_e(s) = N_{e,0} - \sum_{s_0\rightarrow s}\omega_i .
\end{equation}

It is important to set the biasing factors correctly for our simulations. If $\chi$ too low, there will not be any interactions at all. If it is too high, the number of electrons $N_e(s)$ will see significant variations along $s$, therefore inducing an undesired longitudinal bias in the simulation. 
In the outputs of BDSIM simulations, $N_e(s)$ can be evaluated by looking at the number of primary first hits $N_{\rm PFH} (s)$.
In our case, a primary first hit (PFH) is the first interaction between a primary particle and the residual gas as it is tracked in BDSIM.
$N_{\rm PFH}$ corresponds then to number of collisions between the primary electrons and the gas at a given $s$. $N_e$ and  $N_{\rm PFH}$ are proportional and follows the same exponential attenuation law.
This study will primarily rely on the cross section biasing of the PFH only, i.e. the subsequent hits of the primary electrons and every interaction with the secondary particles are not biased.
However, the simulation are also run at a mass density $n_{\rm mass}=1\times10^{-12}\ \rm g.cm^{-3}$, which is roughly two orders of magnitude higher than the real density. This higher density decreases the mean free path $\lambda$ and increases the number of collision along the lattice. In the results, a scaling factor is applied in order to see the effect of the real gas density. 

\subsection{Biasing factors for each process}

The biasing factors are adjusted independently for the three processes. For each of them, the goal is to obtain a uniform distribution of the primary first hits along the lattice. Using BDSIM, \SI{1e6}{} primary electrons were tracked along the beam line for different values of $\chi$.

\subsubsection{Coulomb scattering}

\begin{figure}[!ht]
    \centering
    \includegraphics[width=\textwidth]{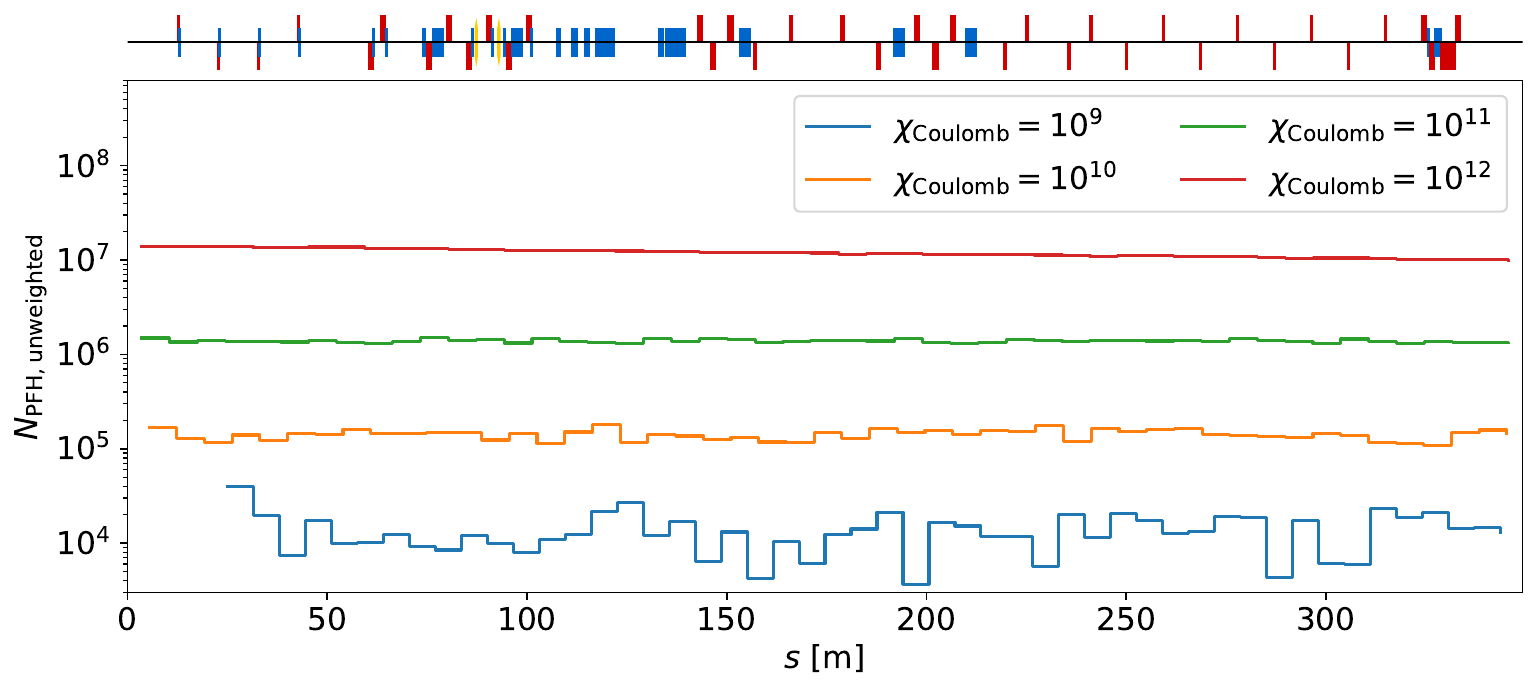}
    \caption{Histogram of unweighted primary first hits along the lattice for different values of $\chi_{\rm Coulomb}$. The number of hits is scaled to one bunch crossing (\SI{2e9}{} electrons).}
    \label{fig:pfh_coulomb}
\end{figure}

\begin{figure}[!ht]
    \centering
    \includegraphics[width=\textwidth]{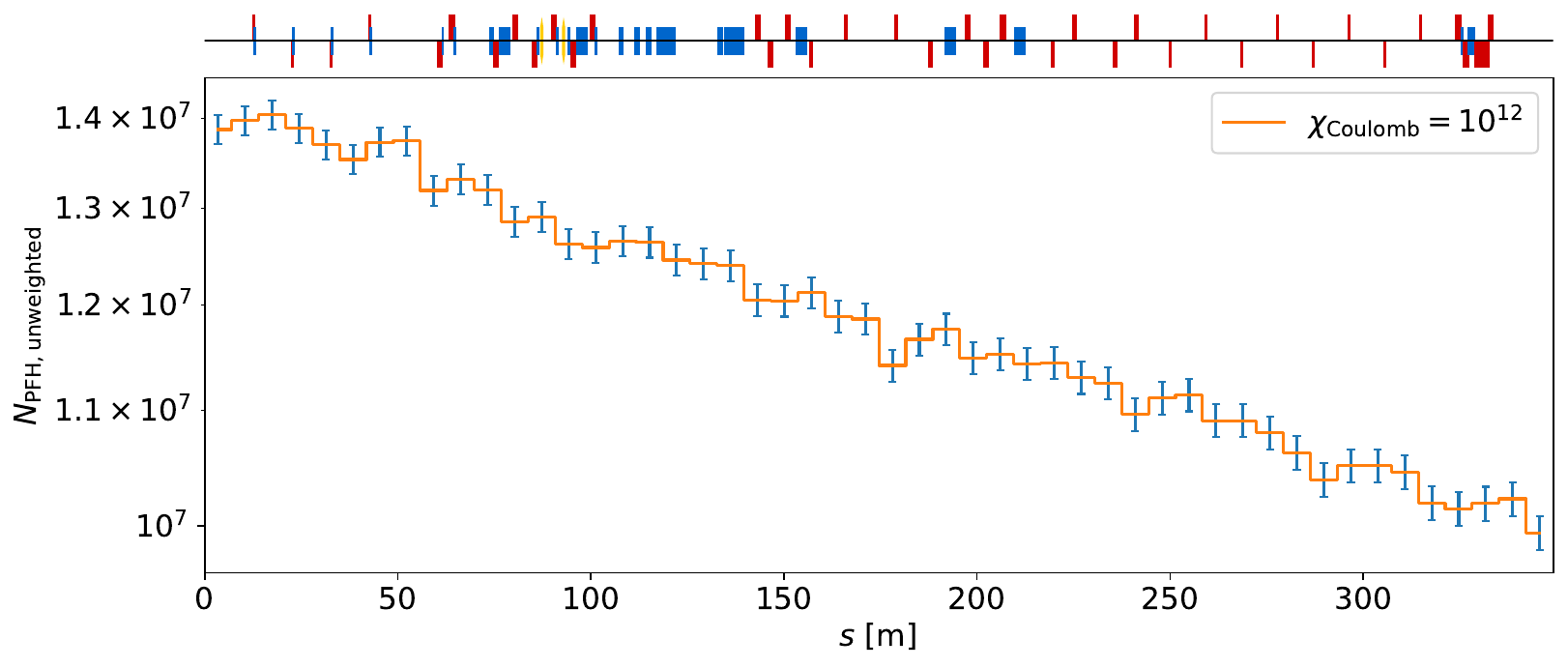}
    \caption{Histogram of unweighted primary first hits $N_{\rm PFH}$ along the lattice for $\chi_{\rm Coulomb} = 10^{12}$ with error bars.}
    \label{fig:pfh_coulomb_error}
\end{figure}

Let us start with the biasing factor for Coulomb scattering $\chi_{\rm Coulomb}$. In Figure~\ref{fig:pfh_coulomb}, the biasing factor was progressively increased until some hits are observed.
Below $\chi_{\rm Coulomb} = 10^8$, barely any Coulomb scattering interactions were simulated on the primary electrons. The average number of hits along $s$, $\langle N_{\rm PFH} (s)\rangle$, is linearly proportional to $\chi_{\rm Coulomb}$. At $\chi_{\rm Coulomb} = 10^{12}$ he number $N_{\rm PFH}$ is decreasing as a function of $s$. This can be more easily seen in Figure~\ref{fig:pfh_coulomb_error}.

Here the biasing factor is high enough to introduce a longitudinal biasing. As previously stated, this is not desired for our background simulations. Looking at those $N_{\rm PFH}$ curves it is not trivial to evaluate what biasing factor is sufficient for our study without longitudinal biasing. 
Suppose $S_{\rm PFH}$ is defined as the normalised standard deviation of $N_{\rm PFH}$ along $s$. It is expressed as a percentage with

\begin{equation}
    S_{\rm PFH}(s) [\%] = \frac{\langle N_{\rm PFH}(s)^2 \rangle}{\langle N_{\rm PFH}(s)\rangle}\times 100 .
\end{equation}

This value should increase at low biasing factors, where the number of interactions is small, and the variation is large. Similarly, it should increase at high biasing factors due to the longitudinal biasing seen in Figure~\ref{fig:pfh_coulomb_error}. 

\begin{figure}[!ht]
    \centering
    \includegraphics[width=\textwidth]{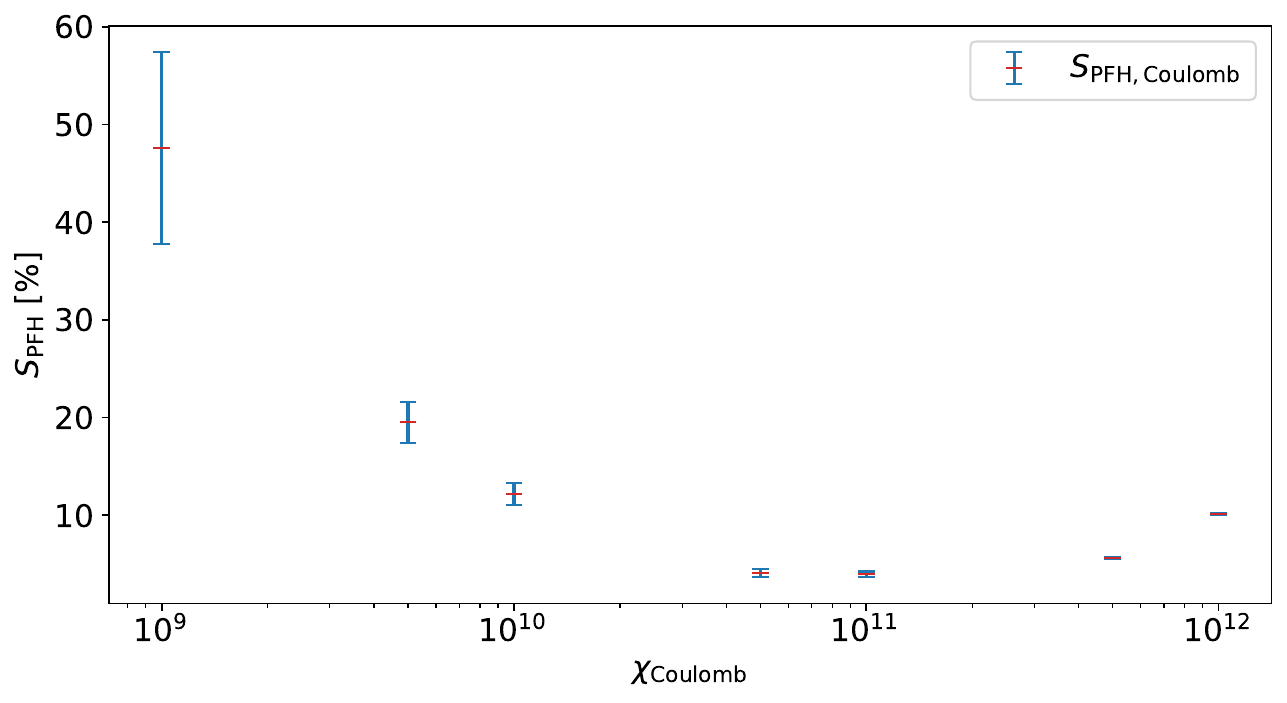}
    \caption{Plot of the normalised standard deviation of number of primary first hits $S_{\rm PFH}$ as a function of the biasing factor $\chi_{\rm Coulomb}$.}
    \label{fig:var_coulomb}
\end{figure}

\begin{figure}[!ht]
    \centering
    \includegraphics[width=\textwidth]{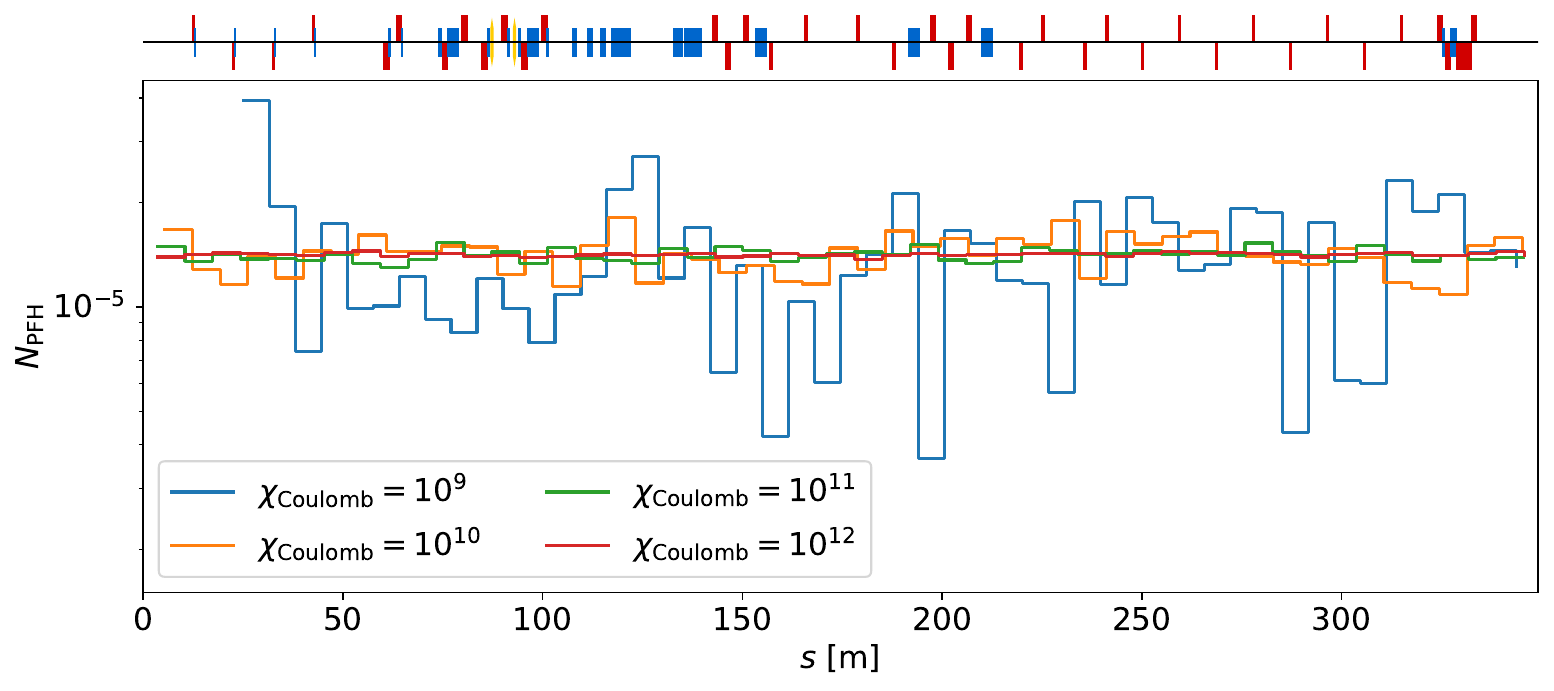}
    \caption{Histogram of weighted primary first hits $N_{\rm PFH}$ along the lattice for different values of $\chi_{\rm Coulomb}$. The number of hits is scaled to one bunch crossing (\SI{2e9}{} electrons).}
    \label{fig:pfh_coulomb_weighted}
\end{figure}

In Figure~\ref{fig:var_coulomb}, $S_{\rm PFH}(s)$ was calculated for different values of $\chi_{\rm Coulomb}$.
Below $\chi_{\rm Coulomb}=5\times10^{10}$, the normalised standard deviation increases up to around \SI{50}{\%} due to the low statistics. On the other end, there is a slight increase in $S_{\rm PFH}$ at $\chi_{\rm Coulomb}=10^{12}$. 
Figure~\ref{fig:pfh_coulomb_error} showed that at this value of biasing, $N_{\rm PFH}$ decreases along $s$. The standard deviation of $N_{\rm PFH}$ then increase, and so does $S_{\rm PFH}$.
Around $\chi_{\rm Coulomb} = 10^{11}$ there is an optimal value of the biasing factor. At this value, $N_{\rm PFH, unweighted}$ is around \SI{1.4}{} per histogram bin, where each bin is \SI{7}{\meter} long. This is equivalent to an average rate of \SI{2e5}{} hits per meter with a bunch of \SI{2e9}{} electrons. 
To get the real, i.e. unbiased, number of primary first hits $N_{\rm PFH}$, the weights needs to be applied to all PFH as shown in Figure~\ref{fig:pfh_coulomb_weighted}.
With the weights, the average value of $N_{\rm PFH}$ is around \SI{1.41}{} per bin for all values of $\chi_{\rm Coulomb}$. This gives an average rate of \SI{2.02e-6}{} hits per meter.

\subsubsection{Bremsstrahlung}

\begin{figure}[!ht]
    \centering
    \includegraphics[width=\textwidth]{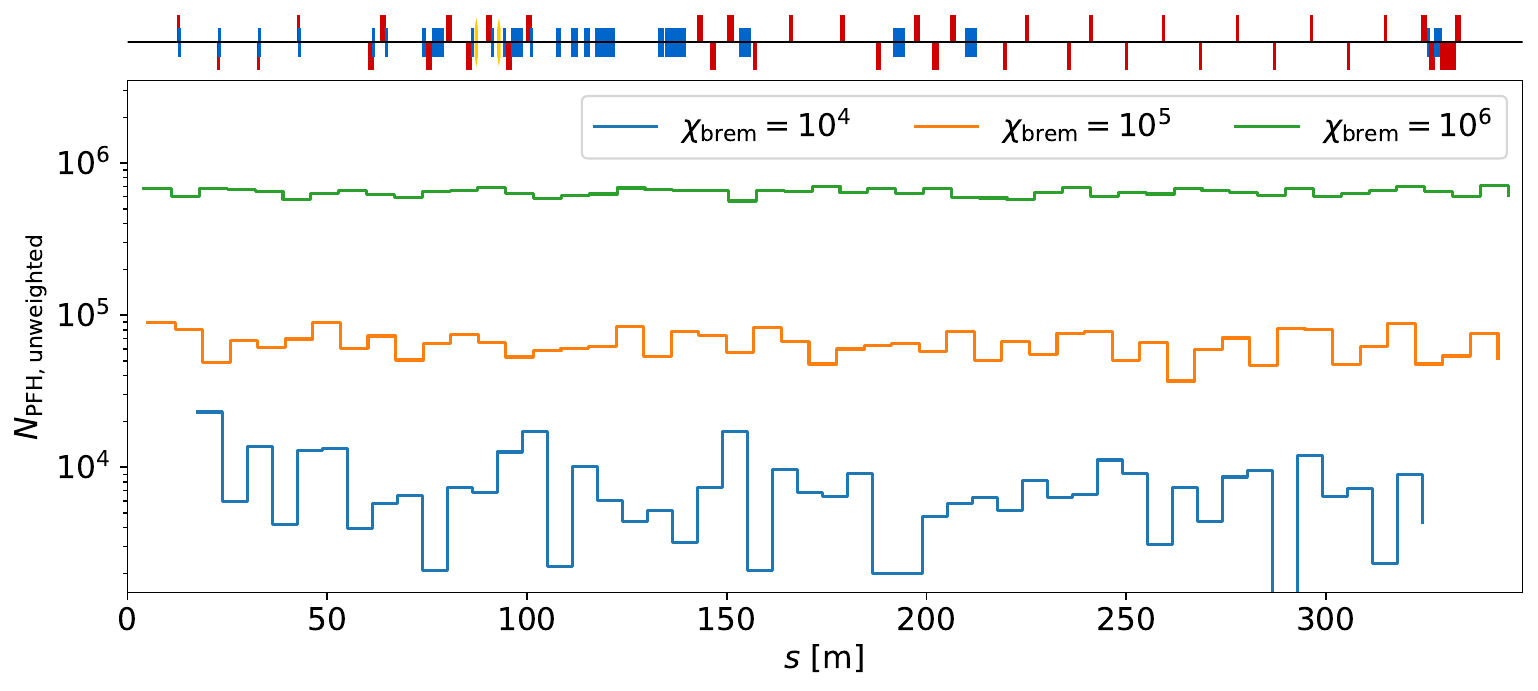}
    \caption{Histogram of unweighted primary first hits along the lattice for different values of $\chi_{\rm brem}$. The number of hits is scaled to one bunch crossing (\SI{2e9}{} electrons).}
    \label{fig:pfh_brem}
\end{figure}

\begin{figure}[!ht]
    \centering
    \includegraphics[width=\textwidth]{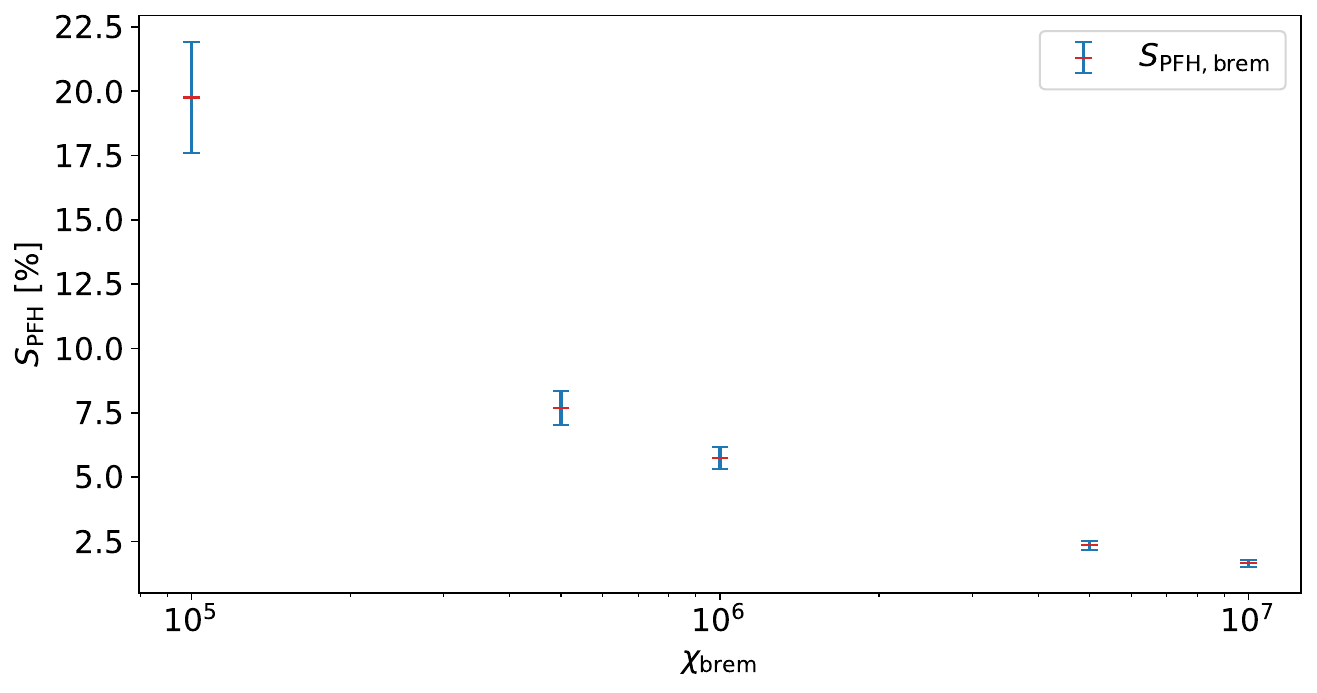}
    \caption{Plot of the normalized standard deviation of number of primary first hits $S_{\rm PFH}$ as a function of the biasing factor $\chi_{\rm brem}$.}
    \label{fig:var_brem}
\end{figure}

In the same way as Coulomb scattering, the biasing factor for bremsstrahlung $\chi_{\rm brem}$ needs to be adjusted. In Figure~\ref{fig:pfh_brem}, the biasing factor is progressively increased to see the impact on $N_{\rm PFH}$ along $s$.
Below $\chi_{\rm brem} = 10^4$ barely any bremsstrahlung interactions could be seen on the primary electrons. As with the Coulomb scattering, the normalised standard deviation $S_{\rm PFH}$ can be calculated for different values of $\chi_{\rm brem}$. The results are shown in Figure~\ref{fig:var_brem}.

As $\chi_{\rm brem}$ increases, $S_{\rm PFH}$ decreases in a similar way as before. To clearly identify the optimum value, simulation above $\chi_{\rm brem}= 10^7$ would be needed. 
However, at those high values of biasing, some errors have been encountered in the simulations. In some particle tracks, a huge number of Geant4 exceptions are raised, which increases the simulation time significantly. It appears that some particle might be stuck in infinite loops and for now the source of this problem has not been clearly identified.
As such, it proved to be difficult to run a million primaries for a biasing factor $\chi_{\rm brem}$ above $10^7$. 
To avoid most of those exceptions, a biasing factor $\chi_{\rm brem}= 10^6$ was chosen for the following simulations. At this value, $S_{\rm PFH}$ is low enough to ensure an even distribution of the bremsstrahlung interactions along the lattice.

\subsubsection{Hadronic inelastic collision}

\begin{figure}[!ht]
    \centering
    \includegraphics[width=\textwidth]{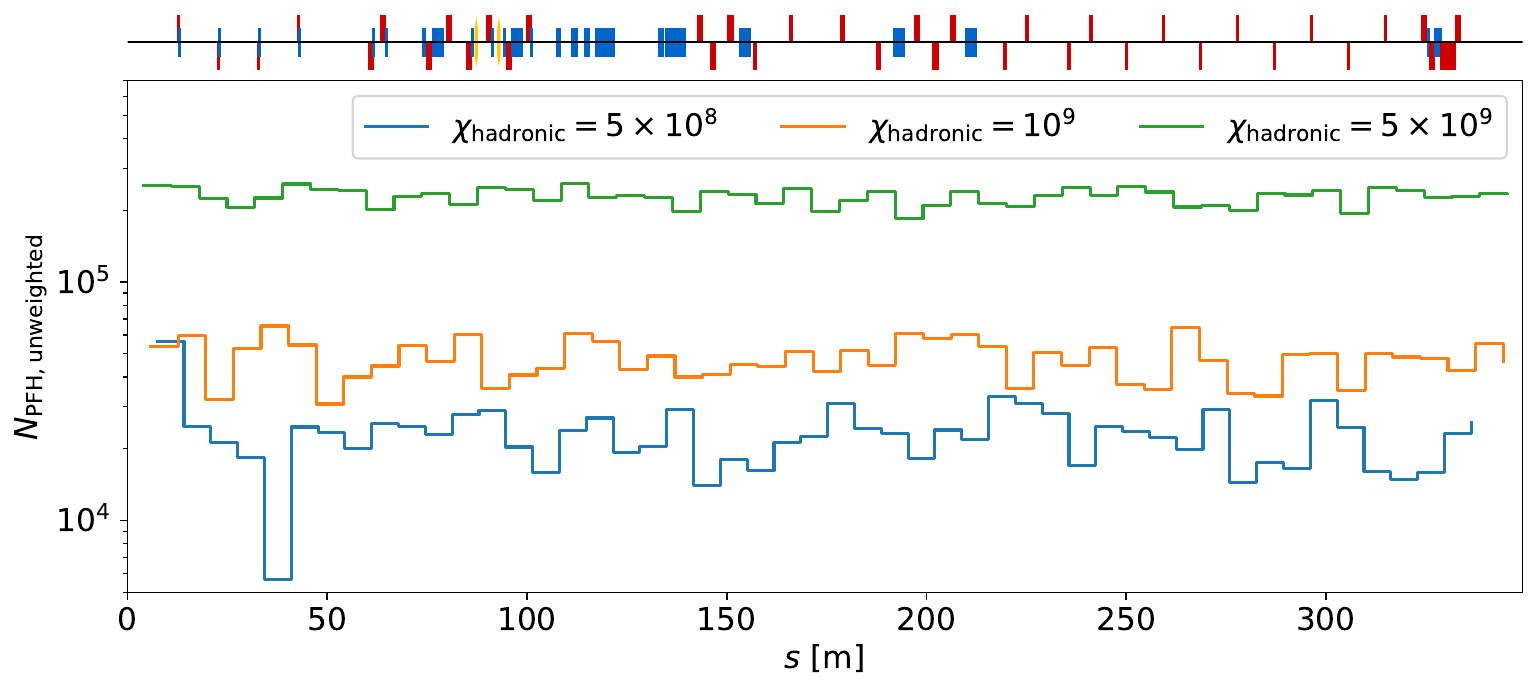}
    \caption{Histogram of unweighted primary first hits along the lattice for different values of $\chi_{\rm hadronic}$. The number of hits is scaled to one bunch crossing (\SI{2e9}{} electrons).}
    \label{fig:pfh_hadronic}
\end{figure}

\begin{figure}[!ht]
    \centering
    \includegraphics[width=\textwidth]{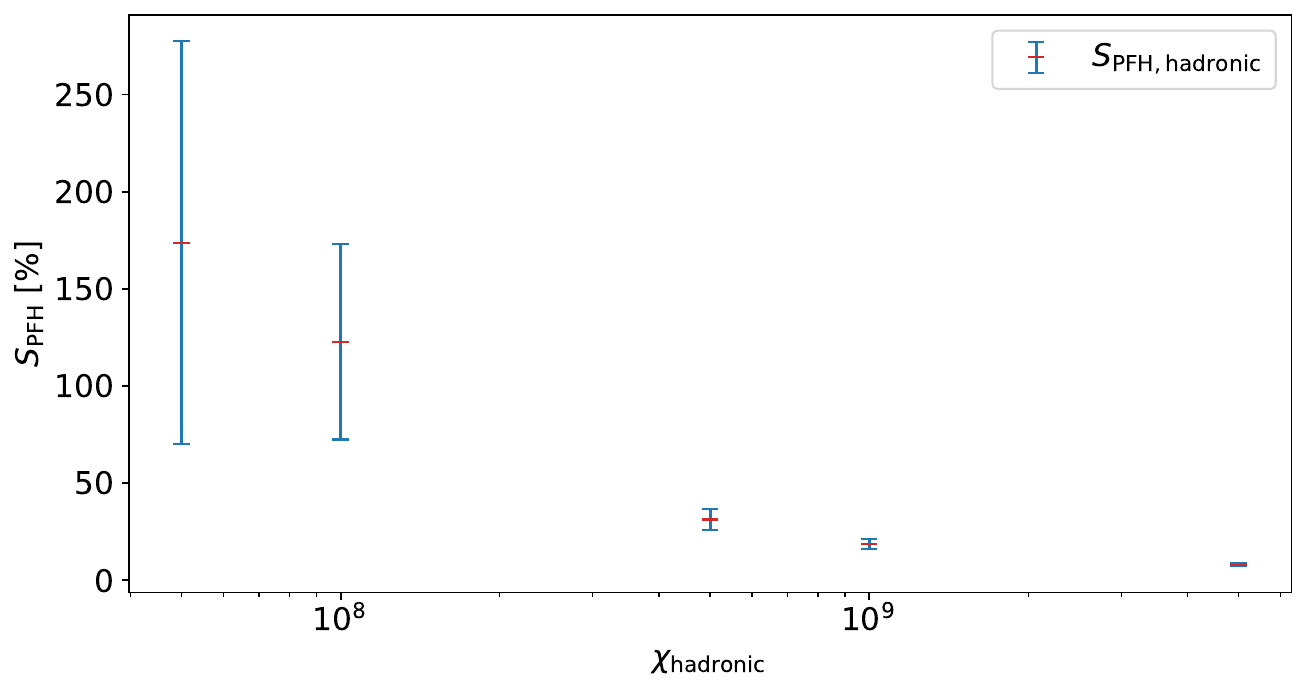}
    \caption{Plot of the normalised standard deviation of number of primary first hits $S_{\rm PFH}$ as a function of the biasing factor $\chi_{\rm hadronic}$.}
    \label{fig:var_hadronic}
\end{figure}

Finally, the biasing factor for hadronic inelastic collision $\chi_{\rm hadronic}$ needs to be adjusted. In Figure~\ref{fig:pfh_hadronic}, the biasing factor is progressively increased to see the impact on $N_{\rm PFH}$ along $s$.
Below $\chi_{\rm hadronic} = 10^7$ there is little to no hadronic inelastic collisions on the primary electrons. As with the other processes, the normalized standard deviation $S_{\rm PFH}$ can be calculated for different values of $\chi_{\rm hadronic}$. The results are shown in Figure~\ref{fig:var_hadronic}.

Similar to the bremsstrahlung variance plot, $S_{\rm PFH}$ decreases as a function of the biasing factor $\chi_{\rm hadronic}$. The hadronic interaction simulations showed an even higher number of error as the biasing factor increased. The maximum biasing value with a properly running simulation was $5\times 10^9$. Since the following background simulation will have an even higher number of primaries, it was decided to keep $\chi_{\rm hadronic}$ at $10^7$ to avoid most of the errors.

\subsection{Impact of the backgrounds at the IP}

\begin{figure}[!ht]
    \centering
    \includegraphics[width=12cm]{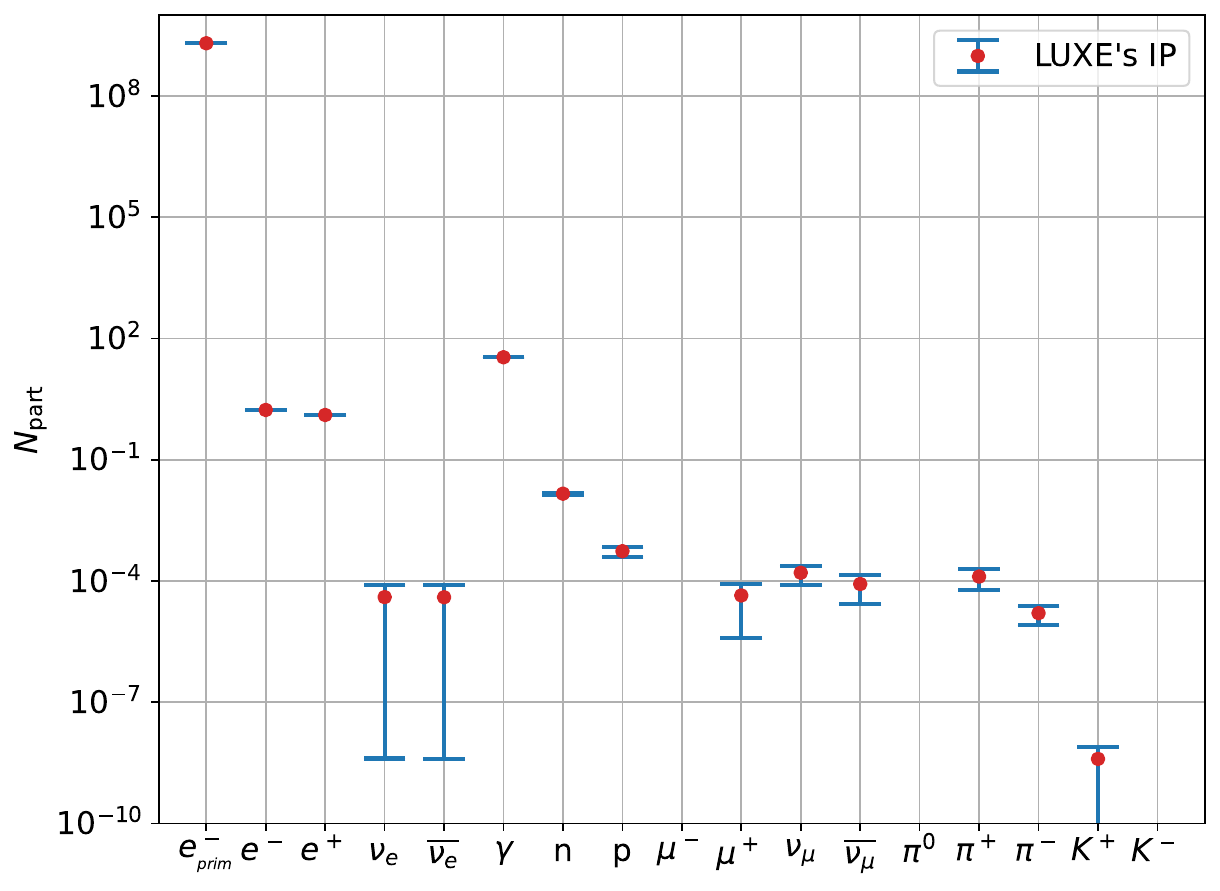}
    \caption{Plot of the beam composition at LUXE's IP}
    \label{fig:beam_composition}
\end{figure}

In Table~\ref{tab:bias_factors} a set of biasing factors for the three processes can be established. Those factors, derived from Figure~\ref{fig:var_coulomb}, Figure~\ref{fig:var_brem} and Figure~\ref{fig:var_hadronic}, have been set to insure that all interaction are produced evenly in $s$.
To properly evaluate the impact of the background on LUXE, 50 million primaries were ran in BDSIM simulations of the XTD8 lattice. The input beam has a Gaussian distribution and is matched to the lattice entrance. 
The particles produces from interactions with the residual gas are then propagated to LUXE's IP.
As with the previous figures, all plots and histograms are then scaled to a bunch crossing (\SI{2e9}{} electrons). 
At the end of the lattice, the interaction point of LUXE is a major point of interest for the impact of the backgrounds. In Figure~\ref{fig:beam_composition} the partition of particle species at LUXE's IP has been plotted.

\begin{table}[!ht]
\centering
\begin{tabular}{| c | c | c |}
\hline
$\chi_{\rm Coulomb}$ & $\chi_{\rm brem}$ & $\chi_{\rm hadronic}$ \\
\hline
$10^{10}$ & $10^{6}$ & $10^{7}$ \\ 
\hline
\end{tabular}
\caption{Biasing factors used for the residual gas background simulations.}
\label{tab:bias_factors}
\end{table}

The $2\times10^9$ primary electrons are almost entirely going through the lattice without interaction. Per bunch crossing, around \SI{34}{} photons are expected as well as \SI{1.5}{} positrons and secondary electron and \SI{0.015}{} neutrons. For the other species, the rate per bunch crossing is less than $10^{-3}$. 
Figure~\ref{fig:beam_composition} shows every particles arriving at the longitudinal position $s$ of LUXE's IP. 
This includes the particles outside the beam pipe, which implies an overestimation of the number of secondary particles that could interact on LUXE's detectors.
At those rates, the interaction at LUXE's IP (hence the interaction a the ICS chamber) should not be significantly impacted by the residual gas induced background.



\section{Background from beam halo interactions}

\begin{figure}[!ht]
    \centering
    \includegraphics[width=14cm]{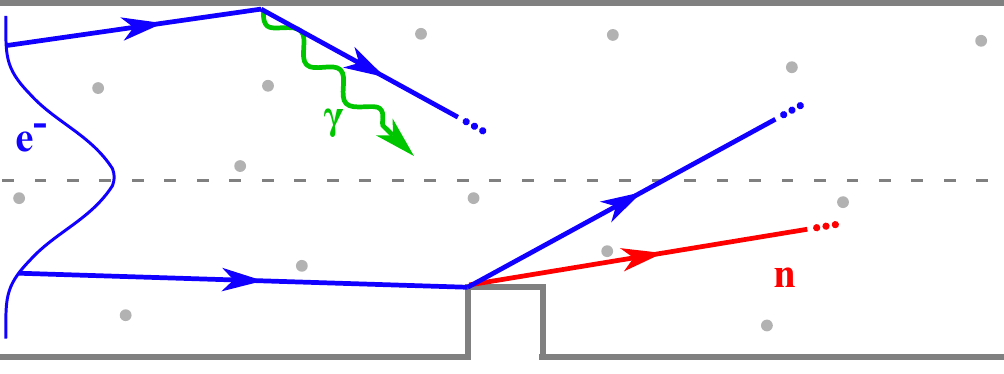}
    \caption{Schematic of the halo interaction in the beam pipe.}
    \label{fig:pipe_halo}
\end{figure}

Another source of background particle comes from the beam halo. As a electron beam travels down the pipe, electrons in the tails of the beam distribution can interact with the beam pipe and produce secondary particles, as shown in Figure~\ref{fig:pipe_halo}. This tail is known as the halo and does not have to be distributed as a Gaussian.

The halo distribution is hard to measure and typically requires dedicated beam instrumentation. To estimate the halo rate, the beam is artificially expanded as in the following description.
Similar to the residual gas induced background, the secondary particles could impact the signal measured at LUXE's detectors. The intensity of this background is directly related to the number of hits from the halo electrons.
Interactions between the beam halo and the pipe depends on the beam sizes $\sigma_{x,y}(s)$ and on the pipe radius $R_{\rm pipe}$. In the XTD8 lattice there is no collimator and $R_{\rm pipe}$ is constant ($R_{\rm pipe} = 0.02413\ \rm m$). From the beam size plot in Figure~\ref{fig:XTD8_sigma} the majority of the hits are expected in the dogleg section.
Contrary to the gas induced background, the distribution of hits is not uniform along $s$ and should vary with a larger beam size. The goal is to evaluate what spectrum of secondaries is expected per hit on the beam pipe. Let us consider an electron bunch with a Gaussian distribution $g$ expressed as

\begin{equation}
    g(x, \sigma)=\frac{1}{\sqrt{2\pi}\sigma}{\rm exp}\left[-\frac{(x-\mu)^2}{2\sigma^2}\right],
\end{equation}

where $\sigma$ is the standard deviation and $\mu$ is the mean. The probability $P$ for an electron to be between $-\infty$ and $x$ is defined as the cumulative distribution function (CDF) $\Phi$ and is expressed as

\begin{equation}
    \Phi(x, \sigma) = P(X \leq x) = \frac{1}{\sqrt{2\pi}\sigma}\int_{-\infty}^{x}{\rm exp}\left[-\frac{(u-\mu)^2}{2\sigma^2}\right]du.
\end{equation}

Because the distribution is symmetric, the probability for an electron to hit the beam pipe $P_{\rm hit}$ is expressed as

\begin{equation}
    P_{\rm hit} = P(X\geq|x|) = P(X\leq -x)+P(X\geq x) = 2\Phi(-x, \sigma).
\end{equation}

\begin{figure}[!ht]
    \centering
    \includegraphics[width=12cm]{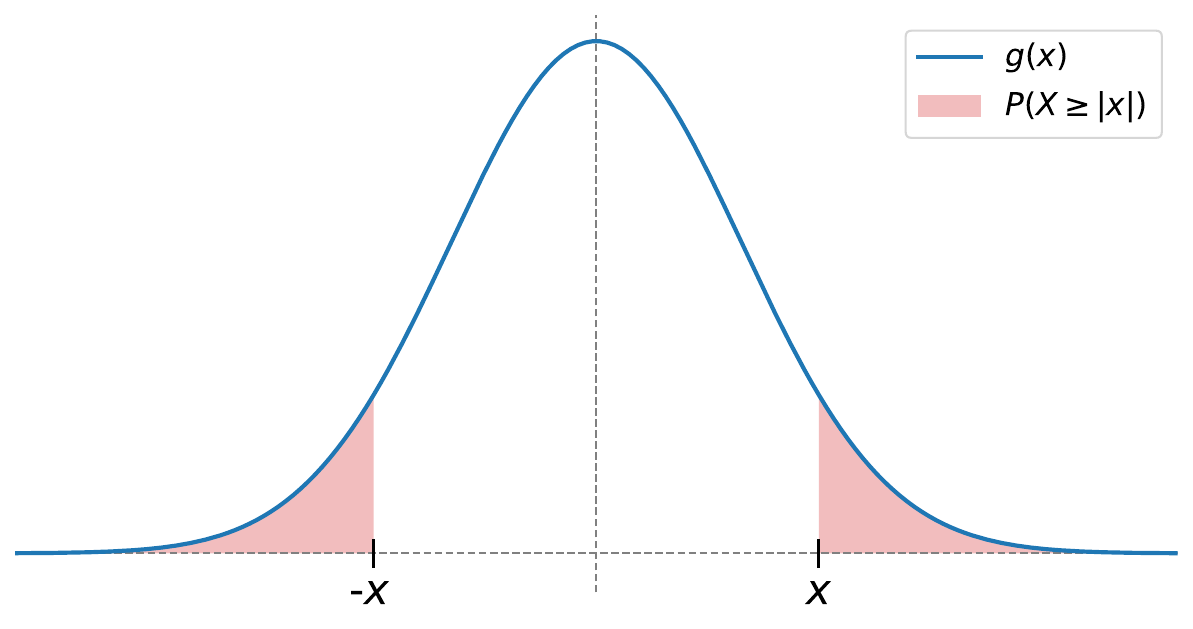}
    \caption{Plot of the probability $P(X\geq|x|)$ under a Gaussian curve.}
    \label{fig:example_halo}
\end{figure}

This probability is represented in Figure~\ref{fig:example_halo} by the area under the tails of the Gaussian curve.

\begin{figure}[!ht]
    \centering
    \includegraphics[width=12cm]{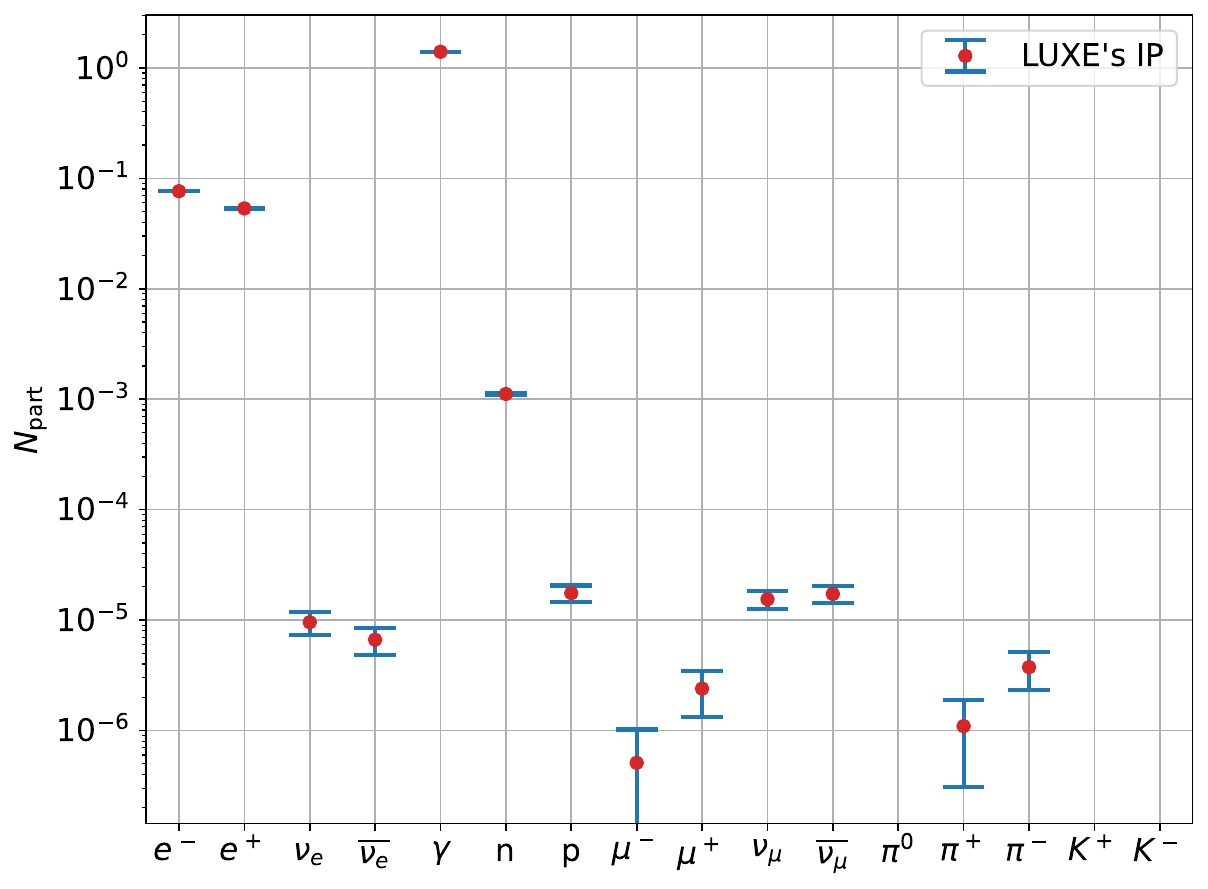}
    \caption{Plot of the expected secondaries at LUXE's IP scaled to one hit on the beam pipe.}
    \label{fig:beam_composition_halo}
\end{figure}

From Figure~\ref{fig:XTD8_sigma}, the maximum horizontal beam size in the XTD8 design is $\sigma_x = 75\ \rm \mu m$. With a pipe radius $R_{\rm pipe}=0.02413\ \rm m$ the probability $P_{\rm hit} = 2\Phi(-R_{\rm pipe}, \sigma_x)$ is too small. Simulating electrons with the designed beam size will virtually never produce any hit. A solution is to increase the horizontal emittance $\varepsilon_x$ of the beam from \SI{3.58e-11}{\meter\radian} to \SI{2.5e-7}{\meter\radian}. The beam size becomes $\sigma_x=6.28\ \rm mm$ and the probability $P_{\rm hit} \approx 1.23\times10^{-4}$. The number of hits expected $N_{\rm hit}$ is then expressed as

\begin{equation}
    N_{\rm hit} = P_{\rm hit} N_{\rm gen},
\end{equation}

where $N_{\rm gen}$ is the total number of primary electrons generated in the Gaussian bunch. With $10^5$ primary electrons, \SI{12}{} hits are expected. At this rate, there is still a majority of primaries in the centre of the Gaussian distribution that travel the lattice without interaction. Generating only the electrons outside a certain number of $\sigma$ will improve the speed of the simulations and make sure a majority of particles will interact with the beam pipe at some point
With BDSIM, the halo distribution type can be used to generate $2\times10^6$ particles between $4.5\sigma$ and $5.5\sigma$. The result spectrum of particles at LUXE's IP is shown in Figure~\ref{fig:beam_composition_halo}.

For one primary electron hit on the beam pipe, the main secondaries expected at the IP are \SI{1.2}{} photons, \SI{0.8}{} electrons, \SI{0.5}{} positrons and $10^{-3}$ neutrons. All other secondaries are expected below $10{-4}$ per hit.
Again this plot is all particles at the s location of the LUXE IP, including the particles which pass safely through the interaction region, so an overestimate of the particles which be observed in the LUXE detectors.
The results of Figure~\ref{fig:beam_composition_halo} will eventually need to be scaled by the actual halo-aperture interaction rate to get a complete picture of the number of particles passing through LUXE.

\section{Alignment at the interaction point}

\begin{figure}[!ht]
    \centering
    \includegraphics[width=14cm]{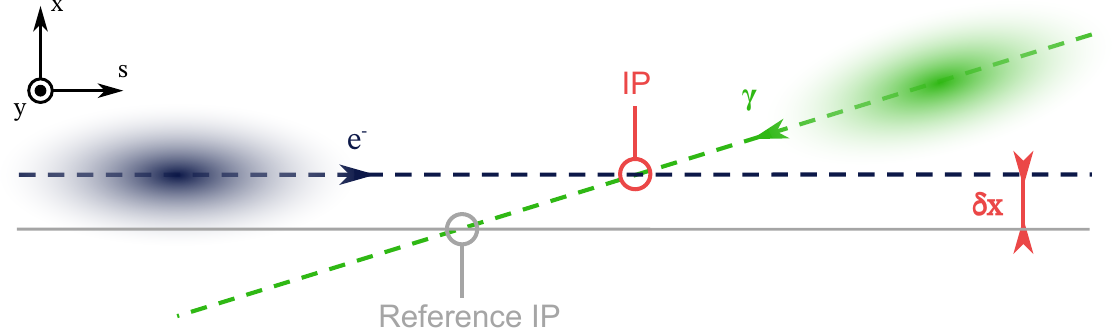}
    \caption{Schematics of the collision between an electron bunch and a laser pulse in the transverse plane. As the electron beam arrives with an horizontal offset $\delta x$, the trajectories intersection is changed.}
    \label{fig:elec_laser_collision}
\end{figure}

In LUXE's interaction chamber, the electron beam coming from the the EuXFEL will collide with a high power laser. A good alignment of the electron beam and the laser is crucial to maximize the number of particles produced by strong field QED processes. Due to the beam position jitter or a misalignment of the average beam position, a electron beam will arrive at the IP with a horizontal position offset $\delta_x$ as shown in Figure~\ref{fig:elec_laser_collision}.

The same is true for a vertical offset $\delta_y$ and for angle offsets $\delta_{x'}$ and $\delta_{y'}$. In an accelerator beam line the transverse position is usually measured using a BPM. It is evident that such a device can not be installed at the IP, so a direct measurement of the electron beam position is not possible at this location. 
However the BPMs in XTD5 and XTD8 can be used to perform a similar virtual measurement as in Chapter~\ref{chap:measurments}. Derived from Eqn~\ref{eq:correl_matrix}, the vector of $N$ horizontal positions at LUXE's IP $\boldsymbol{x}_{\rm IP}$ passing through $M$ BPMs can be written as

\begin{equation}
    \boldsymbol{x}_{\rm IP} = \boldsymbol{X}_{\rm IP} \cdot \boldsymbol{c}_{\rm IP},
    \label{eq:correl_matrix_ip}
\end{equation}

where $\boldsymbol{X}_{\rm IP}$ is the $N\times M$ matrix containing every horizontal positions for every BPM and $\boldsymbol{c}_{\rm IP}$ is the coefficients vector for each BPM. This coefficients vector would be expresses as

\begin{equation}
    \boldsymbol{c}_{\rm IP} = \boldsymbol{X}_{\rm IP}^{-1} . \boldsymbol{x}_{\rm IP}.
    \label{eq:coeff_IP}
\end{equation}

So this solution still requires a measurement of the position at the IP to calibrate the coefficients $\boldsymbol{c}_{\rm IP}$.

\subsection{Position measurement at the interaction point}

\begin{figure}[!ht]
    \centering
    \includegraphics[width=12cm]{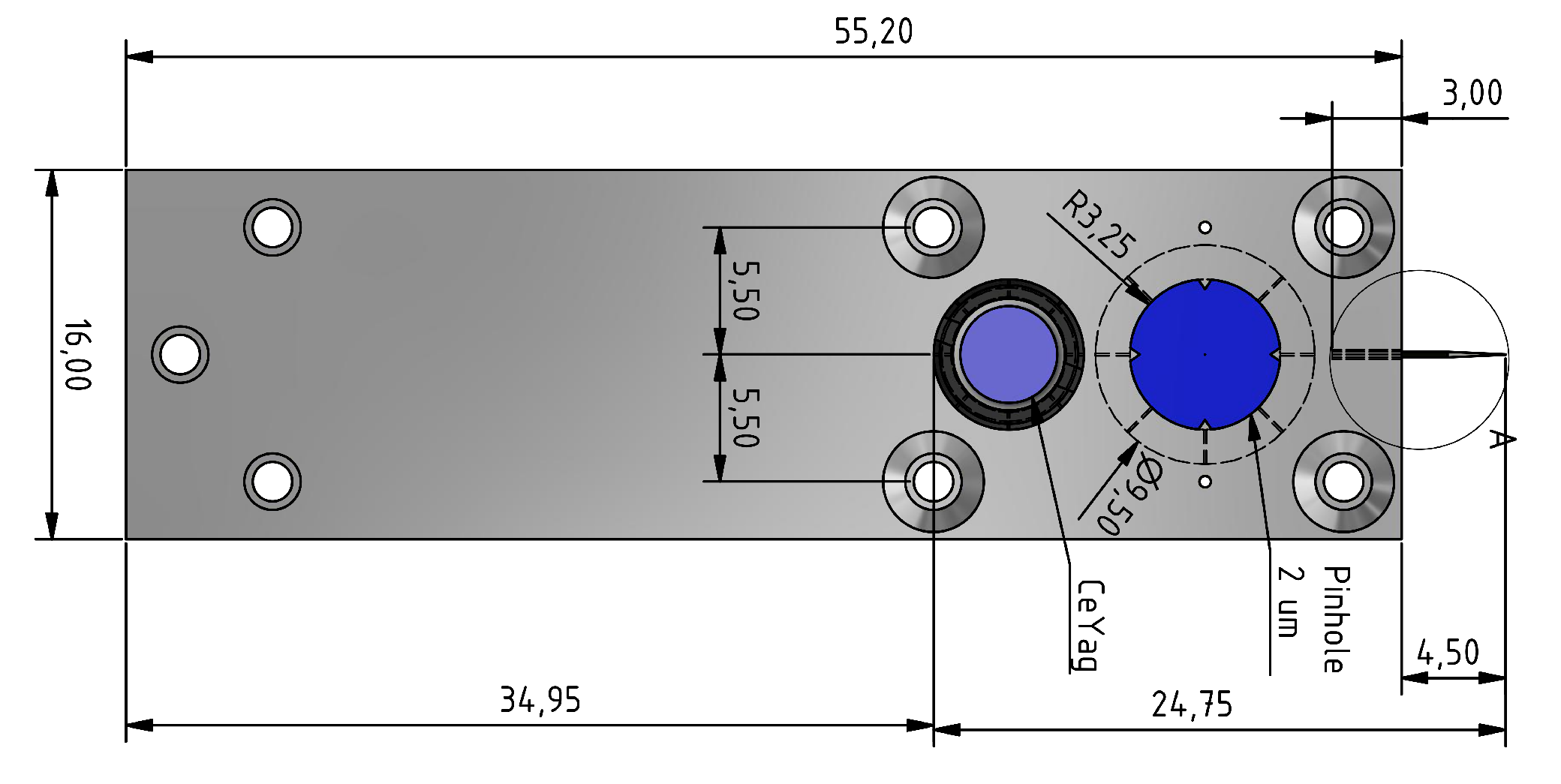}
    \caption{CAD model and dimensions of the alignment-target holder. To determine the transverse position of the electron beam, a needle, a \SI{2}{\micro\meter} pinhole and a Ce:YAG screen are available}
    \label{fig:holder}
\end{figure}

In the interaction chamber, diagnostics of both the electron beam and the laser will be achieved using an alignment-target holder~\cite{luxe_tdr} as shown in Figure~\ref{fig:holder}. 

\begin{figure}[!ht]
    \centering
    \includegraphics[width=5cm]{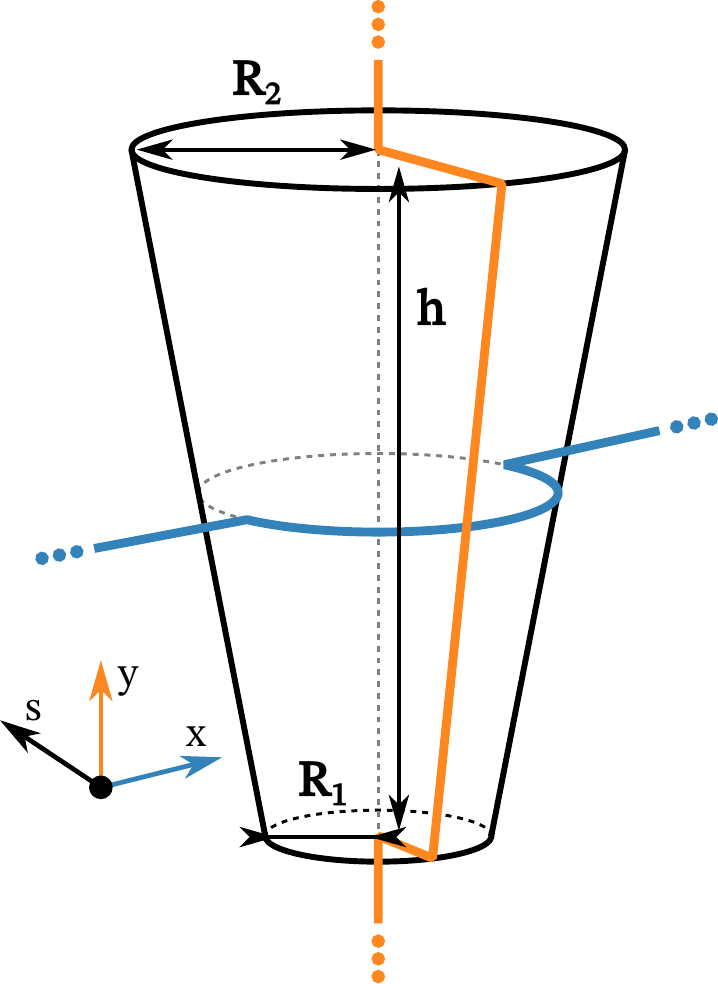}
    \caption{Schematic of the two transverse scans path of the beam on the needle. In blue is the horizontal scan path and in orange the vertical scan path.}
    \label{fig:needle_scans}
\end{figure}

Among the elements on this device there is a thin tungsten needle that can be used to measure the beam position. Similar to a wire scan, this needle can be moved in the transverse plane to make it interact with the electron beam. However, the shape of the needle allow for both a horizontal and a vertical scan as shown in Figure~\ref{fig:needle_scans}.

Let us consider a Gaussian electron beam whose density is expressed as

\begin{equation}
    g(u) = \frac{1}{\sqrt{2\pi}\sigma}{\rm exp}\left(-\frac{(u-\mu)^2}{2\sigma^2}\right),
\end{equation}

where $u$ is either $x$ or $y$. The needle geometry is a truncated cone of height $h$, minimal radius $R_1$ and maximal radius $R_2$. For a horizontal scan, the needle profile $p_{\rm needle}$ is the same as a wire scanner of radius $R$ with

\begin{equation}
    p_{\rm needle, x}(x) = p_{\rm wire}(x) = 2\sqrt{R^2-x^2}.
\end{equation}

\begin{figure}[!ht]
    \centering
    \includegraphics[width=12cm]{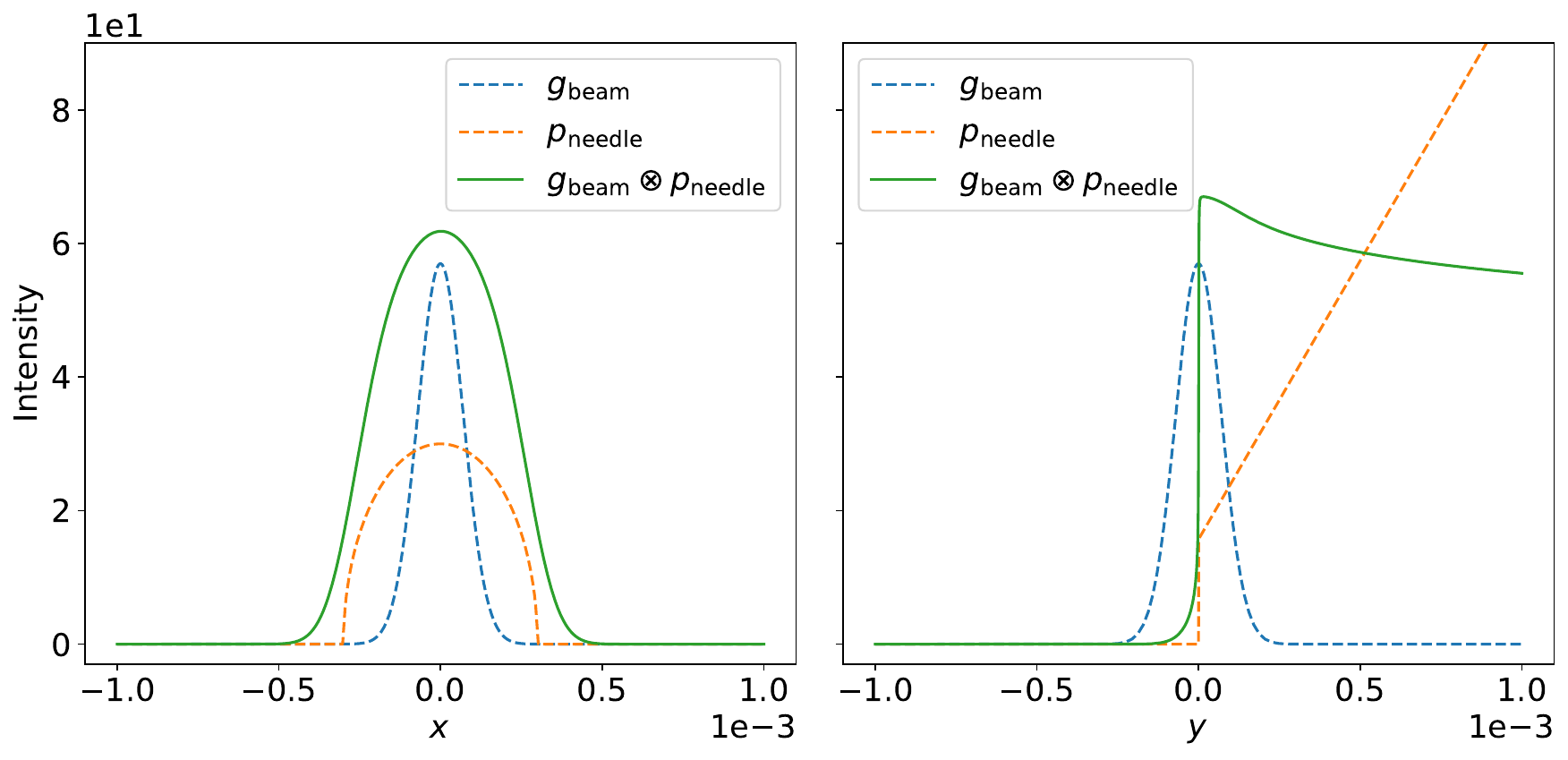}
    \caption{Plots of the convolution curves for the needle. On the left the horizontal scan convolution and on the right the vertical scan convolution.}
    \label{fig:conv_needle_xy}
\end{figure}

For a vertical scan, the needle profile is linear between $R_1$ and $R_2$. It is expressed as

\begin{equation}
    p_{\rm needle, y}(y) = \frac{R_2-R_1}{h}y+\left(R_1-\frac{R_2-R_1}{h}*y_1\right),
\end{equation}

where $y_1$ is the vertical position of the bottom of the needle, i.e. where $R(y)=R_1$. The number of interactions expected for both scan is driven by the convolutions $g(x)\otimes p_{\rm needle, x}(x)$ and $g(y)\otimes p_{\rm needle, y}(y)$. The shape of those convolutions is shown in Figure~\ref{fig:conv_needle_xy}.
Multiple BDSIM simulations with $10^5$ primary electrons were ran for different positions of the needle. As the beam interacts with the needle it will mainly produce bremsstrahlung photons. For both scan in $x$ and $y$ the goal is to find a linear relationship between the position of the needle and the number of photons produced $N_{\gamma}$. 

\subsubsection{Horizontal scan}

\begin{figure}[!ht]
    \centering
    \includegraphics[width=12cm]{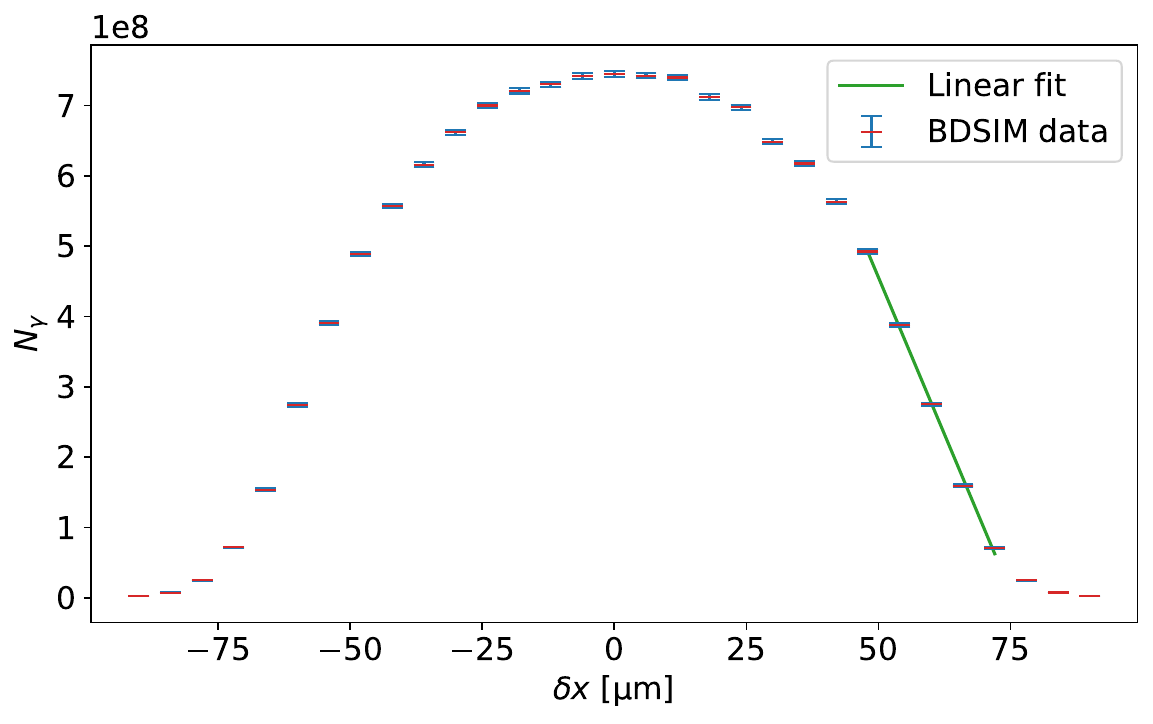}
    \caption{Plot of the number of photons $N_{\gamma}$ as a function of the horizontal position $x$. In red is a linear fit. $N_{\gamma}$ is scaled to one bunch crossing.}
    \label{fig:needle_scan_x}
\end{figure}

In the horizontal plane, the needle scan is analogous to a wire scan. The results of this needle scan are shown in Figure~\ref{fig:needle_scan_x}.

Each side of the bell curve have a quasi-linear section. As long as the side on which the needle approached the beam is known, the position of the needle $x_{\rm n}$ can be deduced by the number of photons $N_{\gamma}$ from the linear equation

\begin{equation}
    N_{\gamma} = ax_{\rm n} + b,
\end{equation}

where $a$ and $b$ are the coefficients obtained with the linear fit. The quasi-linear section of the curve is \SI{24}{\micro\meter} long. The parameters of the linear fit are listed in Table~\ref{tab:fit_param_x}.

\begin{table}[!ht]
\centering
\begin{tabular}{| c c c c c |}
\hline
$a$ & $b$ & ${\rm Cov}(a,b)$ & $\sigma_a$ & $\sigma_b$ \\
\hline
$-1.79\times10^{13}$ & $1.35\times10^{9}$ & $-1.01\times10^{19}$ & $4.10\times10^{11}$ & $2.48\times10^{7}$ \\
\hline
\end{tabular}
\caption{Linear fit parameters for the x scan.}
\label{tab:fit_param_x} 
\end{table}

The needle can be placed in the middle of the linear section, at $x_{\rm n} = 60\ \rm \mu m$. The position of the beam $x_{\rm b}$ is then expressed by 

\begin{equation}
    x_{\rm b} = \frac{N_{\gamma}-b}{a}-x_{\rm n}.
\end{equation}

If $N_{\gamma,\rm meas}$ is the number of photons measured at the Cherenkov detector, the measured position of the beam is

\begin{equation}
    x_{\rm b, meas} = \frac{N_{\gamma,\rm meas}-b}{a}-x_{\rm n}.
\end{equation}

The error on the beam position measurement $\sigma_{x_{\rm b}}$ can be obtained through error propagation as

\begin{equation}
    \sigma_{x_{\rm b}}^2 = 
    \left(\frac{\partial x_{\rm b}}{\partial N_{\gamma}}\sigma_{N_{\gamma}}\right)^2 +
    \left(\frac{\partial x_{\rm b}}{\partial x_{\rm n}}\sigma_{x_{\rm n}}\right)^2 +
    \left(\frac{\partial x_{\rm b}}{\partial a}\sigma_a\right)^2 +
    \left(\frac{\partial x_{\rm b}}{\partial b}\sigma_b\right)^2 +
    2\frac{\partial x_{\rm b}}{\partial a}\frac{\partial x_{\rm b}}{\partial b}{\rm Cov}(a, b),
\end{equation}

where ${\rm Cov}(a,b)$ is the covariance between $a$ and $b$ obtained from the linear fit. After developing all the partial derivatives the error becomes

\begin{equation}
    \sigma_{x_{\rm b}} = \frac{1}{|a|}\sqrt{\sigma_{N_{\gamma}}^2+a^2\sigma_{x_{\rm n}}+x_{\rm b}^2\sigma_a^2+\sigma_b^2+2x_{\rm b}{\rm Cov}(a,b)}.
\end{equation}

Considering the ideal case where there is no error on the photon measurement and on the needle position, i.e. $\sigma_{N_{\gamma}}=\sigma_{x_{\rm n}}=0$, the error becomes

\begin{equation}
    \sigma_{x_{\rm b}} = \frac{1}{|a|}\sqrt{x_{\rm b}^2\sigma_a^2+\sigma_b^2+2x_{\rm b}{\rm Cov}(a,b)}.
\end{equation}

Along the fit range, the average error of the beam horizontal position is $\langle \sigma_{x_{\rm b}}\rangle = 1.39\ \rm \mu m$. The needle can then be considered as a BPM with a measurement noise $\sigma_{\rm N} = \langle \sigma_{x_{\rm b}}\rangle$. In reality the impact of the needle position error $\sigma_{x_{\rm n}}$ and the Cherenkov measurement $\sigma_{N_{\gamma}}$ will increase the noise.



\subsubsection{Vertical scan}

\begin{figure}[!ht]
    \centering
    \includegraphics[width=12cm]{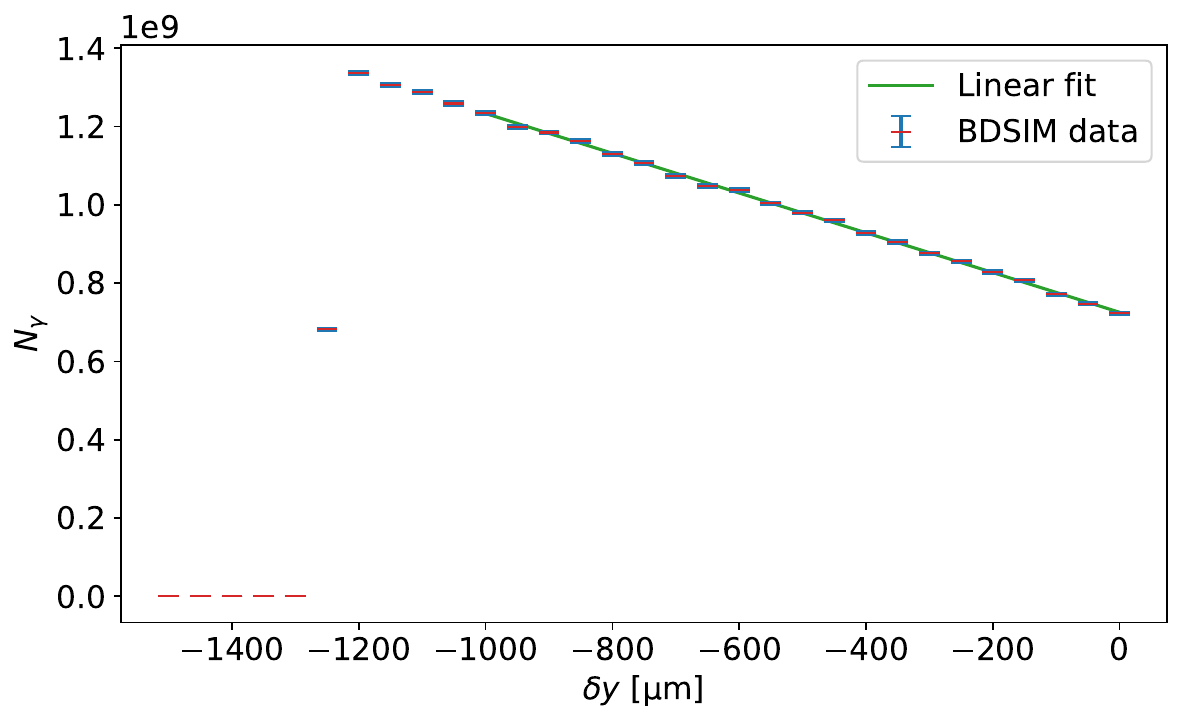}
    \caption{Plot of the number of photons $N_{\gamma}$ as a function of the vertical position $y$. In red is a linear fit. $N_{\gamma}$ is scaled to one bunch crossing.}
    \label{fig:needle_scan_y}
\end{figure}

In the vertical plane a similar simulation scan using BDSIM is performed. The results are shown in Figure~\ref{fig:needle_scan_y}.

A linear fit is computed between $y=-1\ \rm mm$ and $y=0\ \rm mm$ whose parameters are listed in Table~\ref{tab:fit_param_y}.

\begin{table}[!ht]
\centering
\begin{tabular}{| c c c c c |}
\hline
$a$ & $b$ & ${\rm Cov}(a,b)$ & $\sigma_a$ & $\sigma_b$ \\
\hline
$-5.08\times10^{11}$ & $7.25\times10^{8}$ & $5.71\times10^{15}$ & $3.38\times10^{9}$ & $1.98\times10^{6}$ \\
\hline
\end{tabular}
\caption{Linear fit parameters for the y scan.}
\label{tab:fit_param_y} 
\end{table}

The error of the vertical position measurement is then expressed as 

\begin{equation}
    \sigma_{y_{\rm b}} = \frac{1}{|a|}\sqrt{y_{\rm b}^2\sigma_a^2+\sigma_b^2+2y_{\rm b}{\rm Cov}(a,b)}.
\end{equation}

With this linear fit, an average error $\langle \sigma_{y_{\rm b}}\rangle = 4.08\ \rm \mu m$ was found. 
Using the needle scans in both plane, the transverse positions of the electron beam can be obtained. However the absolute position of the beam imports less for LUXE than the relative position between electron beam and laser.
If an attenuated laser can be be used in similar needle scans, the relative position of laser and electron beam can be quickly established, allowing rapid collision setup during LUXE operation.

\subsection{Virtual measurement at LUXE's IP}

At LUXE's IP, the needle scan is a method to measures the transverse position of the electron beam. With all \SI{43}{} BPMs in the XTD8 lattice, a virtual measurement can now be performed. Using BDSIM, $5\times10^4$ macroparticles are tracked and all $x$ and $y$ positions are stored in the $\boldsymbol{M}_{\rm track}$ matrix. From the BDSIM tracking, the vector of $x$ and $y$ positions at the IP $\boldsymbol{v}_{\rm track}$ is extracted as well.
Measuring the horizontal position $x$ at a BPM, it is expected that the major contributions in terms of correlations coefficient will come from the $x$ plane. However, it is possible to have a cross correlation between the two transverse planes. In that case a measurement of $x$ at a BPM will have some contributions from vertical positions $y$ in other BPMs. This is why measurements from both places are kept in the same matrix 
To simulate a real measurement from the BPMs, a noise $\sigma_N$ is applied to all measurements. In chapter~\ref{chap:measurments}, the noise for a \SI{40}{\milli\meter} button BPM was measured at $3\ \mu m$. A set of noise $n_{ij}$ is then generated following a Gaussian distribution of standard deviation $\sigma_N = 3\ \mu m$. The simulated measurement matrix $\boldsymbol{M}_{\rm meas}$ is the expressed as

\begin{equation}
    \boldsymbol{M}_{\rm meas} = \boldsymbol{M}_{\rm track} + 
    \left(
    \begin{matrix}
        n_{11} & ... & n_{i1} & ... & n_{n1} \\
        : & \ddots & & & : \\
        n_{1j} &  & n_{ij} &  & n_{nj}\\
        : & & & \ddots & : \\
        n_{1m} & ... & n_{im} & ... & n_{nm} \\
    \end{matrix}
    \right).
\end{equation}

Similarly, the real measurement vector at the IP $\boldsymbol{v}_{\rm meas}$ is expressed as 

\begin{equation}
    \boldsymbol{v}_{\rm meas} = \boldsymbol{v}_{\rm track} +
    \left(
    \begin{matrix}
        m_{x,1} \\
        : \\
        m_{x,n} \\
        m_{y,1} \\
        : \\
        m_{y,n} \\
    \end{matrix}
    \right),
\end{equation}

where $\{m_{x,i}\}$ and $\{m_{y,i}\}$ are two sets of noise following Gaussian distributions with the standard deviation found in the last section. The correlation coefficients of the virtual measurement are then expressed by 

\begin{equation}
    \boldsymbol{c}_{\rm IP} = \boldsymbol{M}_{\rm meas}^{-1}\cdot \boldsymbol{v}_{\rm meas}.
\end{equation}

With the set of coefficients $\boldsymbol{c}_{\rm IP}$ the position at the IP can be reconstructed with the virtual measurement by following

\begin{equation}
    \boldsymbol{v}_{\rm VM} = \boldsymbol{M}_{\rm meas}\cdot\boldsymbol{c}_{\rm IP}.
\end{equation}

The residuals of the virtual measurement, i.e. the difference between the measured positions and the real positions at LUXE's IP, are then defined as

\begin{equation}
    \left(
    \begin{matrix}
        \boldsymbol{R}_x \\
        \boldsymbol{R}_y \\
    \end{matrix}
    \right) = \boldsymbol{v}_{\rm VM} - \boldsymbol{v}_{\rm meas}.
\end{equation}

\begin{figure}[!ht]
    \centering
    \includegraphics[width=10cm]{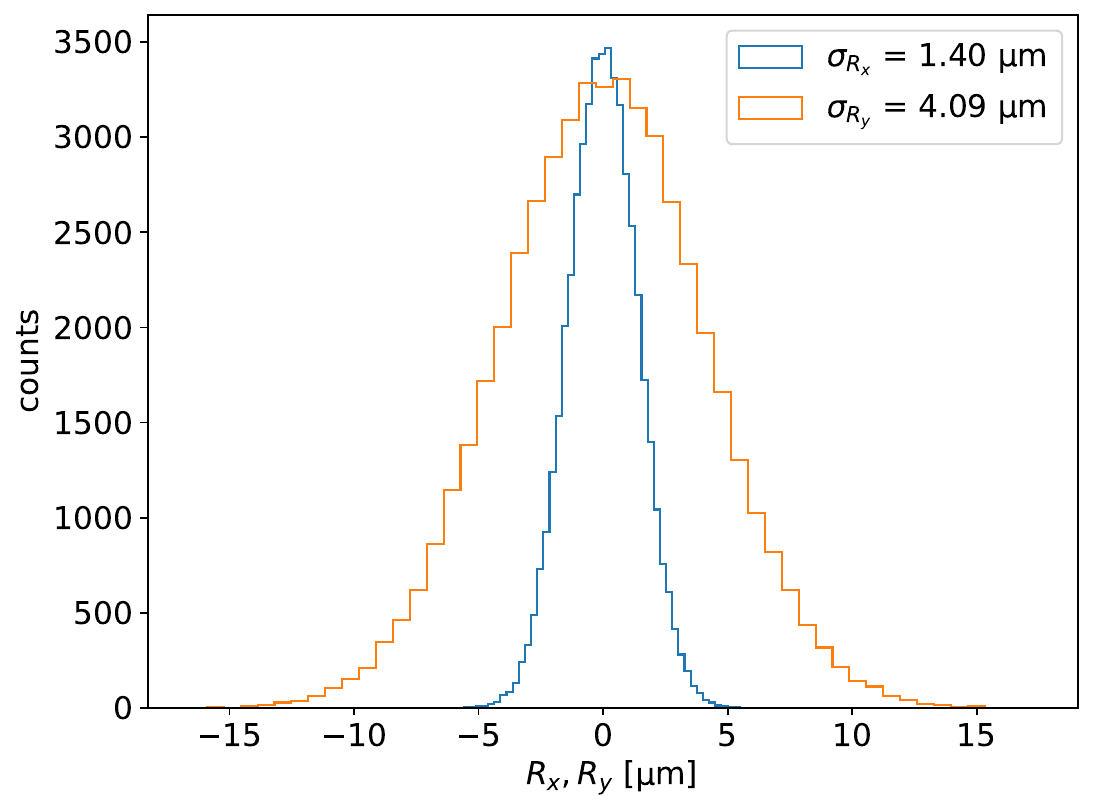}
    \caption{Histogram of the position measurement residual $\boldsymbol{R}$ for both transverse planes at LUXE's IP.}
    \label{fig:resolution}
\end{figure}

\begin{figure}[!ht]
    \centering
    \includegraphics[width=10cm]{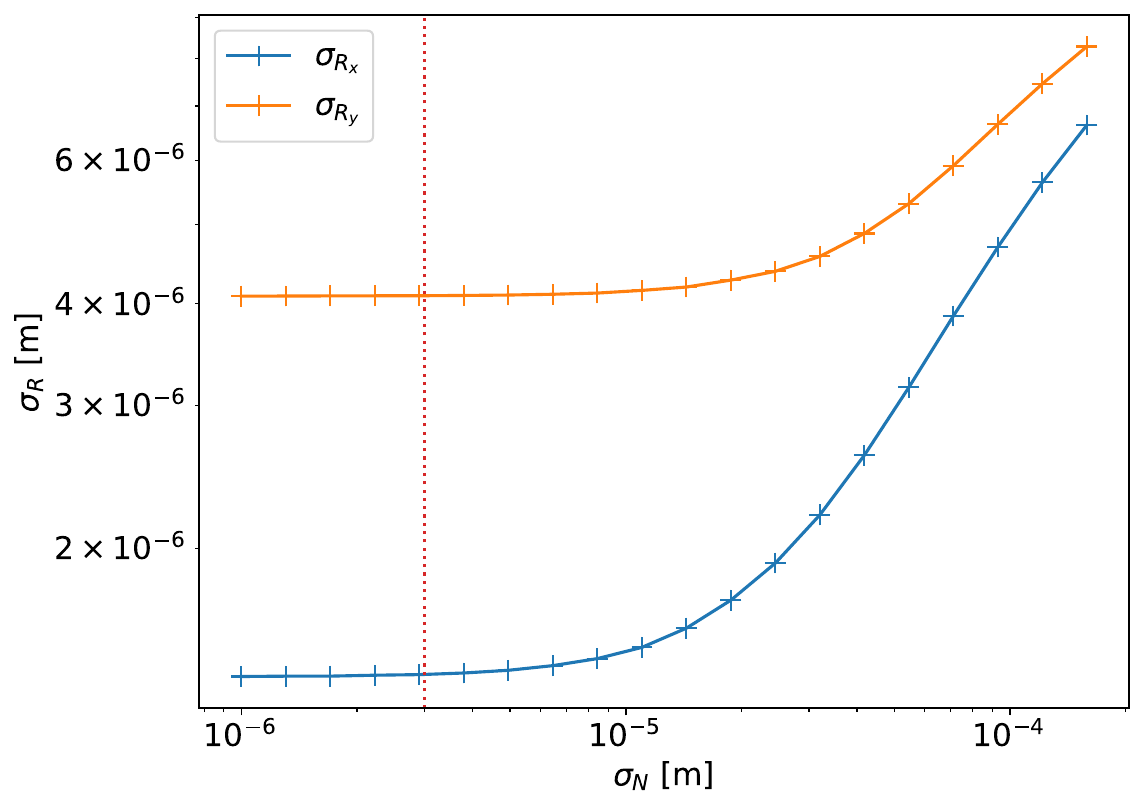}
    \caption{Plot of the virtual measurement resolution as a function of the BPM noise}
    \label{fig:resolution_wrt_noise}
\end{figure}

In Figure~\ref{fig:resolution} is the histogram of the residual for measurements in both planes using the $5\times10^4$ tracked macroparticles.
The resolutions of the virtual measurement in both plane, $\sigma_{R_x}$ and $\sigma_{R_y}$ are close to the resolution of the needle measurement. In Figure~\ref{fig:resolution_wrt_noise}, this virtual measurement is performed for different values of BPM noise.
At low values of BPM noise $\sigma_N$, the virtual measurement resolution is dominated by the needle measurement at the IP. A better measurement at the IP should allow a smaller resolution of the virtual measurement.

\begin{figure}[!ht]
    \centering
    \includegraphics[width=10cm]{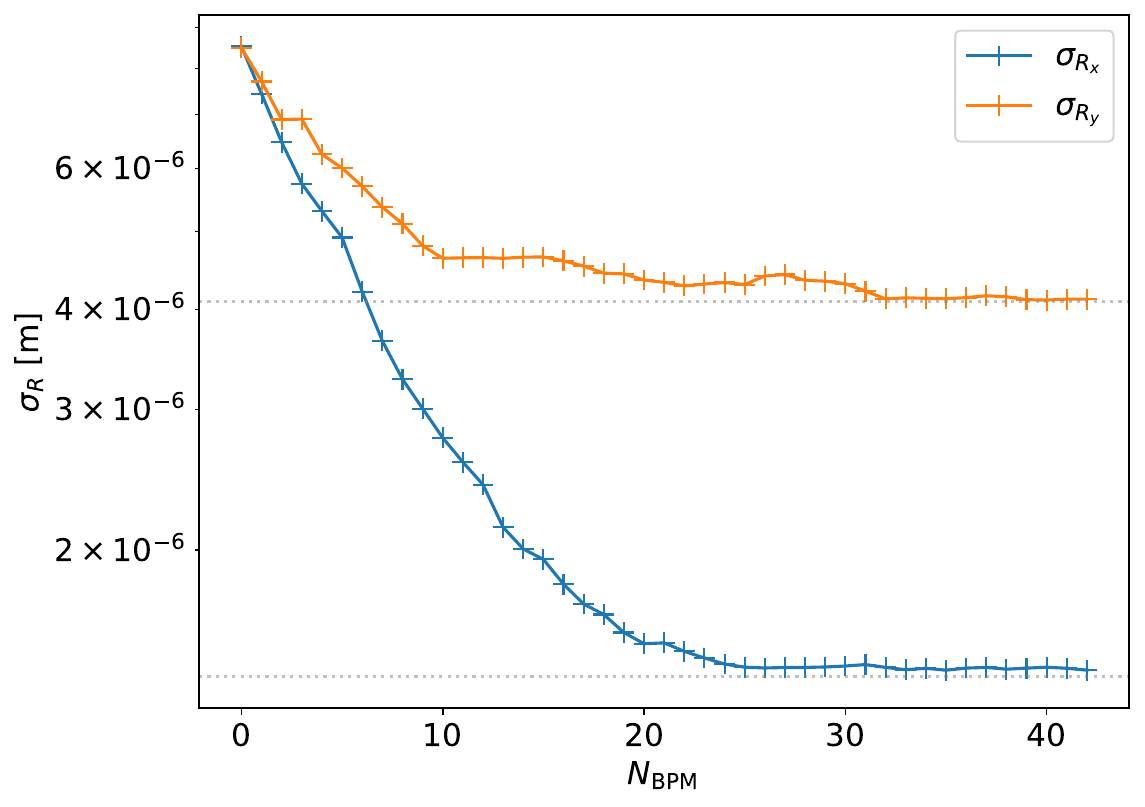}
    \caption{Plot of the virtual measurement resolution as a function of the number of BPM.}
    \label{fig:resolution_wrt_number}
\end{figure}

The coefficient vector $\boldsymbol{c}_{\rm IP}$ indicates which BPMs are correlated the most with the IP. A BPM with a high coefficient $c_i$ will be very sensitive to a variation in position at the IP. Some of the BPMs in the XTD8 have a low coefficient which means a low impact on the virtual measurement. In Figure~\ref{fig:resolution_wrt_number}, the virtual measurement resolution is computed with respect to the number of BPM used. In this plot the BPMs are sorted by their relative coefficients $c_i$.
For the resolution of the horizontal position measurement $\sigma_{R_x}$, only 20 BPMs are needed to get to \SI{1.5}{\micro\meter} of resolution. For the resolution of the vertical position measurement $\sigma_{R_y}$, only 10 BPMs are needed to get to \SI{4.6}{\micro\meter} of resolution. 

\section{Propagation of jitter to the IP}

In Chapter~\ref{chap:measurments}, the jitter of the electron beam was measured for different parameters in the EuXFEL switch yard. The goal is to extrapolate, from those values, the beam variations at LUXE's IP. The propagation of parameters is done using the transfer matrix defined as
\begin{equation}
    \begin{pmatrix}
        x \\ x' \\
        y \\ y' \\
        ct \\ \delta \\
    \end{pmatrix}
     = \boldsymbol{R}_{s_0 \rightarrow s} \cdot
    \begin{pmatrix}
        x_0 \\ x'_0 \\
        y_0 \\ y'_0 \\
        ct_0 \\ \delta_0 \\
     \end{pmatrix},
     \label{eq:rmat_propag}
\end{equation}
where $\boldsymbol{R}_{s_0 \rightarrow s}$ is the matrix propagating the position vector from $s_0$ to $s$. Since there is little to no correlation between $x$, $y$ and $t$, this matrix is bloc diagonal and can be written as
\begin{equation}
    \boldsymbol{R}_{s_0 \rightarrow s} =
    \begin{pmatrix}
        \begin{matrix}
        R_{11} & R_{12} \\
        R_{21} & R_{22} \\
        \end{matrix}
         & 0 & 0 \\ 
         0 & 
        \begin{matrix}
        R_{33} & R_{34} \\
        R_{43} & R_{44} \\
        \end{matrix}
        & 0 \\
        0 & 0 & 
        \begin{matrix}
        R_{55} & R_{56} \\
        R_{65} & R_{66} \\
        \end{matrix}
    \end{pmatrix}.
    \label{eq:rmat_bloc_diag}
\end{equation}

From the Xsuite lattices, the R-matrices can be extracted from a given position $s$ to LUXE's IP. 

\subsection{Transverse position and angle}

From the virtual measurement performed in Chapter~\ref{chap:measurments}, the position jitter in non-dispersive sections is estimated around \SI{5}{\%} of the beam size. At LUXE's IP the beam size requirement is \SI{10}{\micro\meter} in both planes. This correspond to a jitter $\sigma_J = 0.5\ \rm \mu m$ in $x$ and $y$ at the IP. 
To evaluate the actual position jitter at the IP, requires finding the BPMs that have a high correlation with the IP. From Figure~\ref{fig:correl_mux} and Figure~\ref{fig:correl_muy} there are multiple location where this correlation is present. Assuming there is one BPM per quadrupole in the lattice, the goal is to find quadrupole with the highest correlation.
Suppose there is a BPM location for which $x_{\rm BPM}$ is highly correlated to $x_{\rm IP}$. The horizontal position and angle are propagated in the lattice as

\begin{equation}
    \begin{pmatrix}
        x_{\rm IP} \\ x'_{\rm IP} \\
    \end{pmatrix}
    =
    \begin{pmatrix}
        R_{11} & R_{12} \\ R_{21} & R_{22} \\
    \end{pmatrix}
    \cdot
    \begin{pmatrix}
        x_{\rm BPM} \\ x'_{\rm BPM} \\
    \end{pmatrix}.
\end{equation}

The beam position at the IP is then expressed as

\begin{equation}
    x_{\rm IP} = R_{11}x_{\rm BPM} + R_{12}x'_{\rm BPM}.
\end{equation}

A variation of the measurement of $x_{\rm BPM}$ will be scaled at the IP by the factor $R_{11}$. The lower $|R_{11}|$ is, the smaller IP jitter can be detected at the BPM. The same is true for $y, x',y'$ with the respective factors $R_{33}, R_{21}, R_{43}$. From the existing BPMs in the XTD8 lattice, one must find the correlated BPMs that have the smallest respective matrix factor for each coordinate. 
By inspecting the Xsuite R-matrices, a good location for each coordinate can be found. They are listed in Table~\ref{tab:bpm_selec}.

\begin{table}[!ht]
    \centering
    \begin{tabular}{| c || c | c | c |}
        \hline
         & Closest Quad. & S / m & Factor \\
        \hline
        \hline
        x & QF.2906.T8 & 2901.46 & $R_{11}=0.139$ \\
        y & QH.2946.T8 & 2950.84 & $R_{33}=0.212$ \\
        x' & QH.11.T8 & 3068.74 &  $R_{21}=-0.0948$ \\
        y' & QH.2941.T8 & 2941.76 & $R_{43}=-0.0111$ \\
        \hline
    \end{tabular}
    \caption{Correlated BPMs in the XTD8 lattice for each transverse coordinate.}
    \label{tab:bpm_selec}
\end{table}

Now the BPM noise can be propagated using the factors in Table~\ref{tab:bpm_selec}. In the XTD8 a button BPM noise of $\sigma_N=3\ \rm \mu m$ is expected. For a measurement of $x$, this mean that the BPM at \SI{2901.46}{\meter} could detect a jitter of the electron beam $\sigma_J$ if $\sigma_J > |R_{11}|\sigma_N$. Applied to our four selected BPMs, the minimum jitter measurable $\sigma_{J, \rm min}$ are listed in Table~\ref{tab:min_jitter_IP}

\begin{table}[!ht]
    \centering
    \begin{tabular}{| c | c | c |}
        \hline
         Parameter & Unit & $\sigma_{J, \rm min}$ \\
        \hline
        \hline
        x & $\rm \mu m$ & 0.417 \\
        y & $\rm \mu m$ & 0.636 \\
        x' & $\rm \mu rad$ & 0.284 \\
        y' & $\rm \mu rad$ & 0.033 \\
        \hline
    \end{tabular}
    \caption{Minimum values of transverse jitter that can be measured at the BPMs.}
    \label{tab:min_jitter_IP}
\end{table}

\subsection{Transverse beam size}

The beam size follows the same betatron variation as the position. Since the position of the wire scanners is known, the transfer matrix $\boldsymbol{R}$ can be built from the wire scanners to LUXE's IP. In the horizontal plane the twiss propagation matrix can be written in terms of the transfer matrix through Eqn~\ref{eq:twiss_mat_x}.
%
%
But since $\sigma_x = \sqrt{\beta_x \varepsilon_x}$, the beam size can be expressed as

\begin{equation}
    \sigma_x = \sqrt{R_{11}^2\sigma_{x_0}^2 + R_{12}^2(1+\alpha_{x_0}^2)\varepsilon_x^2\sigma_{x_0}^{-2}-R_{11}R_{12}\alpha_{x_0}\varepsilon_x},
\end{equation}

where $\sigma_{x_0}$ is the beam size measured at the wire scanner and $\varepsilon_x$ is the horizontal emittance. From error propagation, the beam size jitter at the IP $\sigma_{J,\sigma_x}$ is given by
\begin{equation}
    \sigma_{J,\sigma_x} = \frac{\partial\sigma_x}{\partial\sigma_{x_0}}\sigma_{J,\sigma_{x_0}} = \frac{R_{11}^2\sigma_{x_0} - R_{12}^2(1+\alpha_{x_0})\varepsilon_x^2\sigma_{x_0}^{-3}}{\sigma_x} \sigma_{J,\sigma_{x_0}},
\end{equation}

where $\sigma_{J,\sigma_{x_0}}$ is the beam size jitter at the wire scanner. In the end the beam size jitter found are $\sigma_{J,\sigma_x} \approx 0.45\ \rm \mu m$ and $\sigma_{J,\sigma_y} \approx 0.35\ \rm \mu m$.

\subsection{Time-energy relationship}

From Eqn~\ref{eq:rmat_bloc_diag} the propagation of time and energy is expressed as

\begin{equation}
    \begin{pmatrix}
     ct_{\rm IP} \\ \delta_{\rm IP} \\
    \end{pmatrix}
    =
    \begin{pmatrix}
        R_{55} & R_{56} \\ R_{65} & R_{66} \\
    \end{pmatrix}
    \cdot
    \begin{pmatrix}
        ct_{0} \\ \delta_{0} \\
    \end{pmatrix}.
    \label{eq:rmat_t_E}
\end{equation}

Looking at the transfer matrix from XTL to XTD8 the factor values are $R_{55}=R_{66}=1$, $R_{65}=0$ and $R_{56}\approx0$. The transfer matrix for time and energy approaches the identity matrix. This mean that the arrival time $t$ and the energy deviation $\delta$ are constant along the lattice and by extension the bunch length $\sigma_t$ as well.

\subsection{Bunch charge}

In an accelerator line, there are three main sources of impact of the bunch charge: background interactions, collimators and RF cavities. Results from the previous residual gas background simulations have showed a low impact on the electron bunch. Furthermore, there is no collimators and no RF cavity in the XTD8 lattice. This mean that the bunch charge measured in the switch yard does not vary significantly before reaching LUXE's IP. As such, the same is true for the charge jitter $\sigma_{J,q}$.

\subsection{Summary of the jitter at the IP}

All the propagated beam jitter and noise values are summarized in Table~\ref{tab:ip_jitter}. With those values of $\sigma_J$, an electron beam with realistic variations can now be simulated at LUXE's IP.
The positions jitter calculated for LUXE, around \SI{0.5}{\micro\meter} is much less that the beam size, around \SI{10}{\micro\meter}. This indicates that the position jitter should not have a huge impact on the signal. 
To investigate this impact as well as the impact of the other parameters, a detailed study is required, which is reported in the next section.

\begin{table}[!ht]
\centering
\begin{tabular}{| c | c || c | c | c|}
\hline
Parameter & Unit & Central value & $\sigma_J$ & $\sigma_N$ \\
\hline
\hline
$x$ & $\rm \mu m$ & 0 & 0.5 & 0.417 \\
\hline
$y$ & $\rm \mu m$ & 0 & 0.5 & 0.636 \\
\hline
$x'$ & $\rm \mu rad$ & 0 & 0.09 & 0.284 \\
\hline
$y'$ & $\rm \mu rad$ & 0 & 0.01 & 0.033 \\
\hline
$\sigma_x$ & $\rm \mu m$ & 10 & - & 0.45 \\
\hline
$\sigma_y$ & $\rm \mu m$ & 10 & - & 0.35 \\
\hline
$E$ & GeV & 16.1 & $1\times10^{-3}$ & $2.53\times10^{-4}$ \\
\hline
$q$ & pC & 250 & 1.25 & 0.5 \\
\hline
$\sigma_t$ & fs & 40 & $\sim 1$ & $<1$ \\
\hline
$t$ & fs & 0 & 25.78 & 5 \\
\hline
\end{tabular}
\caption{Summary of the jitter and noise at LUXE's IP for all the considered parameters.}
\label{tab:ip_jitter} 
\end{table}

\section{Impact of jitter on LUXE's signal}

\subsection{Parameters of the laser}

To evaluate the potential impact on the signal, the collision between an electron bunch and a laser pulse needs to be simulated. At LUXE's "phase-0", the laser installed could achieve powers up to $P_L=40\ \rm TW$. The power of a laser pulse is expressed as

\begin{equation}
    P_L=\frac{\pi I_0 w_0^2}{2},
\end{equation}

where $I_0$ is the peak laser intensity and $w_0$ is the laser waist. $I_0$ is derived from the peak electrical field $\mathcal{E}_0$ as

\begin{equation}
    \mathcal{E}_0 = \sqrt{\frac{2I_0}{c\epsilon_0}},
\end{equation}

where $\epsilon_0$ is the dielectric permittivity of the vacuum. The normalized amplitude $a_0$ used in Ptarmigan is expressed as 

\begin{equation}
    a_0 = \frac{\mathcal{E}_0 e}{c m_e \omega},
\end{equation}

where $e$ is the electron charge, $m_e$ is the electron mass and $\omega = \frac{2 \pi c}{\lambda}$ is the angular frequency. Considering a laser pulse of time $\tau$, the pulse energy is $E_L = P_L \tau$ and the normalized amplitude is finally expressed as

\begin{equation}
    a_0=\frac{e\lambda\sqrt{\frac{P_L}{c^5 \epsilon_0}}}{\pi^{\frac{3}{2}}m_e w_0}.
    \label{eq:a0}
\end{equation}

A list of the laser pulse parameters for LUXE's "phase-0" can be found in Table~\ref{tab:laser_param}.

\begin{table}[!ht]
\centering
\begin{tabular}{| c | c | c |}
\hline
Parameter & Unit & Value \\
\hline
\hline
$a_0$ & 1 & 5 \\
\hline
$E_L$ & J & 1.2 \\
\hline
$P_L$ & TW & 21 \\
\hline
$\lambda$ & nm & 800 \\
\hline
$\tau$ & fs & 30 \\
\hline
$w_0$ & $\rm \mu m$ & 5 \\
\hline
\end{tabular}
\caption{Parameters of the laser for LUXE in phase-0.}
\label{tab:laser_param} 
\end{table}

\subsection{Intrinsic variation of the QED processes}

\begin{figure}[!ht]
    \centering
    \includegraphics[width=13cm]{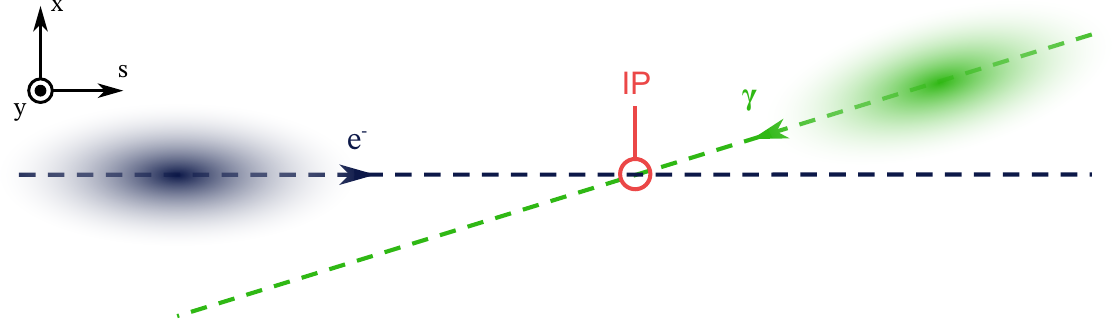}
    \caption{Schematic of the collision between an electron bunch and a laser pulse in the longitudinal plane.}
    \label{fig:pt_elec_laser}
\end{figure}

The collisions between LUXE's laser pulse and an electron bunch are simulated using the QED simulation software Ptarmigan. In an input file for Ptarmigan, the user must define the properties of the laser pulse (Table~\ref{tab:laser_param}) and of the electron bunch (central values in Table~\ref{tab:ip_jitter}). In the bunch definition the user sets the number of electron macroparticles that will collides with the laser. Each macroparticle charge is scaled to the total charge of the bunch, \SI{250}{\pico\coulomb} in this study. 
The first simulation is realized with $5 \times 10^{6}$ primary electrons and with the central values for every bunch parameter. A schematic of the collision is shown in Figure~\ref{fig:pt_elec_laser}.

\begin{figure}[!ht]
    \centering
    \includegraphics[width=12cm]{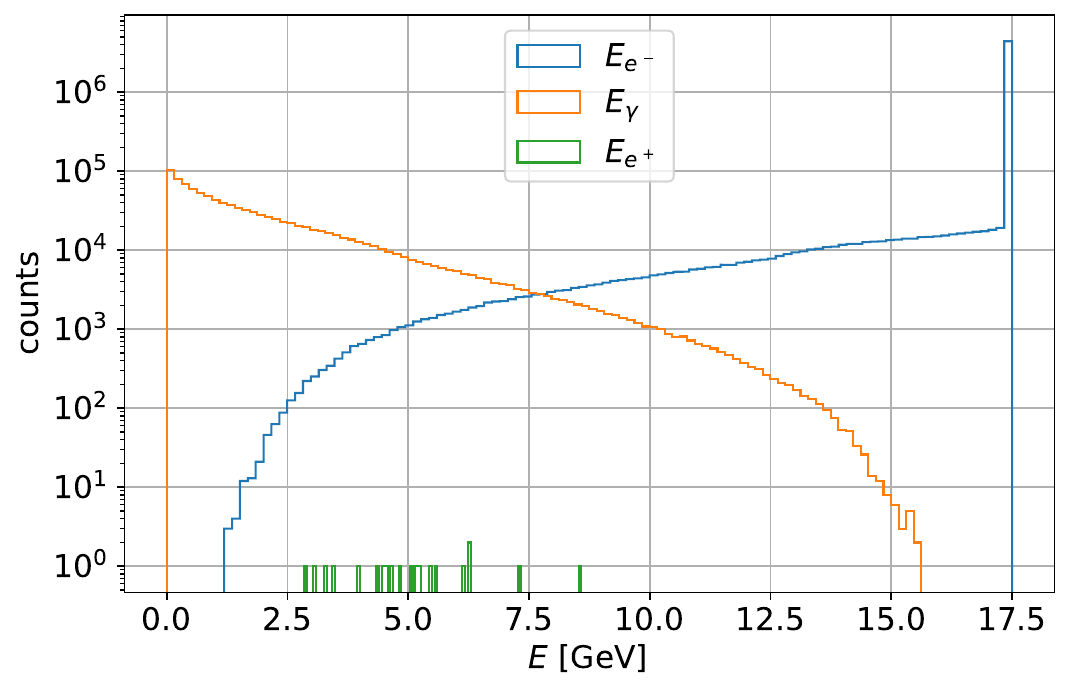}
    \caption{Plot of the energy spectrum for photons, positrons and electrons after the collision.}
    \label{fig:pt_single_energy_unb}
\end{figure}

Through the three QED processes presented in Chapter~\ref{chap:intro}, the collision produced a distribution of particles, mainly photons from inverse Compton, as well as positrons and electrons from Breit-Wheeler and trident interactions. The energy spectrum of those three particles species after the collision is shown in Figure~\ref{fig:pt_single_energy_unb}.
The majority of the output beam is composed of electrons at roughly \SI{17.5}{\giga\electronvolt}, mainly electrons that have not interacted with the laser. Electrons with lower energies have interacted the QED processes, in majority the inverse Compton process. Produced photons are in majority low energy photons with a maximum energy of around \SI{15}{\giga\electronvolt}. With $5 \times 10^{6}$ primary electrons, there are little to no positrons produced due to the lower cross section of the pair production. 

\begin{figure}[!ht]
    \centering
    \includegraphics[width=12cm]{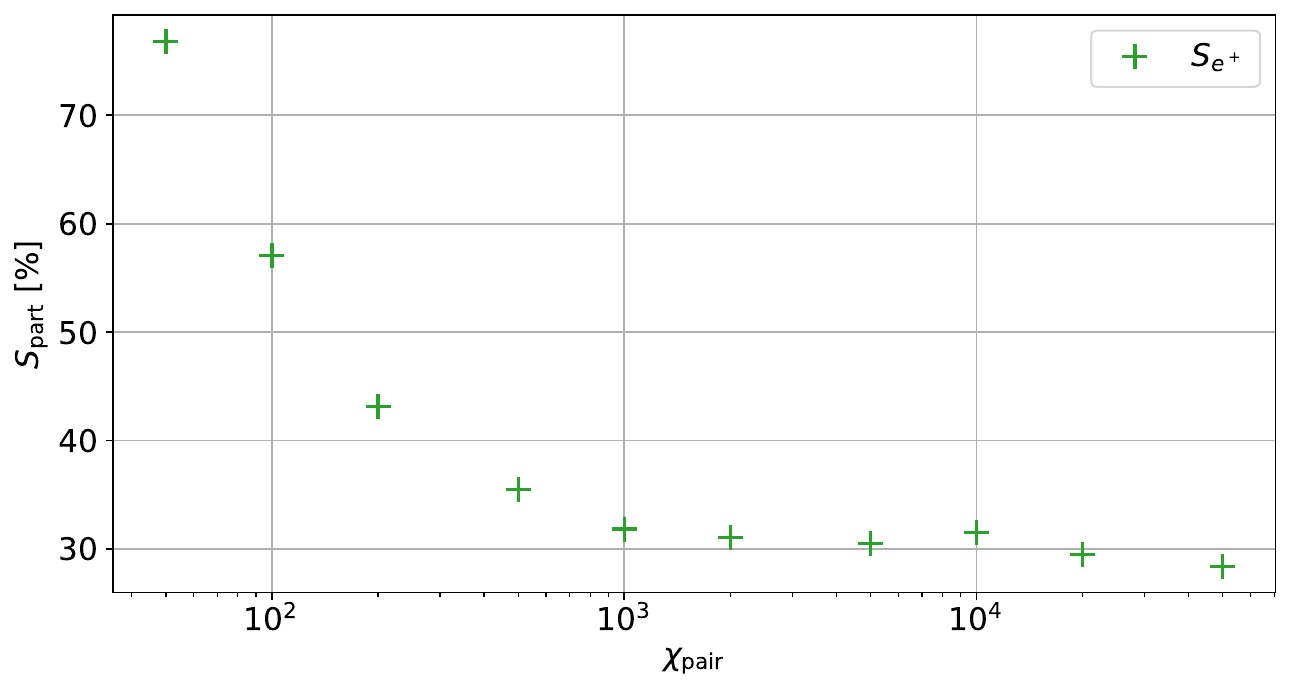}
    \caption{Plot of the normalised standard deviation on the number of positrons $S_{e^+}$ as a function of the pair creation cross section biasing factor $\chi_{\rm pair}$.}
    \label{fig:pt_pair_bias}
\end{figure}

Similar to the background simulations in BDSIM, a cross section bias $\chi_{\rm pair}$ must be applied to the pair production process. As $\chi_{\rm pair}$ increases, the number of positron produced $N_{e^+}$ increases, simultaneously the error on $N_{e^+}$ should decrease. At a given value of $\chi_{\rm pair}$, the variation of $N_{e^+}$ can be calculated by running multiple Ptarmigan simulations. The normalised standard deviation on the number of each particle $S_{\rm part}$ is expressed as

\begin{equation}
    S_{\rm part}[\%] = \frac{\langle {N_{\rm part}}^2\rangle}{\langle N_{\rm part} \rangle}\times 100.
    \label{eq:norm_std}
\end{equation}

\begin{figure}[!ht]
    \centering
    \includegraphics[width=12cm]{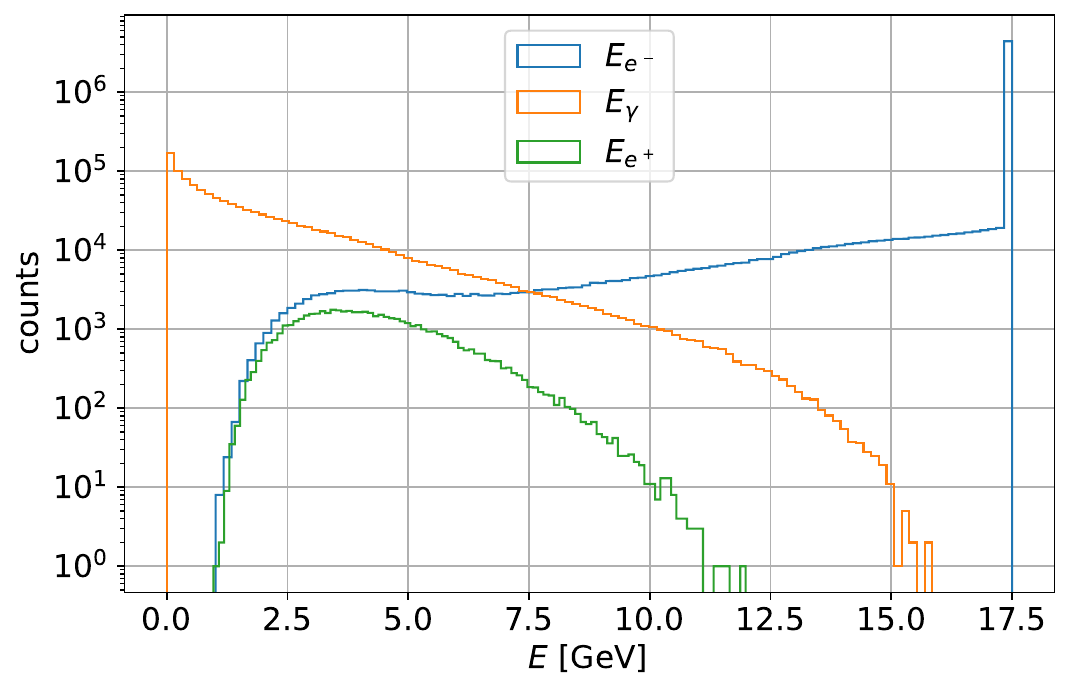}
    \caption{Plot of the energy spectrum for photons, positrons and electrons after the collision with the laser pulse. The pair production cross section is increased by $\chi_{\rm pair}=10^4$}
    \label{fig:pt_single_energy}
\end{figure}

In Figure~\ref{fig:pt_pair_bias}, $S_{\rm part}$ is calculation for different values of $\chi_{\rm pair}$. Each point is set of $5 \times 10^2$ simulations with $10^4$ primary electrons. 
A low values of $\chi_{\rm pair}$ the normalized standard deviation reaches values above \SI{70}{\%}. Above $\chi_{\rm pair}=10^3$, $S_{\rm part}$ is stable at around \SI{30}{\%}. In the following simulations a biasing factor $\chi_{\rm pair}=10^4$ will be used. With this biasing factor the collision with $5 \times 10^{6}$ primary electron can be re-run. The result spectrum can be found in Figure~\ref{fig:pt_single_energy}.

\begin{figure}[!ht]
    \centering
    \includegraphics[width=\textwidth]{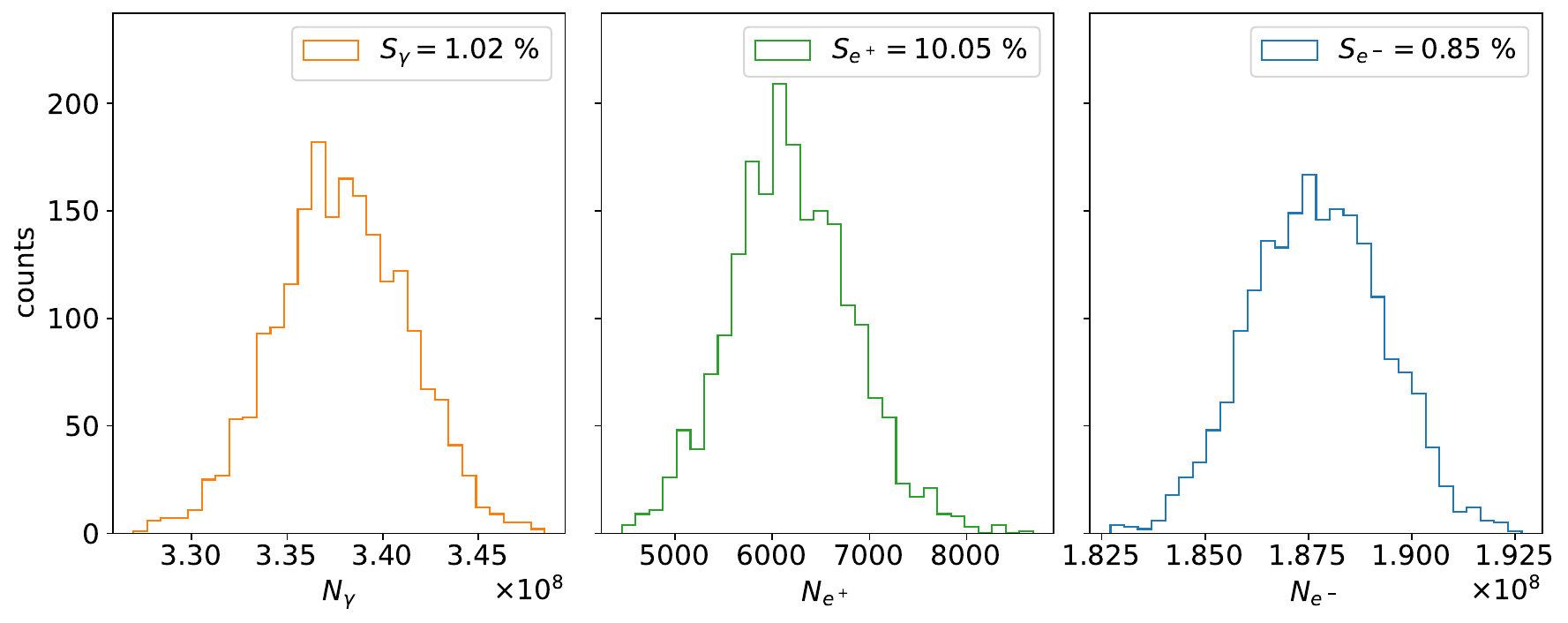}
    \caption{Histograms of the number of particle produced without any variation of the beam parameters. On the left are the photons produced. On the middle the positrons produced. On the right are all the electrons, produced and primaries.}
    \label{fig:pt_ref}
\end{figure}

In the biased spectrum, the spectrum of positrons is now clearly visible. At most, produced positrons can reach energies up to \SI{10}{\giga\electronvolt}. There is also an increase in low energy electrons due to the pair production. 
%
%
Now that the biasing factor is adjusted, a set of $2\times 10^3$ collisions with $10^5$ electron macroparticles can be simulated. The number of particles produced will vary because of the intrinsic statistical variation of the QED processes. This is shown in Figure~\ref{fig:pt_ref}.

Without any jitter in the electron beam, the number of photons and electrons produced varies by \SI{1}{\%} of the mean. For the number of positrons produced, $S_{\rm part}$ is higher than the other two species at around \SI{10}{\%}. This is because, even with the cross section biasing, the Breit-Wheeler pair creation process happens less frequently that the non-linear Compton. Therefore the statistics on number of positions is lower.

\subsection{Impact of the electron beam jitter}

The aim of this section is to simulate the impact of the electron beam jitter on the electron-laser collision. Let us consider a set of $5\times10^2$ Ptarmigan collisions and $10^5$ macroparticles per collision. For every parameters studied, each collision was run with a different values of this parameter. Those values are generated in a Gaussian distribution whose mean is the central value and of standard deviation $\sigma_J$, both taken from Table~\ref{tab:ip_jitter}. With a non-zero $\sigma_J$, the value of $S_{\rm part}$ for each particle should increase. To study more generally the impact of $\sigma_J$ on $S_{\rm part}$, multiple sets of $5\times10^2$ collisions are ran for different values of $\sigma_J$.

\subsubsection{Transverse position}

\begin{figure}[!ht]
    \centering
    \includegraphics[width=\textwidth]{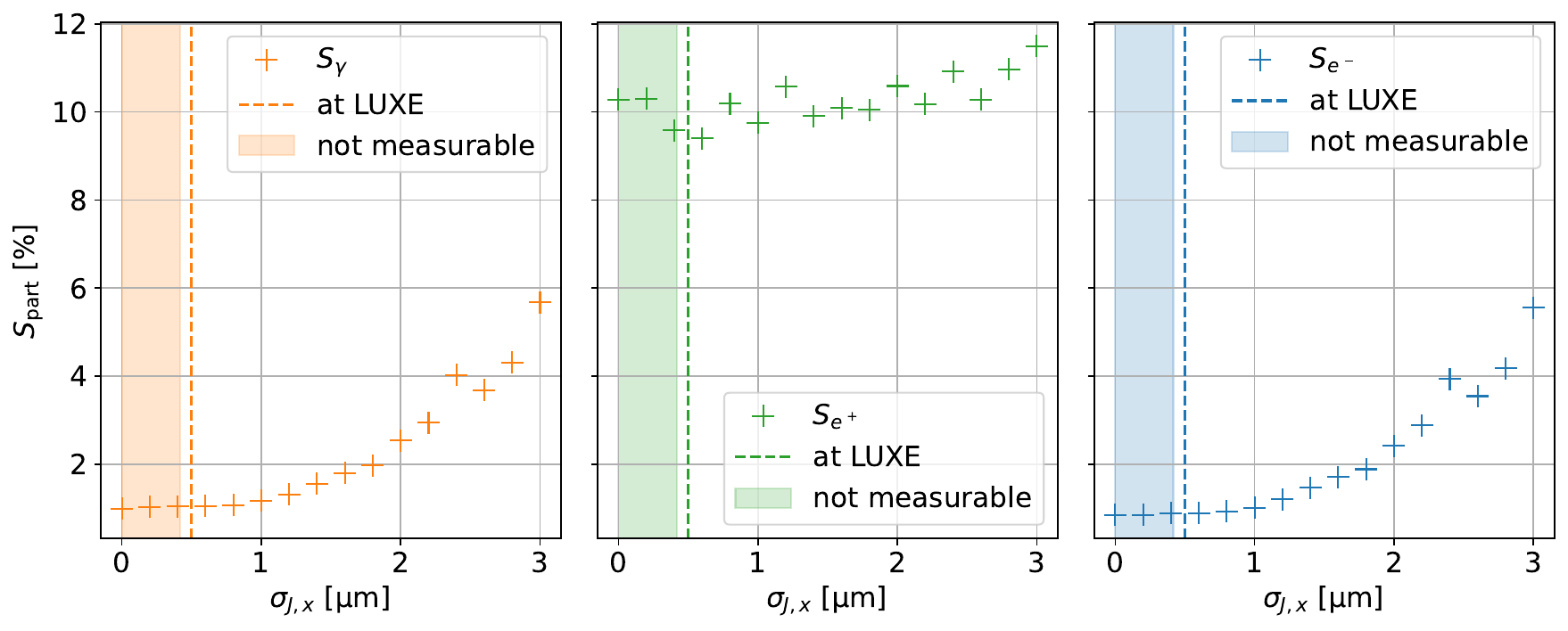}
    \caption{Plots of the normalized standard deviation $S_{\rm part}$ for each particle as a function of the horizontal position jitter $\sigma_{J, x}$. The left plot is the photon signal. The middle plot is the positron signal. The right plot is the electron signal. The vertical dash line indicates the expected value of jitter at LUXE's IP. The vertical bands shows the region where the jitter $\sigma_J$ is below $\sigma_N$.}
    \label{fig:pt_jitter_x}
\end{figure}

The first sets of Ptarmigan simulation ran are with the transverse position jitter. For both $x$ and $y$ the central value is zero as the beam is expected to beam centred in the transverse plane. The jitter $\sigma_{J, x}$ and $\sigma_{J, y}$ have been varies between 0 and \SI{3}{\micro\meter}. The resulting values of $S_{\rm part}$ for each particle type are shown in Figure~\ref{fig:pt_jitter_x}.
A jitter $\sigma_J$ lower than the vertical band in those plots would be below the noise $\sigma_N$, i.e., it would not be detectable from the XTD8 BPMs. In other words, the minimal detectable jitter is $\sigma_J=\sigma_N$. As seen previously, the variation on the number of positrons produced is higher than the two other particle types due to the low statistics on positrons. Below \SI{1}{\micro\meter} there is barely any variation from the configuration without jitter. An increase of $S_{\gamma}$ from \SI{1}{\%} to \SI{3}{\%} requires a jitter above \SI{2}{\micro\meter}. At the expected value $\sigma_J=0.5\ \rm \mu m$ at LUXE's IP, the variation of $S_{\rm part}$ increases by less than \SI{1}{\%} of the reference value.

\begin{figure}[!ht]
    \centering
    \includegraphics[width=\textwidth]{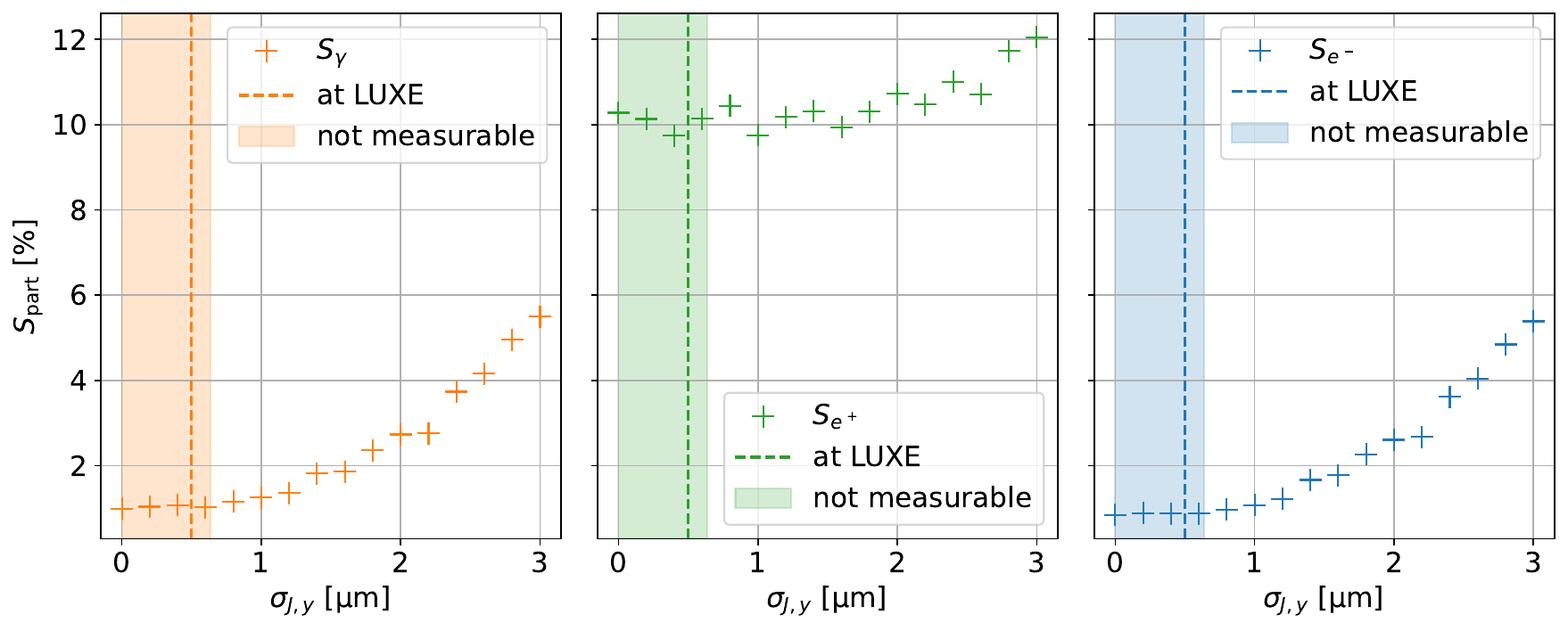}
    \caption{Plots of the normalized standard deviation $S_{\rm part}$ for each particle as a function of the vertical position jitter $\sigma_{J, y}$. The left plot is the photon signal. The middle plot is the positron signal. The right plot is the electron signal. The vertical dash line indicates the expected value of jitter at LUXE's IP. The vertical bands shows the region where the jitter $\sigma_J$ is below $\sigma_N$.}
    \label{fig:pt_jitter_y}
\end{figure}

Similarly, the impact of $\sigma_{J, y}$ can be seen in Figure~\ref{fig:pt_jitter_y}.
Even if the expected $\sigma_{J, y}$ is below the measured $\sigma_{N}$ for BPMs in the XTD8, the induced variations below \SI{1}{\micro\meter} are less than \SI{1}{\%}. In both planes, the expected variation in number of particle produced due to the transverse beam position jitter are negligible. Figures~\ref{fig:pt_jitter_x} and Figure~\ref{fig:pt_jitter_y} should be almost identical as the electron and laser beams are round. The only difference might come from the angle between beam axis and laser axis along the $x$ plane. 

\subsubsection{Bunch charge}

\begin{figure}[!ht]
    \centering
    \includegraphics[width=\textwidth]{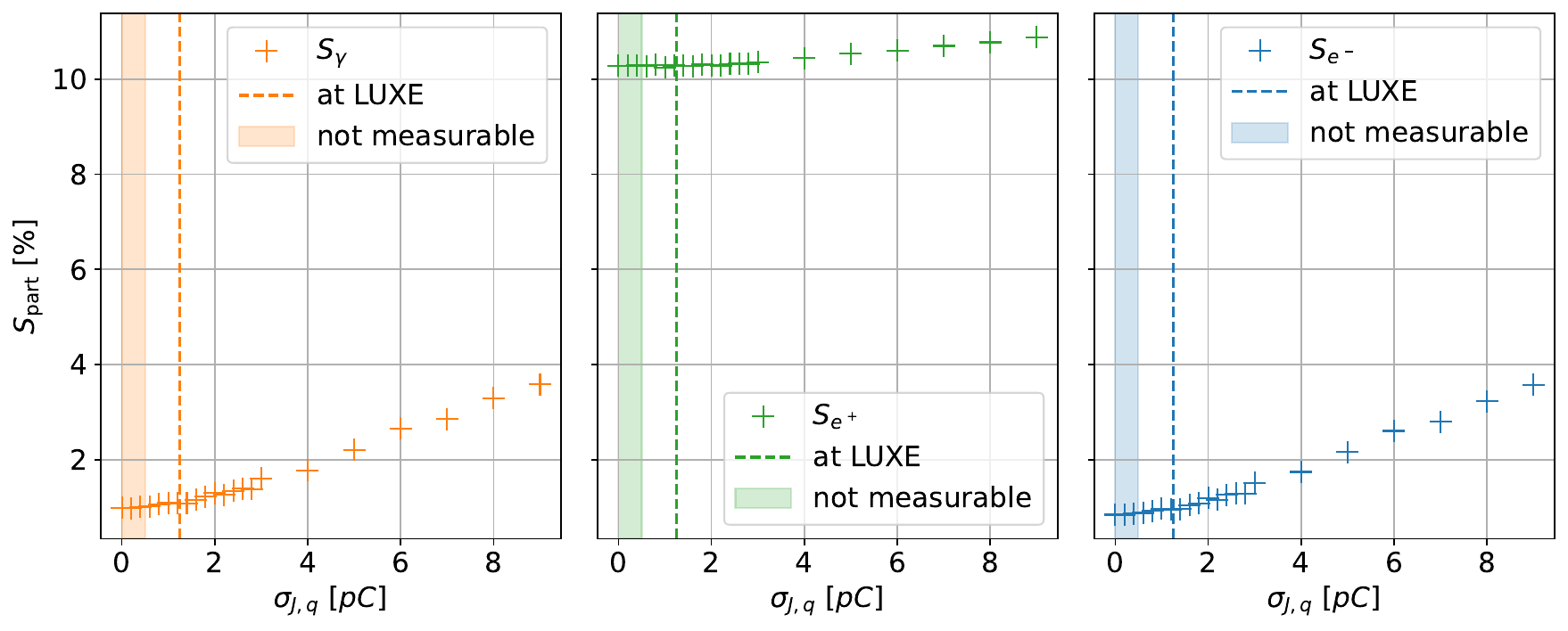}
    \caption{Plots of the normalized standard deviation $S_{\rm part}$ for each particle as a function of the charge jitter $\sigma_{J, q}$. The left plot is the photon signal. The middle plot is the positron signal. The right plot is the electron signal. The vertical dash line indicates the expected value of jitter at LUXE's IP. The vertical bands shows the region where the jitter $\sigma_J$ is below $\sigma_N$.}
    \label{fig:pt_jitter_q}
\end{figure}

The nominal bunch charge from LUXE's interactions is \SI{250}{\pico\coulomb}. A scan of the charge jitter $\sigma_{J, q}$ between 0 and \SI{9}{\pico\coulomb} is performed. The results are shown in Figure~\ref{fig:pt_jitter_q}.
The expected jitter in bunch charge at LUXE is around \SI{1.25}{\pico\coulomb}. At this jitter the variation of signal is way below \SI{1}{\%}. Even a jitter of \SI{4}{\pico\coulomb} produce a variation below \SI{1}{\%}. This shows that the bunch charge variations are negligible for LUXE's signal.

\subsubsection{Arrival time}

\begin{figure}[!b]
    \centering
    \includegraphics[width=\textwidth]{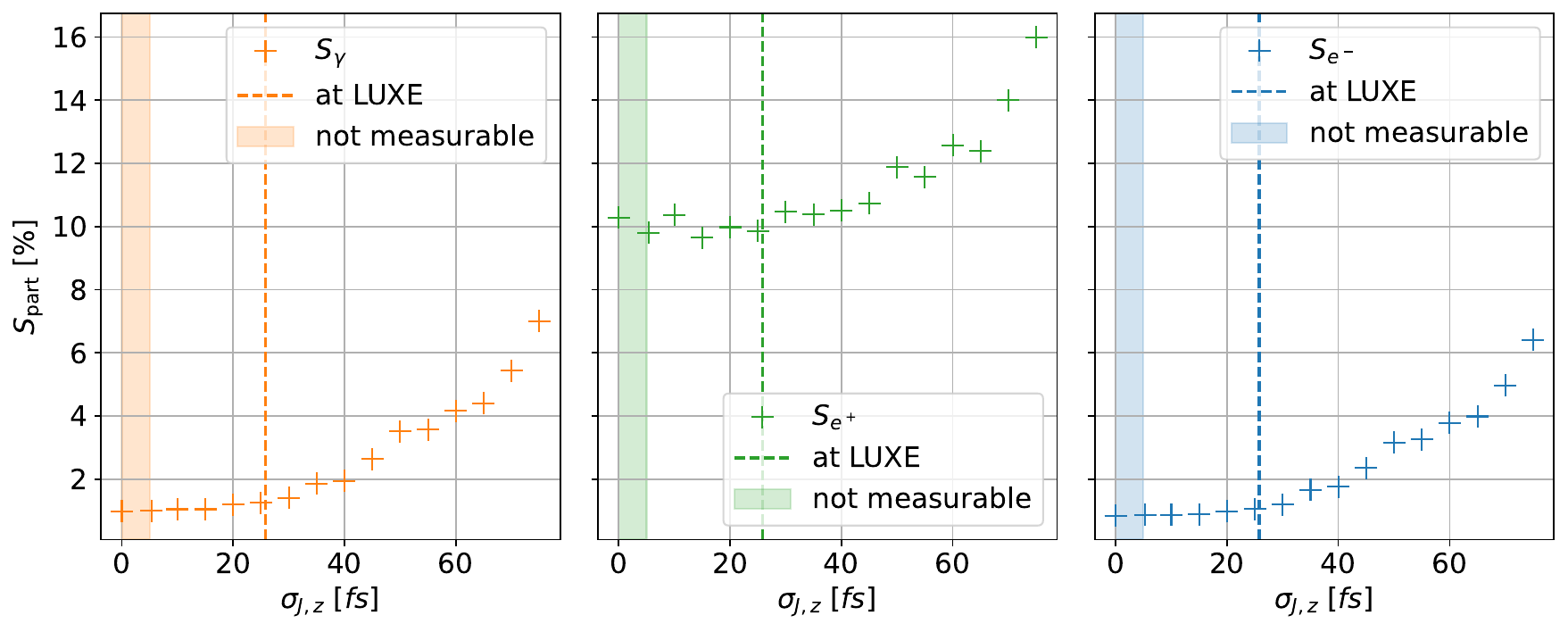}
    \caption{Plots of the normalized standard deviation $S_{\rm part}$ for each particle as a function of the arrival time jitter $\sigma_{J, z}$. The left plot is the photon signal. The middle plot is the positron signal. The right plot is the electron signal. The vertical dash line indicates the expected value of jitter at LUXE's IP. The vertical bands shows the region where the jitter $\sigma_J$ is below $\sigma_N$.}
    \label{fig:pt_jitter_z}
\end{figure}

Since the EuXFEL electron bunch is relativistic, the arrival time $t$ is directly proportional to the longitudinal position $z$ as $z = ct$ with $c$ the speed of light. Ptarmigan uses the position $z$ and the results are converted back in seconds. A scan of the arrival time jitter $\sigma_{J, z}$ between 0 and \SI{80}{\femto\second} is performed. The results are shown in Figure~\ref{fig:pt_jitter_z}.
At the expected time jitter of \SI{26}{\femto\second}, the variation in particle produced is lower than \SI{1}{\%}. Even at $\sigma_J=40\ \rm fs$, the variation in $S_{\rm part}$ is around \SI{1}{\%}. The variations in arrival time are negligible for LUXE's signal.

\subsubsection{Bunch length}

\begin{figure}[!ht]
    \centering
    \includegraphics[width=\textwidth]{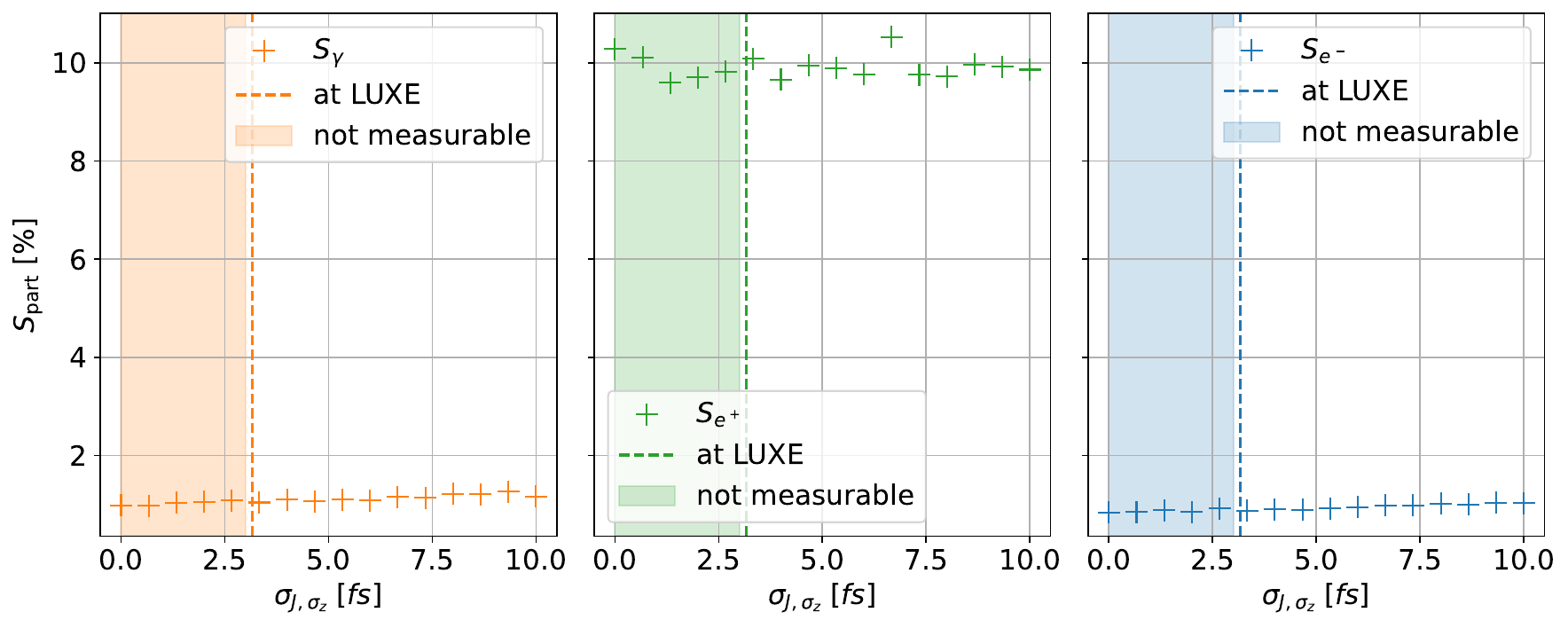}
    \caption{Plots of the normalized standard deviation $S_{\rm part}$ for each particle as a function of the bunch length jitter $\sigma_{J, \sigma_z}$. The left plot is the photon signal. The middle plot is the positron signal. The right plot is the electron signal. The vertical dash line indicates the expected value of jitter at LUXE's IP. The vertical bands shows the region where the jitter $\sigma_J$ is below $\sigma_N$.}
    \label{fig:pt_jitter_length}
\end{figure}

Similar to the arrival time, the bunch length in Ptarmigan is defined is meters and the results are converted back to seconds. A scan of the bunch length jitter $\sigma_{J, \sigma_z}$ was done between 0 and \SI{10}{\femto\second}. The results are shown in Figure~\ref{fig:pt_jitter_length}.
The expected bunch length jitter at LUXE is around \SI{3.17}{\femto\second}. Even at $\sigma_{J, sigma_z}=10\ \rm fs$ the number of particles produced is almost identical to the reference configuration.

\subsubsection{Beam size}

The nominal beam size for LUXE is \SI{10}{\micro\meter} in both transverse planes. A scan of the transverse size jitter $\sigma_{J, \sigma}$ was done between 0 and \SI{3}{\micro\meter}. The results are shown in Figure~\ref{fig:pt_jitter_length}.
The beam size is by far the most impactful parameter in regards to the number of particles produced through QED processes. At a jitter $\sigma_J$ equal to the wire scanner resolution (\SI{0.5}{\micro\meter}), $S_{\gamma}$ and $S_{e^-}$ increases by about \SI{7}{\%}. With the same jitter, $S_{e^+}$ increases by about \SI{2.5}{\%}. As previously stated, the wire scanner measurement method used can not be applied to individual bunches so the actual jitter on the beam size is not known. 

\begin{figure}[!ht]
    \centering
    \includegraphics[width=\textwidth]{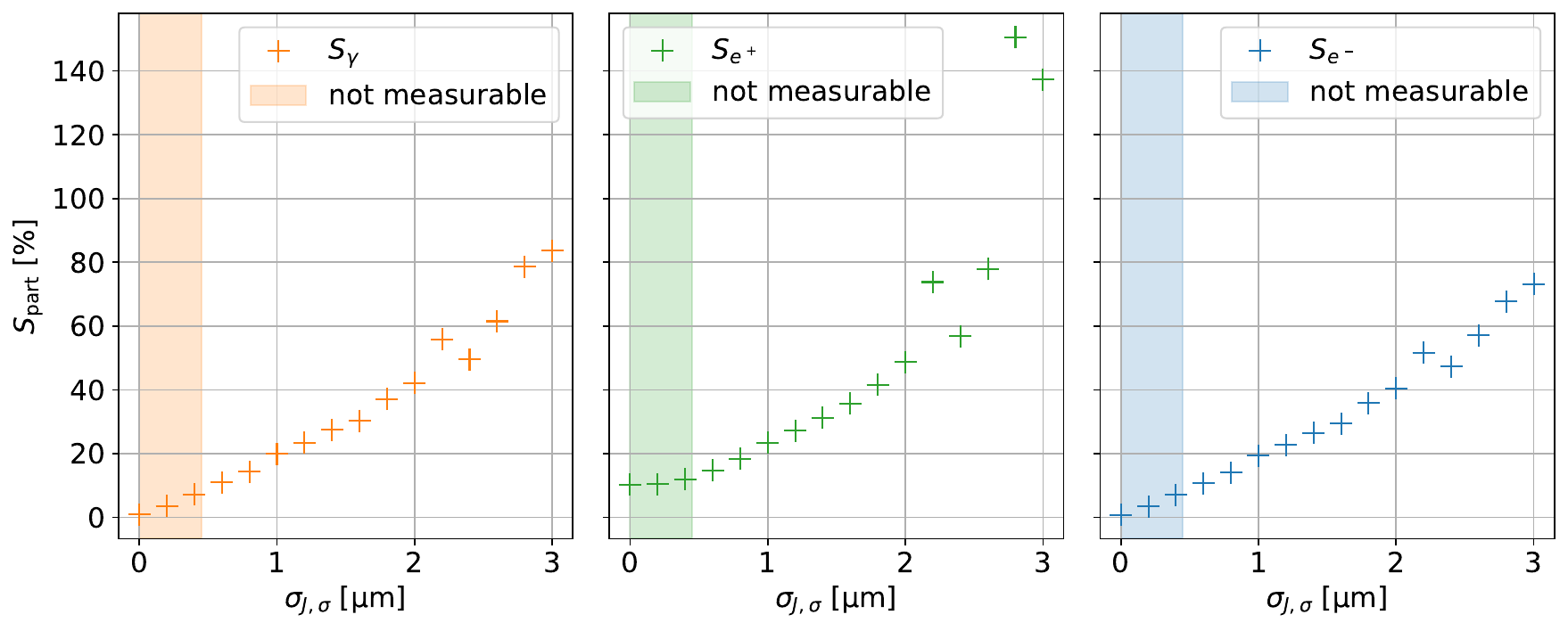}
    \caption{Plots of the normalized standard deviation $S_{\rm part}$ for each particle as a function of the transverse size jitter $\sigma_{J, \sigma}$. The left plot is the photon signal. The middle plot is the positron signal. The right plot is the electron signal. The vertical dash line indicates the expected value of jitter at LUXE's IP. The vertical bands shows the region where the jitter $\sigma_J$ is below $\sigma_N$.}
    \label{fig:pt_jitter_size}
\end{figure}

\subsubsection{Angle and Energy}

At LUXE's IP, the nominal collision angle $\theta$ is \SI{17.2}{\degree} with an expected jitter $\sigma_{J, \theta}=0.1\ \rm \mu rad$. The nominal bunch energy $E$ is \SI{17.5}{\giga\electronvolt} with an expected jitter $\sigma_{J, E}=1\ \rm MeV$. Similar to the other parameters, Ptarmigan scan were perform for both $\theta$ up to \SI{5.5}{\micro\radian} and $E$ up to \SI{3}{\mega\electronvolt}. In both scans the normalised standard deviation $S_{\rm part}$ was not impacted by the jitter. The collision angle and the bunch energy does not have a significant impact on LUXE's signal.

\section{Conclusion on the impact on LUXE}

From the BDSIM simulations, it has been shown that the backgrounds induced from both the residual gas and the halo does not have a significant impact at LUXE's IP. The beam transverse position at the IP can be estimated with a virtual measurement and the calibration at the IP itself can be performed with a thin tungsten needle. Using the transfer matrix on the EuXFEL and the XTD8, the measured jitter were propagated to LUXE's IP. Ptarmigan simulations of electron-laser collisions have shown the impact of each parameters jitter. From all of them, only the beam size has a significant impact on the produced signal. 
It appears as the most important parameter to monitor for LUXE.
Beyond the wire scanner measurements details in Chapter~\ref{chap:measurments}, there are detectors which can (when suitably calibrated) measure beam size variation bunch by bunch, for example coherent diffraction radiation~\cite{codr_size_meas}. These might be considered if there is significant variation in the LUXE beam size during data taking.

\chapter{Strong field QED at CLARA} \label{chap:clara}

\section{Overview}

The compact linear accelerator for research and applications (CLARA)~\cite{clara} is an electron beam line designed at the Daresbury Laboratory. Its main purpose is to test FEL schemes that will be implemented in other facilities. The ultrabright electron beam at CLARA makes it an suitable environment to test a wide range of techniques and ideas in order to prove their viability. This includes plasma wakefield and laser wakefield acceleration, as well as quantum radiation reaction (QRR) experiment~\cite{claraQRR} which has a lot of similarities with LUXE. Both experiments aims to collide electron beam with high power laser. In LUXE, the goal is to observe the difference between linear and non-linear (i.e. strong-field) QED~\cite{sfqed}. At CLARA, the goal is to observe the difference between the classical prediction and quantum prediction or radiation reactions. Although those was recently measured in electron-laser collision experiments, clearly measuring the difference between classical and quantum models has been challenging.
This chapter aims to simulate the radiation reactions predicted by QED and compare it to the classical predictions. In addition, this chapter aims to evaluate the possibility of a strong field QED~\cite{sfqed} experiment in comparison to the LUXE experiment described in previous chapters.

\subsection{Layout of the accelerator}

\begin{figure}[!ht]
    \centering
    \includegraphics[width=\textwidth]{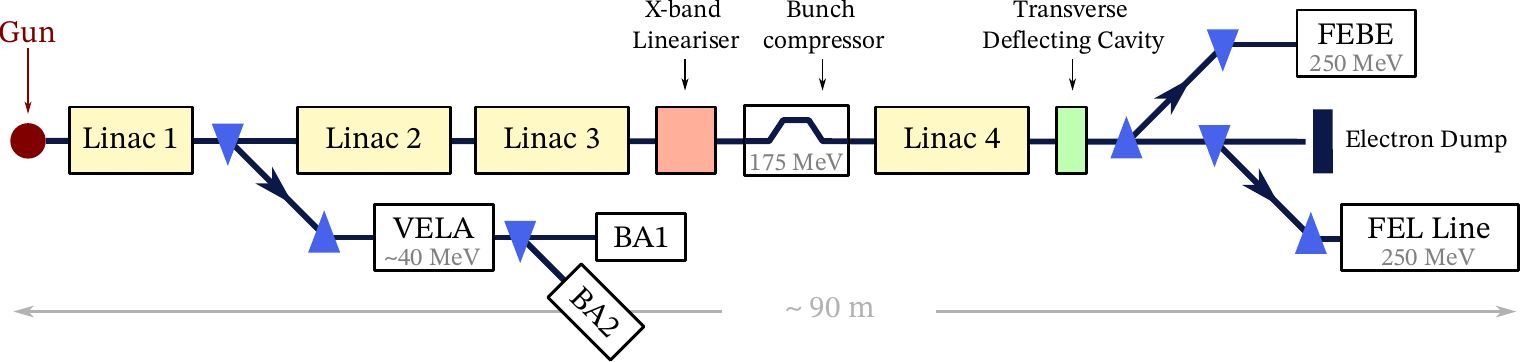}
    \caption{Plan layout of the CLARA accelerator.}
    \label{fig:clara_accelerator}
\end{figure}

CLARA uses a linear accelerator layout to bring electron bunches up to \SI{250}{\pico\coulomb} at around \SI{250}{\mega\electronvolt} with a repetition rate of \SI{400}{\hertz}. The \SI{90}{meter} facility uses four linac sections and the layout is detailed in Figure~\ref{fig:clara_accelerator}
Just after the first linac section, a low energy (40-\SI{50}{\mega\electronvolt}) beam line is extracted toward the versatile electron linear accelerator (VELA)~\cite{vela} and two beam areas (BA1 and BA2). The electron beam delivered to the beam areas is highly customisable and allows for the testing and commissioning of various accelerator equipments. One of the main goals is to develop high-brightness, high-repetition-rate injector technology for future FELs. Downstream of the linac sections, the initial design had only the FEL line and the electron dump. The desired electron bunch properties at this point are listed in Table~\ref{tab:clara_electron_beam}. As of today, the FEL line is not commissioned yet. 

\begin{table}[!ht]
\centering
\begin{tabular}{| c | c || c | c |}
\hline
Parameter & Unit & Low charge & High charge \\
\hline
\hline
$q$ & pC & 5 & 250 \\
\hline
$E$ & MeV & 250 & 250 \\
\hline
$\beta\gamma$ & 1 & 489.24 & 489.24 \\
\hline
$\sigma_{x,y}$ & $\rm \mu m$ & $\sim1$ & 50 \\
\hline
$\sigma_t$ & fs & $\ll 50$ & $\leq 50$ \\
\hline
$\sigma_E/E$ & \% & 0.1 & 1 \\
\hline
$\varepsilon_x$ & m.rad & $< 2.04\times 10^{-9}$ & $< 10.22\times 10^{-9}$ \\
\hline
$\varepsilon_y$ & m.rad & $< 2.04\times 10^{-9}$ & $< 2.04\times 10^{-9}$ \\
\hline
\end{tabular}
\caption{Parameters of the electron beam at the end of CLARA's accelerating section.}
\label{tab:clara_electron_beam} 
\end{table}

\subsection{FEBE and laser collision}

\begin{figure}[!ht]
    \centering
    \includegraphics[width=\textwidth]{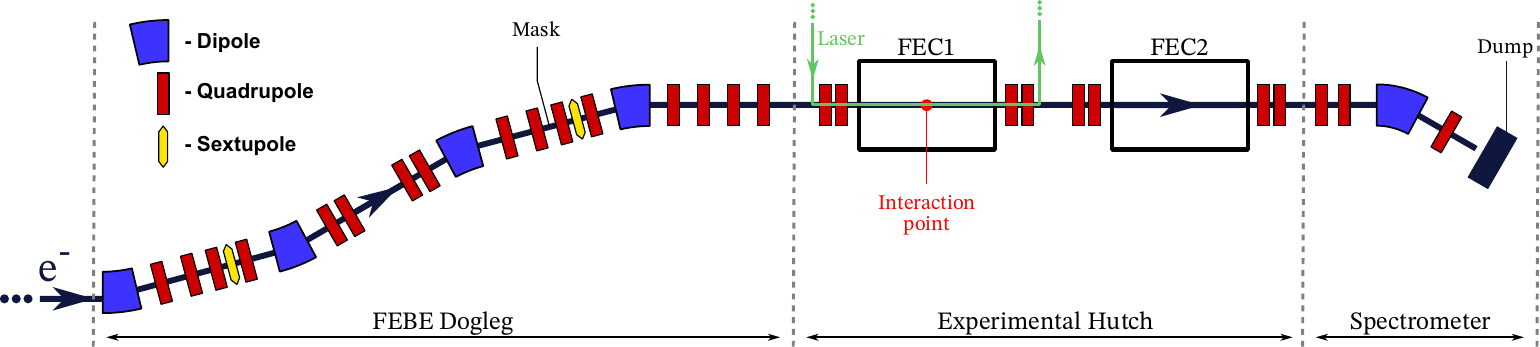}
    \caption{Plan layout of the different sections of FEBE.}
    \label{fig:febe_layout}
\end{figure}

Since the commissioning of phase 2,
many academic and industrial users have accessed the \SI{250}{\mega\electronvolt} electron beam to perform test on advanced accelerator technology, novel particle beam acceleration, medical application, etc. The decision was made to design and build a dedicated line, parallel to the FEL test area, that would be dedicated for user applications. This new line is the full energy beam exploitation (FEBE)~\cite{febe} line. A layout of FEBE can be found in Figure~\ref{fig:febe_layout}

The \SI{250}{\mega\electronvolt} electrons are extracted after the linac via a dogleg composed of two successive DBAs. In the middle of the second DBA, a mask (as shown in Figure~\ref{fig:beam_mask}) can be inserted to separate a drive bunch and a witness bunch for plasma wakefield acceleration. Downstream of the dogleg, the main experimental hutch is composed of two FEBE experimental chambers (FEC). The first chamber, FEC1, contains the main interaction point between the electron bunches and a \SI{100}{\tera\watt} laser. The parameters of this laser can be found in Table~\ref{tab:febe_laser}.

\begin{table}[!ht]
\centering
\begin{tabular}{| c | c | c |}
\hline
Parameter & Unit & Value \\
\hline
\hline
$a_0$ & 1 & 1.77 \\
\hline
$E_L$ & J & 2.5 \\
\hline
$P_L$ & TW & 100 \\
\hline
$\lambda$ & $\rm \mu m$ & 0.8 \\
\hline
$\tau$ & fs & 25 \\
\hline
$w_0$ & $\rm \mu m$ & 30 \\
\hline
\end{tabular}
\caption{Parameters of the laser at FEBE.}
\label{tab:febe_laser} 
\end{table}

The second chamber, FEC2, is a flexible space for diagnostics post-interaction. At the end of this experimental hutch there is a spectrometer for further diagnostics and the final electron dump. In FEC1, the collision setup between electron bunches and laser pulses is similar to LUXE's interaction chamber. As such, a strong field QED experiment could be installed at CLARA. 

\section{Simulating Compton scattering}

\subsection{Electron bunch expectations}

The electron parameters described in Table~\ref{tab:clara_electron_beam} are still in development since the commissioning of phase 2. To efficiently get strong field QED interaction at FEBE's IP, the transverse dimensions of the bunches must be similar the what is planned for LUXE. This means that the transverse beam size must be at most \SI{10}{\micro\meter}. Only the high charge configuration should be used as it will also increase the number of interactions. In Table~\ref{tab:febe_electron_beam} are the minimal beam parameters required to observe strong field QED at CLARA.

\begin{table}[!ht]
\centering
\begin{tabular}{| c | c | c |}
\hline
Parameter & Unit & Value \\
\hline
\hline
$q$ & pC & 250 \\
\hline
$E$ & MeV & 250 \\
\hline
$\sigma_{x,y}$ & $\rm \mu m$ & 10 \\
\hline
$\sigma_t$ & fs & 100 \\
\hline
$\sigma_E/E$ & \% & 1 \\
\hline
$\theta$ & deg & 17.2 \\
\hline
\end{tabular}
\caption{Expected parameters of the electron beam at FEBE's interaction point.}
\label{tab:febe_electron_beam} 
\end{table}

\begin{figure}[!ht]
    \centering
    \includegraphics[width=6cm]{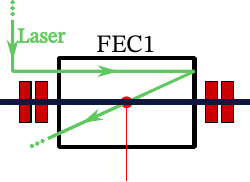}
    \caption{Schematic of the back propagating laser path in FEC1.}
    \label{fig:alt_laser}
\end{figure}

Similar to LUXE's interaction chamber, the laser is planed to collide with the electron beam at an angle $\theta$ of \SI{17.2}{\degree}. The laser setup will be adapted to get the back propagating laser, as shown in Figure~\ref{fig:alt_laser}
The Compton signal should increase with a smaller beam size $\sigma_{x,y}$, a higher charge $q$ or higher energy $E$, a lower energy spread $\delta=\sigma_E/E$, a smaller bunch length $\sigma_t$ or a lower collision angle $\theta$.

\subsection{Laser foci}

\begin{figure}[!ht]
    \centering
    \includegraphics[width=10cm]{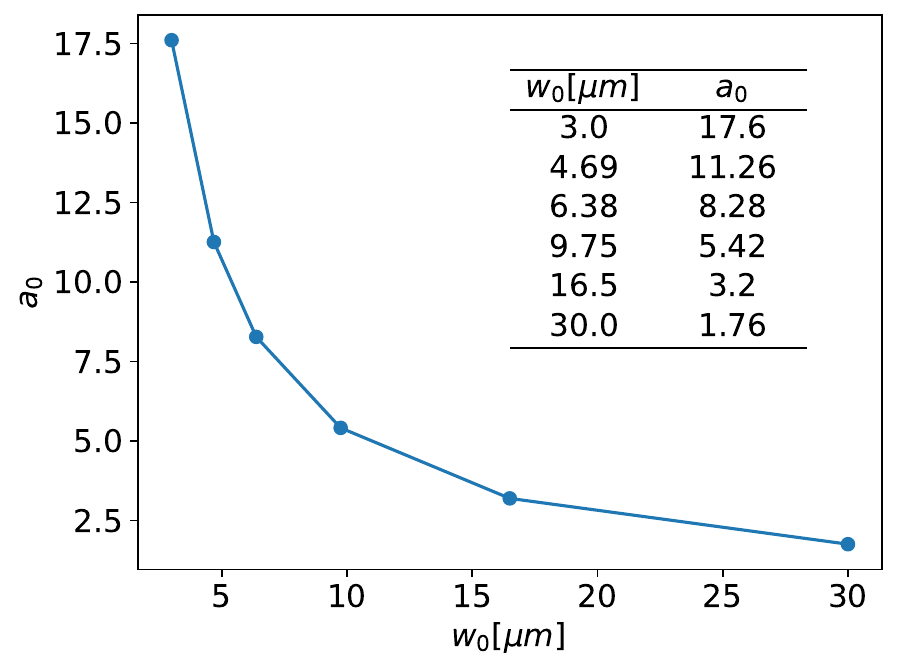}
    \caption{Plot of the normalised amplitude $a_0$ as a function of the laser waist $w_0$.}
    \label{fig:febe_laser_w0}
\end{figure}

The amount of Compton scatters is directly linked to the intensity of the laser. In Chapter~\ref{chap:intro} the factor $\xi$ was defined in Eqn~\ref{eq:xi} to evaluate the laser intensity. It is equivalent to the normalized amplitude $a_0$ defined in Eqn~\ref{eq:a0}. With the nominal laser parameters in power $P_L=100\ \rm TW$ and waist $w_0=30\ \rm \mu m$ the laser reaches $a_0=1.77$. Assuming there is no increase in laser power, the same laser can achieve a higher amplitude by reducing $w_0$.
Figure~\ref{fig:febe_laser_w0} shows different values of $a_0$ calculated from Eqn~\ref{eq:a0} for a range of waists between \SI{3}{\micro\meter} and \SI{30}{\micro\meter}. With a waist $w_0=3\ \rm \mu m$ the laser can reach an amplitude of $a_0=17.7$. The laser waist for a perfectly Gaussian pulse is calculated as~\cite{laser_waist}

\begin{equation}
    w_0 = \frac{4\lambda f}{\pi D},
\end{equation}

where $f$ is the focal length of the mirror and $D$ is the laser diameter at the mirror. At FEBE, the co-propagating focal length is around \SI{3.5}{\meter} and the standard mirrors have a diameter $D\approx 0.18\ \rm m$. In theory this correspond to a minimal waist $w_0\approx15.7\ \rm \mu m$. With the back propagating laser, a focal length of around \SI{0.5}{\meter} would be need to reach $w_0=3\ \rm \mu m$. Even higher values of $a_0$ could be achieved this way but this is limited by the diffraction limit. There are other way to increase $a_0$, either by increasing the laser power $P_L$ or increasing the wavelength $\lambda$, both of which are impossible at CLARA/FEBE.

\subsection{Ptarmigan simulation of Compton at FEBE}

\begin{figure}[!ht]
    \centering
    \includegraphics[width=10cm]{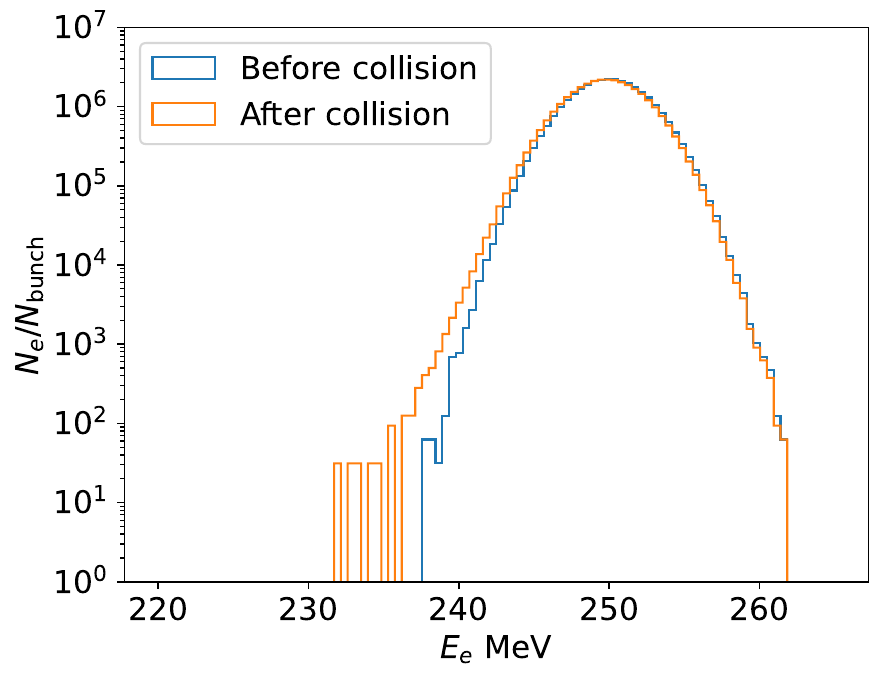}
    \caption{Histogram of the energy spectrum for electrons before and after collision.}
    \label{fig:hist_E_ref}
\end{figure}

Simulating the collision between electron bunches from CLARA and laser pulses in FEC1 is again done using Ptarmigan. Taking the input parameters from Table~\ref{tab:febe_laser} and Table~\ref{tab:febe_electron_beam}, collision of $10^6$ primary electron macroparticles was simulated. The resulting electron spectrum is shown in Figure~\ref{fig:hist_E_ref}.
The increase in low energy electrons shows that some Compton scattering happened in the collision. However, with $a_0=1.77$ the produced signal is quite low.

\begin{figure}[!ht]
    \centering
    \includegraphics[width=\textwidth]{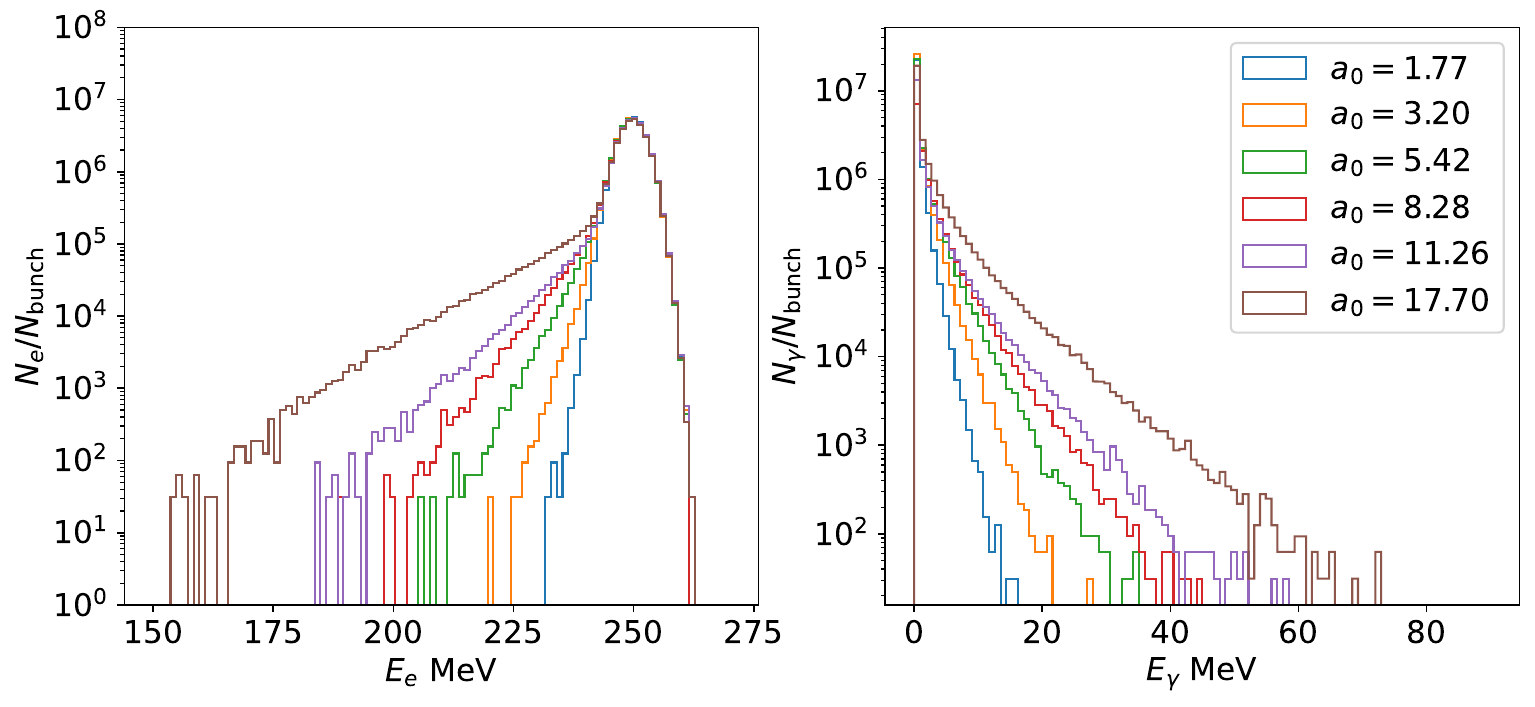}
    \caption{Histograms of the energy spectrum for electrons and photons after electron-laser collision for different values of $a_0$. On the left is the number of electrons as a function of energy. On the right is the number of photos as a function of energy.}
    \label{fig:hist_E_a0}
\end{figure}

In Figure~\ref{fig:hist_E_a0}, energy spectrums are shown for multiple collisions with varying $a_0$, i.e. with varying $w_0$.
As $a_0$ increases, there is a higher number of photons produced through Compton scattering. At $a_0=17.7$ those photons can reach energies up to \SI{70}{\mega\electronvolt} and the scattered electrons can reach energies as low as \SI{160}{\mega\electronvolt}.

\begin{figure}[!ht]
    \centering
    \includegraphics[width=\textwidth]{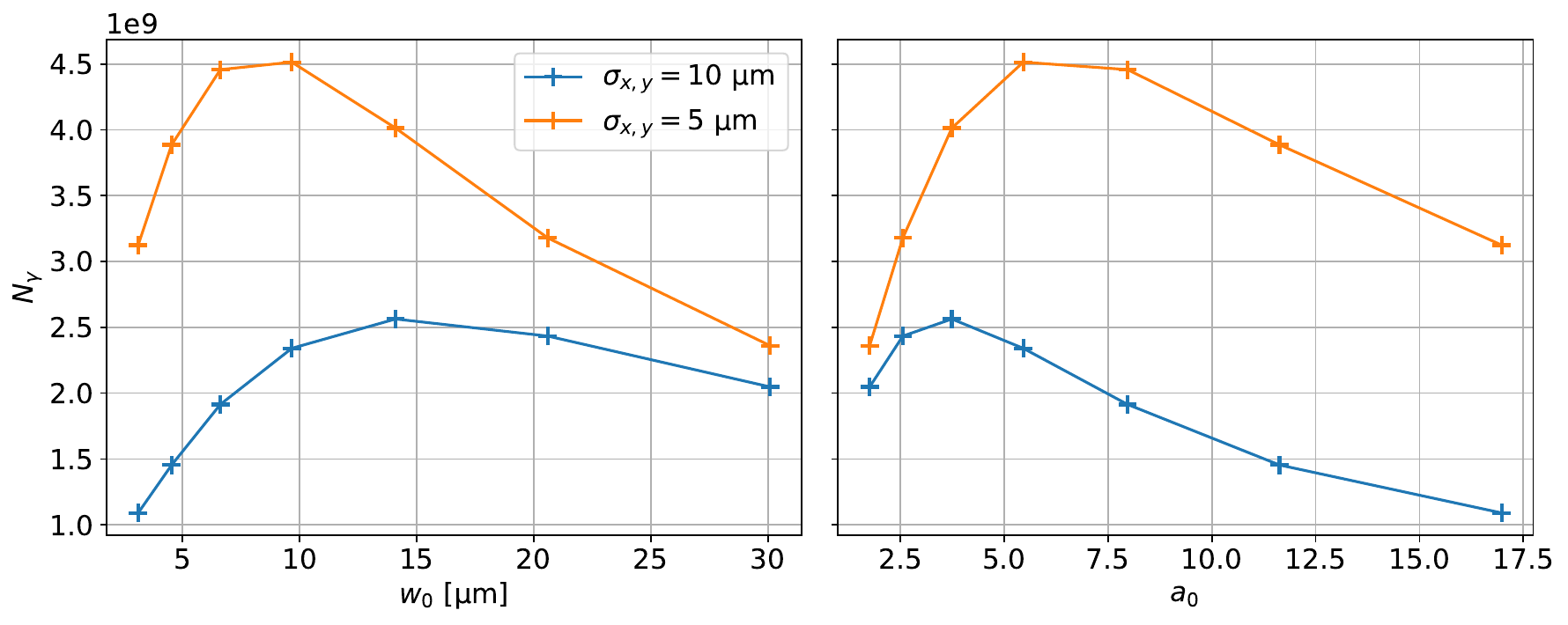}
    \caption{Plots of the number of photons produced $N_\gamma$ for different laser intensities. On the left is $N_\gamma$ as a function of $w_0$. On the right is $N_\gamma$ as a function of $a_0$.}
    \label{fig:nphoton_a0_w0}
\end{figure}

With higher values of $a_0$, an increase in the number of photons produced is expected. However, if this increase is achieved by reducing the waist, it gets below the transverse electron beam size of \SI{10}{\micro\meter}. The number of photons produced $N_\gamma$ for every configuration is shown in Figure~\ref{fig:nphoton_a0_w0}.

\begin{figure}[!ht]
    \centering
    \includegraphics[width=\textwidth]{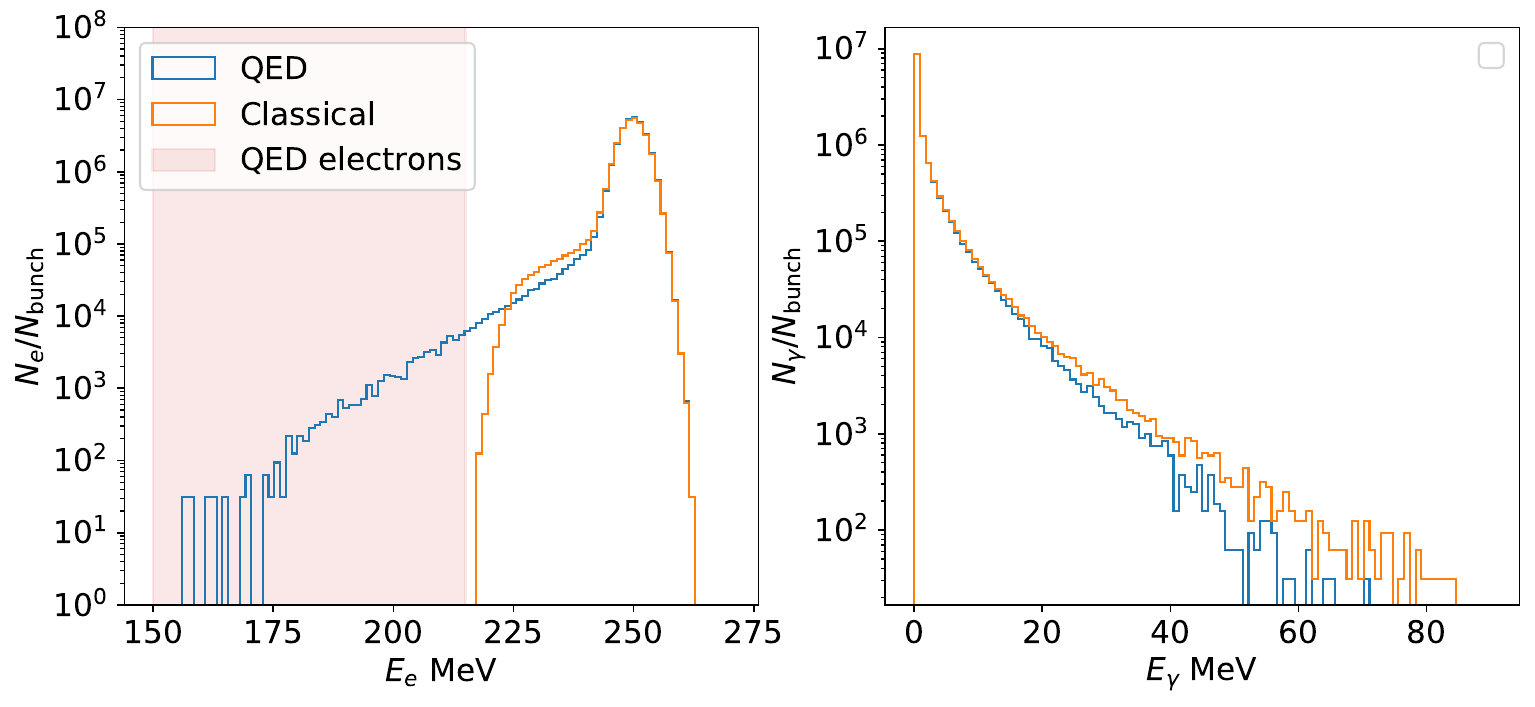}
    \caption{Histograms of the energy spectrum for electrons and photons after electron-laser collision for the classical and QED model. On the left is the number of electrons as a function of energy. On the right is the number of photos as a function of energy.}
    \label{fig:qed_classical_comp}
\end{figure}

With the regular transverse beam size $\sigma_{x,y}=10\ \rm \mu m$, the number of photons is maximized around $w_0\approx15\ \rm \mu m$, i.e around $a_0\approx3$. Doing the same simulations with $\sigma_{x,y}=5\ \rm \mu m$ the maximum $N_\gamma$ is shifted to $w_0\approx9\ \rm \mu m$, i.e. $a_0\approx7$. 
This is because a smaller electron beam size has an higher overlap with a small $w_0$ laser.
As expected, increasing the electron beam size also increases the number of photons produced for every values of $a_0$.
Using Ptarmigan, the electron-laser collision can be simulated with the classical Compton scattering model. The comparison between classical model and QED model can be found in Figure~\ref{fig:qed_classical_comp}

\begin{figure}[!ht]
    \centering
    \includegraphics[width=11cm]{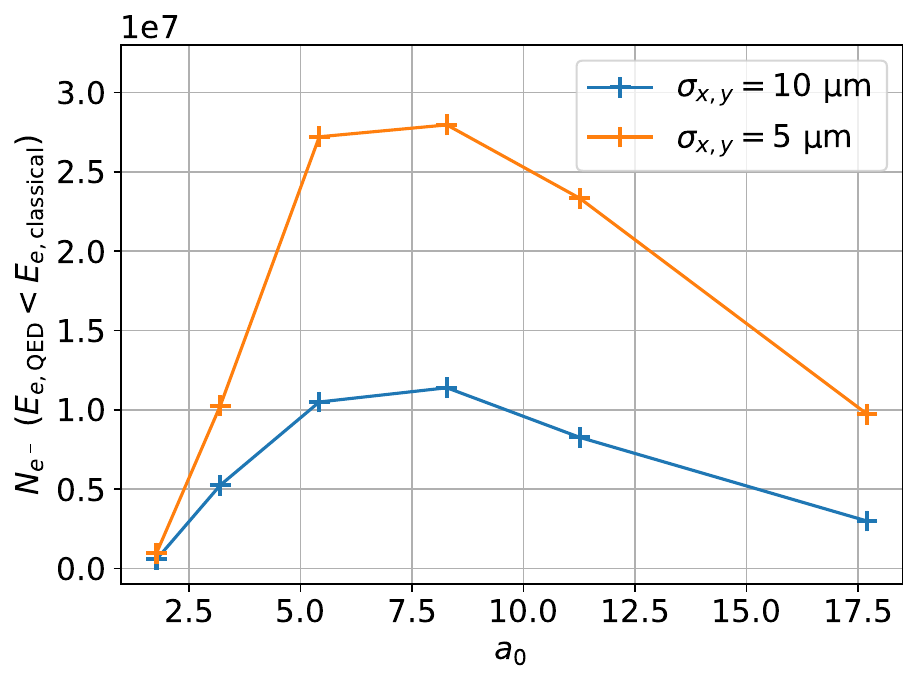}
    \caption{Plot of the number of QED electron $N_{e^-}$ below the minimum energy of classical electrons for different laser intensities.}
    \label{fig:nelec_cut_a0}
\end{figure}

The spectrum of the produced photons is quite similar between the two models. For the scattered electron, the minimum energy in the classical model is around \SI{210}{\mega\electronvolt}. Between \SI{160}{\mega\electronvolt} and \SI{210}{\mega\electronvolt}, only the electrons produced by the QED radiation reaction are present. With a running electron-laser collision experiment, measuring electrons energies below the classical limit would be a strong evidence of the quantum nature of Compton scattering. Similar to the number of photons $N_\gamma$, Figure~\ref{fig:nelec_cut_a0} shows the number of electron whose energy is below the classical limit. For both electron beam size, there is a maximum number of electron produced below this classical limit around $a_0=7.5$. 

\section{Electron beam jitter}

Like with LUXE at the EuXFEL, a Compton experiment at FEBE will rely on the beam stability from CLARA. A too high jitter in transverse position, arrival time, bunch length, etc, could impact the Compton photons significantly. Similar to the previous chapter, a set of scans was perform for different electron beam parameters. Each point of the scans being calculated with $5\times 10^2$ Ptarmigan simulations containing $10^5$ macroparticles. The variation in number of particle produced is still evaluated with the normalised standard deviation, as defined in Eqn~\ref{eq:norm_std}. The Ptarmigan simulations in Chapter~\ref{chap:impact_luxe} showed that some parameters like the bunch charge $q$, the bunch energy $E$ and the collision angle $\theta$ have very little impact on the signal. Those will not be tested in the following sections.

\subsection{Transverse position jitter}

\begin{figure}[!b]
    \centering
    \includegraphics[width=\textwidth]{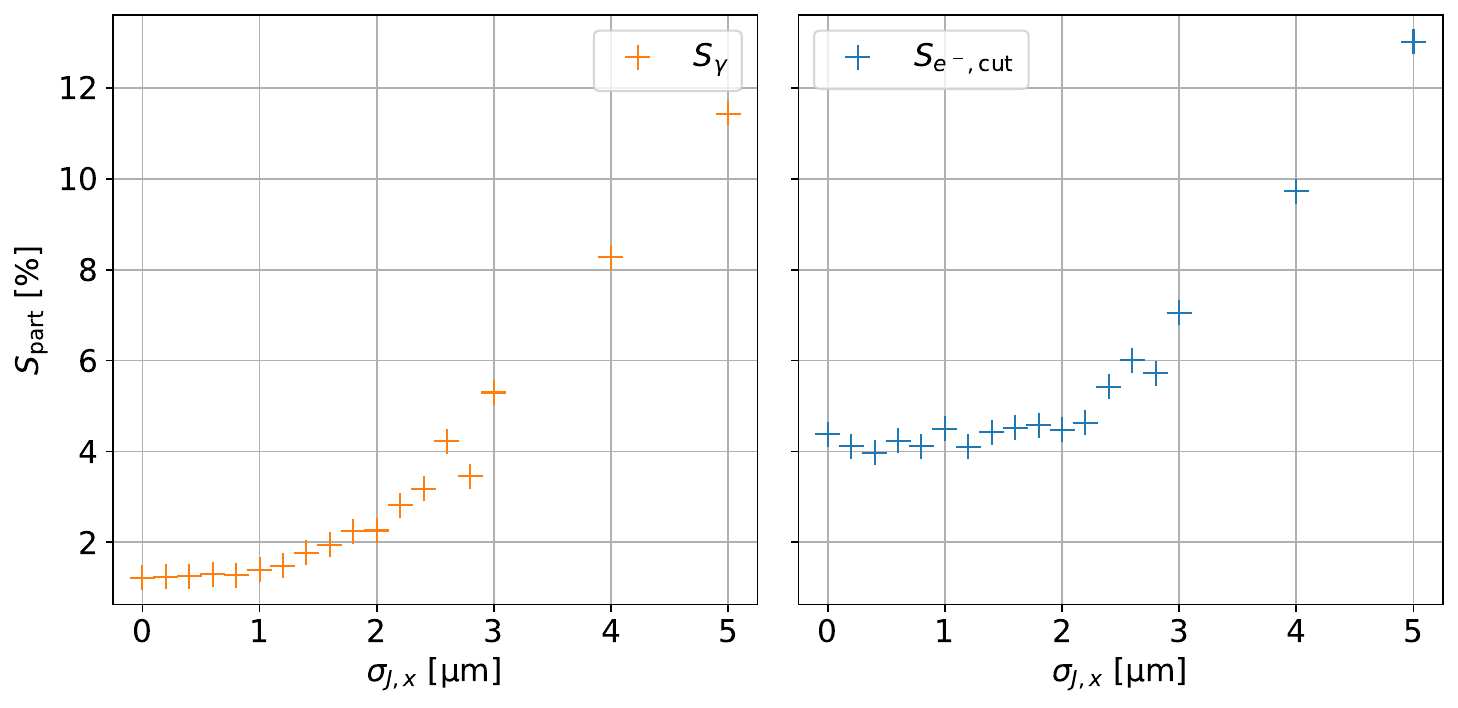}
    \caption{Plots of the normalised standard deviation $S_{\rm part}$ for both photons and electrons as a function of the horizontal position jitter $\sigma_{J, x}$. On the left is plot for photons. On the right is the plot for electrons below the classical limit.}
    \label{fig:clara_jitter_scan_x}
\end{figure}

As for the collision at LUXE's IP, the alignment in the transverse plane is the first concern for the signal stability. Along the CLARA lattice, the main diagnostics used for transverse position are button BPMs with around \SI{20}{\micro\meter} individual resolution in the high charge configuration~\cite{clara}. To evaluate both the position jitter $\sigma_{J, x}, \sigma_{J, y}$ and the combined BPM noise $\sigma_N$, a similar virtual measurement study could be performed. Meanwhile those BPMs are quite similar in resolution with the button BPMs in the EuXFEL. So a \SI{3}{\micro\meter} noise and a relative jitter of about \SI{5}{\%} of the beam size can be expected. In the end, both $\sigma_J$ and $\sigma_N$ had smaller values at the IP but it is expected for the beam at CLARA to be less stable than the EuXFEL beam. 
To evaluate the impact of the position jitter on the normalized standard deviation $S_{\rm part}$ defined in Eqn~\ref{eq:norm_std}, multiple sets of Ptarmigan simulation were ran between $\sigma_J=0\ \rm \mu m$ and $\sigma_J=5\ \rm \mu m$. The results are shown in Figure~\ref{fig:clara_jitter_scan_x} and contain both the variation of the number of photons $S_\gamma$ and the variation on the number of electrons below the classical limit $S_{e^-, \rm cut}$ as defined in Figure~\ref{fig:qed_classical_comp}. 

For LUXE's IP, the jitter was estimated at around \SI{0.5}{\micro\meter}. This correspond to \SI{5}{\%} of the \SI{10}{\micro\meter} beam size which is the requirement for both LUXE and the Compton experiment at CLARA. Such a jitter at CLARA will not produce a noticeable variation in the Compton photons produced. Considering a worst situation where the jitter is of the order of the combined BPM resolution, $\sigma_J=3\ \rm \mu m$, the normalised standard deviation for photons $S_\gamma$ would increase from \SI{1.25}{\%} to around \SI{5.5}{\%}. For the cut electrons, the normalised standard deviation $S_{e^-, \rm cut}$ would increase from \SI{4}{\%} to \SI{7}{\%}. This would be noticeable in the signal but not blocking factor for the experiment.

\subsection{Arrival time and bunch length jitter}

\begin{figure}[!b]
    \centering
    \includegraphics[width=\textwidth]{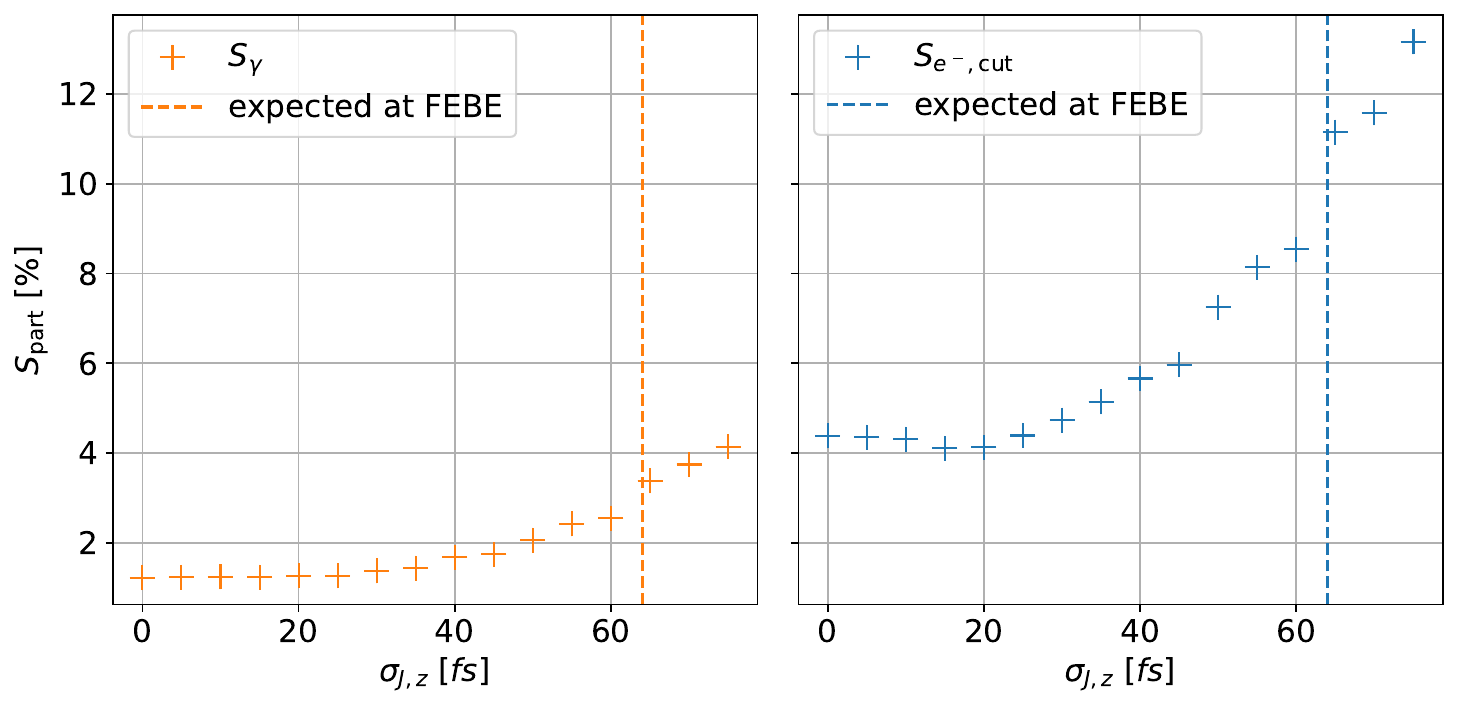}
    \caption{Plots of the normalised standard deviation $S_{\rm part}$ for both photons and electrons as a function of the arrival time jitter $\sigma_{J, z}$. On the left is plot for photons. On the right is the plot for electrons below the classical limit.}
    \label{fig:clara_jitter_scan_z}
\end{figure}

The impact of the electron beam longitudinal profile is crucial as the CLARA beam line does not include a bunch arrival $t$ or bunch length $\sigma_t$ monitors at this moment. Significant variation in both parameters can not be measured and corrected for. Other studies performed simulations on CLARA to evaluate the expected jitter on $t$ and $\sigma_t$~\cite{clara_jitter_sim}. The results found an absolute jitter of \SI{64}{\femto\second} for the arrival time and a relative jitter of \SI{21}{\%} on the bunch length (for a \SI{100}{\femto\second} bunch it correspond to \SI{21}{\femto\second} of absolute jitter). In Figure~\ref{fig:clara_jitter_scan_z}, the jitter scan for arrival time $\sigma_{J, z}$ is shown.
At the expected jitter of \SI{64}{\femto\second}, $S_\gamma$ increases from \SI{1.25}{\%} to \SI{3}{\%} and $S_{e^-, \rm cut}$ increases from \SI{4}{\%} to around \SI{10}{\%}. The electron signal is more sensitive to the arrival time than the transverse position. Still, such a variation seams reasonable for the Compton experiment. 

\begin{figure}[!ht]
    \centering
    \includegraphics[width=\textwidth]{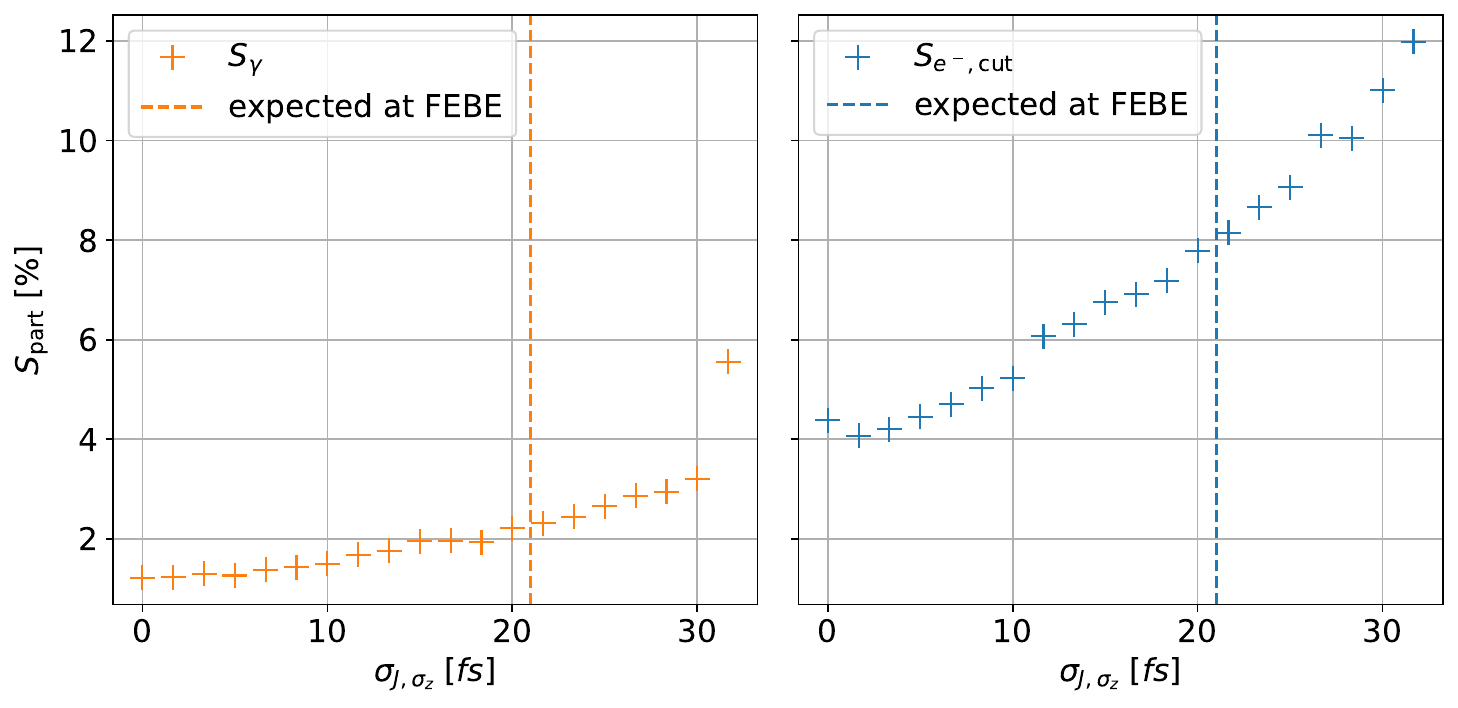}
    \caption{Plots of the normalised standard deviation $S_{\rm part}$ for both photons and electrons as a function of the arrival time jitter $\sigma_{J, \sigma_z}$. On the left is plot for photons. On the right is the plot for electrons below the classical limit.}
    \label{fig:clara_jitter_scan_length}
\end{figure}

In Figure~\ref{fig:clara_jitter_scan_length}, the jitter scan for bunch length $\sigma_{J,\sigma_z}$ is shown.
At the expected jitter of \SI{21}{\femto\second}, $S_\gamma$ increases from \SI{1.25}{\%} to around \SI{2.5}{\%} and $S_{e^-, \rm cut}$ increases from \SI{4}{\%} to \SI{8}{\%}. This is an even lower variation than the arrival time and should also be quite reasonable for the Compton experiment.
Compared to the impact of the EuXFEL jitter shown in Chapter~\ref{chap:impact_luxe}, the beam jitter at CLARA have a higher impact on the Compton signal, especially for the arrival time and the bunch length. However, those jitter are at most \SI{10}{\%} for the electron signal, which should not prevent the success of a Compton experiment.

\section{Possibilities of pair creation}

\begin{figure}[!ht]
    \centering
    \includegraphics[width=12cm]{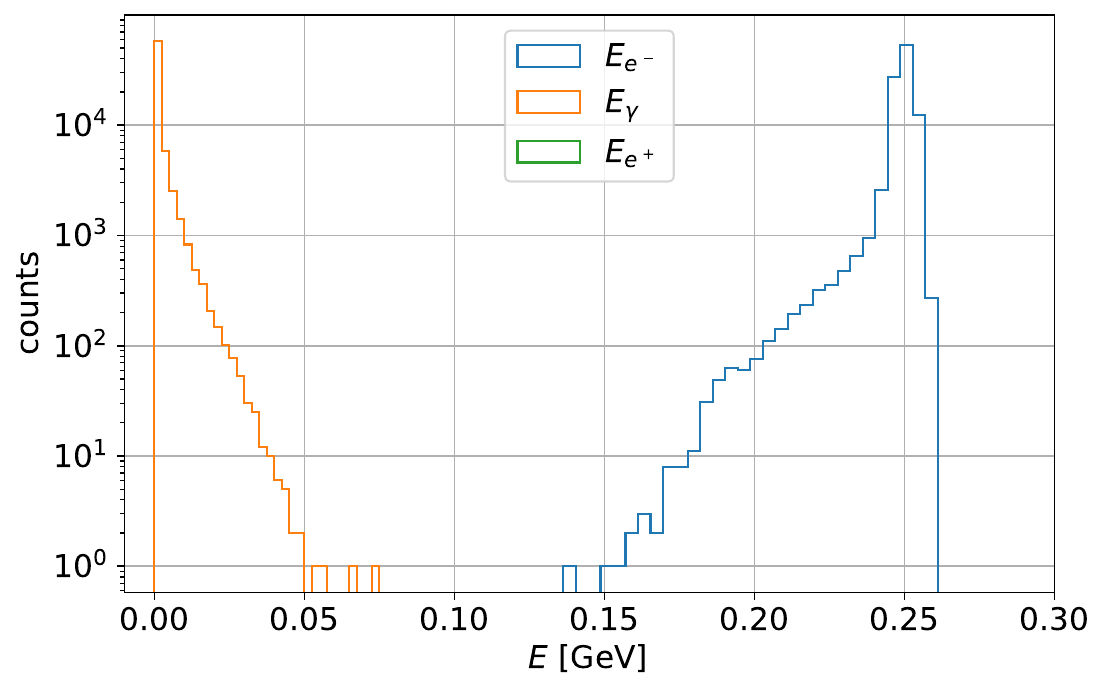}
    \caption{Histogram of the energy spectrum for every produced particles at $a_0=17.7$}
    \label{fig:e_spectrum_17}
\end{figure}

With the current electron beam and laser power, a Compton experiment at FEBE is possible. However, a question remains on the possibilities of Breit-Wheeler pair creation. In the current configuration, the post-collision results shows no positrons even at $a_0=17.7$, which means no pair production interaction. This can be seen in Figure~\ref{fig:e_spectrum_17}.
Expanding on the waist range in Figure~\ref{fig:febe_laser_w0}, additional Ptarmigan simulations were ran at higher values of $a_0$. Moreover, the collisions were also simulated with an electron beam of higher energy. 

\begin{figure}[!ht]
    \centering
    \includegraphics[width=12cm]{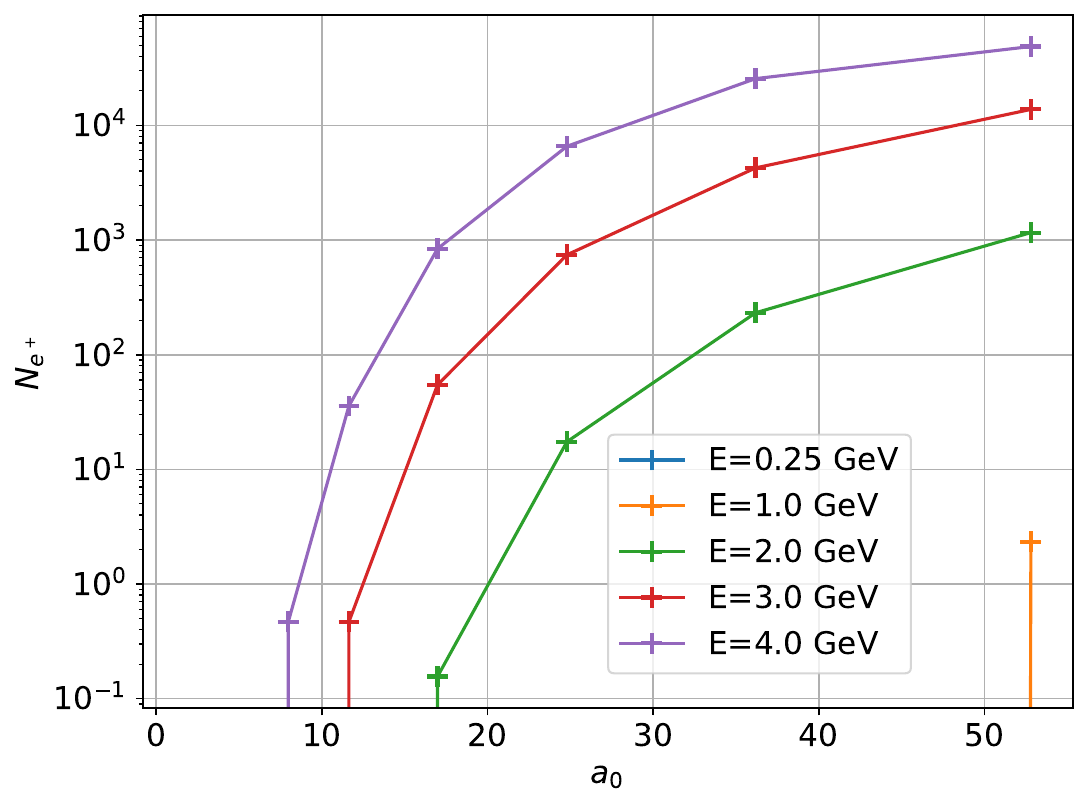}
    \caption{Plot of the number of positron produced $N_{e^+}$ as a function of the normalised amplitude $a_0$ for different beam energies.}
    \label{fig:clara_pair_creation}
\end{figure}

Figure~\ref{fig:clara_pair_creation} shows the positron rate for different beam energies and $a_0$. 
Even with values of $a_0$ above 50, no positrons were produced at the nominal energy $E=250\ \rm MeV$. At $a_0=17.7$, positrons start to appears at $E=2\ \rm GeV$ and a thousand positrons is only reached at $E=4\ \rm GeV$. With the current electron beam energy at FEBE, pair creation is not possible. This would require a significant increase in the beam energy. Such increase could be achieved with plasma wakefield acceleration, which is a planned experiment at FEBE. Although \SI{2}{\giga\electronvolt} electrons beams and $a_0 > 20$ is probably beyond the reach of what is realistically available.

\chapter{Improvements and future work} \label{chap:next}

\section{Lattice improvements}

The XTD8 lattice design presented in Chapter~\ref{chap:design} fulfil the main requirements in electron beam properties for the ELBEX project. This lattice is achromatic, isochronous and produces a \SI{10}{\micro\meter} transverse beam size at LUXE’s IP. However, there are still some improvements that can be made on the design.

\subsection{Smaller beam size at the focus points.}

In Chapter~\ref{chap:design}, LUXE's final focus section is designed to provide the \SI{10}{\micro\meter} required transverse beam size. The Xsuite optimisation presented in Figure~\ref{fig:knob_sigma_IP} and Figure~\ref{fig:knob_sigma_ICSC} showed that a \SI{5}{\micro\meter} beam size is possible on both focus points. Despite some attempts at a lower beam size, our Xsuite optimisation did not found any solution. One explanation on why the software can not find a solution is because of the chromatic effects. Since the electron bunch have an energy spread $\sigma_E>0$, particles of different energies passing through quadrupoles will see a slightly different quadrupole strength. This effects puts a limit on the minimum possible beam size after a quadrupole. The total transverse beam size $\sigma$ at the focus point is the expressed as a function of the linear beam size $\sigma_{\rm linear} = \sqrt{\varepsilon\beta^*}$ and a factor $\sigma_{\rm chrom}$ related to the chromaticity. It can be written as~\cite{chromat}

\begin{equation}
    \sigma_{}^2 = \sigma_{\rm linear}^2 + 
    \sigma_{\rm chrom}^2 = \varepsilon\beta^* \left[1+\left(\frac{L^* \sigma_E}{\beta^*}\right)\right],
\end{equation}

where $\varepsilon$ it the emittance, $\beta^*$ is the beta function at the focus point and $L^*$ is the distance between the quadrupole and the focus point. 
This chromaticity is usually corrected using sextupole magnets~\cites{sext_1, sext_2}. In the centre of the sextupole, a particle perceives no magnetic field. With a transverse position offset, the particle perceives a quadrupole field whose intensity depends on the offset. In a dispersive section, i.e. a section where particles with different energy are spread along a transverse distance, a sextupole applies a variable focusing strength which is used to correct the chromaticity.
At LUXE's IP, $\sigma_E =6\times10^{-4}$, $\beta^*\approx1\ \rm m$ and $L^*\approx9.5\ \rm m$. Running the calculation for $\sigma_{\rm linear}=5\ \mu m$ and even  $\sigma_{\rm linear}=1\ \mu m$, it is clear that $\sigma_{\rm linear} \gg \sigma_{\rm chrom}$, which means than the chromatic effect should be negligible at such beam sizes. As such, it should be possible to redesign the final focus a get to a \SI{1}{\micro\meter} beam size without any sextupole correction. Reaching a smaller beam size at the focus would require a larger beam size in the final focus. A future final focus, to get a \SI{1}{\micro\meter} focus would need much larger quadrupole adjustment along $s$ during optimisation.

\subsection{Additional focus points for the other experiments}

Along the simulation on the XTD8 lattice, the strong field QED at LUXE was the principal guide for the studies presented in this thesis. As such, the current lattice only contains the final focus magnets for LUXE. At least one other final focus section should be included for the upstream plasma experiment. Otherwise, a suggestion was raised about having one final focus section for both experiments. This would imply some adjustments in the LUXE layout to include the plasma cell between the ICS target and the first dump. In both cases, with one final focus or two, the longitudinal locations ot the experiments must be clearly defined. In the current XTD8, the placement of LUXE, hence the final focus and the other experiments, was done as a rough approximation since it had little impact on the final focus design and no impact on both the extraction and dogleg design. As the correct placement of the experiment is given, it is the FODO cell that will be adjusted to match the optics functions.

\subsection{Integration in the CAD}

\begin{figure}[!ht]
    \centering
    \includegraphics[width=\textwidth]{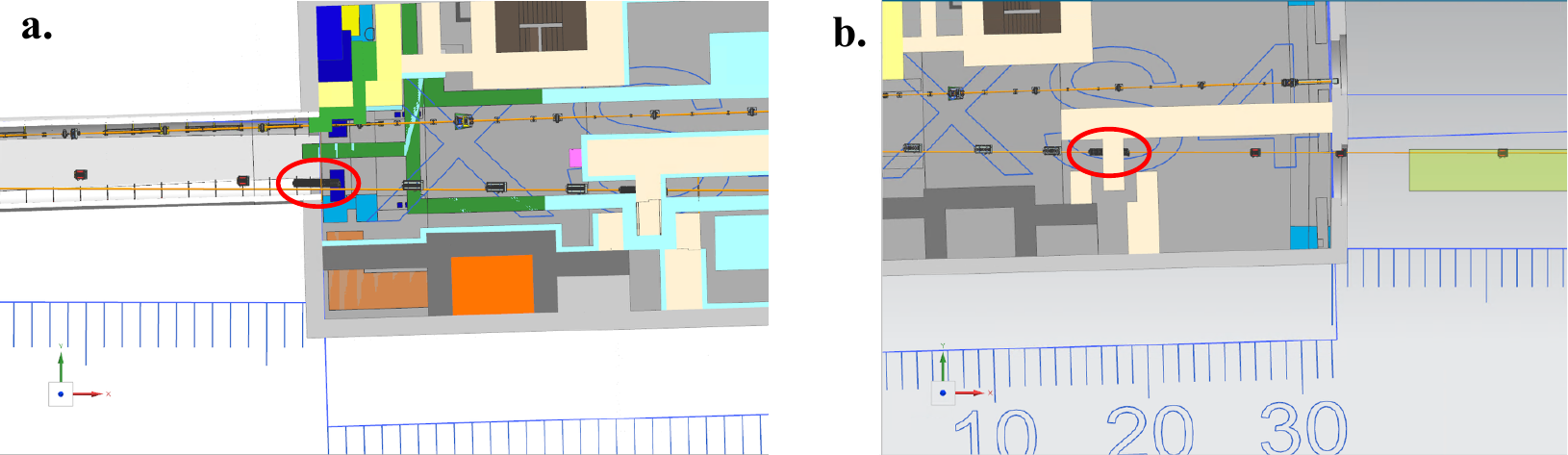}
    \caption{CAD rendering of the XTD8 design in the EuXFEL tunnels. a) The bend, indicated by a red circle, overlaps with the wall between the XTD3 tunnel and the XS4 building. b) The bend, indicated by a red circle, overlaps with the concrete chicane before the XTD8 tunnel.}
    \label{fig:xtd8_mag_ovelap}
\end{figure}

The goal of the XTD8 design done in Xsuite is to provide a solution lattice with matched optics function, the correct beam size at the focus point and both the required achomaticity and isochronicity. In that sense, the solution presented in Chapter~\ref{chap:design} fulfil all the above requirements. However, this design does not consider the real geometry of the XS4 hall and the XTD8 tunnel. Through some exchanges with the team in charge of CAD integration for ELBEX, some of the magnets appeared to overlap with already present walls and equipment. Some examples of overlapping magnets are shown in Figure~\ref{fig:xtd8_mag_ovelap}.
To avoid these constraints, some adjustments must be made on the XTD8 design. The displacement required are of the order of a few meters. So, the easiest way to adjust the lattice is to move the overlapping elements and then re-run the Xsuite optimisation only using the magnet strengths as variables. Some preliminary tests have been conducted and the beam line optics functions do not vary significantly. 

\section{Background from the dumps}

In Chapter~\ref{chap:impact_luxe}, the impact of both the residual gas background and the background from halo were thoroughly studies through BDSIM simulations. There is however another type of background that have not been addressed: the background induced by the beam dumps. In the gamma-laser configuration of LUXE, the electron beam is focused at an inverse Compton scattering target to create a photon beam. After the target, the electrons are deflected by a bending magnet to a dump surrounded by shielding. The interaction chamber is behind this dump-shield assembly. The dump is designed to absorb the electron beam entirely. But among the particle shower created inside the dump, there is a probability for some high energy particles to exit the dump and the shielding. Like with the other background types, those residual particles could impact the detected signal. Using BDSIM, the particle showers created by the \SI{17.5}{\giga\electronvolt} electron beam can be simulated. However, such simulation is particularly computation intensive due to the material density and the number of interaction created from on single macroparticle. In addition, the amount of particles exiting the dump per bunch crossing is very small. 

\subsection{Final state splitting}

The cross section biasing is not the most suited method for the dump background problem as the objective would be to bias the interaction based on if a produced particle at the end of a shower manages to escape the dump. Such a method is more akin to a final state splitting~\cite{fss}. In a more general approach, a final state splitting applies a biasing not on the physics processes but based on the properties of the final excaping particle. It can be a particle type, a certain energy or in our case the longitudinal position $s$. When a particle escapes the dump, i.e. when a particle is detected beyond a given $s$ in the BDSIM simulation, the event that led to this particle is repeated $n$ times and a weight of $1/n$ is applied to the products. BDSIM currently have an option for muon splitting but a more general option for final state splitting could be a useful tool for dump simulations.

\section{Plasma boosting simulations for CLARA}

The Ptarmigan simulations in Chapter~\ref{chap:clara} have shown the possibilities of a Compton scattering experiment. However, such an experiment could not show Breit-Wheeler pair creation with a \SI{250}{\mega\electronvolt} electron beam. A solution considered is to increase the beam energy at FEBE using plasma wakefield acceleration. In the dogleg lattice upstream of FEBE, there is already a location assigned to insert a mask in order to create both a drive bunch and a witness bunch. The main question is then: what energies can the witness bunch reach after passing through a plasma cell in FEC1. Answering this question requires a simulation in which both the drive bunch and the witness bunch are modelled and propagated through a plasma. None of the software previously described in this paper are suited to perform such simulation. A good candidate to run preliminary PWFA simulations is Wake-T~\cite{wakeT}. Wake-T is a python base plasma acceleration code that simulate the propagation of a drive bunch and a witness bunch in a cylindrical plasma cell. As other codes rely on a computationally heavy 3D particle-in-cell~\cite{pic} (PIC) model, Wake-T uses two approximations to reduce the simulation time : the model is considered axially symmetric and quasi static. Running a Wake-T simulation should clarify how much energy can be expected with the initial \SI{250}{\mega\electronvolt} electron beam. It must be noted that, because the electron beam is separated to create the drive and witness bunch, the bunch charge $q$ will decrease. From an initial \SI{250}{\pico\coulomb} bunch the majority will go in the drive bunch and the witness bunch will have a charge of the order of \SI{10}{\pico\coulomb}~\cite{clara_pwfa}. A decreased bunch charge will impact the strong field QED processes rate so the threshold energy for pair creation should be higher than the previously found \SI{4}{\giga\electronvolt}.

\chapter{Conclusions} \label{chap:conclu}

The perturbative regime of quantum electrodynamic has now been thoroughly tested. It describes accurately the scattering interactions like Coulomb scattering and Compton scattering in a relativistic case. However, the theory of QED predicts a non-perturbative regime with electromagnetic field higher than the Schwinger limit. Reaching this strong field regime was not possible until recently. It requires both a high EM field provided by a terawatt laser and electrons above \SI{1}{\giga\electronvolt} to allow a sufficient Lorentz boost of the laser field in the particles rest frame. LUXE is one experiment aiming to reach the strong field and observe the non-linear QED processes for the first time. It proposes to collide the \SI{17.5}{\giga\electronvolt} electron of the EuXFEL on a \SI{350}{\tera\watt} laser pulse. 

Initially planned in the XTD20 tunnel of the EuXFEL, LUXE is now associated with the ELBEX project and will be installed in the XTD8 tunnel. As there is currently no beam line in this tunnel, a new lattice has been designed to extract electrons from an existing line. This lattice recovers the \SI{10}{\micro\meter} transverse beam size required for LUXE. Going down to \SI{5}{\micro\meter} was proven possible with the current final focus layout. The lattice is achromatic, as the dispersion introduced by the extraction is closed at the end of the dogleg. The downstream beam size is then independent from the energy spread. The lattice is also isochronous, as the matrix element $R_{56}$ of the total transfer matrix of the line can be set to zero. In that case there is no compression of the electron bunch. Although the lattice optics are matched and optimised, collective effects like coherent synchrotron radiation can alter the electron beam emittance and consequently the beam size. Ocelot simulations of the XTD8 lattice have shown a global increase of the horizontal emittance by a factor 2 as well as a dispersion leak. Overhaul the horizontal beam size increases by \SI{10}{\%} due to CSR.

Measurements at the EuXFEL have shown that the electron bunches are produced with a high level of stability. The train by train jitter on the transverse position is measured at \SI{3}{\micro\meter} in button BPMs and at \SI{1}{\micro\meter} in cavity BPMs. In straight sections, the position jitter stays at around \SI{5}{\%} of the beam size. The jitter on the bunch charge is at most \SI{1.5}{\pico\coulomb} for a \SI{250}{\pico\coulomb} bunch. The bunch arrival time jitter is around \SI{26.25}{\pico\second} between trains. The bunch length jitter is around \SI{3.33}{\femto\second}. The bunch energy jitter is below \SI{1}{\mega\electronvolt} for a \SI{16.15}{\giga\electronvolt} electron beam. Using those measurements and the transfer matrix of the new line, it was possible to propagate the election beam jitter and estimate their value at LUXE's interaction point. Running Ptarmigan simulations while varying each parameters, it is clear that the electron beam jitter should have a low impact on the signal measured at LUXE.

As an electron bunch travels down the beam pipe, some electrons will produce backgrounds particles. It could be from the interactions with the residual gas or from the interaction of electrons in the halo of the distribution with the beam pipe. Using BDSIM simulations with cross section biasing, a spectrum of the expected secondary particles has been established. At one bunch crossing ($2\times10^9$ electrons), about \SI{34}{} secondary photons are expected at LUXE's IP as well as \SI{1.5}{} secondary electrons and \SI{1.5}{} secondary positrons. All other species have a rate below $10^{-3}$ per bunch. At those rates, the backgrounds will not have a significant impact on the collision at LUXE's IP.

Measuring the transverse position in LUXE's interaction chamber can not be achieved with a regular BPM. However, the upstream BPMs in the XTD8 can be used to perform a virtual measurement of the position at the IP. This method still requires a calibration measurement at the IP. In the design of the interaction chamber, an alignment-target holder has been designed to get diagnostics of both the electron beam and the laser. The thin tungsten needle on this holder can be used like a wire scanner in both transverse plane. As it interacts with the electron beam, photons will be produce as a function of the overlap between the beam and the needle. Through BDSIM simulations, this method is able to measure the beam position with a \SI{1.40}{\micro\meter} error on the horizontal position and a \SI{4.09}{\micro\meter} error on the vertical position.

The LUXE experiment aims to observe three strong-field QED processes : non-linear Compton scattering, non linear Breit-Wheeler pair creation and non-linear trident. The collisions between an electron bunch and a laser pulse, hence those processes, can be simulated using Ptarmigan,  a Monte-Carlo simulation code. Without any jitter of the electron beam, the intrinsic statistics of the QED processes induces a variation in the number of particles produced. With $5\times10^2$ collision, each having $10^5$ electron macroparticles, this intrinsic variation expressed as the normalised standard deviation $S_{\rm part}$ is around \SI{1}{\%} for electrons and photons and around \SI{10}{\%} for positrons. Using the measurement at the EuXFEL, the impact of each electron beam jitter on $S_{\rm part}$ can be evaluated. For almost every parameters (transverse position, bunch charge, arrival time, bunch length, energy and collision angle) the variation $\sigma_J$ expected at LUXE's IP and the minimum detectable variations $\sigma_N$ produces less than \SI{1}{\%} variation on $S_{\rm part}$ from the perfectly aligned configuration. The only exception is with the beam size jitter, even with $\sigma_{J, \sigma} = 1\ \rm \mu m$, the number of particle produced varies by \SI{20}{\%}. This confirms that the transverse electron beam size is crucial to get a stable signal at LUXE.

At the CLARA linear accelerator, the FEBE beam line is planned to host a large variety of experiments using the \SI{250}{\mega\electronvolt} electron beam. Among those experiments, an electron-laser collision similar to the LUXE experiment could produce Compton scattering photons. The available \SI{100}{\tera\watt} laser can be focused down to \SI{3}{\micro\meter} to reach a normalised amplitude $a_0=17.7$. Ptarmigan simulations of the electron-laser collisions showed produced photons with energies up to \SI{70}{\mega\electronvolt}. Coincidently, the scattered electron have energy as low as \SI{160}{\mega\electronvolt}. When compared to the classical model of Compton scattering, those low energy electron are clearly distinct from the classical Compton tail. Measuring electron energies way below the classical spectrum would be an evidence that this interaction is dictated by quantum mechanics. The stability of both the electron and photon signal after the collision is dependant on the electron beam stability as it arrives at FEBE. Similar to the study of LUXE, Ptarmigan is used to run jitter scans of different electron beam parameters, mainly the transverse position, the arrival time and the bunch length. Without a virtual measurement on the BPMs at CLARA, the estimated position jitter is extrapolated from the results at LUXE. In the worst case of a jitter $\sigma_{J, x}$ around \SI{3}{\micro\meter}, the variation in the number of photons produced $S_\gamma$ is about \SI{5}{\%} which would be noticeable but reasonable. Prior simulations were ran to evaluate the jitter on arrival time and bunch length. The estimated arrival time jitter of \SI{64}{\femto\second} would impact the signal by around \SI{3}{\%}, whereas the estimated bunch length jitter of \SI{21}{\femto\second} would impact the signal by around \SI{2.5}{\%}. Both jitter should not be a blocking factor for the experiment. In addition to the Compton scattering, the question was raise of the possibilities of Breit-Wheeler pair creation at FEBE. Ptarmigan simulations have shown that even at $a_0=50$, the electron beam energy is too low to see any positron. It would require energies of at least \SI{2}{\giga\electronvolt} to see a significant amount of pair creation.

This study opens of a few obvious next steps. 
The XDT8 lattice could feature a smaller focus size for LUXE with a final focus upgrade. Dedicated magnets will be added for the plasma experiment and the test beam experiment. In coordination with the CAD integration team, small adjustments on the magnet placements should be made to insure no overlap with the walls. While the residual gas background and halo background are well studied, the background induced by the dump is missing to complete the expected background at LUXE's detectors. To avoid a computation heavy simulation of the particle showers, implementing a final state splitting method in BDSIM is a solution. Finally, if a pair creation experiment is not currently possible at CLARA, it could be achieved with an electron beam of higher energy. Plasma wakefield acceleration simulations must confirm if it is possible to reach the required GeV electron bunches.

\printbibliography[title={References},heading=bibintoc] 

%

\end{document}